\documentclass[a4paper,fleqn,usenatbib,useAMS]{mnras}

\usepackage{mathptmx}
\usepackage[T1]{fontenc}
\usepackage{ae,aecompl}

\usepackage{graphicx}	% Including figure files
\usepackage{amsmath}	% Advanced maths commands
\usepackage{amssymb}	% Extra maths symbols
\usepackage{color}
\usepackage{multicol}
\usepackage{longtable}
\usepackage{rotating}
\usepackage{pdflscape}
\usepackage{natbib}
\usepackage{enumitem}
\usepackage{appendix}
\usepackage{url}
\usepackage{float}
\usepackage[caption = false]{subfig}
\def \simless {\mathbin{\lower 3pt\hbox{$\rlap{\raise 4pt
              \hbox{$\char'074$}}\mathchar"7218$}}}
\def \simgreat {\mathbin{\lower 3pt\hbox{$\rlap{\raise 4pt
              \hbox{$\char'076$}}\mathchar"7218$}}}
\def\ie{{\it i.e.}}

\def\eg{{\it e.g.}}
\title[Velocity dispersion profiles of BCGs]{Diversity in the stellar velocity dispersion profiles of a large sample of Brightest Cluster Galaxies $z\leq0.3$}

\author[Loubser et al.]{S. I. Loubser$^{1}$\thanks{E-mail:Ilani.Loubser@nwu.ac.za (SIL).}, H. Hoekstra$^{2}$, A. Babul$^{3}$, E. O'Sullivan$^{4}$\\
$^{1}$Centre for Space Research, North-West University, Potchefstroom 2520, South Africa\\
$^{2}$Leiden Observatory, Leiden University, PO Box 9513, 2300 RA, Leiden, The Netherlands\\
$^{3}$Department of Physics and Astronomy, University of Victoria, Victoria, BC, V8W 2Y2, Canada\\
$^{4}$Harvard-Smithsonian Center for Astrophysics, 60 Garden Street, Cambridge, MA 02138, USA}

\date{Accepted 2018 February 16. Received 2018 February 8; in original form 2017 October 17.}

\pubyear{2018}

\begin{document}
\label{firstpage}
\pagerange{\pageref{firstpage}--\pageref{lastpage}}
\maketitle

\begin{abstract}
We analyse spatially-resolved deep optical spectroscopy of Brightest Cluster Galaxies (BCGs) located in 32 massive clusters with redshifts of 0.05 $\leq z \leq$ 0.30, to investigate their velocity dispersion profiles. We compare these measurements to those of other massive early-type galaxies, as well as central group galaxies, where relevant. This unique, large sample extends to the most extreme of massive galaxies, spanning M$_{K}$ between --25.7 to --27.8 mag, and host cluster halo mass M$_{500}$ up to 1.7 $\times$ 10$^{15}$ M$_{\sun}$. To compare the kinematic properties between brightest group and cluster members, we analyse similar spatially-resolved long-slit spectroscopy for 23 nearby Brightest Group Galaxies (BGGs) from the Complete Local-Volume Groups Sample (CLoGS). We find a surprisingly large variety in velocity dispersion slopes for BCGs, with a significantly larger fraction of positive slopes, unique compared to other (non-central) early-type galaxies as well as the majority of the brightest members of the groups. We find that the velocity dispersion slopes of the BCGs and BGGs correlate with the luminosity of the galaxies, and we quantify this correlation. It is not clear whether the full diversity in velocity dispersion slopes that we see is reproduced in simulations.
\end{abstract}


%%%%%%%%%%%%%%%%%%%%%%%%%%%%%%%%%%%%%%%%%%%%%%%%%%
%%%%%%%%%%%%%%%%% BODY OF PAPER %%%%%%%%%%%%%%%%%%%%%%%

\section{Introduction}
\label{introduction}

% Introduction
Brightest Cluster Galaxies (BCGs) reside predominantly in the dense cores, in the deep gravitational potential well, of rich galaxy clusters. Because of this location, they are the sites of interesting evolutionary phenomena, \eg\ dynamical friction, mergers, galactic cannibalism, and cooling flows. BCGs have many well-known, unique properties such as high luminosities and diffuse stellar envelopes. It is also known that (some) BCGs have rising velocity dispersion profiles with increasing radius \citep{Loubser2008, Newman2013a}, may contain secondary nuclei \citep{Laine2003} or large flat cores in their surface brightness profiles \citep{Lauer2007}, may have experienced AGN activity and recent star formation episodes \citep{Bildfell2008, Loubser2013, Donahue2015, Loubser2016}, or may have a mass-to-light ratio (M/L) that is different from other massive early-type galaxies \citep{VonderLinden2007}. The observable properties of BCGs are shaped by the baryonic processes that are fundamental to our understanding of galaxy and cluster formation, \eg\ AGN feedback, star formation and stellar feedback, and chemical enrichment. 

Additionally, observed BCG velocity dispersion profiles, as presented for a representative sample here, directly relate to the dynamical mass profiles, and is an important step towards the full resolution of cluster mass profiles, which in turn, is necessary to constrain galaxy formation and evolution models (\eg\ \citealt{Newman2013a}). Outside the central regions of clusters, X-ray observations and weak-lensing measurements provide good mass estimates of the host halo, but can not probe the innermost region of the cluster. The fact that BCGs are at the bottom of the cluster potential in regular, non-interacting systems, means the dynamics of the stellar component offers a valuable route to resolving this problem. 

%Velocity dispersion profiles
\citet{Dressler1979} first showed that the velocity dispersion profile of the BCG in Abell 2029 (IC 1101) rises with increasing radius from the galaxy centre, and it was interpreted as evidence that the diffuse stellar halo consists of accumulated debris of stars stripped from cluster members by tidal encounters and by dynamical friction against the growing halo. \citet{Fisher1995} found that, with the exception of the BCG in Abell 2029, the velocity dispersion gradients of their sample of 13 nearby BCGs are all negative, \ie\ decreasing outwards. More mixed results followed: \eg\ \citet{Carter1999} found one out of their sample of three BCGs (NGC 6166 in Abell 2199) to have a positive velocity dispersion gradient, and \citet{Brough2007} found one significant negative velocity dispersion gradient, and five velocity dispersion gradients consistent with zero from their observations of three BCGs and three Brightest Group Galaxies (BGGs). \citet{Loubser2008}, investigating the first large sample of spatially-resolved kinematics of BCGs, found at least five of the 41 nearby BCGs ($z<0.07$) studied to have flat to rising velocity dispersion profiles. These rising velocity dispersion profiles have also been interpreted as evidence for the existence of high M/L components in these galaxies \citep{Dressler1979, Carter1985}. In contrast, the velocity dispersion profiles of normal (\ie\ non-central) early-type galaxies either remain flat or decrease with radius \citep{Kronawitter2000}, with the exception of the most massive early-types \citep{Veale2017}. At present, it is not yet clear what the increasing or decreasing velocity dispersion tells us about the galaxy, especially for BCGs in the cluster potential well. It can be a reflection of the gravitational potential of the galaxy, the centre of the cluster, or a snapshot of a dynamical system which has not yet reached equilibrium \citep{Murphy2014, Bender2015}.

%This paper compared to previous
Up to now studies that included detailed velocity dispersion profiles of massive early-type galaxies contained a very small number of BCGs, \eg\ ATLAS$^{\rm 3D}$ \citep{Cappellari2011}, or focused specifically on the most nearby massive early-types, \eg\ MASSIVE \citep{Ma2014, Veale2017}. This also translates into limited coverage of the galaxy property parameter space, \eg\ MASSIVE includes early-type galaxies between M$_{K}$ = --25.7 to --26.6 mag, limiting their ability to characterise any strong trends with mass or luminosity. \citet{Loubser2008} limited their study to nearby BCGs/BGGs below $z \sim 0.07$. \citet{Newman2013a} presented a detailed study of the dynamical modelling of seven cluster mass profiles, and the velocity dispersion profiles of their seven BCGs were very homogeneous (their figure 11), as we discuss in Section \ref{Velocityprofiles}. 

Here, we present a study of a complimentary, large sample of 32 BCGs, up to a redshift of $z \sim 0.3$, from the well-characterised Multi--Epoch Nearby Cluster Survey (MENeaCS) and Canadian Cluster Comparison Project (CCCP) cluster samples (as studied in \eg\ \citealt{Bildfell2008, Sand2011, Sand2012, Bildfell_thesis, Mahdavi2013, Mahdavi2014, Hoekstra2015, Sifon2015, Loubser2016}). Our 32 BCGs span M$_{K}$ = --25.7 to --27.8 mag, with host cluster halo masses M$_{500}$ from 1.6 $\times$ 10$^{14}$ to 1.7 $\times$ 10$^{15}$ M$_{\sun}$. To compare the kinematics between the brightest group and cluster members, we also analyse 23 BGGs in the Complete Local-Volume Groups Sample (CLoGS, \citealt{O'Sullivan2017}), thereby extending our M$_{K}$ range to a lower limit of --24.2 mag. In clusters of galaxies, the evolution of gas is governed by thermal processes (cooling) and because of these systems' deep gravitational potential wells, by AGN feedback. On the other hand in groups, due to their relatively shallower gravitational wells, the evolution of the gas can be impacted by large-scale galactic flows powered by SNe and stellar winds in addition to radiative cooling and AGN feedback \citep{Liang2016}. 

%This paper content
Section \ref{data} presents the MENeaCS and CCCP samples of BCGs, and the CLoGS sample of BGGs, as well as the spectroscopic data. Section \ref{profiles} contains the details of the stellar kinematic measurements. Section \ref{Calproperties} contains the calculation and discussion of the BCG kinematic profiles, the comparison to those measured for the BGGs, and the correlations to host cluster/group properties. The conclusions are summarised in Section \ref{conclusions}. In the second paper of this series (Loubser et al., in prep, hereafter Paper II), we use the data and measurements presented here, as well as the measurements of the central higher order velocity moments (Gauss-Hermite $h_{3}$ and $h_{4}$), $r$-band surface brightness profiles, stellar population modelling and predicted stellar M/L ratios of the BCGs, to do detailed dynamical modelling. 

We use H$_{0}$ = 73 km s$^{-1}$ Mpc$^{-1}$, $\Omega_{\rm matter}$ = 0.27, $\Omega_{\rm vacuum}$ = 0.73 throughout, and make cosmological corrections where necessary.

\section{Data}
\label{data}

%Summary
We summarise the overall sample, and describe each of the three sub-samples, together with their optical spectroscopic observations, below. We use spatially-resolved long-slit spectroscopy for 14 MENeaCS and 18 CCCP BCGs, taken on the Gemini North and South telescopes. In addition, we use \textit{Chandra/XMM-Newton} X-ray, and weak lensing properties of the host clusters themselves \citep{Mahdavi2013, Hoekstra2015, Herbonnet_thesis}. The BCG sample then consists of 32 BCGs in X-ray luminous clusters between redshifts of 0.05 $\leq z \leq$ 0.30. In addition, to compare the derived kinematic properties between the central galaxies in clusters and groups, we include a sub-sample of 23 nearby BGGs from the CLoGS sample ($D<80$ Mpc). For these galaxies we use archival spatially-resolved long-slit spectroscopy from the Hobby-Eberly Telescope (HET).

\subsection{MENeaCS sample and spectroscopic data}

The MENeaCS sample \citep{Sand2011, Sand2012} was initially designed to measure the cluster supernovae rate in a sample of 57 X--ray selected clusters at 0.05 $< z <$ 0.15, and to utilize galaxy-galaxy lensing to measure the dark matter content of early-type galaxies as a function of clustercentric distance \citep{Sifon2017}. The BCGs of 14 of these clusters were also observed with the Gemini North and South telescopes using GMOS long--slit mode during the 2009A (from February to June 2009) and 2009B (two nights in November 2009) semesters (PI: C. Bildfell). Table \ref{objects_BCGs} lists the observations and the relevant exposure times, and  we follow a similar spectroscopic data reduction method as for the Gemini observations of the CCCP BCGs, described in detail in \citet{Loubser2016}.

\subsection{CCCP sample and spectroscopic data}

The full CCCP sample, as well as the sub-sample selection for the spectroscopic observations, and the reduction thereof, are discussed in detail in \citet{Loubser2016}. Briefly, we target 19 BCGs in X-ray luminous galaxy clusters in the redshift range 0.15 $< z <$ 0.30, where the BCGs reside within a projected distance of 75 kpc of their host cluster's X-ray peak (see Table \ref{objects_BCGs} for the list of objects). After careful analysis of the choice of BCG in the clusters, we found that the choice of the `BCG' in Abell 209 could be ambiguous, and to avoid uncertainty it is excluded from further analysis. This exclusion does not influence any conclusions made here or in \citet{Loubser2016}. 

\subsection{CLoGS sample and spectroscopic data}
\label{sub:CLoGS}

A detailed discussion of the CLoGS sample selection is presented in \citet{O'Sullivan2017}, and is only briefly summarised here. The CLoGS sample starts from the shallow, all-sky Lyon Galaxy Group (LGG) catalogue of \citet{Garcia1993}, which is complete to m$_{B}=14$ mag and $v_{\rm rec}=5500$ km s$^{-1}$ (equivalent to $D <$ 80 Mpc, correcting for Virgocentric flow). The groups are then selected, and the group members determined, as detailed in \citet{O'Sullivan2017}. The sample is divided into two subsamples, based on their richness parameter $R$, which is the number of galaxies with $\log L_{B} \ge 10.2$ within 1 Mpc and 3$\sigma$ of the brightest member. $R > 10$ systems are known clusters and excluded. The CLoGS high-richness subsample contains the 26 groups with $R$ = 4 -- 8, and the low-richness subsample contains the 27 groups with $R$ = 2 -- 3.

\citet{Vandenbosch2015} conducted an optical long-slit spectroscopic survey, HETMG, of 1022 galaxies using the 10m HET at McDonald Observatory, originally motivated by the search for nearby massive galaxies that are suitable for black hole mass measurements. The spectra cover 4200 -- 7400 \AA{}, and have a default 2 $\times$ 2 binning. This setup provides an instrumental resolution of 4.8 \AA{}, or a dispersion of 108 km s$^{-1}$. When practical, the slit was aligned on the major axis and centred on the galaxy, and single 15 minute exposures were obtained. The typical spatial resolution of the observations is 2.5$\arcsec$ FWHM. 

We use the CLoGS sample and select the groups for which the brightest members were observed by \citet{Vandenbosch2015}. In cases where there is more than one spectral exposure, we choose the exposure with the highest S/N. We do not combine the exposures due to different (sometimes poor) observing conditions. The objects are listed in Table \ref{objects_BGGs}, and consist of 14 high richness, and 9 low richness BGGs.

\section{Measurements}
\label{profiles}

\subsection{Spatial binning and stellar template fitting} 
\label{binning}

The BCG spectra were binned into fixed spatial bins from the centre of the galaxy outwards. The number of bins was chosen so that they are sufficiently small to detect rotation and possible substructure in the kinematic profile measurements, whilst still having S/N high enough ($\ge 5$) to maintain acceptable errors on the velocity and velocity dispersion measurements. As a result, the spatial bins become wider with increasing radius from the centre of the galaxy, typically reaching 15 kpc to each side of the CCCP and MENeaCS BCGs. 

The CCCP spectra were binned into nine fixed spatial bins (one central bin, and four bins on each side of the central bin). The MENeaCS BCG spectra were generally higher S/N and typically binned into 11, 13 or 15 fixed spatial bins (one central bin, and 5, 6 or 7 on each side of the central bin) depending on the S/N. In addition, the velocity and velocity dispersion measurements, were also measured within a 5 kpc circular aperture and a 5 to 15 kpc aperture for direct comparison to the CCCP stellar population aperture measurements as described in \citep{Loubser2016}. 

For the BGG spectra, we use the binning by \citet{Vandenbosch2015}, who combined spatial rows into bins with a minimum S/N of 25. The lowest number of bins is 14 (for NGC 5846) and the highest is 68 (for NGC 5353), and the bins typically reach 10 kpc to each side of the BGG.

The central velocity dispersion ($\sigma_{0}$) was measured within an aperture of 5 kpc from the centre of the galaxy to each side (\ie\ 10 kpc in total, the inner bin as described above) for the BCGs and within an aperture of 1 kpc for the CLoGS BGGs. The radii of the apertures (in arcsec) where the central velocity dispersion measurements ($\sigma_{0}$) are made, are large enough to avoid being significantly affected by seeing.

We implement the penalised pixel-fitting (pPXF) measurement method \citep{Cappellari2004} to measure the relativistically-corrected recession velocities and the physical velocity dispersion of the BCGs/BGGs. For the velocity dispersion, we use
\begin{equation}
\sigma^2_{\rm BCG} = \sigma^2_{\rm M}  - \sigma^2_{\rm I} -  \sigma^2_{\rm T}, 
\end{equation}
where $\sigma_{\rm BCG}$ is the physical velocity dispersion of the galaxy, $\sigma_{\rm M}$ is the velocity dispersion as measured from the broadened spectra, $\sigma_{\rm I}$ is the instrumental broadening and $\sigma_{\rm T}$ is the resolution of the stellar templates used to measure the kinematics. For the Gemini BCG data, the instrumental broadening, $\sigma_{\rm I} = 71$ km s$^{-1}$, was measured using the standard star spectra at every 200 \AA{} interval. For the HET BGGs data, the instrumental broadening, $\sigma_{\rm I} = 108$ km s$^{-1}$, was taken from \citet{Vandenbosch2015}.

All 985 stars of the MILES stellar library \citep{Sanchez2006} were used to construct linear combinations of stars that form the optimal stellar absorption templates. The MILES stellar library covers a very large stellar parameter space which enables an accurate fit of the stellar continuum, and has a fixed instrumental resolution, $\sigma_{\rm T}$, of $\sim$ 2.3 \AA{} ($\sim$ 125 km s$^{-1}$, FWHM). 

We firstly fit only the velocity and velocity dispersion in every spatial bin, as we are interested in the spatially-resolved profiles. In a second, separate process, we fit $V, \sigma_{0}, h_{3}, h_{4}$ simultaneously in just the central bin (\ie\ 10 kpc for the BCGs, and 2 kpc for the BGGs). We have tested that the measurements of velocity and velocity dispersion (only), and velocity and velocity dispersion (simultaneously with $h_{3}$ and $h_{4}$) are consistent in the centres where $h_{3}$ and $h_{4}$ are measured. The central measurements of $h_{3}$ and $h_{4}$ are not used in this paper, and will be presented in Paper II, alongside the other dynamical modelling ingredients $\eg$ the stellar populations and stellar mass profiles. We allow free fitting of the entire template stellar library in each bin. 

%\begin{landscape}
\begin{table*}
\caption{MENeaCS and CCCP BCGs observed for this study. In all cases the slit position angle (PA) is given as clockwise from North. We use the ellipticities, $\epsilon$, of the BCGs as measured from the 2MASS isophotal $K$-band, and obtained through the NASA Extragalactic Database (NED). The M$_{K}$ absolute luminosities were also obtained from 2MASS measurements and corrected as described in Section \ref{Kband}. R$_{500}$ and M$_{500}$ are from \citet{Herbonnet_thesis} for MENeaCS, and \citet{Hoekstra2015} for CCCP, and M$_{500}$ is given in $10^{13}$ M$_{\sun}$ to be directly compared to the corresponding values for CLoGS in Table \ref{objects_BGGs}. The $\star$ next to the object name indicates whether optical emission lines were present in the spectra analysed here.} 
\label{objects_BCGs}
\centering
\begin{tabular}{l c c c c c c c c l l}
\hline
Name & $z$ & $\alpha_{J2000}$ &  $\delta_{J2000}$ & Exp.  & Slit  		& Telescope &  $\epsilon$ &  M$_{K}$ & M$_{500}$ & R$_{500}$  \\
            &         &                                  &                                 & time (s)                         & PA ($\degr$) 	& semester   &   & (mag)  & $10^{13}$ M$_{\sun}$  & (kpc)  \\                                      
\hline
\multicolumn{11}{c}{MENeaCS}\\
\hline
Abell 780$^{\star}$ & 0.054 & 09:18:05.65 & --12:05:43.5 & 3600 &145  & GS09A& 0.24 & --25.78 $\pm$ 0.06 & 15.73	$\pm$ 6.71 & 786 $\pm$ 105 \\ 
Abell 754 & 0.054 &  09:08:32.37 & --09:37:47.2 &3600 &294  & GS09A& 0.30 & --26.24 $\pm$ 0.05 & 52.93	$\pm$14.19 & 1179 $\pm$ 96  \\ 
Abell 2319 & 0.056 &  19:21:10.00  & +43:56:44.5 &7200 &191  & GN09B& 0.24 & --26.41 $\pm$ 0.06 & -- & -- \\ 
Abell 1991 & 0.059 &  14:54:31.48 & +18:38:33.4 &3600 &192 & GS09A& 0.28 & --25.83 $\pm$ 0.08 & 17.93 $\pm$ 12.95 & 815 $\pm$ 192  \\ 
Abell 1795$^{\star}$ & 0.063 &  13:48:52.49 & +26:35:34.8 &3600 &19   & GN09A& 0.12 & --26.34 $\pm$ 0.08 & 57.25	$\pm$ 16.11 & 1208 $\pm$ 105  \\ 
Abell 644 & 0.070 &  08:17:25.61 & --07:30:45.0 &3600 &12 & GS09A& 0.14 & --26.03 $\pm$ 0.13 &  -- & --  \\ 
Abell 2029 & 0.077 &  15:10:56.09  & +05:44:41.5 &5400 &205  & GS09A& 0.50 & --27.17 $\pm$ 0.05 &  82.95 $\pm$ 17.16 & 1352 $\pm$ 77 \\ 
Abell 1650 & 0.084 & 12:58:41.49  & --01:45:41.0 &4702 & 161 & GS09A& 0.30 & --25.77 $\pm$ 0.10 &  46.89 $\pm$ 9.11 & 1122 $\pm$ 58 \\ 
Abell 2420 & 0.085 &  22:10:18.76  & --12:10:13.9 &2257 &237  & GS09A& 0.18 & --26.51 $\pm$ 0.13 &  50.92 $\pm$ 20.52 & 1151 $\pm$ 153    \\ 
Abell 2142 & 0.091 &  15:58:19.99  & +27:14:00.4 &7200 &313  & GN09A& 0.16 & --25.91 $\pm$ 0.10 &  70.67 $\pm$ 19.27 & 1275 $\pm$ 105    \\ 
Abell 2055$^{\star}$ & 0.102 &  15:18:45.72 & +06:13:56.4 &5400 &139  & GS09A& 0.20 & --25.68 $\pm$ 0.12 &  16.11 $\pm$ 8.05 & 777 $\pm$ 125    \\ 
Abell 2050 & 0.118 &  15:16:17.92 & +00:05:20.9 &5400 &227  & GS09A& 0.26 & --25.78 $\pm$ 0.12 &  23.59 $\pm$ 9.21 & 882 $\pm$ 115   \\ 
Abell 646 & 0.129 &  08:22:09.53  & +47:05:53.3 &3600 &61   & GN09A& 0.24 & --25.93 $\pm$ 0.11 & 18.03 $\pm$ 11.89 & 805 $\pm$ 173    \\ 
Abell 990 & 0.144 & 10:23:39.91  & +49:08:38.8 & 7200 &250  & GN09A& 0.27 & -- &  72.49 $\pm$ 17.45 & 1266 $\pm$ 96    \\ 
\hline
\multicolumn{11}{c}{CCCP}\\
\hline
Abell 2104 & 0.153 & 15:40:07.94 & --03:18:16.3& 7200 & 239 & GS08A& 0.50 & --26.31 $\pm$ 0.14 & 85.92 $^{+\ 17.16}_{-\ 16.40}$ & 1333 $\pm$ 0 \\ 
Abell 2259 & 0.164 & 17:20:09.66 & +27:40:08.3& 3600 & 286 & GS08B& 0.38 & --26.40 $\pm$ 0.10 & 44.40	$^{+\ 13.23}_{-\ 12.37}$ & 1064 $\pm$ 0 \\ 
Abell 586 & 0.171 & 07:32:20.31 & +31:38:01.1& 14400 & 136 & GN08B& 0.42 & --27.00 $\pm$ 0.10 &  26.47 $^{+\ 11.03}_{-\ 10.16}$ & 901 $\pm$ 0 \\  
MS 0906+11 & 0.174 & 09:09:12.76 & +10:58:29.1& 7200 & 208 & GS07B& 0.31 & --26.72 $\pm$ 0.13 &  -- & -- \\ 
Abell 1689 & 0.183 & 13:11:29.52 & --01:20:27.9& 7200 & 163 & GN08B& -- & --  &  166.27	$^{+\ 24.16}_{-\ 23.40}$ & 1649 $\pm$ 0 \\ 
MS 0440+02 & 0.187 & 04:43:09.92 & +02:10:19.3& 7200 & 270 & GS07B& 0.26 & --27.79 $\pm$ 0.10 &  20.14 $^{+\ 9.49}_{-\ 9.49}$  & 815 $\pm$ 0 \\  
Abell 383$^{\star}$ & 0.190 & 02:48:03.38 & --03:31:44.9& 12600 & 2 & GS07B& 0.16 & --26.84 $\pm$ 0.12 &  32.79	$^{+\ 13.33}_{-\ 12.56}$ & 959 $\pm$ 0 \\ 
Abell 963 & 0.206 & 10:17:03.63 & +39:02:49.7 & 7200 & 353 & GN08B& 0.28& --27.25 $\pm$ 0.11 &  68.27 $^{+\ 15.05}_{-\ 15.05}$ & 1218 $\pm$ 0 \\ 
Abell 1763 & 0.223 & 13:35:20.12 & +41:00:04.3 & 7200 & 86 & GN08A& 0.43& --27.33 $\pm$ 0.11 &  92.92 $^{+\ 17.36}_{-\ 17.36}$ & 1342 $\pm$ 0  \\ 
Abell 1942 & 0.224 & 14:38:21.88 & +03:40:13.3& 7200 & 149 & GS08A& 0.28& --27.40 $\pm$ 0.17 & 74.99	$^{+\ 13.90}_{-\ 13.04}$ & 1247 $\pm$ 0 \\ 
Abell 2261 & 0.224 & 17:22:27.23 & +32:07:57.7& 7200 & 174 & GN08A& 0.02& --27.37 $\pm$ 0.10 & 133.19	$^{+\ 20.23}_{-\ 19.47}$ & 1505 $\pm$ 0 \\ 
Abell 2390$^{\star}$ & 0.228 & 21:53:36.84 & +17:41:44.1& 7200 & 315 & GS08A& -- & --27.10 $\pm$ 0.17 & 126.48	$^{+\ 18.70}_{-\ 17.93}$ & 1477 $\pm$ 0 \\ 
Abell 267 & 0.231 & 01:52:41.95 & +01:00:25.9& 7200 & 201 & GS08B& 0.40& --26.82 $\pm$ 0.13  & 44.78	$^{+\ 12.47}_{-\ 11.70}$ & 1045 $\pm$ 0 \\  
Abell 1835$^{\star}$ & 0.253 & 14:01:02.10 & +02:52:42.7& 7200 & 340 & GS08A& 0.20& --27.50 $\pm$ 0.14 & 109.79	$^{+\ 18.51}_{-\ 17.74}$ & 1400 $\pm$ 0 \\ 
Abell 68 & 0.255 & 00:37:06.85 & +09:09:24.5& 7200 & 310 & GS08B& 0.36 &  --26.98 $\pm$ 0.18 &  71.82 $^{+\ 13.52}_{-\ 13.52}$ & 1218 $\pm$ 0 \\ 
MS 1455+22$^{\star}$ & 0.258 & 14:57:15.12 & +22:20:34.5& 7200 & 39 & GS08A& -- & --  &  73.74	$^{+\ 12.37}_{-\ 13.14}$ & 1227 $\pm$ 0  \\ 
Abell 611 & 0.288 & 08:00:56.83 & +36:03:23.8& 7200 & 46 & GN08B& 0.27 & --27.08 $\pm$ 0.15  &  52.93 $^{+\ 14.19}_{-\ 14.19}$ & 1084 $\pm$ 0 \\ 
Abell 2537 & 0.295 & 23:08:22.22 & --02:11:31.7& 7200 & 124 & GS08B& 0.38 & --26.48 $\pm$ 0.23 &  115.74	$^{+\ 20.14}_{-\ 19.37}$ & 1400 $\pm$ 0 \\ 
\hline
\end{tabular}
\end{table*}
%\end{landscape}

\begin{table*}
\caption{CLoGS BGGs with long-slit spectroscopy in HETMG. In all cases the position angle (PA) is given as clockwise from North. LGG is the identification number used in the catalogue of \citet{Garcia1993}. Similar to the BCGs, we use the ellipticities, $\epsilon$, of the BGGs as measured from the 2MASS isophotal $K$-band, and M$_{K}$ measured as described in Section \ref{Kband}. R$_{500}$ and M$_{500}$ are from \citet{O'Sullivan2017}, and M$_{500}$ is in $10^{13}$ M$_{\sun}$. We note that the R$_{500}$ and M$_{500}$ for the groups are scaled from system temperature, as thoroughly discussed in \citet{O'Sullivan2017}, and have smaller uncertainties than the clusters where the values are derived from the mass profiles. The $\star$ next to the object name indicates whether optical emission lines were present in the spectra analysed here.}
\label{objects_BGGs}
\begin{tabular}{r l c c c r r r c c}
\hline
LGG & Name & $z$ & $\alpha_{J2000}$ & $\delta_{J2000}$ & Slit PA & $\epsilon$ & M$_{K}$ & M$_{500}$ & R$_{500}$ \\
            &         &                                  &                          &       & ($\degr$) &    & (mag)   & $10^{13}$ M$_{\sun}$  & (kpc)   \\                                      
\hline
\multicolumn{10}{c}{High-richness Groups}\\
\hline
393 & NGC5846 & 0.0057 & 15:06:29.30 & +01:36:20.0 & 90& 0.05	 & --25.11 $\pm$ 0.02   &  2.65	$^{+\ 0.05}_{-\ 0.05}$ & 452 $^{+\ 3}_{-\ 3}$ \\ 
27 & NGC0584 & 0.0060 & 01:31:20.70 & --06:52:05.0 & 242& 0.38 &	--24.22 $\pm$ 0.03  &  -- & --  \\ 
278 & NGC4261$^{\star}$ & 0.0073 & 12:19:23.22 & +05:49:29.7 & 173& 0.14 & 	--25.47 $\pm$ 0.03   & 4.83	$^{+\ 0.18}_{-\ 0.12}$ & 552$^{+\ 7}_{-\ 5}$ \\ 
363 & NGC5353$^{\star}$ & 0.0073 & 13:27:54.32 & --29:37:04.8 & 308& 0.52 & 	--25.06 $\pm$ 0.02   &1.67 $^{+\ 0.04}_{-\ 0.04}$ & 387 $^{+\ 3}_{-\ 3}$ \\ 
402 & NGC5982 & 0.0095 & 15:38:39.78 & +59:21:21.2 & 105& 0.30 & --24.92 $\pm$ 0.02   &1.20 $^{+\ 0.03}_{-\ 0.03}$ & 346 $^{+\ 3}_{-\ 3}$  \\ 
117 & NGC1587$^{\star}$ & 0.0120 & 04:30:39.92 & +00:39:42.2 & 225& 0.22 & 	--25.00 $\pm$ 0.03    &0.55	$^{+\ 0.21}_{-\ 0.14}$ & 267$^{+\ 31}_{-\ 25}$ \\ 
421 & NGC6658 & 0.0124 & 18:33:55.68 & +22:53:17.9 & 185&  0.76 & 	--24.25 $\pm$ 0.03    &0.36	$^{+\ 0.18}_{-\ 0.12}$ & 233$^{+\ 34}_{-\ 28}$ \\ 
473 & NGC7619 & 0.0125 & 23:20:14.52 & +08:12:22.6 & 133& 0.20 & --25.28 $\pm$ 0.02    &2.88	 $^{+\ 0.05}_{-\ 0.05}$ & 464 $^{+\ 3}_{-\ 3}$ \\ 
103 & NGC1453$^{\star}$ & 0.0128 & 03:46:27.27 & --03:58:07.6 & 199& 0.22 & 	--25.48 $\pm$ 0.02   &1.74 $^{+\ 0.12}_{-\ 0.12}$ & 392 $^{+\ 9}_{-\ 9}$ \\ 
61 & NGC0924$^{\star}$ & 0.0147 & 02:26:46.84 & +20:29:50.7 & 55& 0.40	& --24.37 $\pm$ 0.03   &-- & -- \\ 
158 & NGC2563 & 0.0147 & 08:20:35.68 & +21:04:04.3 & 250& 0.22 &	--25.02 $\pm$ 0.02  & 4.18 $^{+\ 0.06}_{-\ 0.06}$ & 525 $^{+\ 2}_{-\ 2}$  \\ 
42 & NGC0777 & 0.0162 & 02:00:14.93 & +31:25:45.8 & 145& 0.16	 & --25.61 $\pm$ 0.02  &   2.37 	$^{+\ 0.09}_{-\ 0.09}$ & 434 $^{+\ 5}_{-\ 5}$  \\ 
72 & NGC1060$^{\star}$ & 0.0167 & 02:43:15.05 & +32:25:30.0 & 70& 0.16	&	--25.97 $\pm$ 0.02  &   2.97		$^{+\ 0.15}_{-\ 0.14}$ & 468 $^{+\ 8}_{-\ 8}$ \\ 
18 & NGC0410$^{\star}$ & 0.0172 & 01:10:58.87 & +33:09:07.3 & 262& 0.26	 &	--25.76 $\pm$ 0.02  &   2.78		$^{+\ 0.10}_{-\ 0.09}$ & 458 $^{+\ 5}_{-\ 5}$ \\ 
\hline
\multicolumn{10}{c}{Low-richness Groups}\\
\hline
167 & NGC2768$^{\star}$ & 0.0043 & 09:11:37.50 & +60:02:13.9 & 93& 0.54	&	--24.54 $\pm$ 0.03     & --  & -- \\ 
236 & NGC3665$^{\star}$ & 0.0066 & 11:24:43.63 & +38:45:46.1 & 25&  0.24 &		--24.84 $\pm$ 0.02    & --  & -- \\ 
232 & NGC3613 & 0.0068 & 11:18:36.10 & +58:00:00.0 & 100&  0.52 &		--24.35 $\pm$ 0.02    &  -- & -- \\ 
23 & NGC0524 & 0.0078 & 01:24:47.71 & +09:32:19.7 & 235&  0.10 &		--25.09 $\pm$ 0.01    & --  & -- \\ 
126 & NGC1779$^{\star}$ & 0.0108 & 05:05:18.03 & --09:08:50.1 & 130&  0.42 &		--24.55 $\pm$ 0.02    &  -- & -- \\ 
383 & NGC5629 & 0.0147 & 14:28:16.36 & +25:50:55.7 & 110& 0.10 &		--24.79 $\pm$ 0.02     & --  & -- \\ 
350 & NGC5127$^{\star}$ & 0.0160 & 13:23:44.98 & +31:33:56.9 & 260& 0.26 & 		--24.81 $\pm$ 0.03     &  -- & -- \\ 
376 & NGC5490 & 0.0160 & 14:09:57.33 & +17:32:43.5 & 184& 0.22 &		--25.20 $\pm$ 0.02     &  -- & -- \\ 
14 & NGC0315$^{\star}$ & 0.0162 & 00:57:48.88 & +30:21:08.8 & 45&  0.22	 & --26.02 $\pm$ 0.02 &  -- & -- \\ 
\hline
\end{tabular}
\end{table*}  	

For the CLoGS BGGs, the available data products for the HET massive galaxies measured by \citet{Vandenbosch2015} include the stellar kinematics, measured with the pixel-fitting code (pPXF; \citealt{Cappellari2004}) using template stars from MILES \citep{Sanchez2006}. Their stellar kinematic extraction is obtained from the stellar continuum in an observed window of 5000 -- 6100 \AA{}, selected to minimise instrumental resolution changes across the slit. We remeasure the velocity and velocity dispersion profiles, as well as a central velocity dispersion within a 1 kpc aperture, and find excellent agreement, within the 1$\sigma$ errors with \citet{Vandenbosch2015}. 

All the spatially-resolved velocity and velocity dispersion profiles of the MENeaCS and CCCP BCGs are presented in Figures \ref{fig:kin1} to \ref{fig:kin11} in Appendix \ref{plots}, with spatial radii indicated in both arcsec and kpc. We compare our central velocity dispersion ($\sigma_{0}$) measurements, for galaxies in common, with \citet{Cappellari2013} and \citet{Veale2017}, and the velocity dispersion profiles for galaxies in common with \citet{Fisher1995, Loubser2008, Newman2013a}, and find excellent agreement in all cases as described in Appendix \ref{kinematics}.

\subsection{$K$-band luminosity}
\label{Kband}

We use the 2MASS extended source catalogue (XSC) to determine each galaxy's absolute $K$-band luminosity. Similar to \citet{Ma2014} and \citet{Veale2017} (for MASSIVE) we use the total extrapolated $K$-band magnitude (XSC parameter $k\_m\_ext$), which is measured in an aperture consisting of the isophotal aperture plus the extrapolation of the surface brightness profile based on a single S\'{e}rsic fit to the inner profile \citep{Jarrett2003}. We then make three corrections to accurately compare the luminosities with each other: foreground and internal extinction, an evolutionary correction, and a k-correction. 

Differential extinction in the $K$-band is an order of magnitude smaller than in the visible bands. Nevertheless, we correct for foreground (line-of-sight) extinction by using the \citet{Schlafly2011} recalibration of the infrared-based dust map by \citet{Schlegel1998}. The average foreground extinction correction for all 32 BCG and 23 BGGs is only 0.018 magnitude. 

Internal extinction by gas and dust only applies to the active, star forming BCGs and is also generally negligibly small. \citet{Oonk2011} find that for the BCG in Abell 2597, a known star forming BCG at 4 -- 5 M$_{\sun}$/yr \citep{Donahue2007}, the Br $\gamma$/Pa $\alpha$ ratio measurements indicate that extinction in the $K$-band is unimportant. Deep optical spectroscopy in \citet{Voit1997} find a $V$-band extinction of A$_{V} \sim 1$ across the Abell 2597 BCG nebulosity, which translates to A$_{K} \sim 0.1$. Since the internal extinction of individual star forming BCGs is difficult to determine, we take this into account by making a correction of A$_{K} \sim 0.05$ mag to the luminosities (for all the star forming BCGs and BGGs, \ie\ those with emission lines in their optical spectra).

To fairly compare all the luminosities at $z = 0$, we make an evolutionary correction to all the BCGs and BGGs by using the photometric predictions generated by the \citet{Vazdekis2010} stellar population models based on the MILES \citet{Sanchez2006} stellar library, and a Salpeter Initial Mass Function (IMF, \citealt{Salpeter1955}) as used in the stellar population fitting (presented in \citealt{Loubser2016} for CCCP, and will be described in Paper II for MENeaCS). We used a metal-rich stellar population typical to BCGs, and the largest adjustment in the $K$-band was 0.5 magnitudes for the $z \sim 0.3$ BCG, and the adjustment for the BGGs was $<$ 0.1 mag.

Similarly, we perform a k-correction to eliminate the redshift effect on the $K$-band luminosity measurements. k-corrections are independent of galaxy type up to $z \sim 2$ \citep{Glazebrook1995}. We use the M$_{J}$ -- M$_{K}$ colours from 2MASS, and the \citet{Chilingarian2010} k-correction algorithms. 

As mentioned above, we have four BGGs in common with the MASSIVE study. We compare our absolute $K$-band luminosities with those listed in \citet{Ma2014}, as both our studies have used the 2MASS XCS to obtain the $K$-band measurements, and find that ours are on average 0.06 mag fainter then MASSIVE. As described in the captions of Tables \ref{objects_BCGs} and \ref{objects_BGGs}, we use the ellipticities, $\epsilon$, of the BCGs/BGGs as measured from the 2MASS isophotal $K$-band, and obtained through the NED. 
We also compare the absolute values of the differences in our ellipticities (with ATLAS$^{\rm 3D}$ in \citealt{Krajnovic2011} and MASSIVE in \citealt{Ma2014}), and find that they differ only 0.04 on average, which is well within the typical error on ellipticities derived from 2MASS.   

\section{Analysis: kinematics}
\label{Calproperties}

\subsection{Rotation} 
\label{rotation}

\begin{figure*}
   \centering
  \subfloat{\includegraphics[scale=0.36]{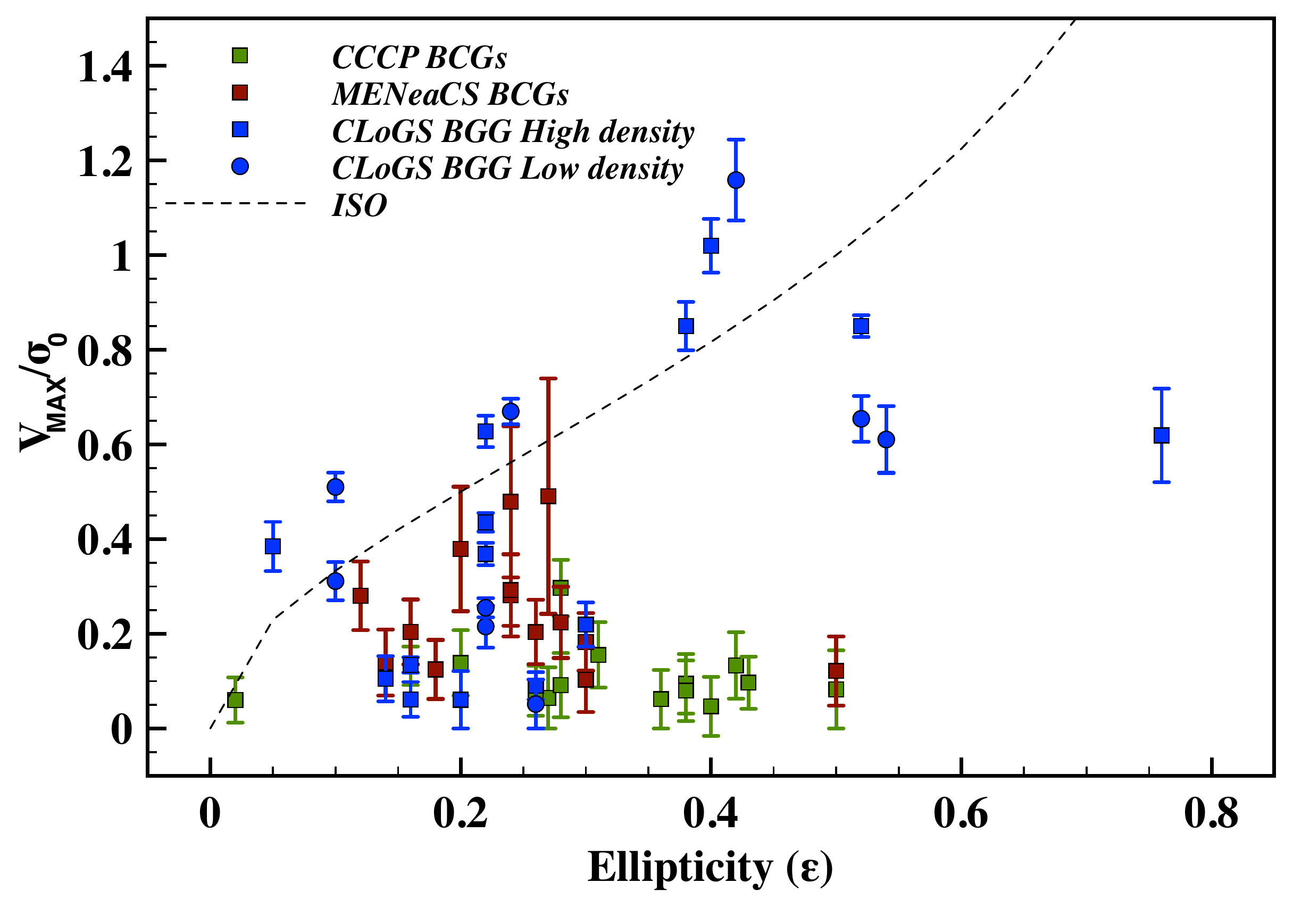}}
  \subfloat{\includegraphics[scale=0.36]{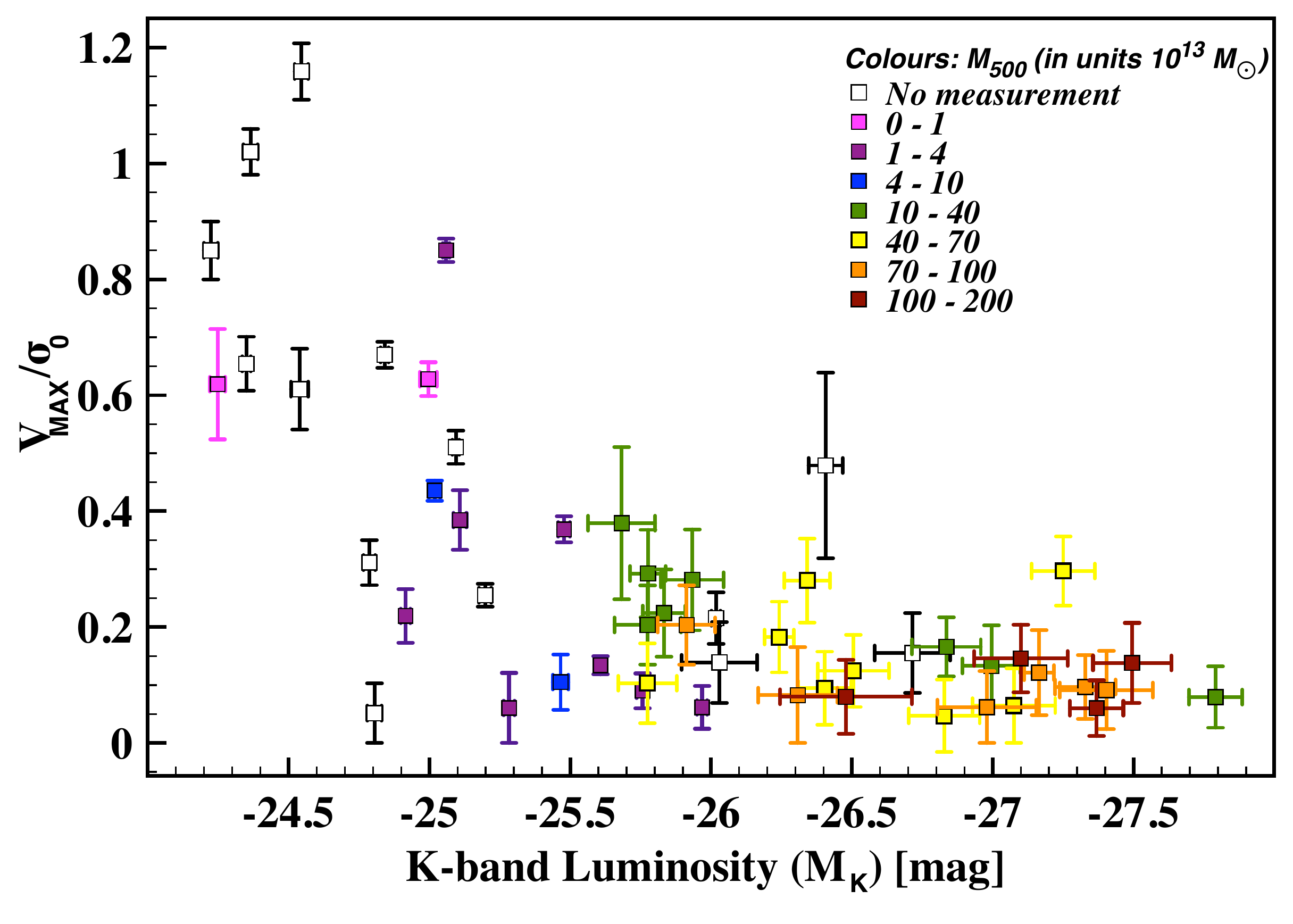}}\\
   \caption[]{Left: Anisotropy parameter $(V_{\rm max}/\sigma_{0}$) vs. ellipticity ($\epsilon$) for the CCCP and MENeaCS BCGs, and the high and low density samples of the CLoGS BGGs. The predicted rotation for isotropic oblate spheroids is shown by the `oblate line', labelled ISO. Right: Anisotropy parameter $(V_{\rm max}/\sigma_{0}$) vs. M$_{K}$ luminosity, coloured by M$_{500}$ in units of $10^{13}$ M$_{\sun}$. Some CLoGS groups, primarily in the low density sample, do not have M$_{500}$ measurements.}
   \label{fig:Rotation}
\end{figure*}

The `anisotropy parameter', $V_{\rm max}/\sigma_{0}$ \citep{Kormendy1982} compares the global dynamical importance of rotation and random motions of stars in a galaxy. Figure \ref{fig:Rotation} shows the anisotropy parameter vs. ellipticity ($\epsilon$) of the CCCP (green) and MENeaCS (red) BCGs, as well as the CLoGS BGGs (blue) for comparison. The predicted rotation for isotropic oblate spheroids is shown by the `oblate line', labeled ISO in the figure, and approximated as $V_{\rm max}/\sigma_{0} = \sqrt{\frac{\epsilon}{1 - \epsilon}}$ \citep{Bender1992}. The ISO curve plotted in Figure \ref{fig:Rotation} is corrected for projection effects, but is specifically for edge-on models with constant ellipticity \citep{Binney1980}. Isotropic oblate spheroid models viewed at other inclinations all fall close to the line, giving rise to very little scatter \citep{Illingworth1977}. The BCG data points that fall well below the ISO line can therefore be interpreted as isotropic prolate spheroids, or much more likely for these massive ellipticals, as anisotropic systems \citep{Binney1978}.  

Three galaxies in the CCCP sample (Abell 1689, Abell 2390 and MS 1455+22) do not have measured ellipticities in 2MASS. However, if we were to use the average ellipticity of the rest of the CCCP sample 0.29 ($\pm$ 0.14), then these three BCGs are located in the same plane as the other CCCP BCGs, \ie\ well below the standard ISO curve. 

The rotational velocity $V_{\rm max}$ was estimated as half the difference between the minimum and maximum of the rotation (Figures \ref{fig:kin1} to \ref{fig:kin11}). As BCGs generally do not have well defined rotation curves, the $V_{\rm max}$ measurements are subject to large uncertainties. In contrast, the BGGs generally have well-defined rotation curves (see figure 5 in \citealt{Vandenbosch2015}). All the spatially-resolved kinematic profiles for galaxies observed on HET by \cite{Vandenbosch2015}, including for the 23 CLoGS galaxies investigated here, are available online\footnote{\url{http://www2.mpia-hd.mpg.de/~bosch/hetmgs/}}.

The majority of the BCGs have velocity curves consistent with being flat (\ie\ no rotation), whilst some BCGs show marginal rotation (\eg\ Abell 780 and Abell 963 shown in Figure \ref{fig:kin3} and \ref{fig:kin4}). None of the BCGs are supported by rotation and above the isotropic oblate spheroids rotation curve in Figure \ref{fig:Rotation}. It should be noted that this standard ISO curve is the parabola $V_{\rm max}/\sigma_{0} = \epsilon^{1/2}$ as $\epsilon \to 0$ \citep{Binney1982}, and that the BCG at $\epsilon = 0.02$ (Abell 2261) is not rotating, even though it appears to be close to the curve at $\epsilon \sim 0$. Seven of the 23 BGGs (three of the low density, and four of the high density sample) are rotating and above the standard ISO curve in Figure \ref{fig:Rotation}.
 
Many additional factors complicate the dynamical interpretation of individual points, \ie\ subjective $V_{\rm max}$ measurements, and that the observed ellipticity is a global property of the galaxy, whereas the kinematic (long-slit) measurements taken here only reflect the kinematics along the axis where the slit was placed, and only close to the centre of the galaxy. For example, a disk component may dominate the measured kinematics but will have little effect on the ellipticity, making the galaxy appear to rotate faster than its global ellipticity would suggest \citep{Merrifield2004}. However, for our purpose, we can conclude that the BCGs studied here do not show any significant rotation, consistent with their high stellar masses and presumably rich merger histories.

We also plot the anisotropy parameter ($V_{\rm max}/\sigma_{0}$) against the luminosity (M$_{K}$) in the right panel of Figure \ref{fig:Rotation}, colour-coded by host cluster halo mass M$_{500}$ (in units of $10^{13}$ M$_{\sun}$). This is qualitatively consistent with the finding of \citet{Veale2017} that the fraction of slow- or non-rotators (measured from a global angular momentum parameter) increases as a function of luminosity, as measured in $K$-band, for their 41 MASSIVE galaxies as well as the ATLAS$^{\rm 3D}$ sample (from 10$\%$ at M$_{K}$ = --22 to 90$\%$ at M$_{K}$ = --26, their figure 4). Similarly, our result is also qualitatively consistent with \citet{OlivaAltamirano2017}, who showed a weak (not statistically-significant) trend in that the probability of a BCG being a slow- or non-rotator increases with cluster mass (their figure 7). 

In rotating galaxies, rotation can contribute a non-negligible amount to the second order velocity moment $v_{rms} \equiv \sqrt{V^{2} + \sigma^{2}}$. For our BCGs, none of which show significant rotation, we find negligible differences between the velocity dispersion ($\sigma$) slope and the $v_{rms}$ slope (we show this in Figure \ref{sigma_second}). 

\subsection{Scaling relations}
\label{scaling}

Early-type galaxies are tightly correlated via three parameters, the effective radius $R_{e}$, effective surface brightness $I_{e}$ and velocity dispersion $\sigma$, that define a three dimensional parameter space called the Fundamental Plane (FP) \citep{Dressler1997, Djorgovski1987, Bender1992}. Projections of this plane are the Faber-Jackson relation (FJR, \citealt{Faber1976}), luminosity $M$ vs. $\sigma$, and the Kormendy relation (KR, \citealt{Kormendy1977}), $R_{e}$ vs $I_{e}$. 

\begin{figure}
   \centering
    \subfloat{\includegraphics[scale=0.35]{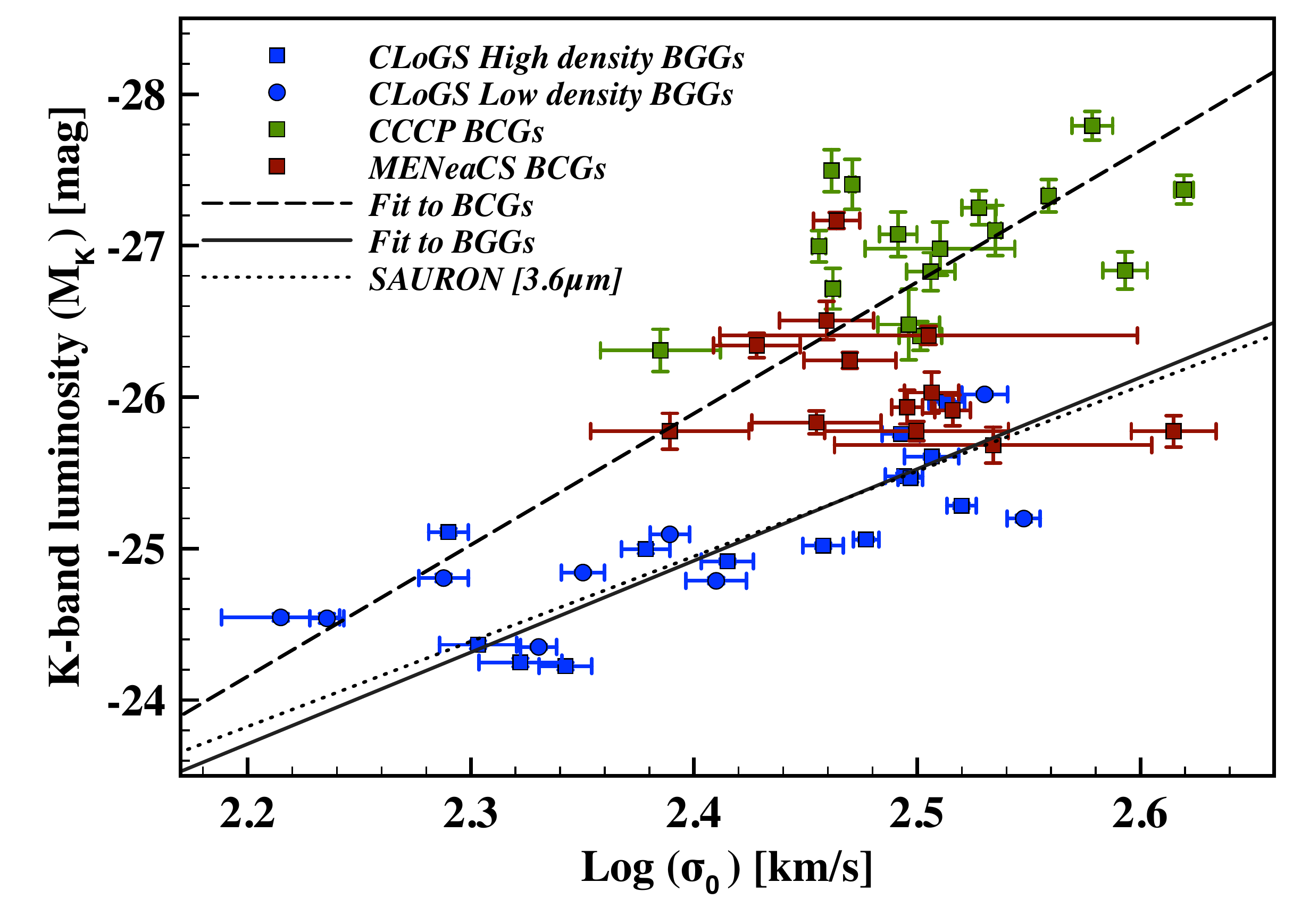}}\\
    \subfloat{\includegraphics[scale=0.35]{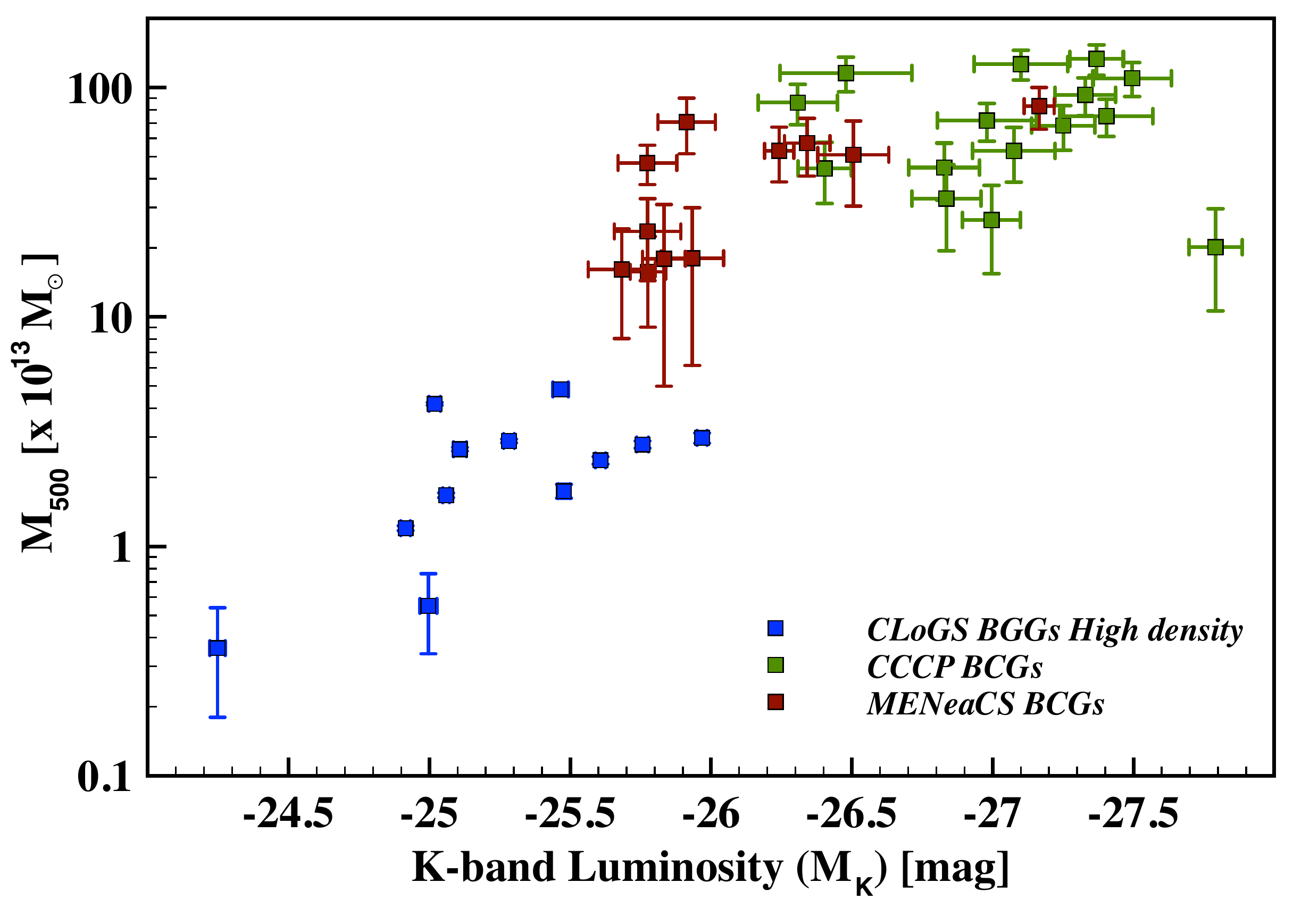}}\\
   \caption[]{Scaling relations: The Faber-Jackson relation (FJR) \citep{Faber1976} (top). The halo mass, M$_{500}$ (in units of $10^{13}$ M$_{\sun}$) vs rest-frame $K$-band luminosity (bottom).}
   \label{checks}
\end{figure}

We plot the $K$-band FJR for our BCGs and BGGs in Figure \ref{checks} (top panel), and find that the best fit to the BGGs is M$_{K} \propto \sigma_{0}^{6.50 \pm 0.21}$ (measured in the range $\log \sigma_{0} = 2.21 - 2.55$), and to the BCGs is M$_{K} \propto \sigma_{0}^{8.68 \pm 0.46}$ (measured in the range $\log \sigma_{0} = 2.38 - 2.62$). Note that these fits take the errors on $x$ and $y$ into account, and is inversely weighted by the errors. From the virial theorem follows $M \propto \sigma^{4}$ \citep{Faber1976}, and others have shown that the slope of the relation can vary from approximately two for low mass galaxies, to approximately eight for the most massive early-types, dependent on band measured, environment and luminosity range in which relationship is measured (see \eg\  \citealt{Gallazzi2006, Lauer2007, Desroches2007}). We also plot the FJR-relation for $Spitzer$/IRAC 3.6$\micron$ for the SAURON E/S0 sample presented in \citet{FalconBarroso2011} in Figure \ref{checks}. Their relation ($M_{3.6\micron} \propto \sigma^{5.62 \pm 0.69}$) agrees remarkably well with our best fit to our BGGs.

These steep deviations from the canonical FJR slope (for the BCGs more so than the BGGs) can be caused by radial changes in the stellar M/L ratio ($\Upsilon_{*}$) of the central galaxy, as would be the expected if, for example, recent star formation in the galaxies is localized, or when the ratio of stellar mass to dynamical mass within the effective radius is not constant (\ie\ scales with either the dynamical or the stellar mass; \citealt{Boylan2005, Boylan2006}). Variation in the ratio of the stellar mass to the dynamical mass (within the effective radius) of galaxies depends on a galaxy's assembly history. Violent relaxation in dissipationless mergers tends to mix dark matter and stars. As a result, the dynamical mass within a physical radius increases more than the stellar mass within the same radius, and the net effect is that the remnants of mergers are more dark matter dominated than their progenitors (as illustrated in the simulations of \citealt{Boylan2005, Boylan2006}). The mixing, and the resulting increase in dark matter fractions, scale with the dynamical or stellar mass. In addition to a non-constant $\Upsilon_{*}$ and dissipationless mergers leading to steep deviations from the canonical FJR slope, the slope also depends on velocity structure. The simulations and analysis by \citet{Boylan2005, Boylan2006} show how the locations of dissipationless merger remnants on the projections of the fundamental plane (but not the fundamental plane itself) depend strongly on the merger orbit, and the relations steepen significantly from the canonical scalings for mergers on radial orbits. 

In the follow-up paper where we present and discuss the surface brightness profiles, we also use the derived structural parameters ($I_{e}$, $R_{e}$ and $\sigma_{e}$) in order to construct the Fundamental Plane and the Kormendy relation, in addition to the Faber-Jackson relation presented here. This will form a more complete picture of the deviations of the CCCP BCGs, MENeaCS BCGs, and CLoGS BGGs from the Fundamental Plane and its projections, as well as the differences between the three samples.

Lastly, we also plot the host cluster halo mass, M$_{500}$ vs. BCG/BGG $K$-band luminosity in Figure \ref{checks} (bottom panel), and recover the known correlation between BCG luminosity and host cluster mass (\eg\ \citealt{Lin2004}).  

%--------------------------------------------------------------------------------------------------------------------------------------------------------------------------------------------------------------------------------------------------------------------------------------------------------

\subsection{Velocity dispersion profiles}
\label{Velocityprofiles}

We measured the velocity dispersion profiles, and normalised them with the central velocity dispersion, $\sigma_{0}$, measured as described in Section \ref{binning}, for each BCG/BGG. We then fitted power laws 
\begin{equation}
\sigma_{R} = \sigma_{0}\Big[\frac{R}{R_{0}}\Big]^{\eta},
\end{equation}
where $R_{0}$ is 5 kpc for BCGs and 1 kpc for BGGs. We excluded the very central bin which may be affected by seeing\footnote{The central bin, possibly affected by seeing, for the BCGs has width of $<$0.8\arcsec, which is much smaller than the physical radius of 10 kpc where $\sigma_{0}$ is measured for all the BCGs. Similarly, the central bin for the BGGs has width of $<$0.5\arcsec, which is much smaller than the physical radius of 2 kpc where $\sigma_{0}$ is measured for all the BGGs. Thus, the central velocity dispersion measurements should not be significantly affected by seeing.}, before we measured the velocity dispersion slope and error ($\eta \pm \Delta \eta$). These fits are also shown in Figures \ref{fig:kin1} to \ref{fig:kin11} for the BCGs, and Figures \ref{fig:kin31} and \ref{fig:kin32} for the BGGs. \citet{Graham1996} measured half-light radii, R$_{e}$, for 119 Abell cluster BCGs, and  found an average R$_{e} \sim 16.7$ kpc. The kinematic profiles of our BCGs are typically measured to 15 kpc (to each side), which is therefore close to the typical half-light radius of a BCG. The average R$_{e}$ for galaxies in the MASSIVE sample (comparable to our BGGs) is R$_{e} \sim 10.1$ kpc, if measured from the NASA Sloan Archive \citep{Ma2014}. The kinematic profiles of our BGGs are typically measured to 10 kpc (to each side), which is close to the typical half-light radius of a BGG.

\subsubsection{Variety in the velocity dispersion profiles}

\begin{figure*}
   \centering
   \includegraphics[scale=0.45]{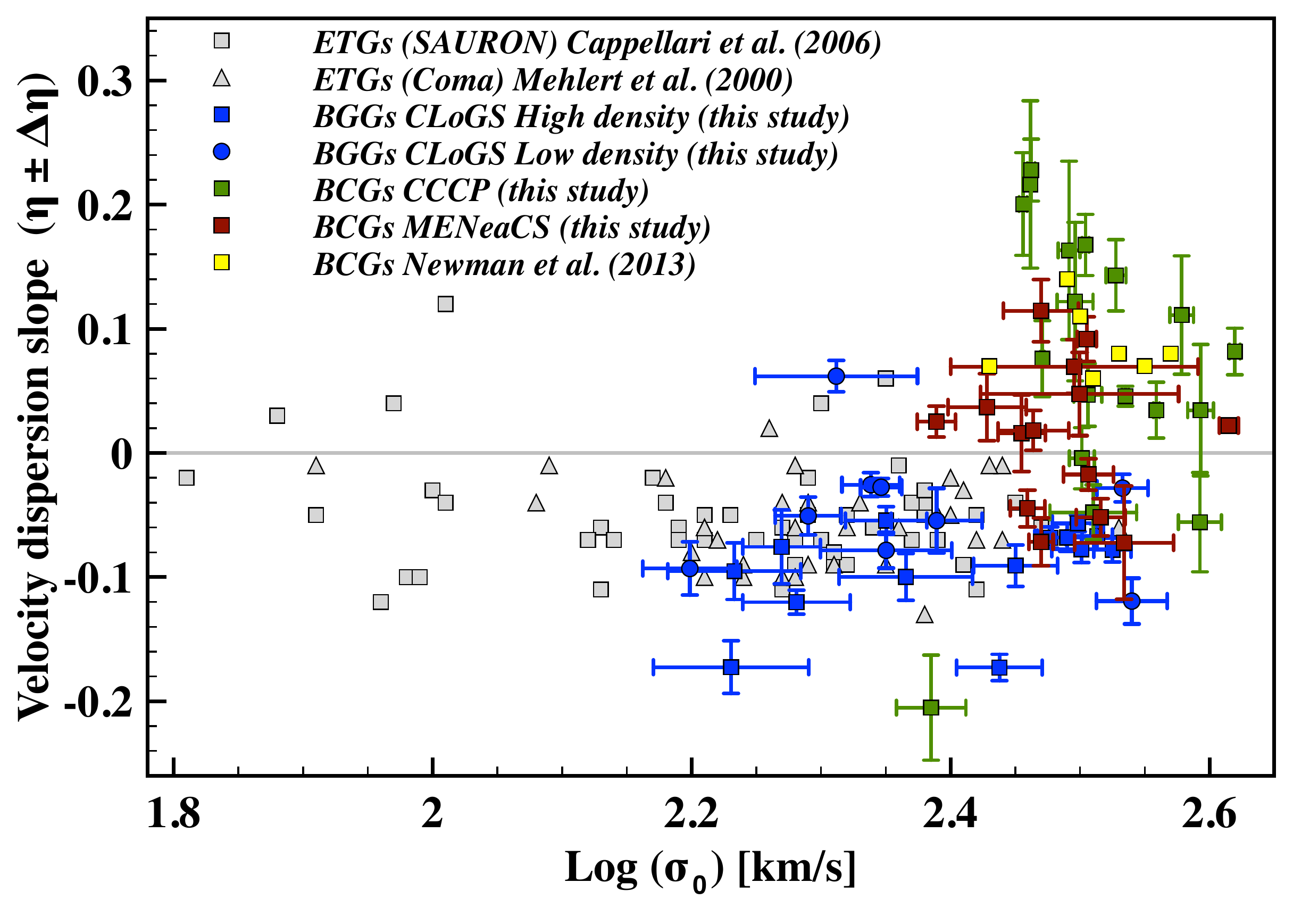}\\
   \caption[]{Velocity dispersion slopes ($\eta \pm \Delta \eta$) against the central velocity dispersion ($\sigma_{0}$). The CCCP and MENeaCS BCGs are indicated with green and red squares, respectively. For comparison we also plot the seven BCGs analysed in \citep{Newman2013a} (with yellow squares), as well as field and cluster early-type galaxies from \citet{Cappellari2006} (grey squares), and early-type galaxy members of the Coma cluster from \citet{Mehlert2000} (grey triangles). We also add the CLoGS high and low density sample BGGs (blue squares and circles, respectively). The CCCP BCG (green square) with a velocity dispersion slope of --0.2 is the BCG in Abell 2104 and a clear exception as discussed in Section \ref{Discussion}.} \label{fig:Sigma}
\end{figure*}

We plot the velocity dispersion slope, $\eta \pm \Delta \eta$, against the central velocity dispersion, $\sigma_{0}$, in Figure \ref{fig:Sigma}. The CCCP and MENeaCS BCGs are indicated with green and red squares, respectively. For comparison we also plot the seven BCGs analysed in \citet{Newman2013a} (with yellow squares), as well as field and cluster early-type galaxies from \citet{Cappellari2006} (grey squares), and early-type galaxy members of the Coma cluster from \citet{Mehlert2000} (grey triangles)\footnote{The literature data is taken from the compilation in \citet{Chae2014} and the velocity dispersion slopes are normalised at half of half-light radii $0.5R_{e}$, however this choice for normalisation has negligible effect on the slope measurements.}. The sample of \citet{Cappellari2006} is a sub-sample of 25 out of the 48 Sauron E/S0 galaxies, which is representative of nearby bright early-type galaxies ($cz \le 3000$ km s$^{-1}$; M$_{K}$ = --21.5 to --25.5 mag\footnote{If their least massive galaxy, M32, at M$_{K}$ = --19.5 mag is excluded.}), but does not include any BCGs. The Coma spectroscopic sample is described in detail in \citet{Mehlert2000}, and contains the three `cD' galaxies, the four most luminous galaxies of type E and S0, and a selection of galaxies drawn from the luminosity function. 

We repeat exactly the same velocity dispersion slope measurements for the CLoGS BGGs, but normalise with 1 kpc instead of 5 kpc (corresponding to the apertures where $\sigma_{0}$ was measured). The CLoGS BGGs are indicated with blue squares and circles, for the high density and low density sample, respectively. The velocity dispersion slopes of the BCGs are clearly much more scattered, with a significantly larger fraction of positive slopes, compared to other (non-central) early-type galaxies as well as the brightest members of the CLoGS groups. We present the velocity dispersion slopes of the BCGs and BGGs in Tables \ref{properties} and \ref{propertiesBGGs}, respectively.

In Figures \ref{fig:kin1} to \ref{fig:kin11} there are four BCGs (Abell 267, Abell 383, Abell 2055, and MS 0440+02) where the velocity dispersion show a pronounced dip in the centre of the profile and a power law fit may not be the most accurate description. We therefore follow the methodology in \citet{Veale2017}, and fit broken power laws with a break radius at 5 kpc to investigate the outer slopes of the velocity dispersion profiles. As emphasised in \citet{Veale2017}, this fitting function is simply a convenient choice for quantifying the overall rise or fall of velocity dispersion with radius and is not motivated by any physical reasoning. In all four cases, we find that when all the points are included in the power law fits, the sign of the overall slope (+ or --) retrieved is the same as the sign of the `outer' slope (further than 5 kpc). A small number of BCGs do not have enough spatial bins beyond 5 kpc to ensure an accurate power law fit to the outer slopes alone. In all the cases where the BCG velocity dispersion outer profiles could be accurately fit, we do however find that the sign of the outer slopes are the same as the sign of the single power law fits. We also compare the four galaxies from our CLoGS sub-sample to those in common with MASSIVE \citep{Veale2017}, and find comparable results (\ie\ NGC0410 negative, NGC0777 negative, NGC1060 negative, and NGC0315 slightly negative/flat, and all best fit by single power laws). We also test the influence on our conclusions when the four BCGs where a single power law may not be the best description are removed (see \ref{variety}). \citet{Veale2017b} find 64/85 of their MASSIVE galaxies are best-fit by a single power law. 

\begin{figure}
   \centering
\subfloat{\includegraphics[scale=0.42]{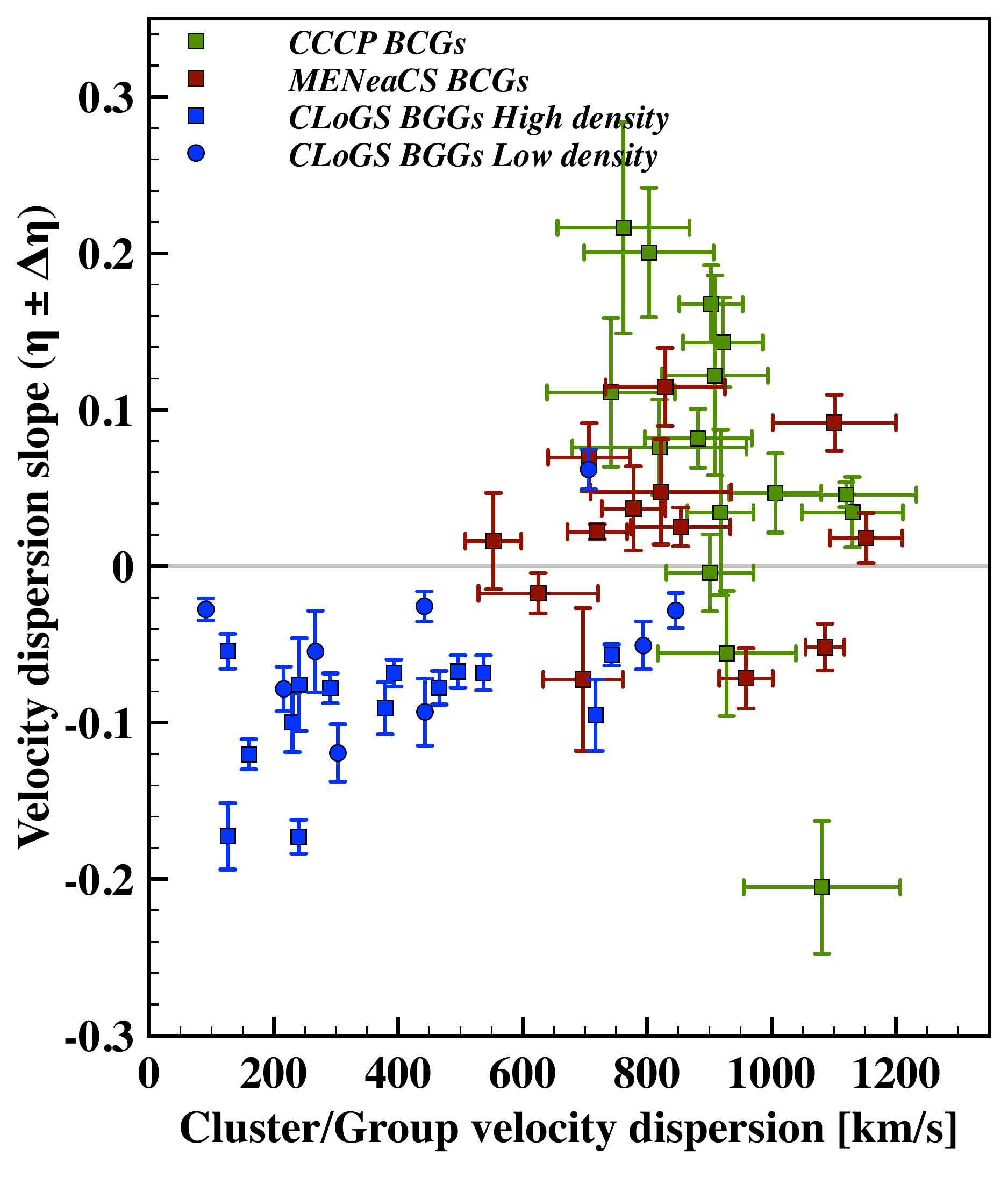}}\\
\caption[]{Velocity dispersion slopes ($\eta \pm \Delta \eta$) against host cluster/group velocity dispersion. We find similar correlations for M$_{500}$ and R$_{500}$. The CCCP BCG (green square in the bottom right-hand corner) with a velocity dispersion slope of --0.2 is the BCG in Abell 2104 and a clear exception as discussed in Section \ref{Discussion}. The cluster velocity dispersions are described in \citet{Bildfell_thesis}, and the group velocity dispersions in \citet{O'Sullivan2017}.}
   \label{fig:Sigmaslope}
\end{figure}

We further plot the velocity dispersion slope against group/cluster velocity dispersion in Figure \ref{fig:Sigmaslope}. If stellar velocity dispersion traces mass directly, then a rising velocity dispersion at large radius is to be expected for galaxies in rich clusters or groups, as it increases towards the cluster or group velocity dispersion. We do see this general trend of increasing velocity dispersion slope with increasing group/cluster velocity dispersion, albeit with very large scatter. We note that we find similar correlations for M$_{500}$ and R$_{500}$.

Our findings are comparable, and complimentary, to \citet{Veale2017} for the 41 most massive nearby galaxies ($M_{\star} \ge 10^{11.8}$ M$_{\sun}$) in the MASSIVE survey. The 12 brightest cluster/group galaxies in their sample have rising or nearly flat velocity dispersion profiles, whereas the less luminous ones show a wide variety of shapes, and the majority (5/7) of their isolated galaxies have falling velocity dispersion profiles. Their study has a smaller range in galaxy luminosity, M$_{K}$ = --25.7 to --26.6 mag, limiting their ability to characterise any strong trends with mass or luminosity as discussed in Section \ref{introduction}, but already suggests that the velocity dispersion profile slopes correlate with galaxy environment and luminosity. We investigate the latter correlation, for all 52 BCGs/BGGs (excluding the three BCGs lacking measurements in 2MASS), from M$_{K}$ = --24.2 to --27.8 mag, in the next subsection.

\begin{table}
\centering
\caption{Derived kinematic properties for the BCGs: central velocity dispersion $\sigma_{0}$, rotation $V_{\rm max}/\sigma_{0}$, and velocity dispersion slope ($\eta \pm \Delta \eta$).}
\label{properties}
\begin{tabular}{l c c c r c c}
\hline
Name & $z$ & $\sigma_{0}$ & $V_{\rm max}/\sigma_{0}$ & $\eta \pm \Delta \eta$ \\
 &  & (km s$^{-1}$) &  &  \\
\hline
\multicolumn{5}{c}{MENeaCS}\\
\hline
Abell 780 & 0.054 & 316 $\pm$ 30&0.29 $\pm$ 0.08&	0.047 $\pm$ 0.034 \\
Abell 754 & 0.054 & 295 $\pm$ 14&	0.18 $\pm$ 0.06&	--0.072 $\pm$ 0.019 \\ 
Abell 2319 & 0.056 &  320 $\pm$ 69&	0.48 $\pm$ 0.16&	0.092 $\pm$ 0.018 \\
Abell 1991 & 0.059 &  285 $\pm$ 19&	0.22 $\pm$ 0.08&	0.016 $\pm$ 0.031 \\
Abell 1795 & 0.063 &  268 $\pm$ 12&	0.28 $\pm$ 0.07&	0.037 $\pm$ 0.027 \\
Abell 644 & 0.070 &  321 $\pm$ 9&	0.14 $\pm$ 0.07&	--0.017 $\pm$ 0.013 \\
Abell 2029 & 0.077 &  291 $\pm$ 7&	0.12 $\pm$ 0.07&	0.018 $\pm$ 0.016 \\
Abell 1650 & 0.084 & 412 $\pm$ 18&	0.10 $\pm$ 0.07&	0.022 $\pm$ 0.005 \\
Abell 2420 & 0.085 &  288 $\pm$ 14&	0.12 $\pm$ 0.06&	--0.045 $\pm$ 0.015 \\
Abell 2142 & 0.091 &  328 $\pm$ 6&	0.20 $\pm$ 0.07&	--0.052 $\pm$ 0.015 \\
Abell 2055 & 0.102 & 342 $\pm$ 56&	0.38 $\pm$ 0.13&	--0.072 $\pm$ 0.046	\\
Abell 2050 & 0.118 &  245 $\pm$ 20&	0.20 $\pm$ 0.07&	0.025 $\pm$ 0.012 \\
Abell 646 & 0.129 &  313 $\pm$ 5&	0.28 $\pm$ 0.09&	0.070 $\pm$ 0.022	\\
Abell 990 & 0.144 & 295 $\pm$ 8&	0.49 $\pm$ 0.25&	0.115 $\pm$ 0.025 \\
\hline
\multicolumn{5}{c}{CCCP}\\
\hline
Abell 2104 & 0.153 & 243 $\pm$ 15&	0.08 $\pm$ 0.08&	--0.205 $\pm$ 0.042 \\
Abell 2259 & 0.164 & 317 $\pm$ 7&	0.09 $\pm$ 0.06&	--0.004 $\pm$ 0.025 \\
Abell 586 & 0.171 & 286 $\pm$ 2&	0.13 $\pm$ 0.07&	0.201 $\pm$ 0.041 \\
MS 0906+11 & 0.174 & 290 $\pm$ 2&	0.16 $\pm$ 0.07&	0.228 $\pm$ 0.025 \\
Abell 1689 & 0.183 & 319 $\pm$ 2&	0.22 $\pm$ 0.06&	0.168 $\pm$ 0.025 \\
MS 0440+02 & 0.187 & 379 $\pm$ 8&	0.08 $\pm$ 0.05&	0.111 $\pm$ 0.048 \\ 
Abell 383 & 0.190 & 392 $\pm$ 9&	0.17 $\pm$ 0.05&	0.034 $\pm$ 0.053 \\
Abell 963 & 0.206 &  337 $\pm$ 6&	0.30 $\pm$ 0.06&	0.143 $\pm$ 0.029 \\
Abell 1763 & 0.223 & 362 $\pm$ 2&	0.10 $\pm$ 0.06&	0.034 $\pm$ 0.023 \\
Abell 1942 & 0.224 & 296 $\pm$ 1&	0.09 $\pm$ 0.07&	0.076 $\pm$ 0.031 \\
Abell 2261 & 0.224 &416 $\pm$ 4&	0.06 $\pm$ 0.05&	0.082 $\pm$ 0.019 \\
Abell 2390 & 0.228 & 343 $\pm$ 2&	0.15 $\pm$ 0.06&	0.046 $\pm$ 0.008 \\
Abell 267 & 0.231 & 321 $\pm$ 8&	0.05 $\pm$ 0.06&	0.047 $\pm$ 0.025 \\
Abell 1835 & 0.253 & 289 $\pm$ 2&	0.14 $\pm$ 0.07&	0.216 $\pm$ 0.067 \\
Abell 68 & 0.255 & 324 $\pm$ 25&	0.06 $\pm$ 0.06&	--0.048 $\pm$ 0.022 \\
MS 1455+22 & 0.258 & 391 $\pm$ 15&	0.11 $\pm$ 0.05&	--0.056 $\pm$ 0.040 \\
Abell 611 & 0.288 & 310 $\pm$ 6&	0.06 $\pm$ 0.06&	0.163 $\pm$ 0.072 \\
Abell 2537 & 0.295 & 313 $\pm$ 10&	0.08 $\pm$ 0.06&	0.122 $\pm$ 0.064 \\
\hline
\end{tabular}
\end{table}

\begin{table}
\centering
\caption{Derived kinematic properties for the BGGs: central velocity dispersion $\sigma_{0}$, rotation $V_{\rm max}/\sigma_{0}$, and velocity dispersion slope ($\eta \pm \Delta \eta$).}
\label{propertiesBGGs}
\begin{tabular}{c c l r r}
\hline
Name & $z$ & $\sigma_{0}$ & $V_{\rm max}/\sigma_{0}$ & $\eta \pm \Delta \eta$ \\
&  & (km s$^{-1}$) &  &  \\                          
\hline
\multicolumn{5}{c}{High-richness Groups}\\
\hline
NGC5846 & 0.0057 & 195 $\pm$ 4 & 0.38 $\pm$ 0.05 &--0.095 $\pm$ 0.023\\ 
NGC0584 & 0.0060 &  220 $\pm$ 6&0.85 $\pm$ 0.05&--0.120 $\pm$ 0.010\\
NGC4261 & 0.0073 & 314 $\pm$ 4&0.11 $\pm$ 0.05	&--0.057 $\pm$ 0.007\\ 
NGC5353 & 0.0073 & 300 $\pm$ 4&0.85 $\pm$ 0.02		&--0.173 $\pm$ 0.011\\ 
NGC5982 & 0.0095 & 260 $\pm$ 7&0.22 $\pm$ 0.05&--0.100 $\pm$ 0.019\\ 
NGC1587 & 0.0120 & 239 $\pm$ 6&0.63 $\pm$ 0.03&--0.054 $\pm$ 0.011\\ 
NGC6658 & 0.0124 & 210 $\pm$ 9&0.62 $\pm$ 0.10&--0.076 $\pm$ 0.030 \\ 
NGC7619 & 0.0125 & 331 $\pm$ 5&0.06 $\pm$ 0.06&--0.078 $\pm$ 0.011 \\ 
NGC1453 & 0.0128 & 312 $\pm$ 6&0.37 $\pm$ 0.02	&--0.068 $\pm$ 0.011 \\ 
NGC0924 & 0.0147 &  201 $\pm$ 8&1.02 $\pm$ 0.04&--0.173 $\pm$ 0.021\\
NGC2563 & 0.0147 &  287 $\pm$ 6&0.44 $\pm$ 0.02	&--0.091 $\pm$ 0.017\\ 
NGC0777 & 0.0162 &   321 $\pm$ 9&0.13 $\pm$ 0.02&--0.078 $\pm$ 0.010\\
NGC1060 & 0.0167 & 326 $\pm$ 6&0.06 $\pm$ 0.04	&--0.067 $\pm$ 0.010\\
NGC0410 & 0.0172 & 311 $\pm$ 6&0.09 $\pm$ 0.03&--0.068 $\pm$ 0.009\\
\hline
\multicolumn{5}{c}{Low-richness Groups}\\
\hline
NGC2768 & 0.0043 & 172 $\pm$ 3&0.61 $\pm$ 0.07&0.062 $\pm$ 0.012\\ 
NGC3665 & 0.0066 &224 $\pm$ 5&0.67 $\pm$ 0.02&--0.028 $\pm$ 0.007 \\ 
NGC3613 & 0.0068 & 214 $\pm$ 4&0.65 $\pm$ 0.05&--0.026 $\pm$ 0.010 \\ 
NGC0524 & 0.0078 & 245 $\pm$ 5&0.51 $\pm$ 0.03&--0.078 $\pm$ 0.014\\ 
NGC1779 & 0.0108 & 164 $\pm$ 10&1.16 $\pm$ 0.05&--0.093 $\pm$ 0.022\\ 
NGC5629 & 0.0147 & 257 $\pm$ 8&0.31 $\pm$ 0.04&--0.055 $\pm$ 0.026 \\ 
NGC5127 & 0.0160 & 194 $\pm$ 5&0.05 $\pm$ 0.05&--0.051 $\pm$ 0.015\\ 
NGC5490 & 0.0160 &  353 $\pm$ 6&0.25 $\pm$ 0.02&--0.119 $\pm$ 0.018\\ 
NGC0315 & 0.0162 & 339 $\pm$ 8	&0.22 $\pm$ 0.04	&--0.028 $\pm$ 0.011\\ 
\hline
\end{tabular}
\end{table}	

\subsection{Velocity dispersion profiles: correlations with other properties}
\label{variety}

\begin{figure}
   \centering
   \includegraphics[scale=0.42]{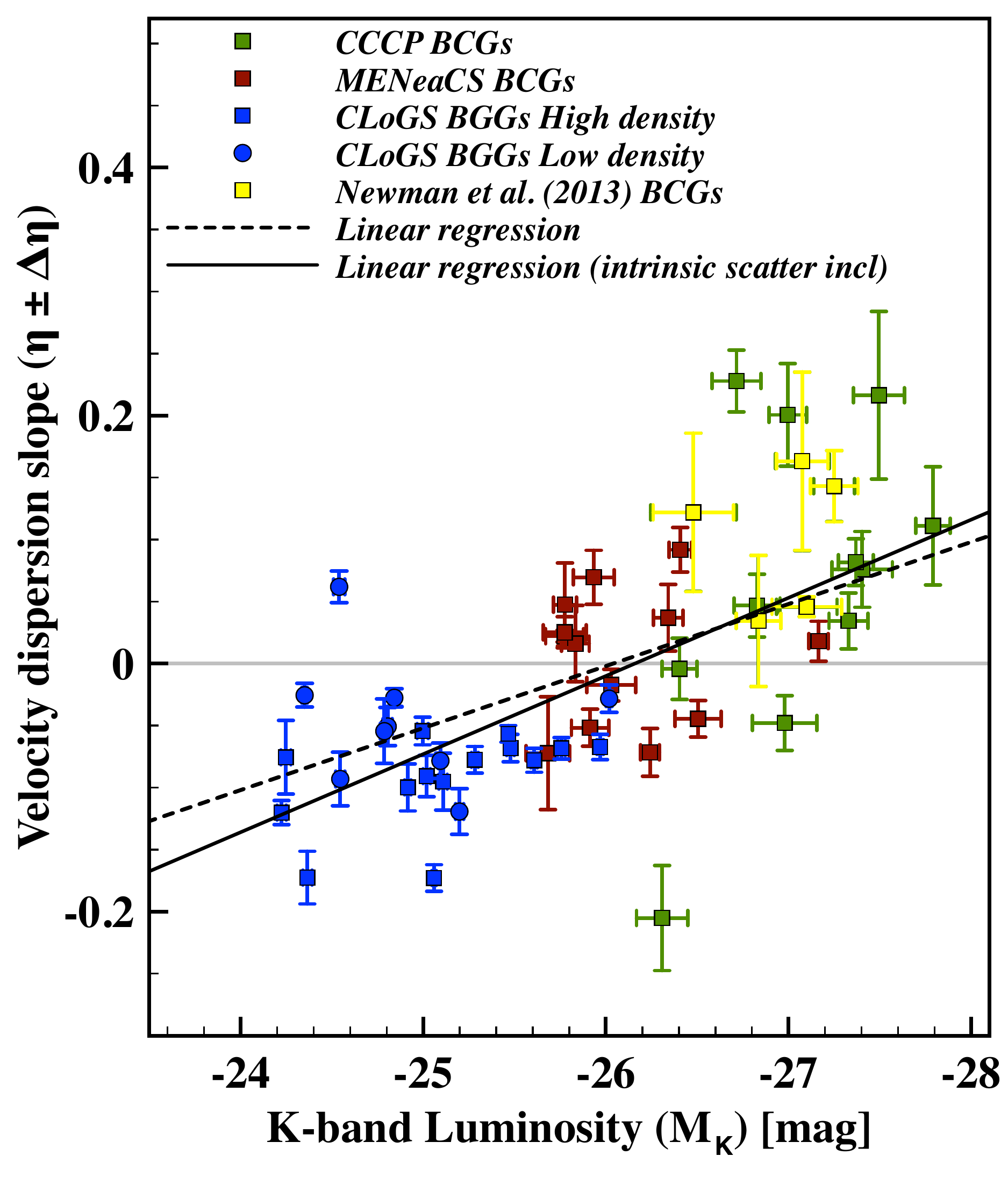}\\
   \caption[]{$K$-band luminosity vs velocity dispersion slope ($\eta \pm \Delta \eta$). The yellow points are the five galaxies in common with \citet{Newman2013a}. The dashed line indicates the best fit to the data points where intrinsic scatter was not taken into account in the linear regression. Similarly, the solid line indicates the best fit to the points where intrinsic scatter was taken into account in the linear regression Section \ref{variety}.}
   \label{fig:variety}
\end{figure}

We plot the velocity dispersion slopes against the $K$-band luminosity of all the central galaxies (BCGs and BGGs) in Figure \ref{fig:variety}. These two parameters form a linear correlation with slope = --0.050 $\pm$ 0.002 (indicated by the dashed line, and with a zero point  = --1.302 $\pm$ 0.064). As mentioned above, there are four BCGs where a single power law fit may not be the most accurate description. When these four BCGs are removed, the linear regression yields: slope = --0.050 $\pm$ 0.002 (zero point = --1.290 $\pm$ 0.064), thus negligibly different from when all 52 BCGs/BGGs with $K$-band luminosities are used. 

In addition, we fit another linear regression to the velocity dispersion slope--luminosity correlation, but taking intrinsic scatter into account. We assume the intrinsic (random) scatter to be normally distributed, and we use the Gibbs sampler implemented in the multivariate Gaussian mixture model routine \texttt{linmix$\_$err} \citep{Kelly2007} with the default of three Gaussians. We use 5000 random draws of the sampler and take the fitted parameters as the posterior mode and the error as the 68 per cent highest posterior density credible interval. We find a slope = --0.063 $\pm$ 0.010, with an intrinsic scatter of 0.067 ($\sigma_{\rm intrinsic}$), and correlation coefficient 0.70 (indicated by the solid line, with a zero point = --1.648 $\pm$ 0.270). 

We are interested in the remarkable diversity of the velocity dispersion profiles, and in particular whether the positive or negative slopes of the velocity dispersion profiles correlate with any of the other derived properties of the BCG, or with those of the host cluster. From the above correlation between velocity dispersion slope and $K$-band luminosity (taking intrinsic scatter into account, \ie\ the solid line in Figure \ref{fig:variety}), we also calculate the residuals and plot it against central velocity dispersion, group/cluster velocity dispersion, ellipticity ($\epsilon$), and M$_{500}$ in Figure \ref{sigma_residuals} in the appendix -- and find no correlations. 

We have also investigated whether there are possible biases in the BCG velocity dispersion slope measurements because of sub-structure, \eg\ multiple nuclei, objects in the line-of-sight or possible misalignment of the slit with the major axis. We show the $r$-band imaging with fitted isophotes for all 32 BCGs in Paper II, and use it to model stellar masses. The BCGs in Abell 780, Abell 990 and Abell 1835 have objects in the line-of-sight that affects the last two velocity dispersion measurements (furthest from the centre). From Figures \ref{fig:kin1} to \ref{fig:kin11} it follows that the last two measurements do not significantly affect the velocity dispersion slope in these three cases. Abell 586, MS 0440+02 and MS 0906+11 has substructure (multiple nuclei) in the centre, and from the plots in Figures \ref{fig:kin1} to \ref{fig:kin11} it can be seen that the velocity dispersion slope of MS 0440+02 may be affected in the central bins, possibly responsible for the non-uniform velocity dispersion profile as described in Section \ref{Velocityprofiles}. However, as shown in the first paragraph of this Section, if this galaxy is removed (along with the three other galaxies where the velocity dispersion profile shape changes in the centre), there is no significant affect on the correlation with luminosity. We do not find any significant misalignment between the placement of the slit and the major axis that could have influenced the measurements. 

\subsection{Implications of the variety in the velocity dispersion profiles, and exceptions}
\label{Discussion}

A rising velocity dispersion profile can be interpreted as evidence for an increasing mass contribution from the dark matter halo, and therefore an increasing dynamical M/L. However, the well-known degeneracy between mass and velocity anisotropy \citep{Binney1982} complicates the interpretation. This degeneracy implies that a low velocity dispersion can be a low enclosed mass, or a radial velocity anisotropy. As a result, a falling velocity dispersion profile can not be interpreted as an absence of a massive dark matter halo, without further information on the velocity anisotropy. If the orbital distribution of the galaxy is unknown, then the detailed shape of the line-of-sight velocity distribution must be measured (the deviation from an exact Gaussian which is produced by an isotropic population that generates an isothermal potential). Dynamical models can be used to disentangle the effects of orbital anisotropy and gravitational potential gradient if the higher order velocity moments (Gauss-Hermite $h_{3}$ and $h_{4}$) are incorporated \citep{Gerhard1993, Merritt1993, Gerhard1998}. Similarly to other samples of the most massive early-type galaxies \citep{Newman2013a, Veale2017}, we likely have anisotropic profiles and variations from isothermal profiles present in the BCGs, and therefore full dynamical modelling is needed. We present the measurements of $h_{3}$ and $h_{4}$, and the dynamical modelling of the BCGs in Paper II.

From Figures~\ref{fig:Sigma} (yellow points) and \ref{fig:variety} (yellow points) it can be seen that the seven BCGs with velocity dispersion profiles presented in \citet{Newman2013a} form a very homogeneous BCG sample. \citet{Laporte2015} use N-body re-simulations and compare their projected line-of-sight velocity dispersion profiles of the simulated BCGs to observations from \citet{Newman2013a} (figure 8 in \citealt{Laporte2015}). Their simulations predict a similar rise in velocity dispersion profiles from $\sim$300 km s$^{-1}$ in the centre to $\sim$400 km s$^{-1}$ in the outskirts. The authors achieve this by assuming that at late times the assembly of the inner regions of clusters is entirely dominated by collisionless merger processes. Similarly, \citet{Schaller2015} also match their simulations to the observations from \citet{Newman2013a} (figure 7 in \citealt{Schaller2015}). These simulations do not produce any decreasing velocity dispersion profiles for their BCGs, and it is not clear whether we will see the same diversity in slopes reproduced in simulations, if the simulations are for host clusters with a similar range of halo masses than those presented here.

Furthermore, the wide variety, and large fraction of positive velocity dispersion profiles for BCGs have implications for power law velocity dispersion aperture correction schemes (\eg\ \citealt{Jorgensen1995}), that assumes a higher velocity dispersion in the centre of the galaxy than in the outer regions.

From Figures \ref{fig:Sigma}, \ref{fig:Sigmaslope}, and \ref{fig:variety}, it can be seen that one BCG is a clear exception in the BCG sample. The BCG in Abell 2104 has been unambiguously identified as the cD galaxy 2MASX J15400795-0318162 \citep{Crawford1999, Liang2000, Bildfell2008, Hoffer2012}. However, we find that the BCG is very unusual in that it has a very negative velocity dispersion slope of --0.2 (see Figure \ref{fig:kin8}), similar to a typical isolated early-type galaxy. \citet{Liang2000} construct a galaxy velocity distribution for $\sim$47 cluster galaxies within 3000 km s$^{-1}$ of this central galaxy and show that this galaxy is at rest at the bottom of the cluster potential. Their X-ray imaging shows significant substructure in the centre of the cluster and an overall elliptical appearance, and it appears that the cluster has not yet reached dynamical equilibrium. Neither the cluster nor the galaxy 2MASX J15400795-0318162 seem unusual in any other property, with the exception that \citet{Martini2002} presented deep \textit{Chandra} observations that revealed a significant X-ray point-source excess over the expectations of blank fields. Their spectroscopy show that all six X-ray sources associated with red counterparts are cluster members and their X-ray properties are consistent with all of them being AGNs. The presence of these AGNs indicate that supermassive black holes have somehow retained a fuel source.

A further exception, although less unusual, can be seen in the velocity dispersion profile of BGG NGC2768, which shows a clear positive slope in contrast to the rest of our brightest group galaxies (the only BGG above the y = 0 line in Figures \ref{fig:Sigma}, \ref{fig:Sigmaslope}, \ref{fig:variety}). Our kinematic results confirm the SAURON kinematic results for this galaxy, which also show strong rotation and a lower central velocity dispersion \citep{McDermid2006}. As shown in \citet{Veale2017} (for their MASSIVE sample), brightest group galaxies with positive slopes do exist, and NGC 2768 is not that unusual, just an outlier in our sample which consists of generally less-massive brightest group galaxies than the MASSIVE sample. It is thus not inconceivable to find other massive BGGs with steeper positive velocity dispersion slopes, whereas the steep negative velocity dispersion slope of the BCG in Abel 2104 is a truly intriguing exception.

\section{Conclusions}
\label{conclusions}

The stellar velocity dispersion profile of a galaxy is a standard indication of the gravitational potential of a galaxy, and is often used as a proxy for a galaxy's dynamical mass. Accurate measurements of velocity dispersion profiles of early-type galaxies is, therefore, a key step towards estimating their dark matter content, necessary to ultimately constrain galaxy formation and evolution models. 

In this paper, we investigate and quantify the intrinsic variety in the (often rising) velocity dispersion profiles of BCGs (0.05 $\leq z \leq$ 0.30). We use optical spectroscopy of 32 MENeaCS and CCCP BCGs, with the advantage that the host clusters themselves are well-characterised \eg\ carefully measured halo masses, etc. (as studied in \citealt{Bildfell2008, Sand2011, Sand2012, Mahdavi2013, Bildfell_thesis, Hoekstra2015, Sifon2015, Loubser2016}, and others). Our 32 BCGs span M$_{K}$ = --25.7 to --27.8 mag, with host cluster halo masses M$_{500}$ up to 1.7 $\times$ 10$^{15}$ M$_{\sun}$. For comparison, we also analyse similar spectra for 23 brightest group members, thereby extending our M$_{K}$ range to a lower limit of --24.2 mag. This sample therefore probes a larger central galaxy luminosity range (and thus also host cluster/group halo mass range) compared to the complimentary, detailed analysis in \citet{Newman2013a, Veale2017}. This now allows us to probe possible correlations between the velocity dispersion profile slopes of the BCGs/BGGs and other properties of the galaxies, and those of their host clusters/groups. We summarise our main findings below:

\begin{enumerate}
\item Whilst some BCGs show marginal rotation, none of them are supported by rotation. The plot of the anisotropy parameter ($V_{\rm max}/\sigma_{0}$) vs. M$_{K}$ luminosity in Figure \ref{fig:Rotation}, coloured by halo mass M$_{500}$, is qualitatively consistent with the recent findings by \citet{Veale2017, OlivaAltamirano2017}, who found that the fraction of slow- and non-rotators increase as a function of luminosity and cluster mass, respectively. 

\item From Figure \ref{fig:Sigma}, it is clear that the velocity dispersion slopes of the BCGs show a much larger variety, with a significantly larger fraction of positive slopes, compared to other (non-central) early-type galaxies as well as the brightest members of the CLoGS groups. A rising velocity dispersion profile can be interpreted as evidence for an increasing dynamical M/L, but the well-known degeneracy between mass and velocity anisotropy \citep{Binney1982} complicates the interpretation, and a falling velocity dispersion profile does not necessarily imply the absence of a massive dark matter halo. Similarly to other samples of the most massive early-type galaxies \citep{Newman2013a, Veale2017}, we likely have anisotropic profiles and variations from isothermal profiles present in the BCGs, and therefore full dynamical modelling is needed.

\item $K$-band luminosity vs. velocity dispersion slopes for BCGs and BGGs show a tight correlation. The residuals of this correlation do not correlate with any other properties, \eg\ central velocity dispersion, group/cluster velocity dispersion, ellipticity of the BCG, M$_{500}$ (or R$_{500}$). 

\item We see a general trend of increasing velocity dispersion slope with increasing group/cluster velocity dispersion in Figure \ref{fig:Sigmaslope}, albeit with large scatter. We find similar correlations for M$_{500}$ and R$_{500}$.

\item From Figures~\ref{fig:Sigma} (yellow points) and \ref{fig:variety} (cyan points) it can be seen that the seven BCGs with velocity dispersion profiles presented in \citet{Newman2013a} form a very homogeneous BCG sample, and in Section \ref{Discussion} we discuss how simulations by \eg\ \citet{Laporte2015, Schaller2015} match these observed profiles. It is not clear whether the same diversity in slopes present in our sample would be reproduced in the simulations, if the simulations are for host clusters with a similar range of halo masses than those presented here.

\item The wide variety, and large fraction of positive velocity dispersion profiles for BCGs have implications for power law velocity dispersion aperture correction schemes (\eg\ \citealt{Jorgensen1995}), that assume a higher velocity dispersion in the centre of the galaxy than in the outer regions.

\item Lastly, we recover the Faber-Jackson relation as well as the host cluster halo mass, M$_{500}$ vs. BCG/BGG $K$-band luminosity relation in Section \ref{scaling}. 

\end{enumerate}

This sample has well-characterized gravitational potentials from lensing analysis, and in the follow-up paper (Paper II), we extend the characterisation of these gravitational potentials inside the radius constrained by lensing using the velocity dispersion of the stars presented here, together with the surface brightness profiles and stellar population analysis. 

\section*{Acknowledgements}

We sincerely thank the anonymous reviewer for the constructive and thoughtful comments which were of great help in revising the manuscript. This research was enabled, in part, by support provided by the bilateral funding agreement between the National Research Foundation (NRF) of South Africa, and the Netherlands Organisation for Scientific Research (NWO) to HH and SIL. SIL is aided by a Henri Chr\'etien International Research Grant administered by the American Astronomical Society. AB acknowledges support from NSERC (Canada) through the Discovery Grant program and to the Pauli Center for Theoretical Studies ETH UZH. He would also like to thank University of Zurich's Institute for Computational Sciences, and especially the members of the Institute's Center for Theoretical Astrophysics and Cosmology, for their hospitality during his recent extended visit. EOS acknowledges support from the National Aeronautics and Space Administration (NASA) through Chandra Awards GO6-17121X and GO6-17122X, issued by the Chandra X-ray Observatory Center, which is operated by the Smithsonian Astrophysical Observatory on behalf of NASA under contract NAS8-03060. We thank Ricardo Herbonnet for providing the CCCP and MENeaCS M$_{500}$ and R$_{500}$ values prior to publication, David Gilbank for useful discussions, and Ando Ratsimbazafy for help with Gemini data reduction.

Based, in part, on observations obtained at the Gemini Observatory, which is operated by the Association of Universities for Research in Astronomy, Inc., under a cooperative agreement with the NSF on behalf of the Gemini partnership: the National Science Foundation (United States), the National Research Council (Canada), CONICYT (Chile), Ministerio de Ciencia, Tecnolog\'{i}a e Innovaci\'{o}n Productiva (Argentina), and Minist\'{e}rio da Ci\^{e}ncia, Tecnologia e Inova\c{c}\~{a}o (Brazil). Based, in part, on observations obtained at the Canada-France-Hawaii Telescope (CFHT) which is operated by the National Research Council of Canada, the Institut National des Sciences de l'Univers of the Centre National de la Recherche Scientifique of France, and the University of Hawaii. This research used the facilities of the Canadian Astronomy Data Centre operated by the National Research Council of Canada with support from the Canadian Space Agency. 

Any opinion, finding and conclusion or recommendation expressed in this material is that of the author(s) and the NRF does not accept any liability in this regard.

%%%%%%%%%%%%%%%%%%%%%%%%%%%%%%%%%%%%%%%%%%%%%
%%%%%%%%%%%%%%%%%%%% REFERENCES %%%%%%%%%%%%%%%%%

\bibliographystyle{mnras}
\bibliography{References} 

%%%%%%%%%%%%%%%%%%%%%%%%%%%%%%%%%%%%%%%%%%%%%%
%%%%%%%%%%%%%%%%% APPENDICES %%%%%%%%%%%%%%%%%%%%%%

\appendix

\section{Comparison of velocity dispersion profiles to previous measurements}
\label{kinematics}

All the spatially-resolved velocity and velocity dispersion profiles of the MENeaCS and CCCP BCGs are presented in Figures \ref{fig:kin1} to \ref{fig:kin11} with spatial radii indicated in both arcsec and kpc. The BCGs in Abell 383, Abell 611, Abell 963, Abell 2390 and Abell 2537 overlap with the sample studied by \citet{Newman2013a}, and their velocity dispersion measurements are indicated with grey circles in the relevant figures. Some of the \citet{Newman2013a} observations were done in multi-object slit mode, and aligned close to the major axis of the BCG, although with some minor deviations tolerated to include gravitational arcs. Both sides of the galaxy velocity dispersion profiles were averaged together in their measurements. We also compare the velocity dispersion profile of the BCG in Abell 2029 (IC 1101) with that by \citet{Fisher1995}, and also indicate it in Figure \ref{fig:kin6} with grey diamonds. Furthermore, we compare the velocity dispersion measurements of the BCGs in Abell 780 and Abell 2029 with those made using different observations and an independent method in \citet{Loubser2008}, and these are indicated with grey squares in Figures \ref{fig:kin4} and \ref{fig:kin6}. The profiles generally agree very well with previous measurements, given the different instruments and setups (\eg\ position angle), spatial binning and measurements methods.

We also compare our central velocity dispersion measurements for some of our CLoGS BGGs (as described in Section \ref{binning}) with the seven galaxies in common with ATLAS$^{\rm 3D}$ \citep{Cappellari2013} and the four in common with MASSIVE \citep{Veale2017}, and we find that the average deviation is 17 km s$^{-1}$, and in general a remarkable agreement given the different observations (long-slit vs IFU) and central apertures. 

\section{Residuals from the $\sigma$-slope -- M$_{K}$ relation}

\begin{figure*}
   \centering
   \subfloat{\includegraphics[scale=0.34]{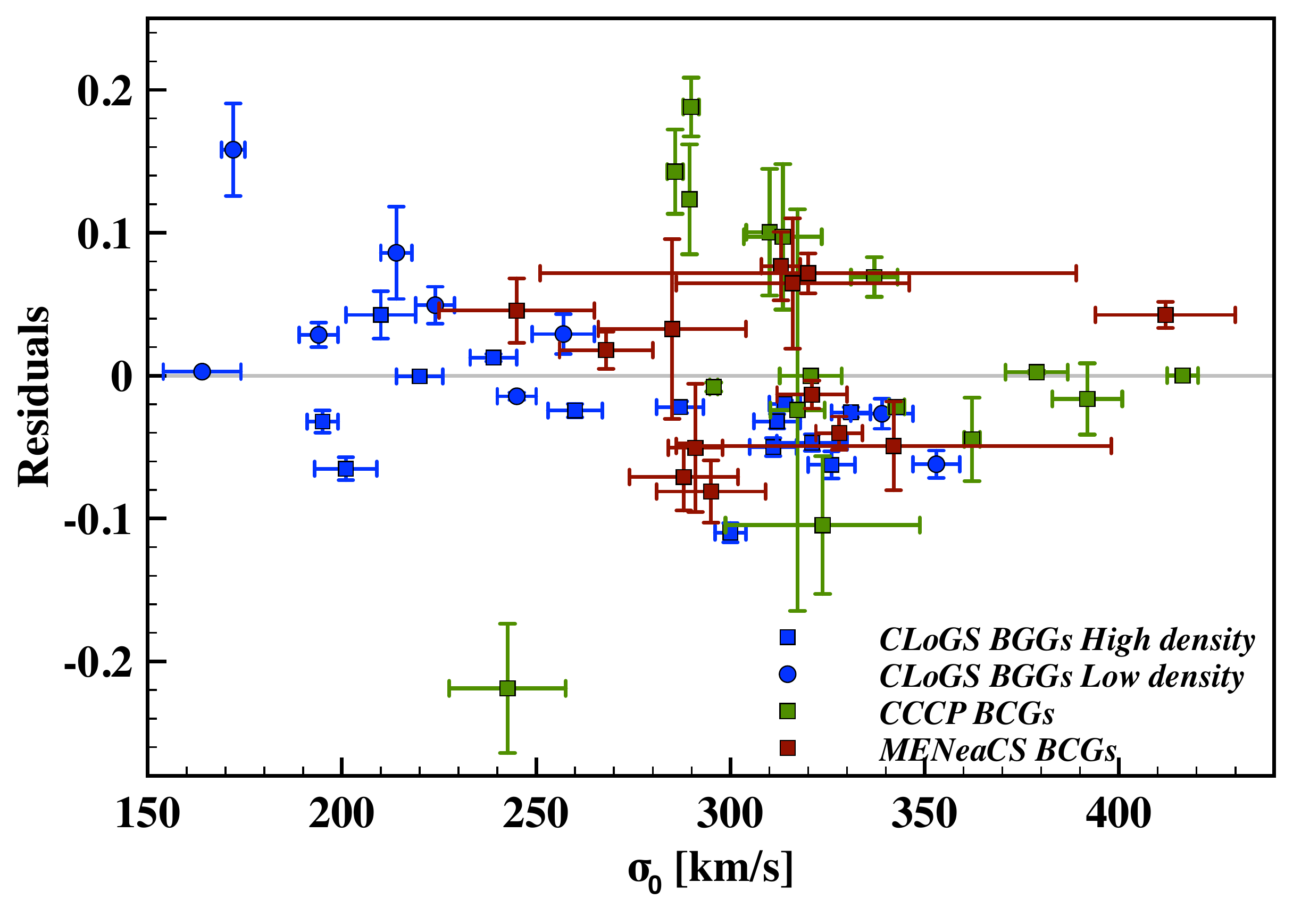}}
   \subfloat{\includegraphics[scale=0.34]{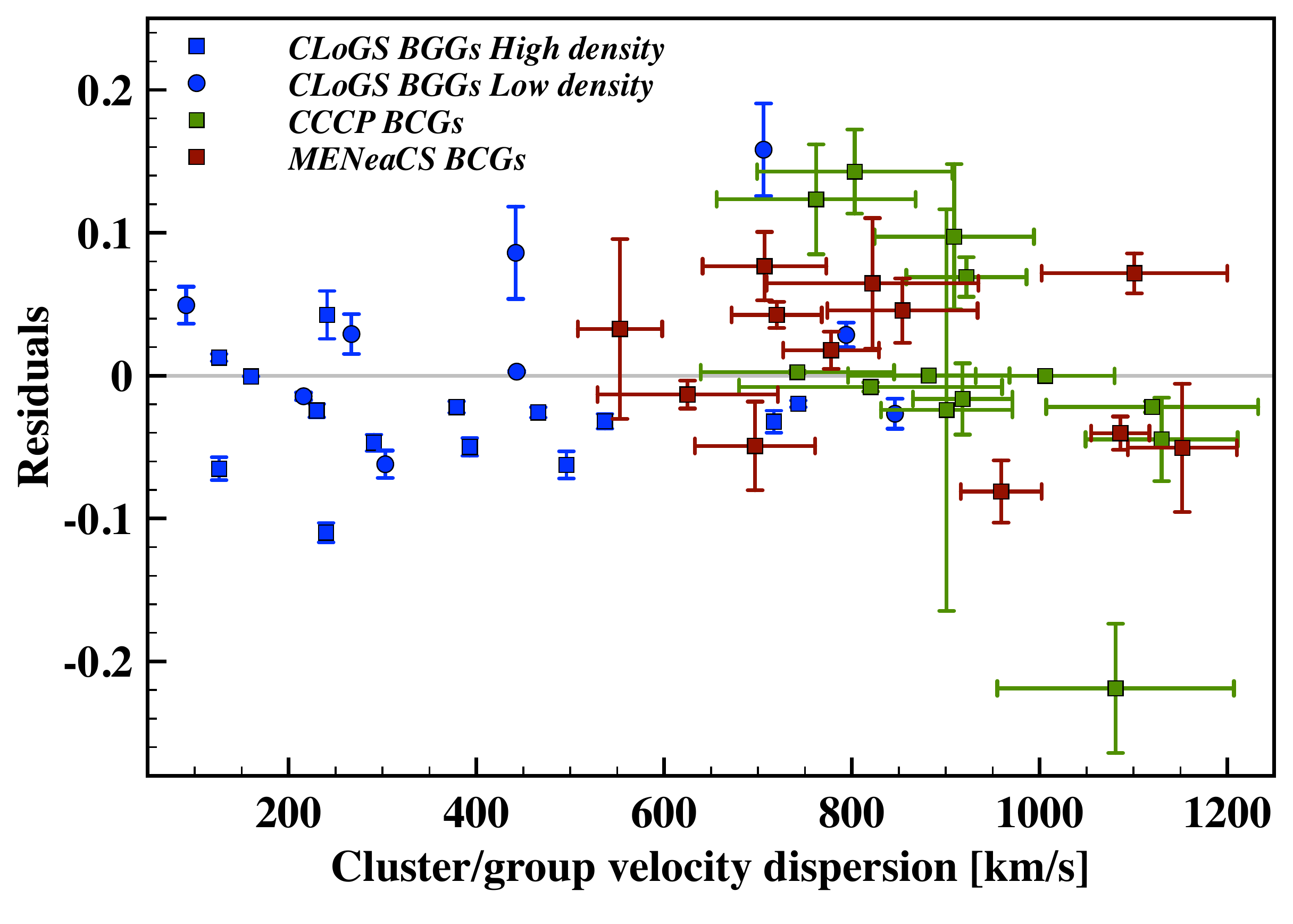}}\\
   \subfloat{\includegraphics[scale=0.34]{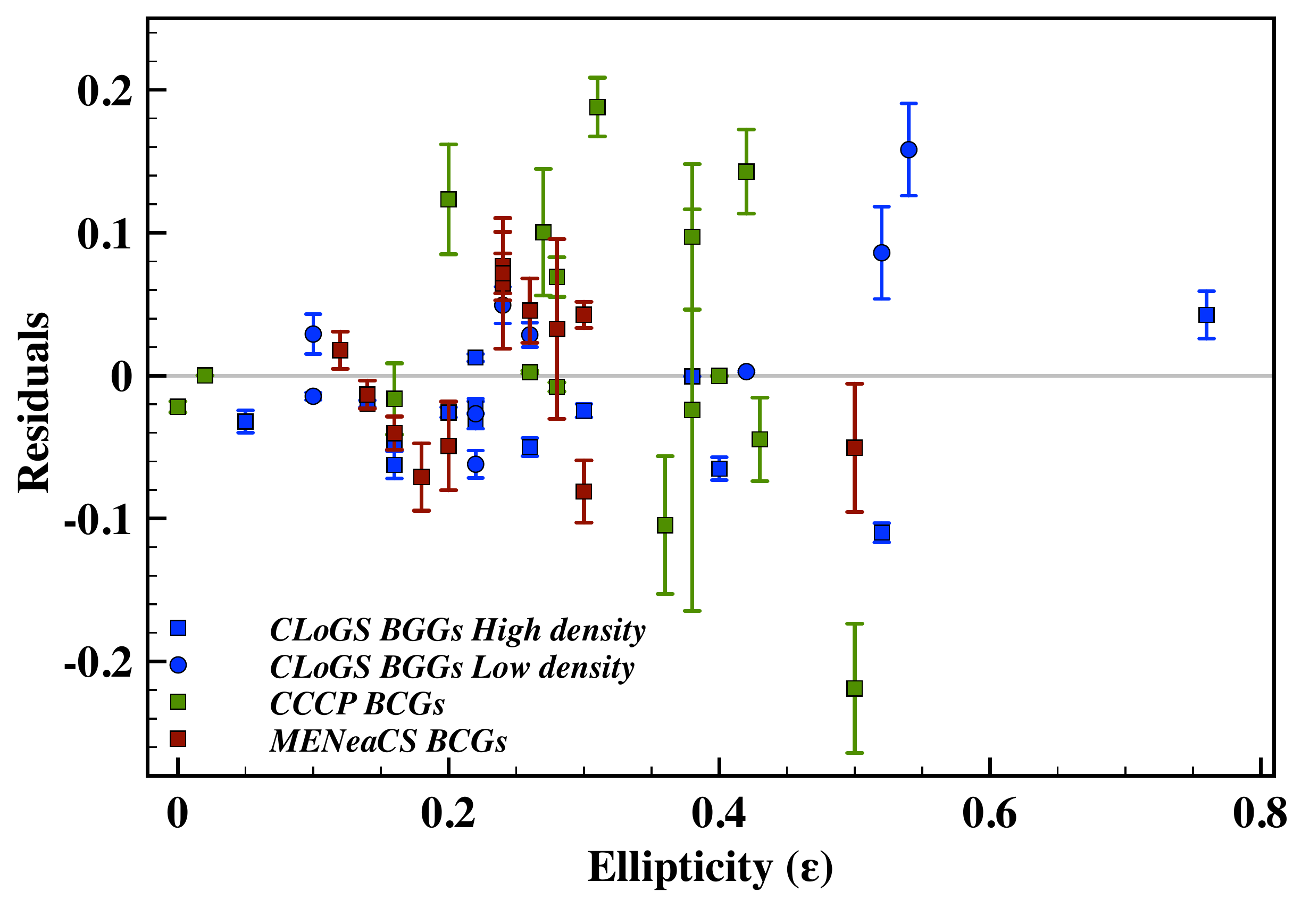}}
   \subfloat{\includegraphics[scale=0.34]{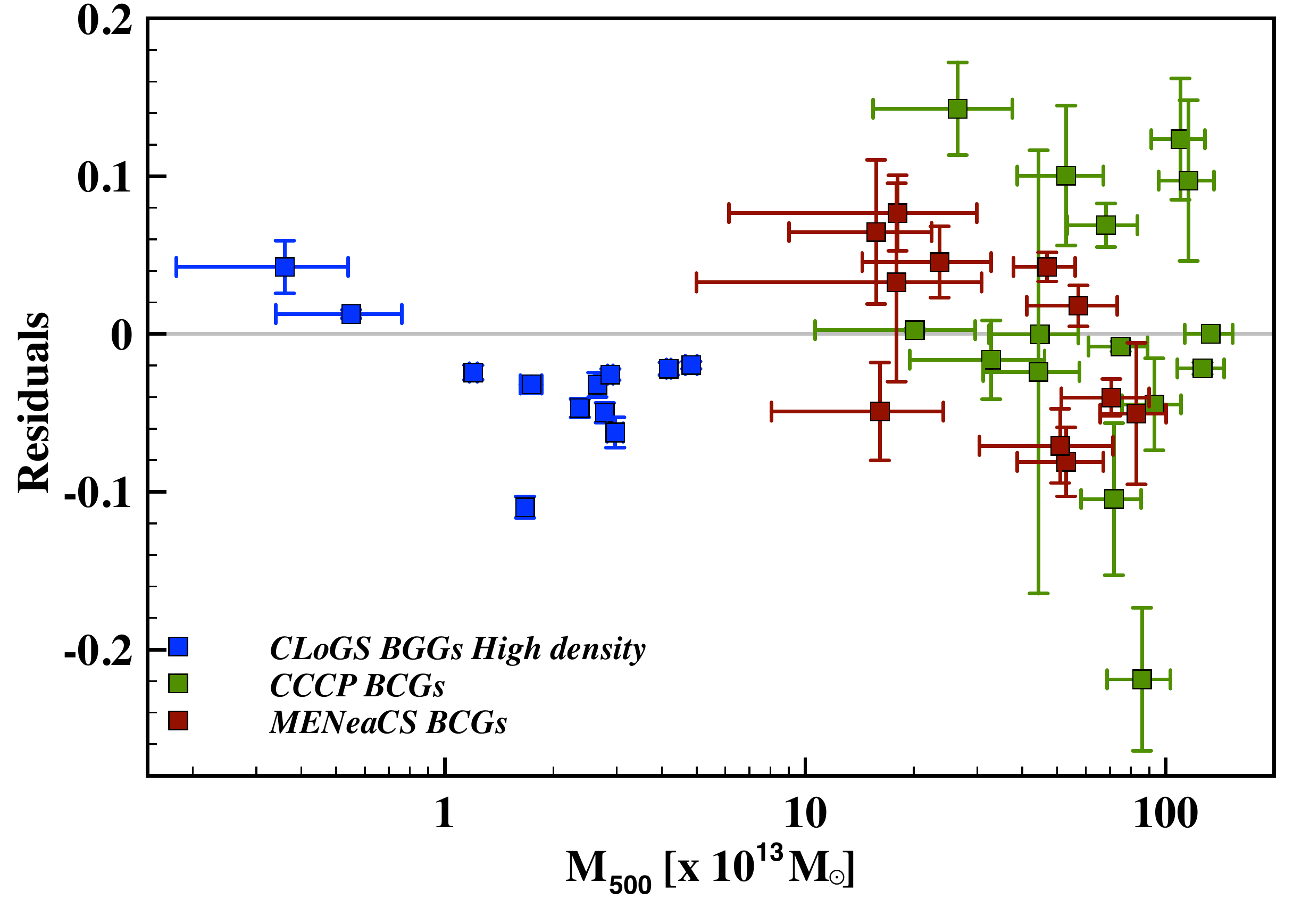}}\\
   \caption[]{Velocity dispersion slope residuals (from correlation with $K$-band luminosity) against central velocity dispersion, group/cluster velocity dispersion, ellipticity, and M$_{500}$ (in units of $10^{13}$ M$_{\sun}$).}
   \label{sigma_residuals}
\end{figure*}

We are interested in the diversity of the velocity dispersion profiles, and in particular whether the positive or negative slopes of the velocity dispersion profiles correlate with any of the other derived properties, or with those of the host clusters. From the correlation between velocity dispersion slope and $K$-band luminosity in Section \ref{variety} (the solid line, where intrinsic scatter was taken into account in Figure \ref{fig:variety}), we calculate the residuals in the velocity dispersion slope and plot these against central velocity dispersion, group/cluster velocity dispersion, ellipticity, and M$_{500}$ in Figure \ref{sigma_residuals}, and do not find any other significant correlations for the BCGs.

Inspecting the top panel in Figure \ref{sigma_residuals}, there appears to be a correlation between the residuals from the M$_{K}$ vs. $\eta \pm \Delta \eta$ correlation and $\sigma_{0}$ for the BGGs (blue data points). We fit the M$_{K}$ vs. $\eta \pm \Delta \eta$ correlation for BCGs and BGGs as if they are one continuous population, as the range in M$_{K}$ is too small to reliably fit two separate populations. It is likely that the overall correlation is driven by the BCGs, and that the correlation should have a slightly flatter slope at the low mass (BGG) end, which gives rise to the correlation in the residuals for the BGG data points.

\section{Second order velocity moment}
\label{second_moment}

\begin{figure}
   \centering
    \subfloat{\includegraphics[scale=0.34]{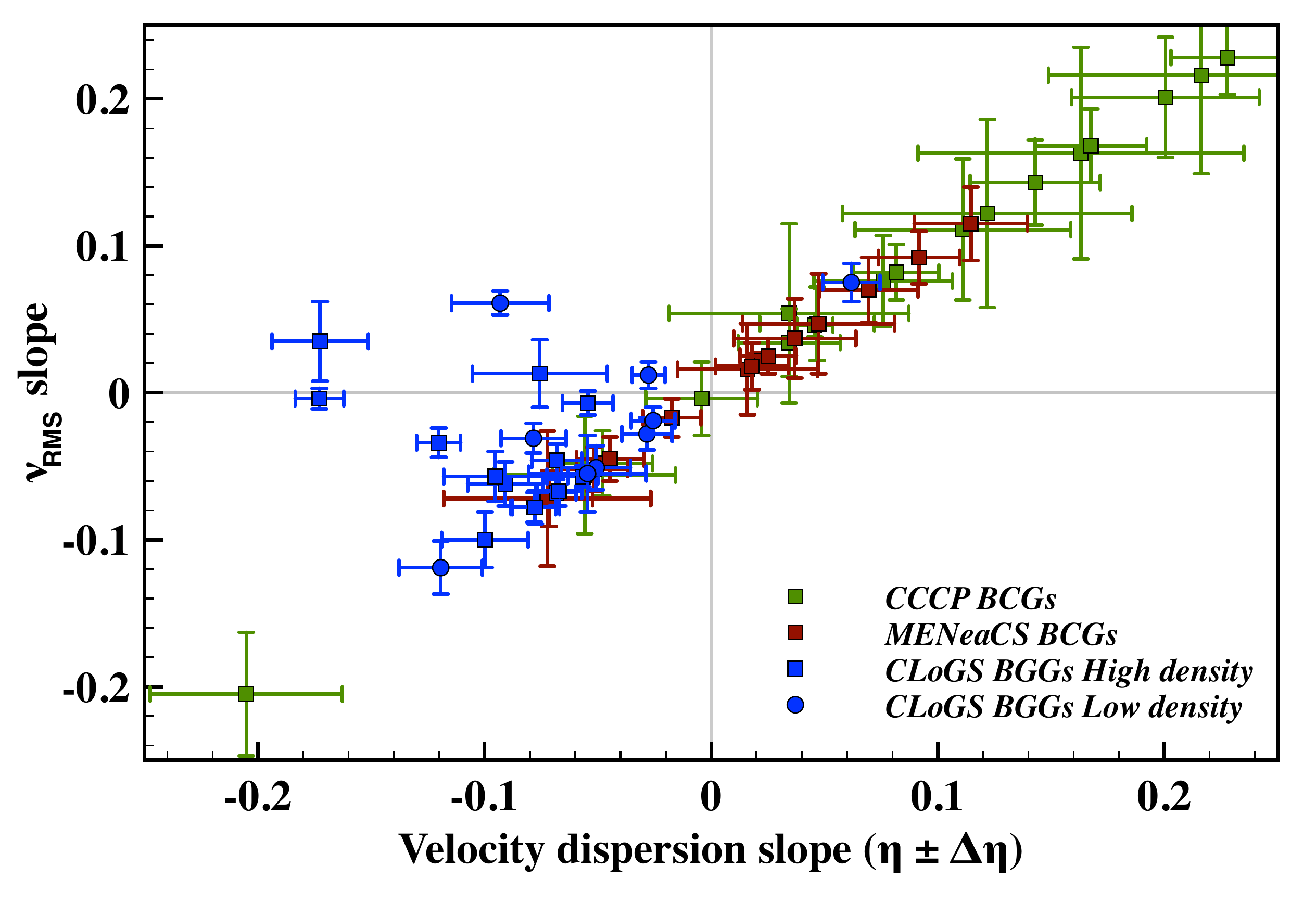}}\\
   \caption[]{Velocity dispersion slopes against second order velocity moment slopes.}
   \label{sigma_second}
\end{figure}

We compare our velocity dispersion slope measurements against similar measurements of the second moment of velocity, $v_{\rm rms} = \sqrt{V^{2} + \sigma^{2}}$, in Figure \ref{sigma_second}. For the BCGs, all of which are non/slow-rotators, velocity dispersion is a good approximation to the second order velocity moment. 

\section{Individual galaxies -- plots}
\label{plots}

\begin{figure*}
\captionsetup[subfigure]{labelformat=empty}
   \subfloat{\includegraphics[scale=0.25]{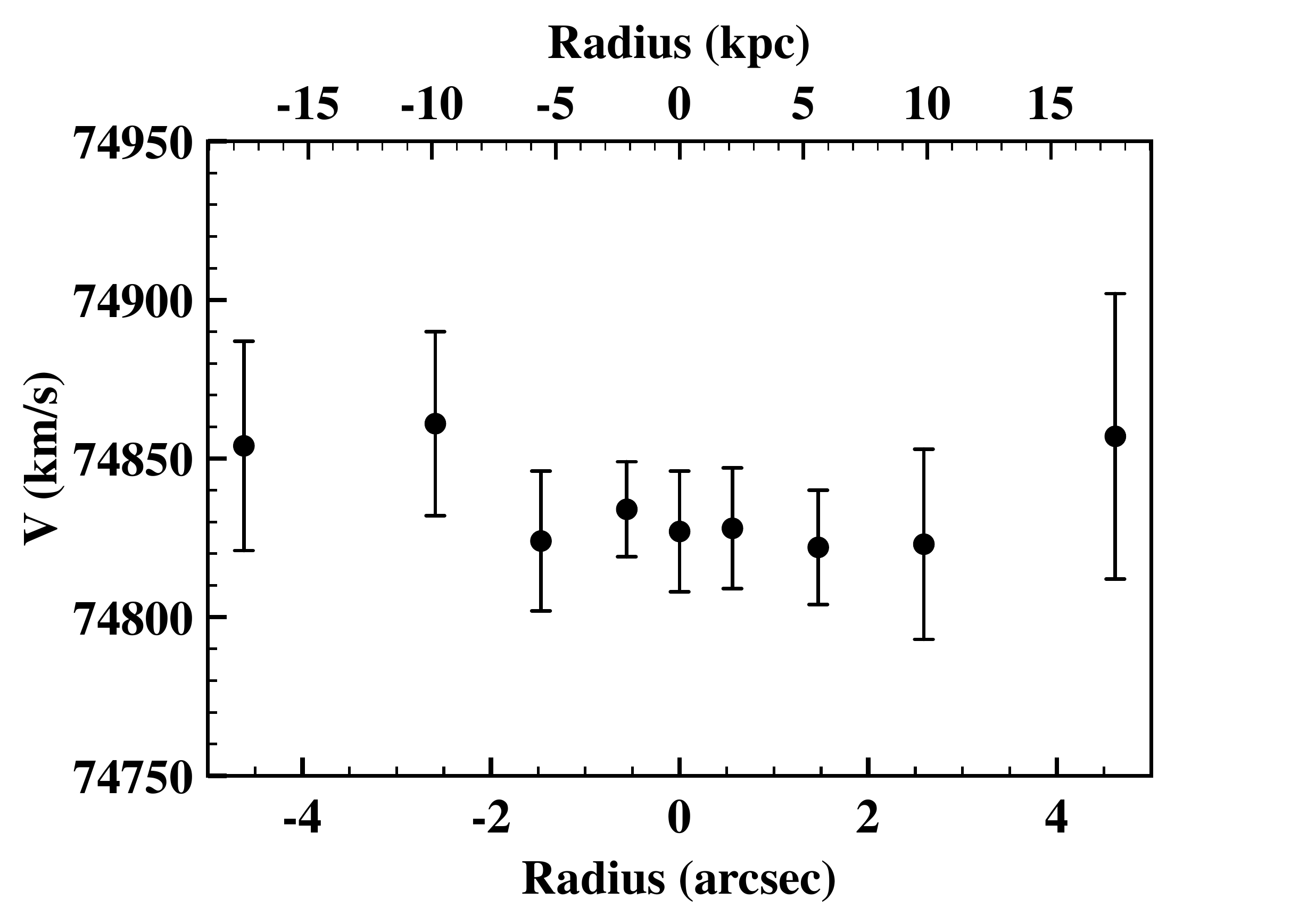}}
         \subfloat[Abell 68]{\includegraphics[scale=0.25]{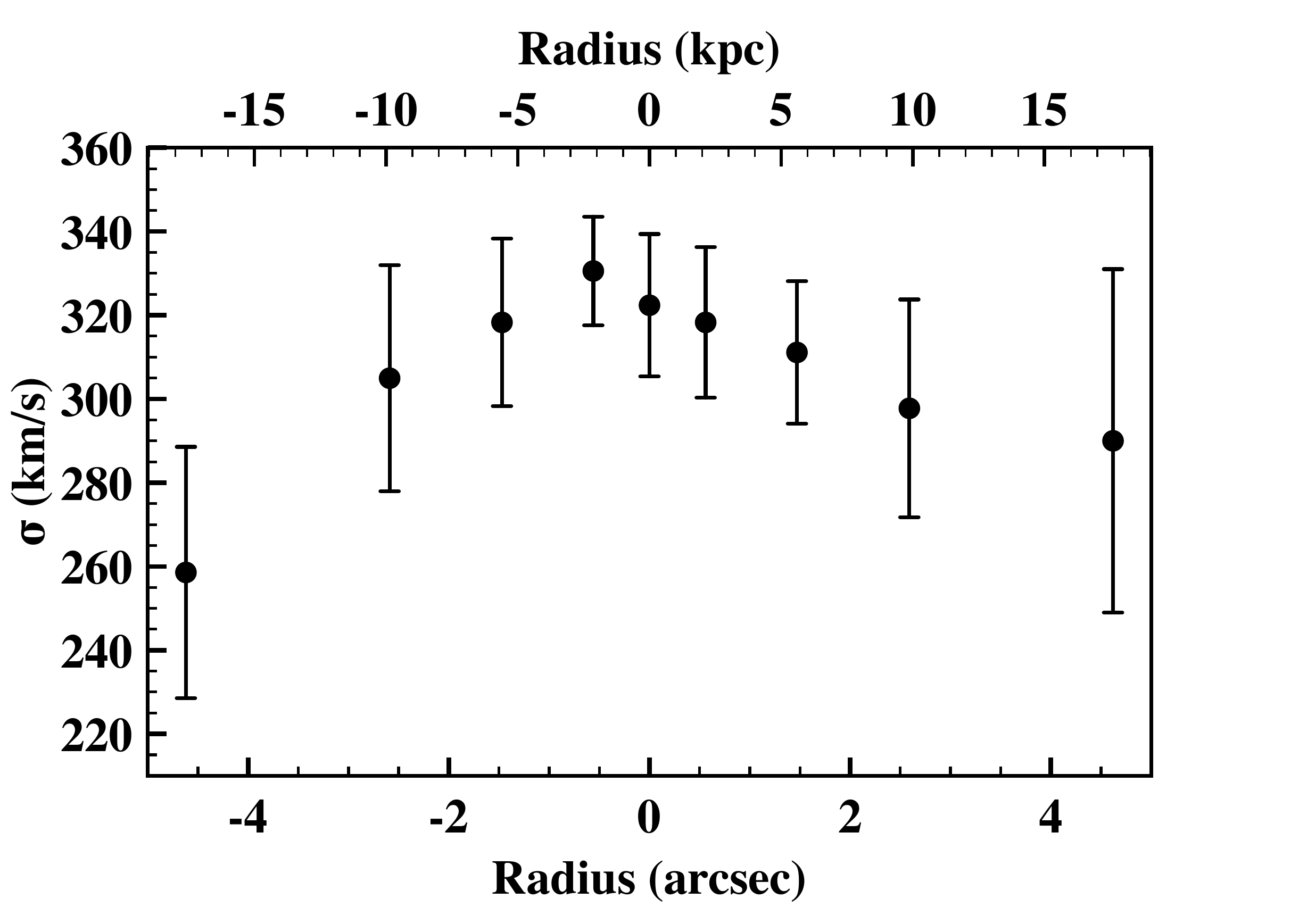}}
   \subfloat{\includegraphics[scale=0.25]{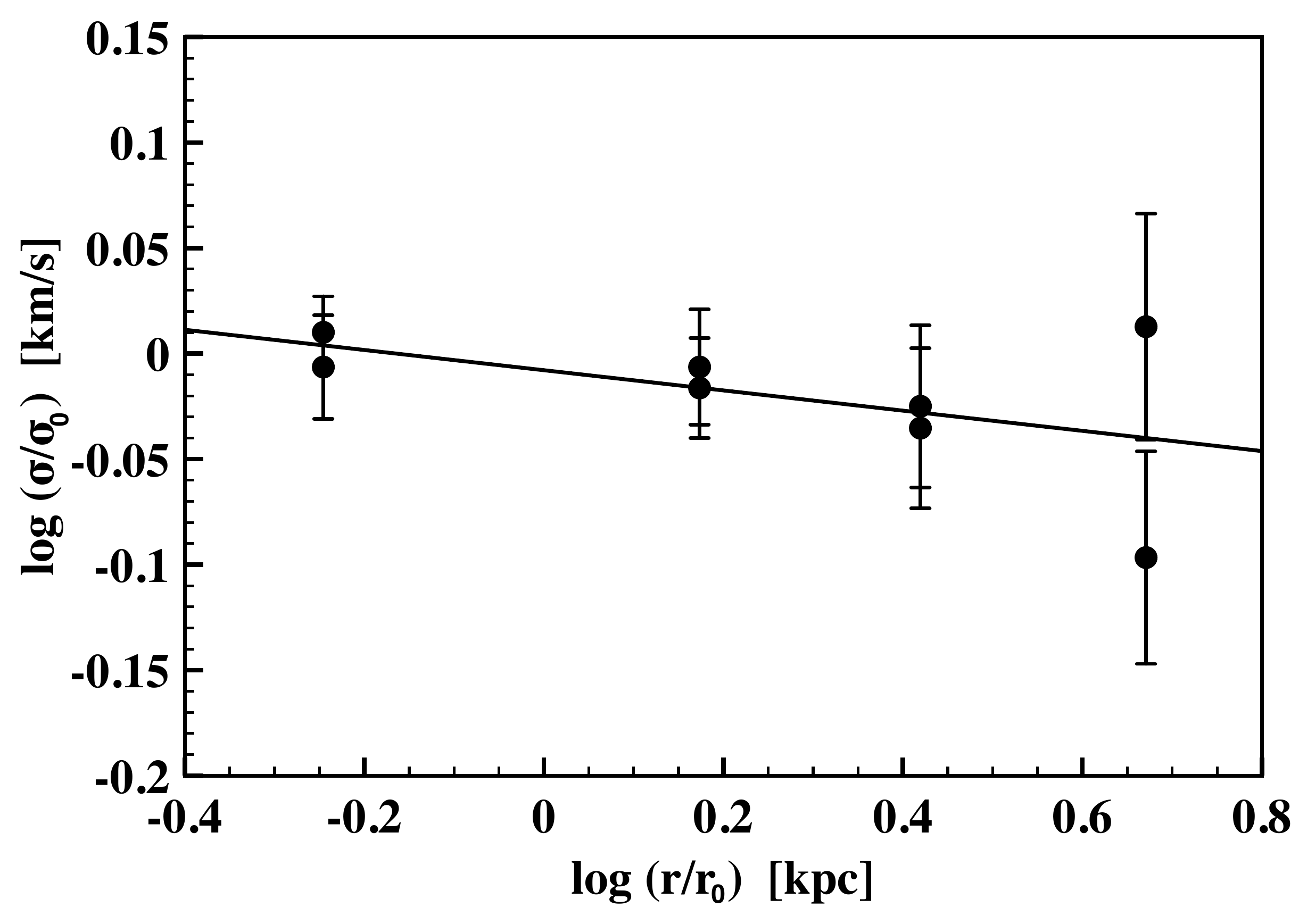}} \\    
     \subfloat{\includegraphics[scale=0.25]{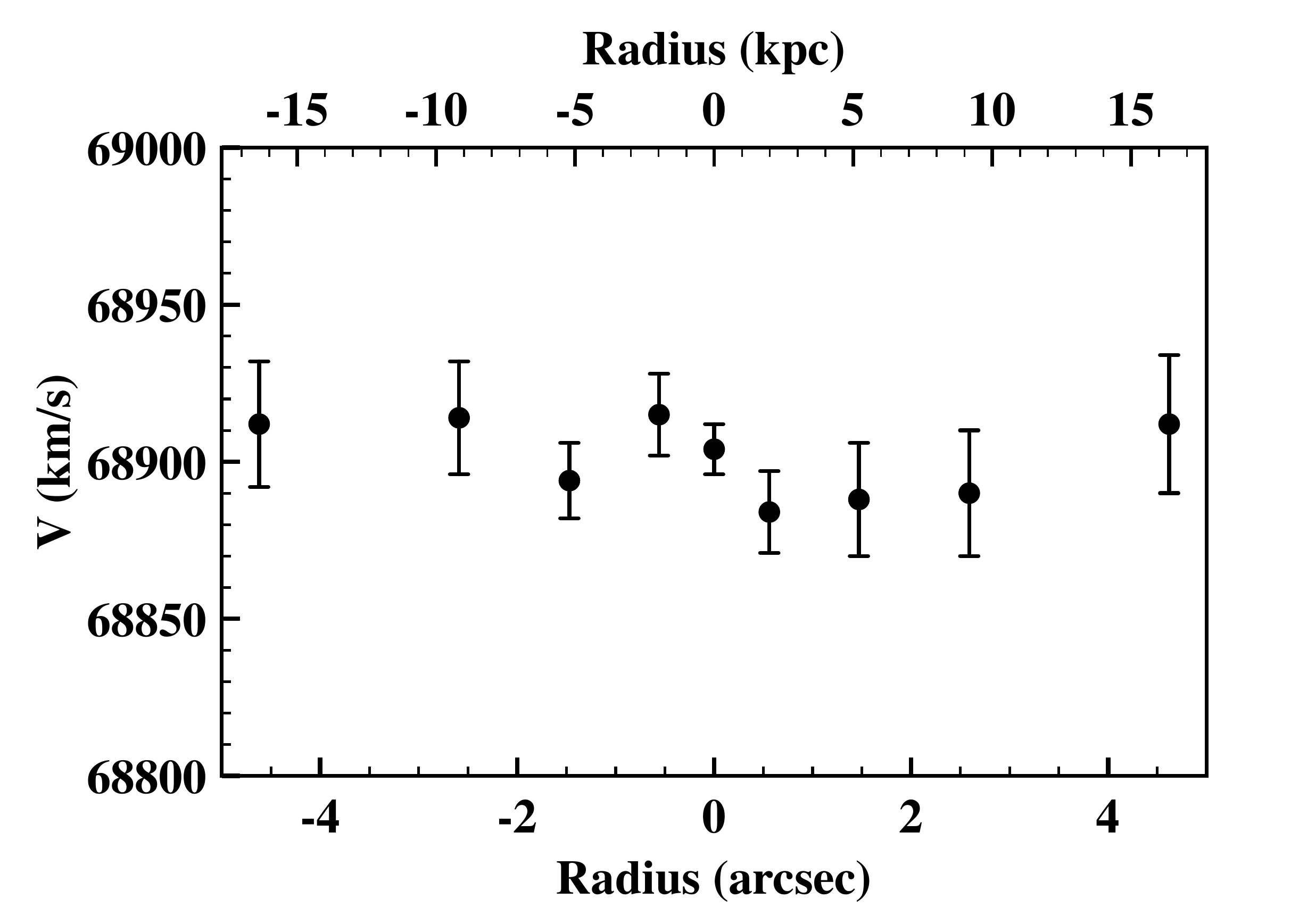}}
         \subfloat[Abell 267]{\includegraphics[scale=0.25]{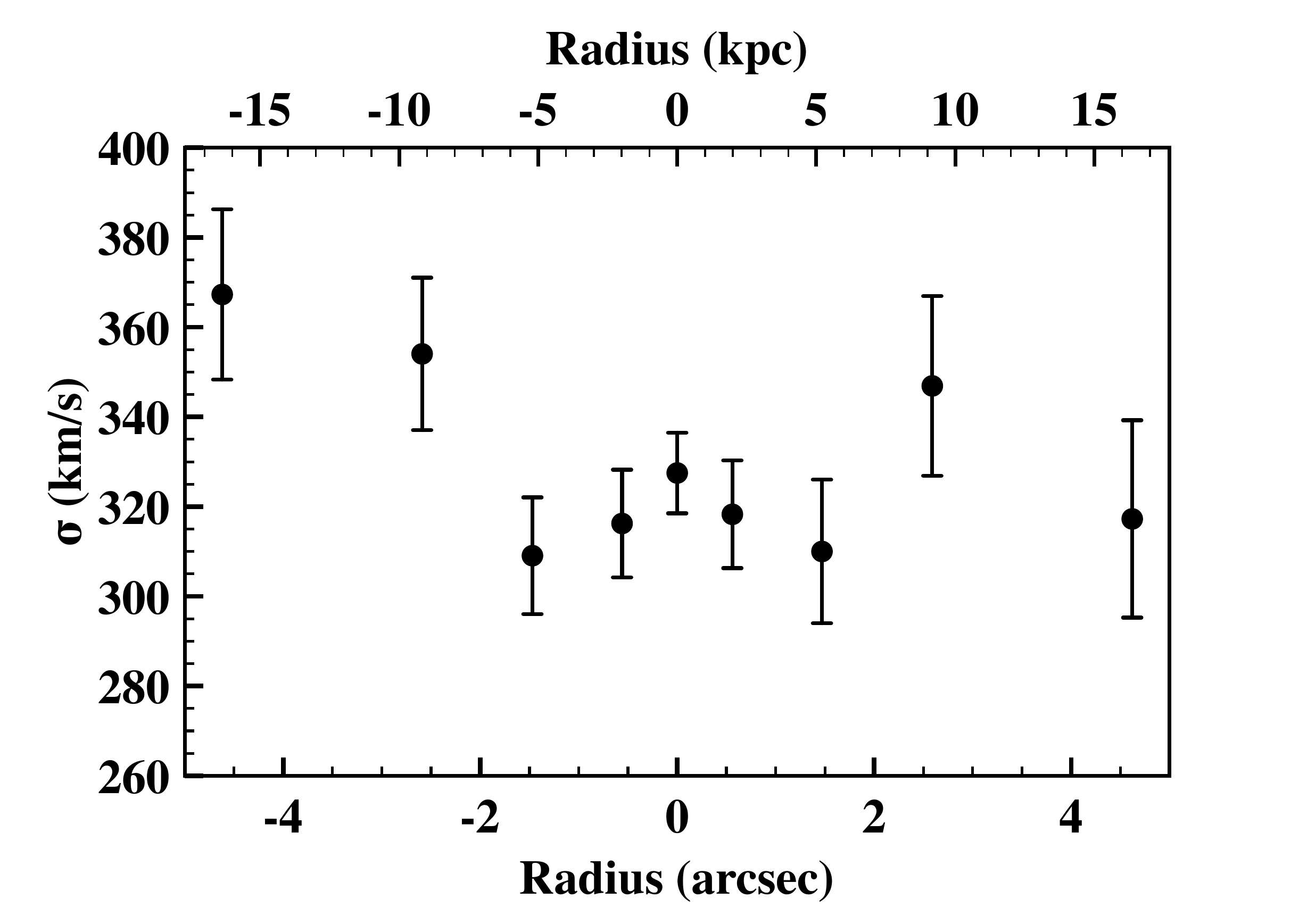}}
   \subfloat{\includegraphics[scale=0.25]{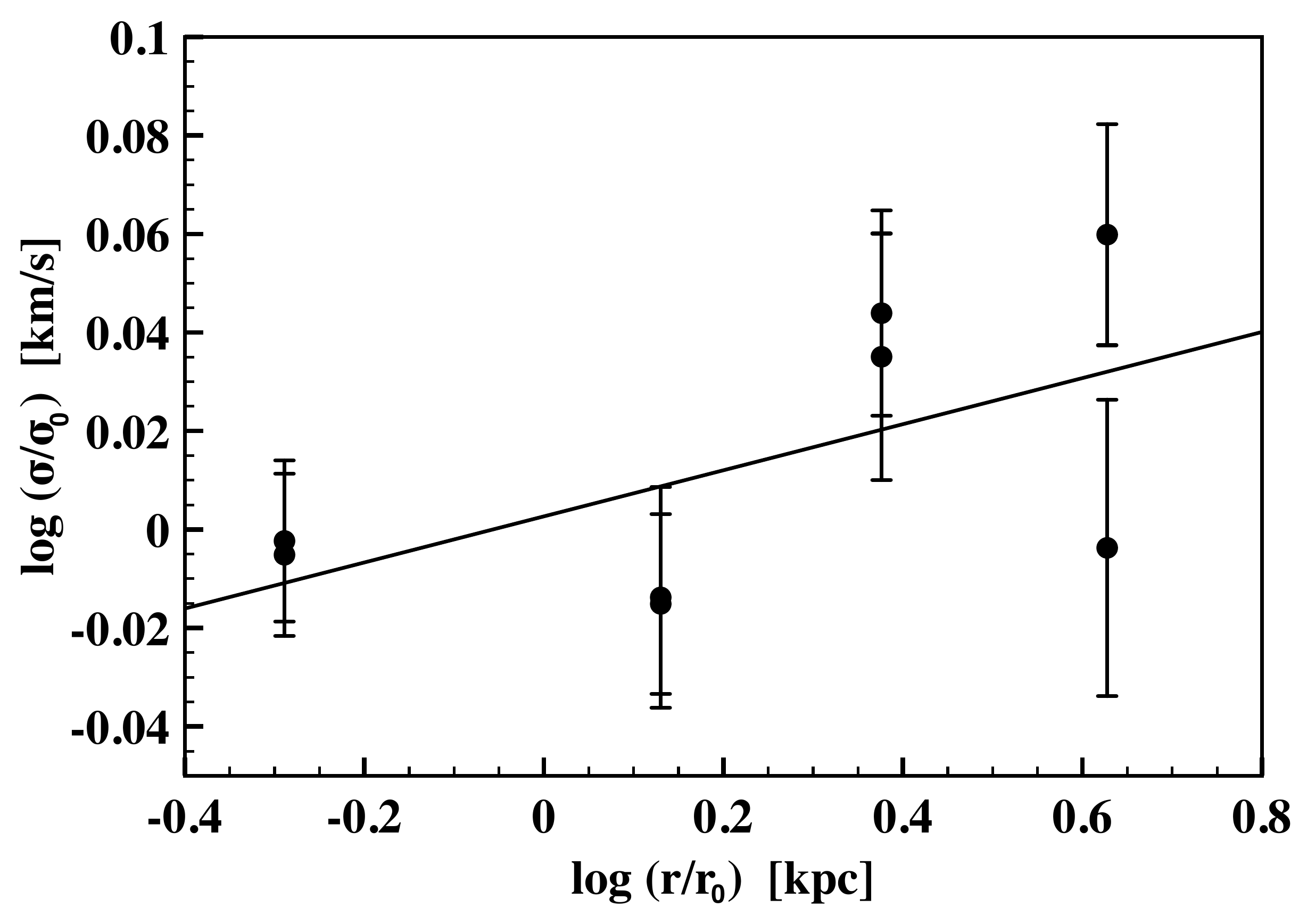}} \\
    \subfloat{\includegraphics[scale=0.25]{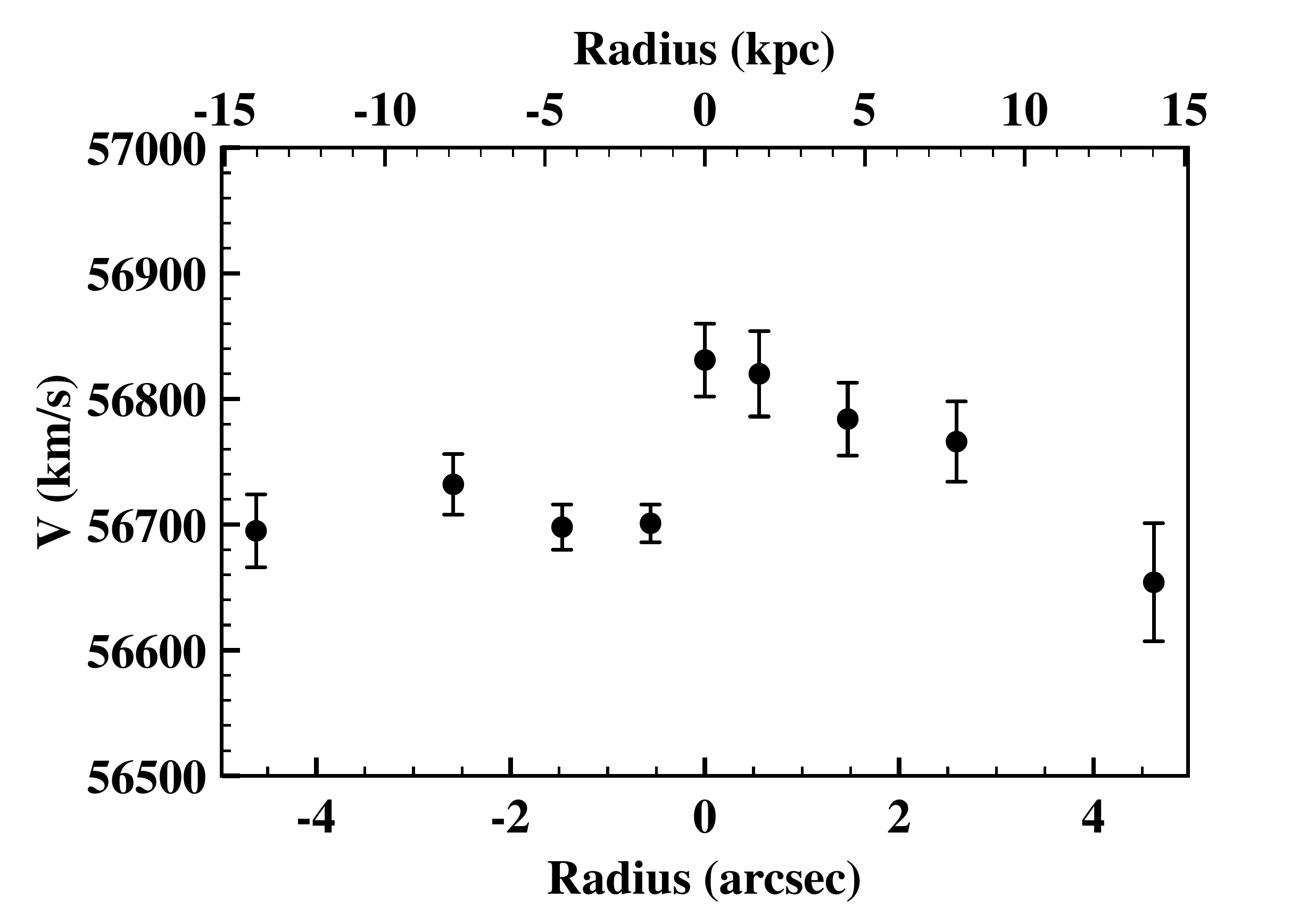}}
         \subfloat[Abell 383]{\includegraphics[scale=0.25]{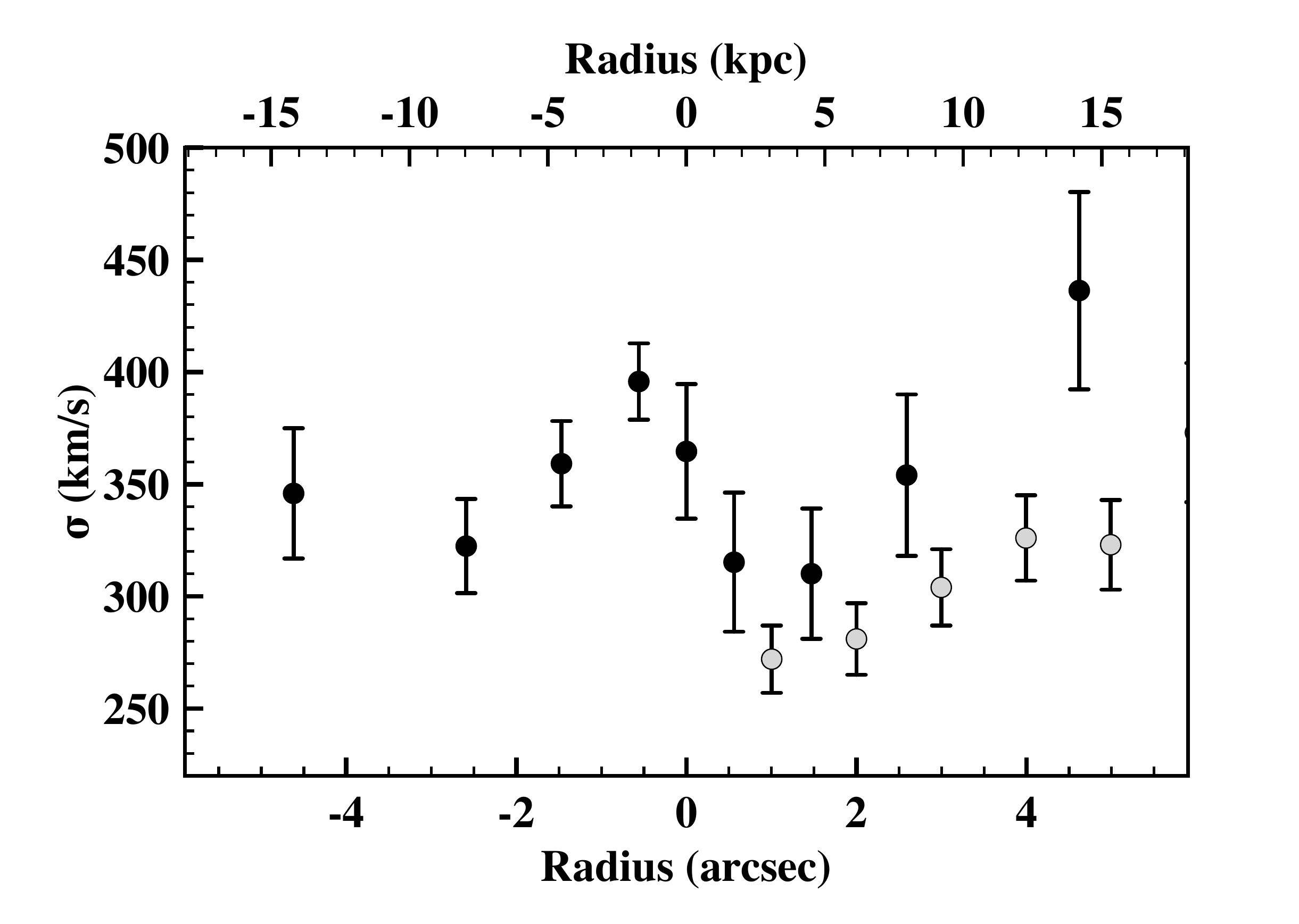}}
   \subfloat{\includegraphics[scale=0.25]{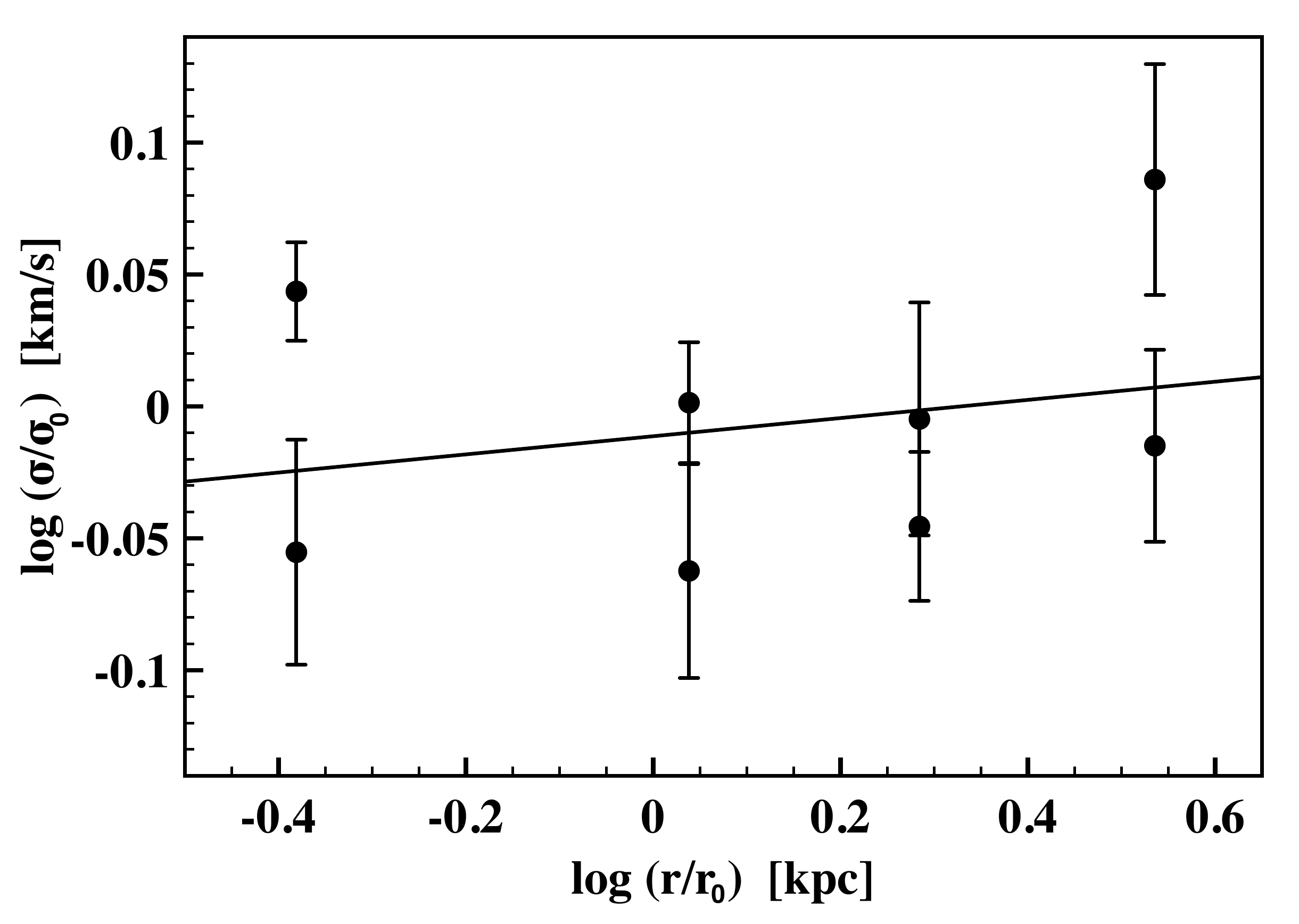}} \\
   \subfloat{\includegraphics[scale=0.25]{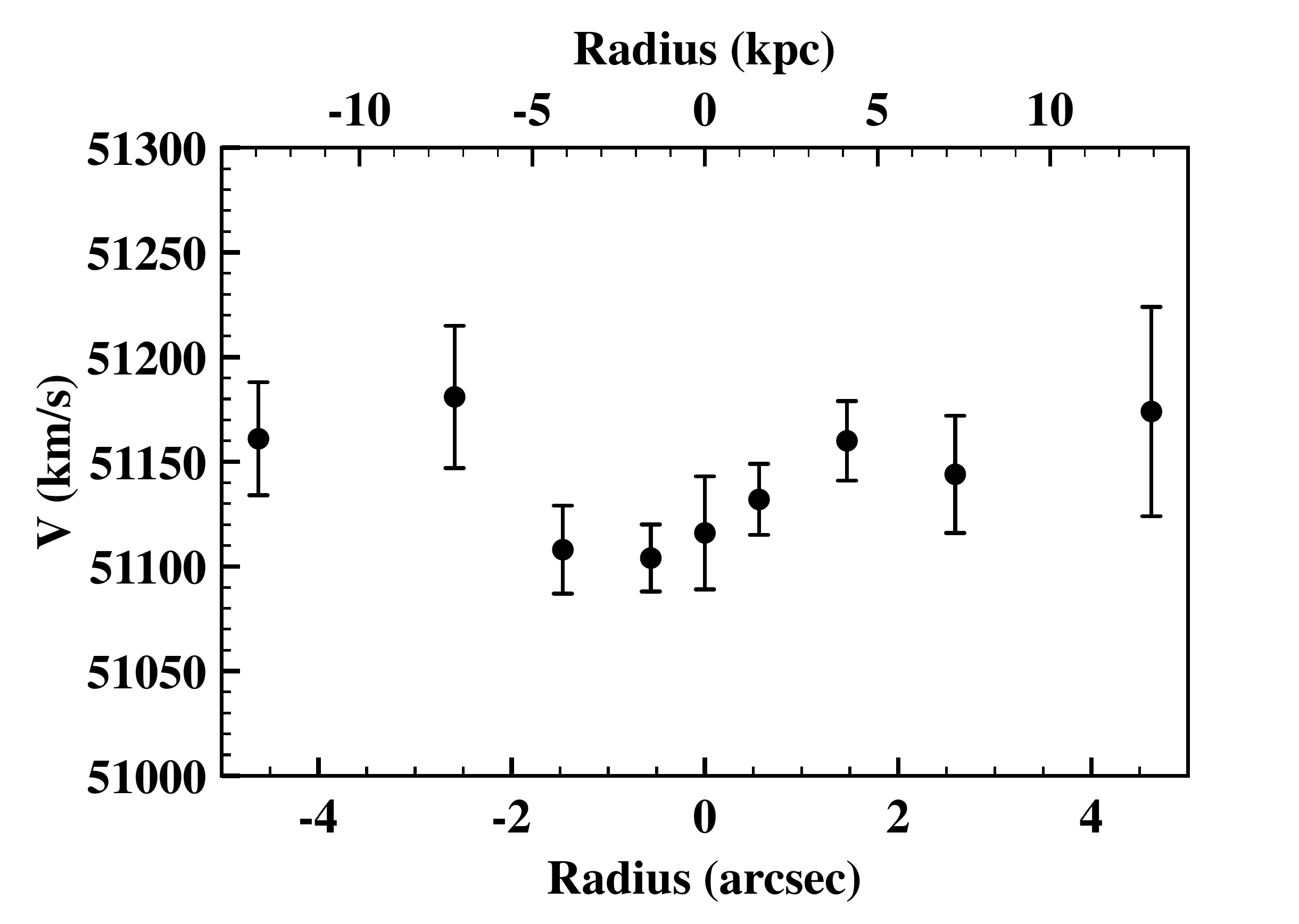}}
         \subfloat[Abell 586]{\includegraphics[scale=0.25]{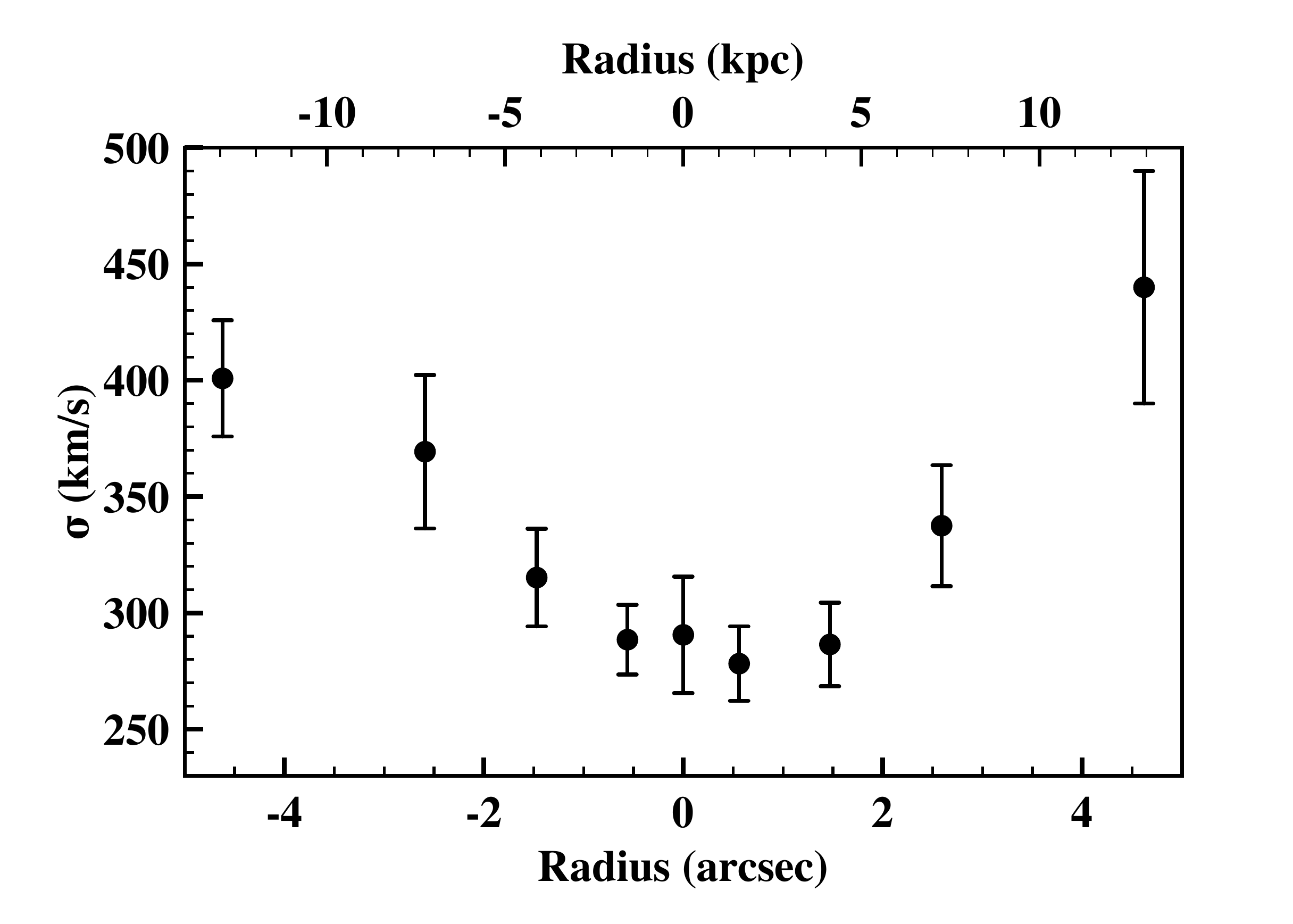}}
   \subfloat{\includegraphics[scale=0.25]{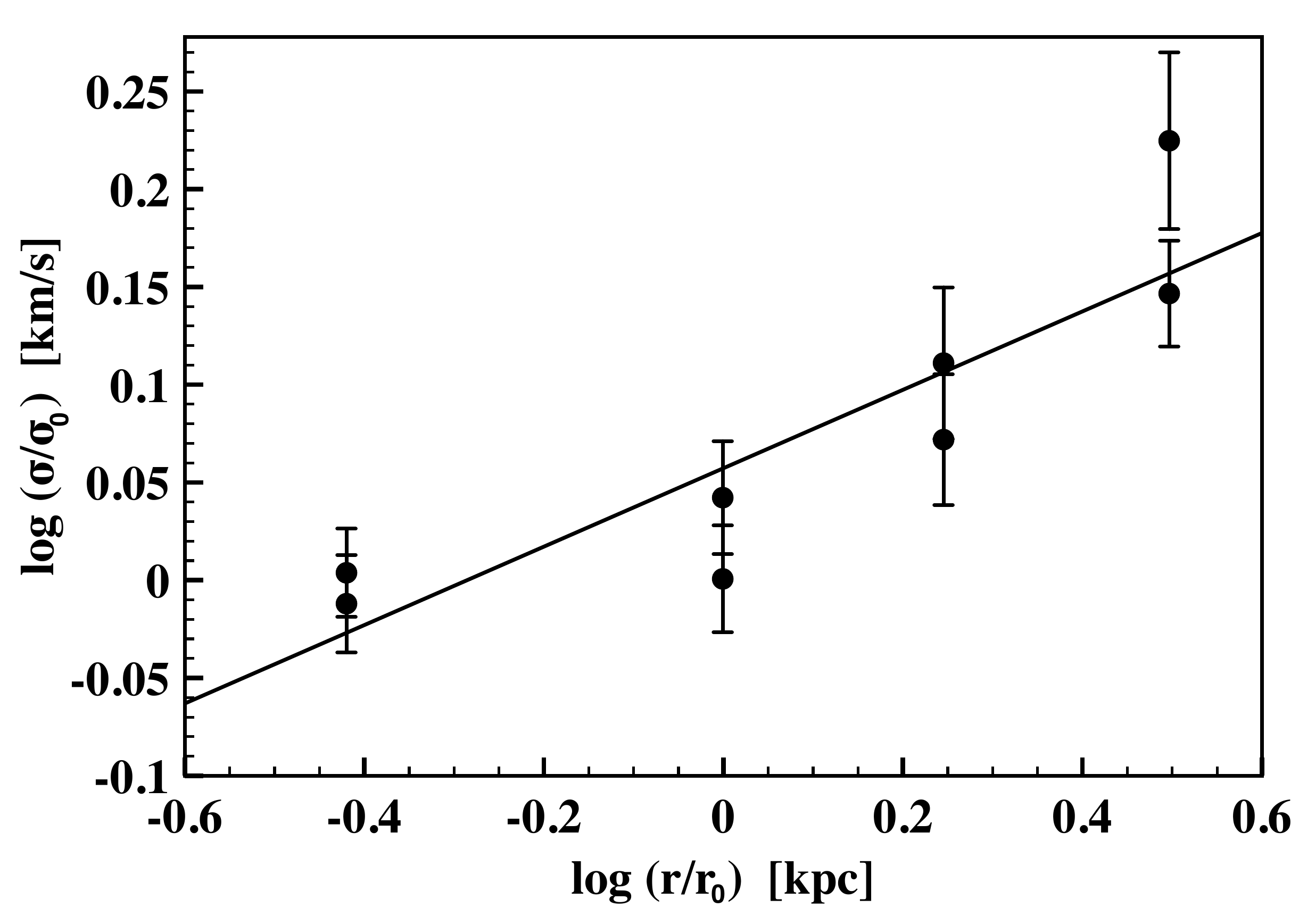}} \\
  \caption{[a] Radial profiles of velocity (V), [b] velocity dispersion ($\sigma$, and [c] power law fit). The grey circles in the velocity dispersion profile of Abell 383 indicate the measurements from \citet{Newman2013a} (see Appendix \ref{kinematics}).}
\label{fig:kin1}
\end{figure*}

\begin{figure*}
\captionsetup[subfigure]{labelformat=empty}
      \subfloat{\includegraphics[scale=0.25]{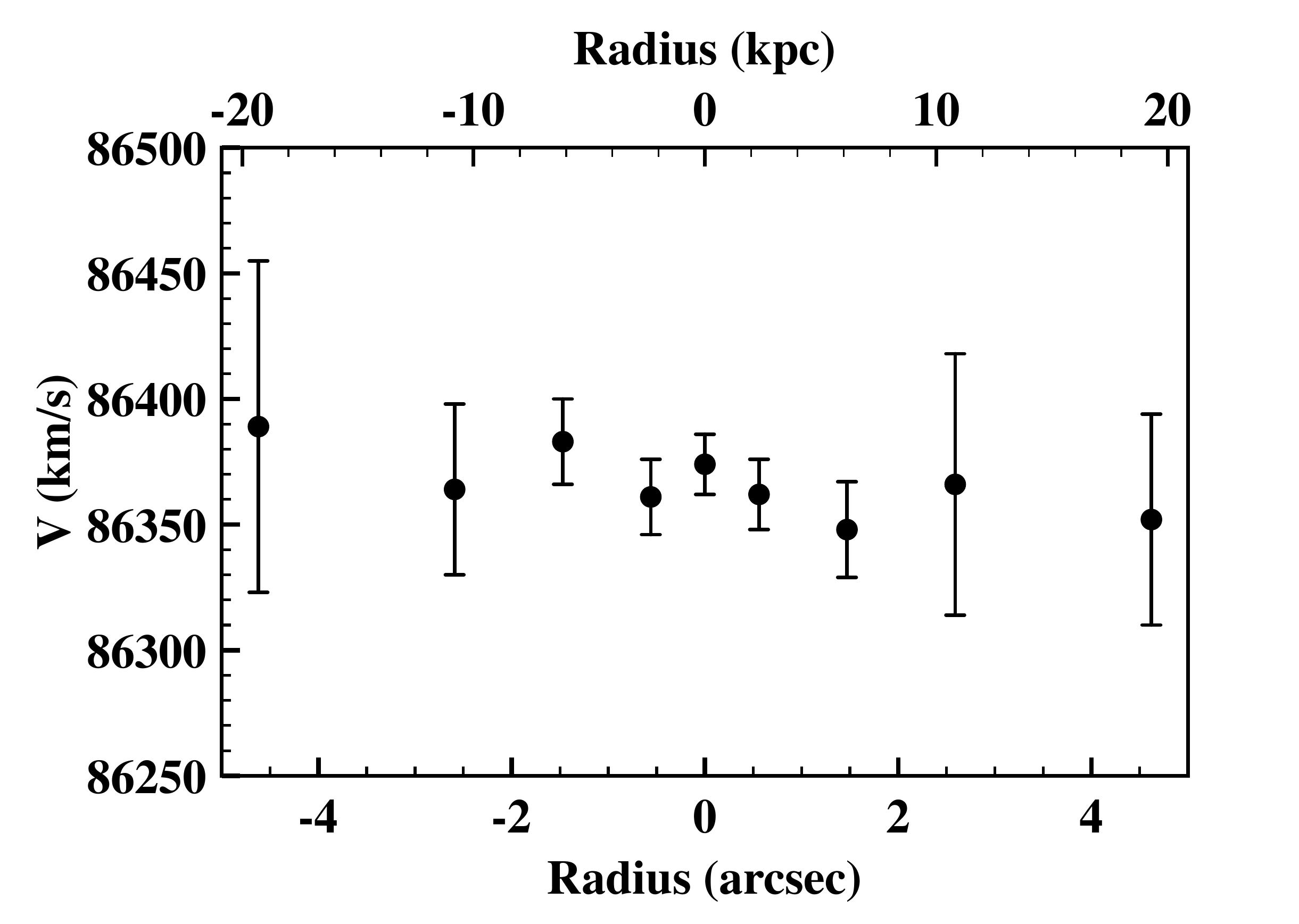}}
         \subfloat[Abell 611]{\includegraphics[scale=0.25]{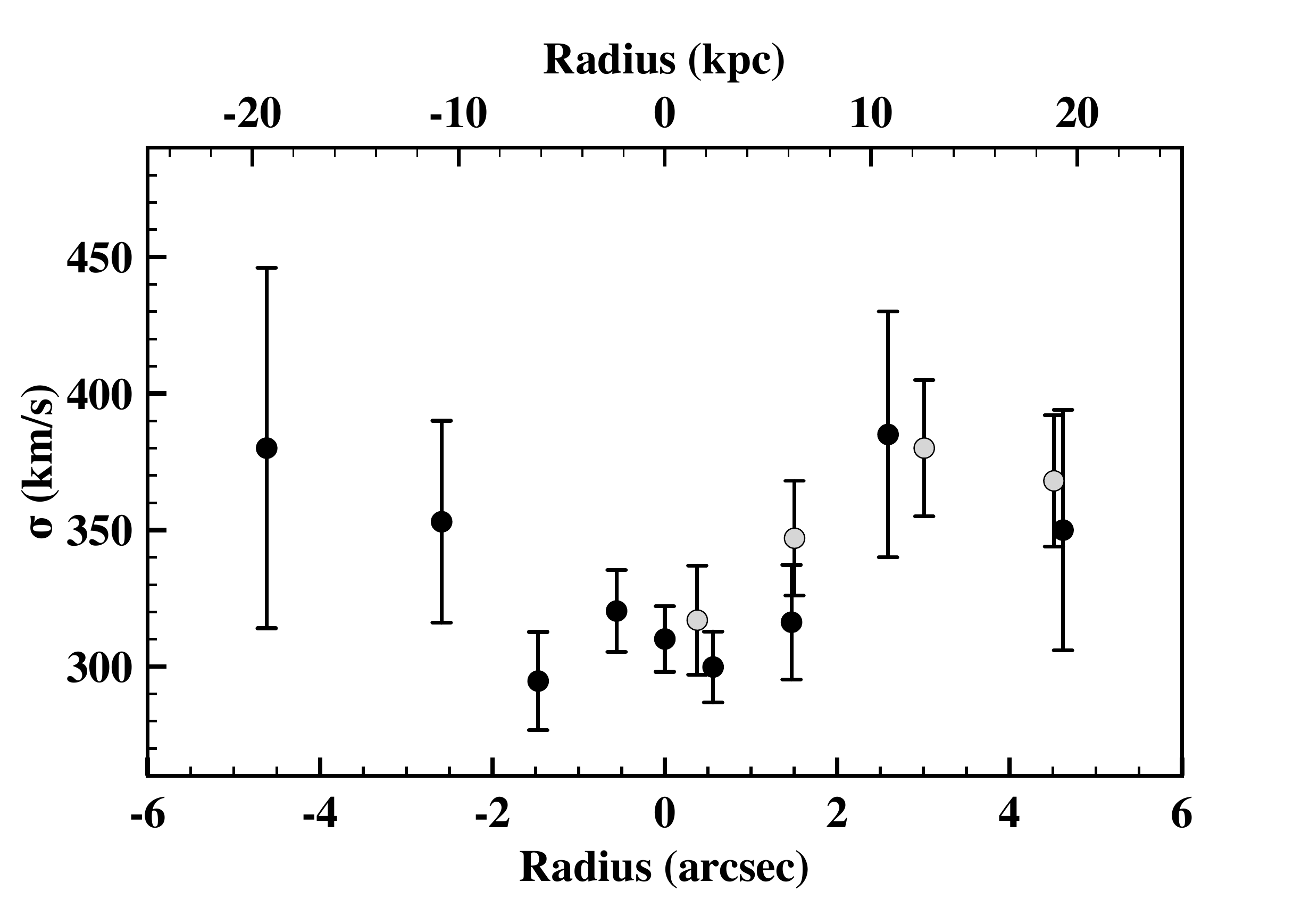}}
   \subfloat{\includegraphics[scale=0.25]{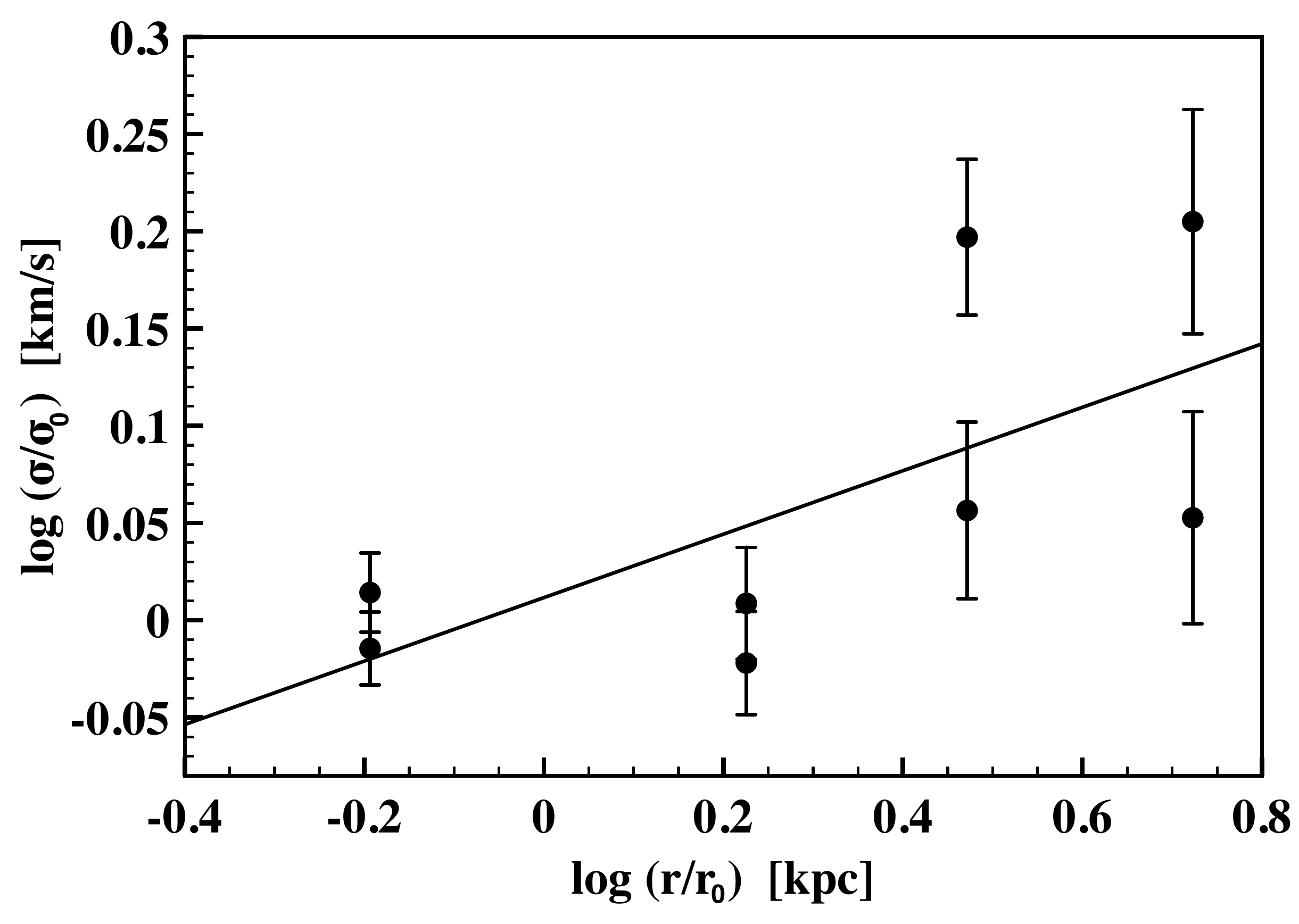}} \\
\subfloat{\includegraphics[scale=0.25]{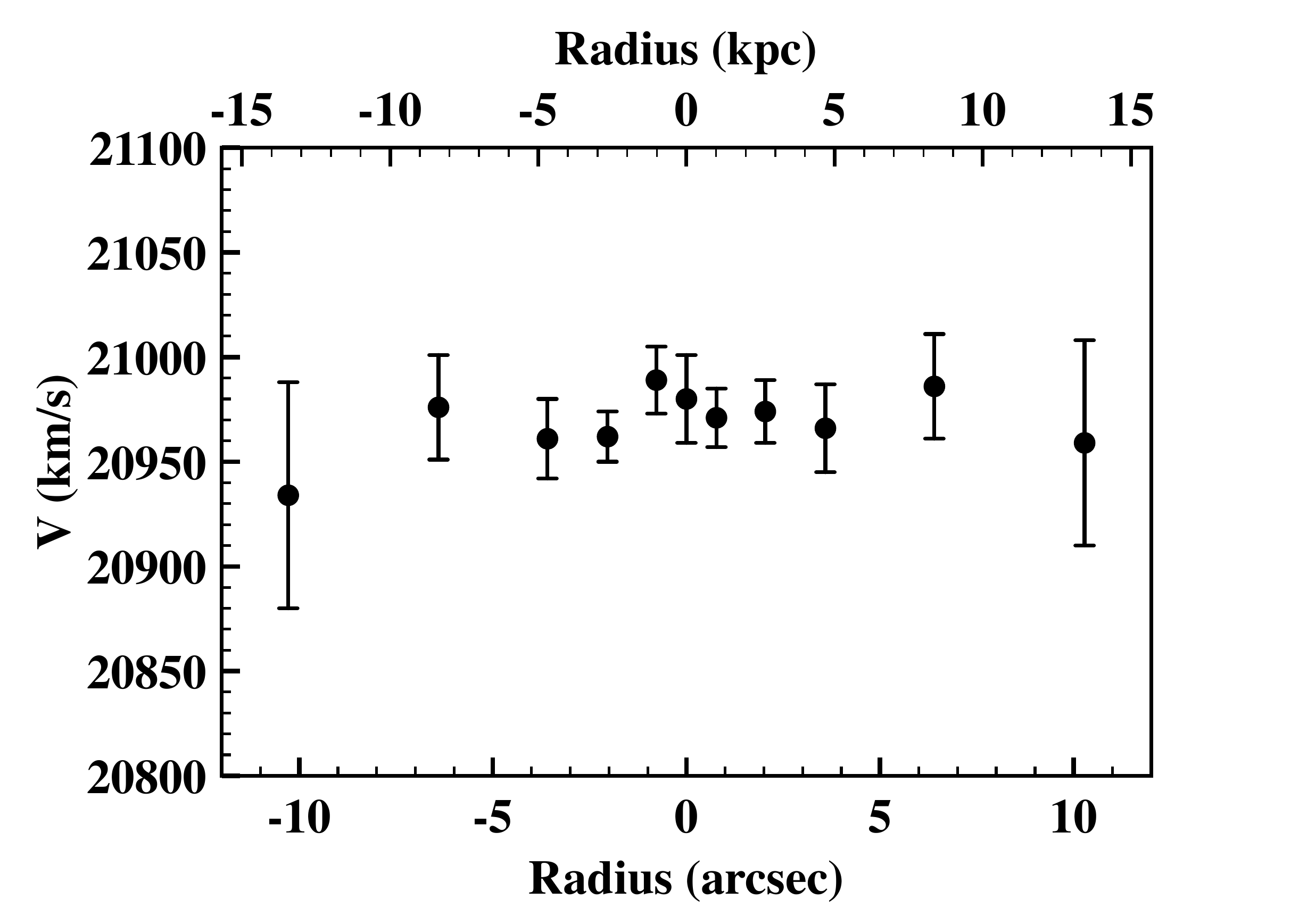}}
         \subfloat[Abell 644]{\includegraphics[scale=0.25]{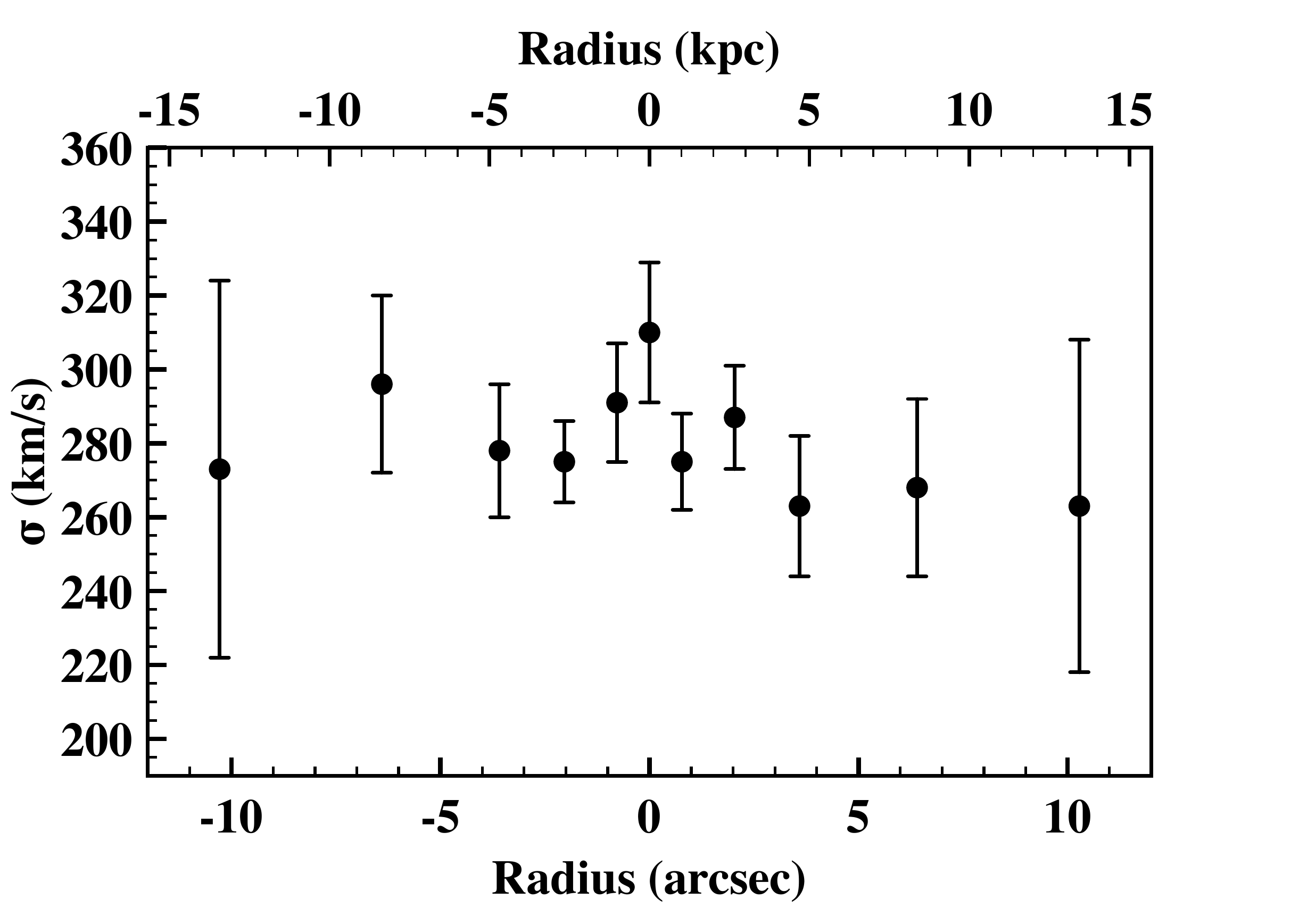}}
   \subfloat{\includegraphics[scale=0.25]{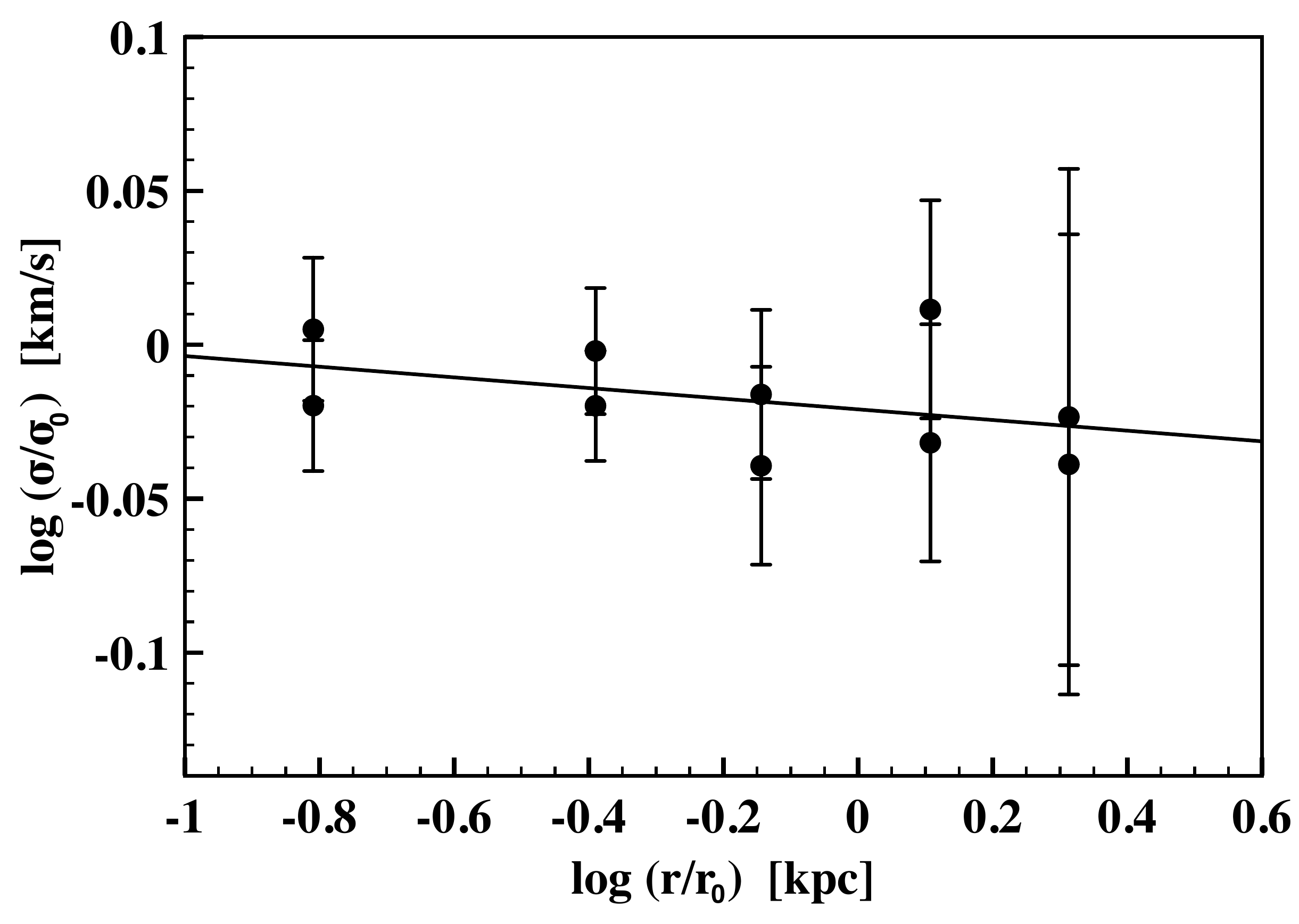}} \\
   \subfloat{\includegraphics[scale=0.25]{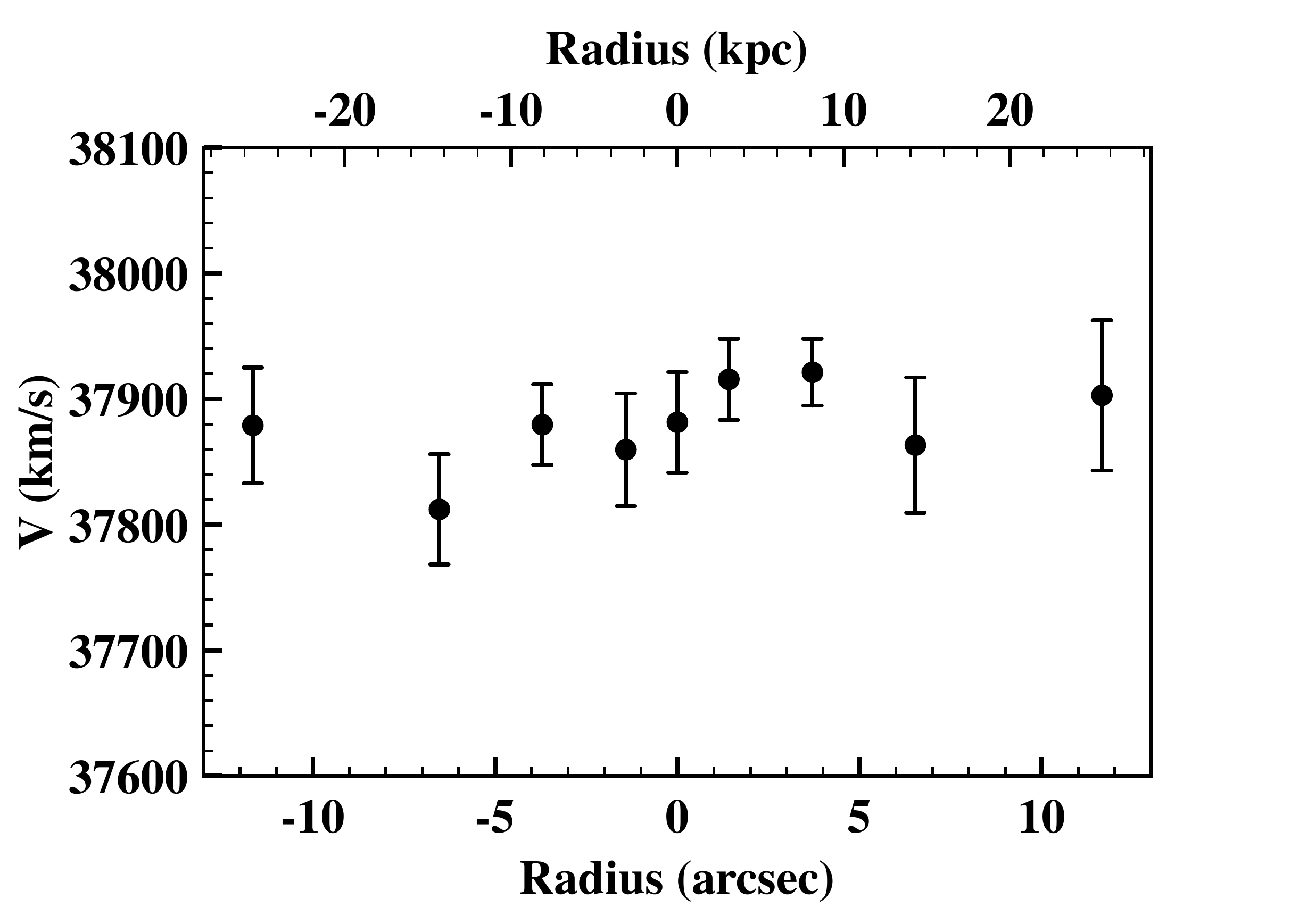}}
         \subfloat[Abell 646]{\includegraphics[scale=0.25]{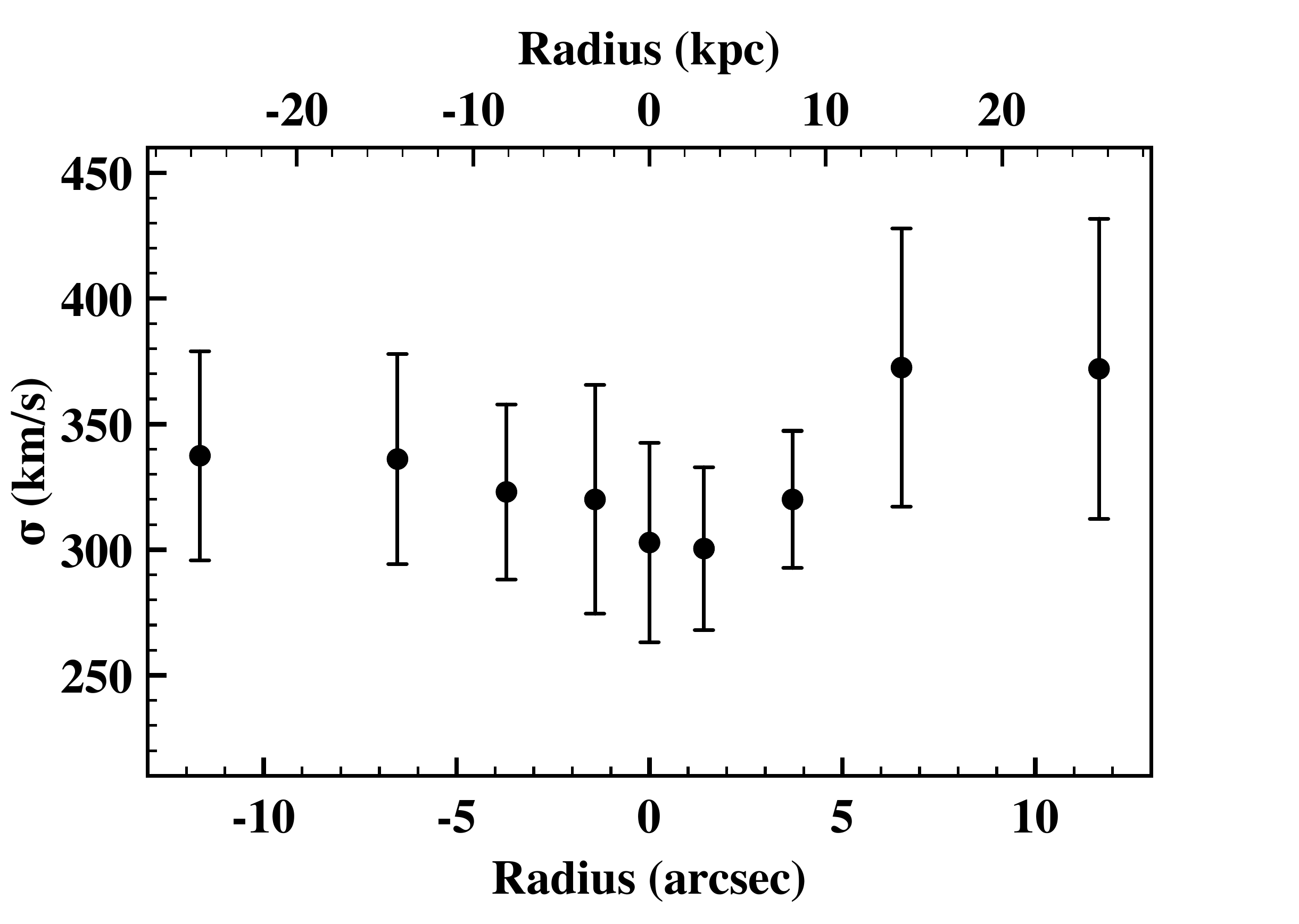}}
   \subfloat{\includegraphics[scale=0.25]{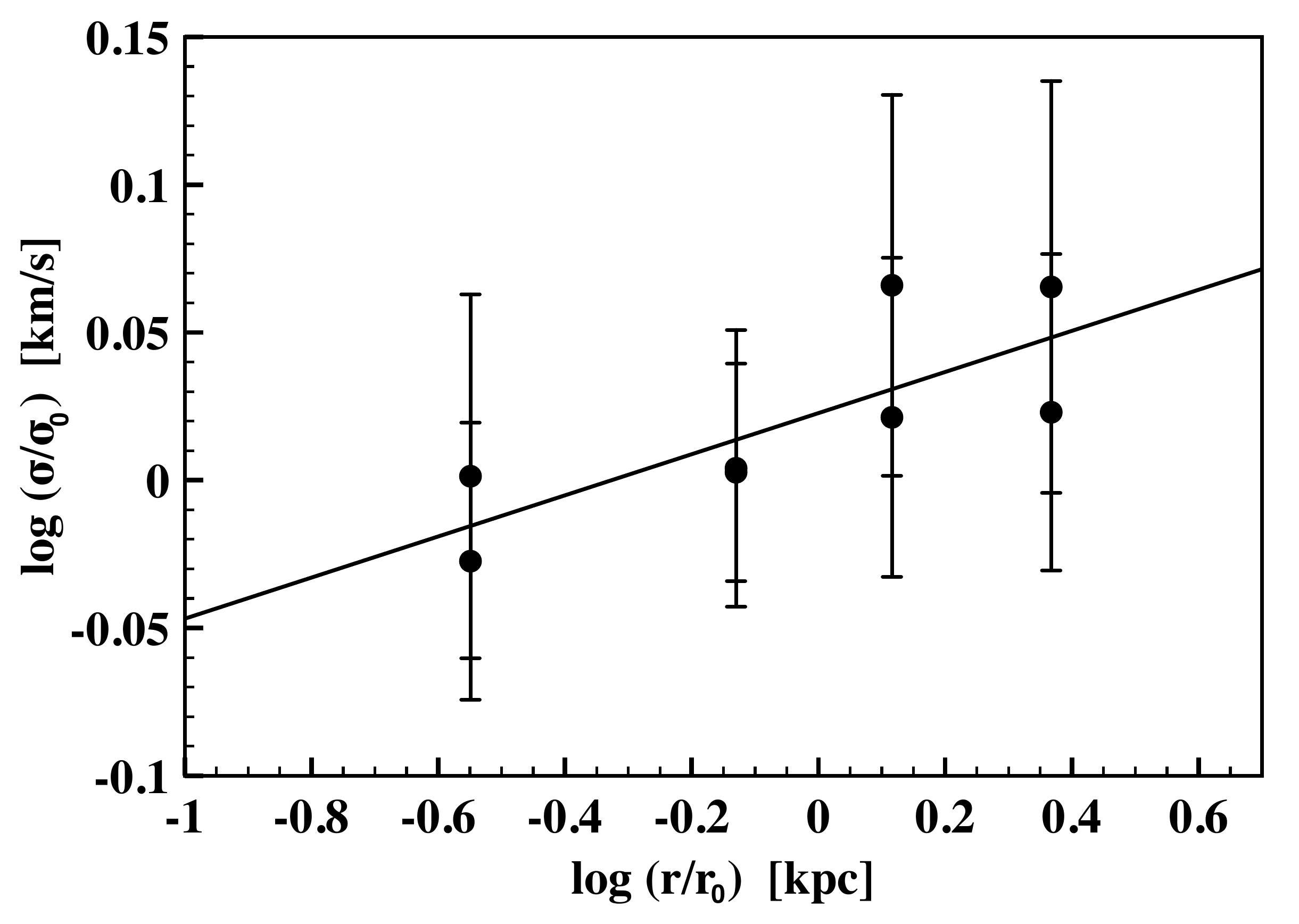}} \\
     \subfloat{\includegraphics[scale=0.25]{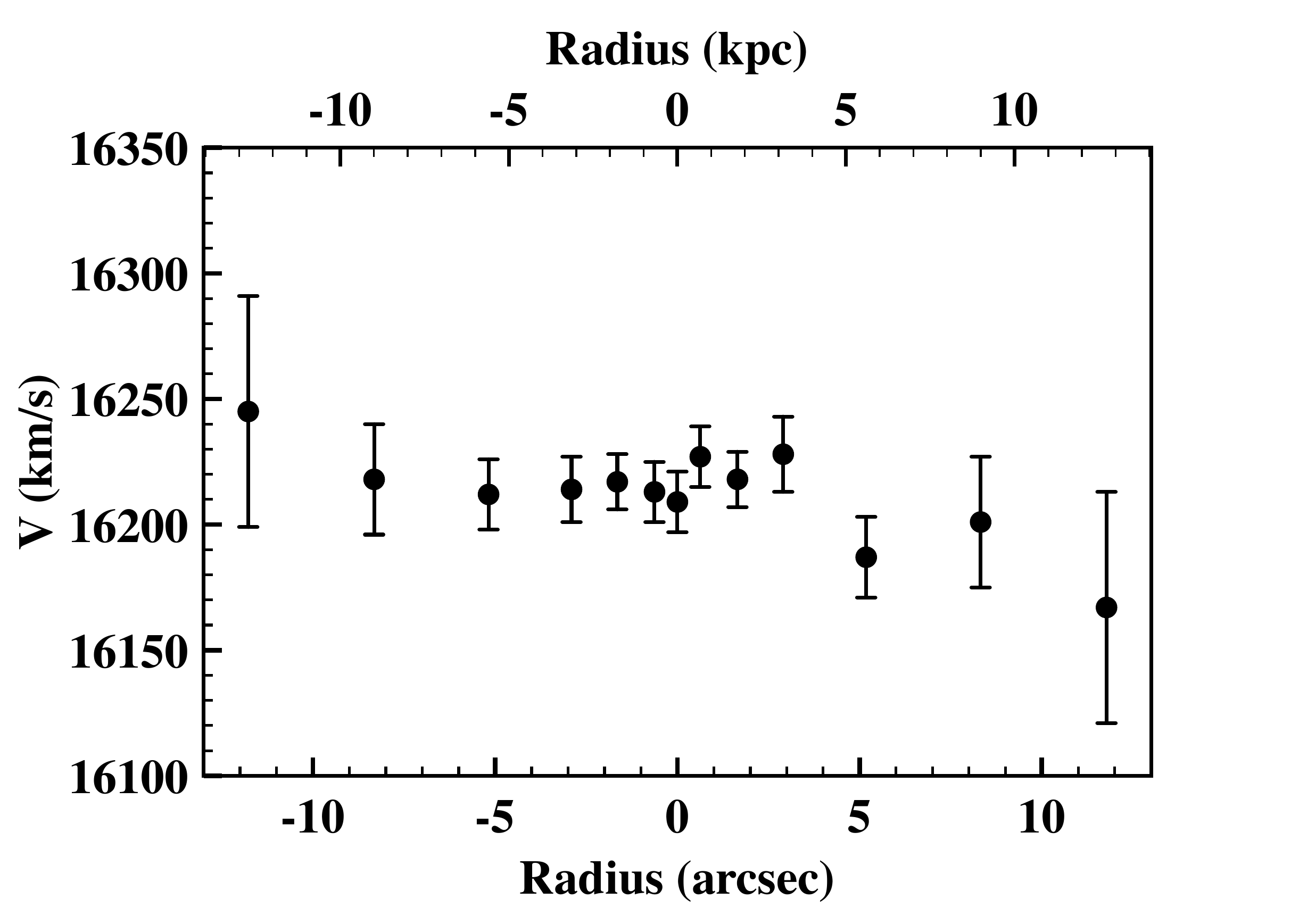}}
         \subfloat[Abell 754]{\includegraphics[scale=0.25]{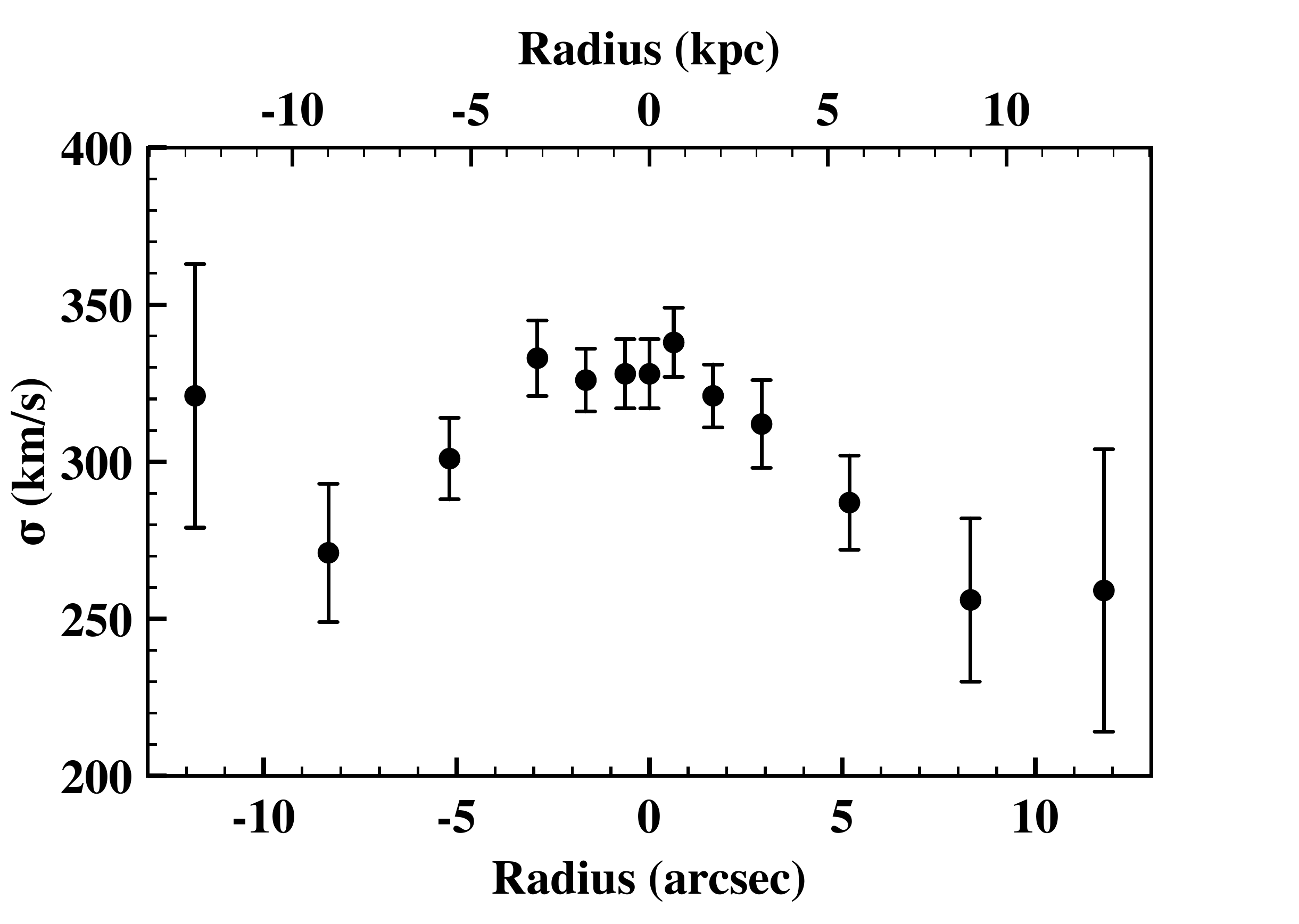}}
   \subfloat{\includegraphics[scale=0.25]{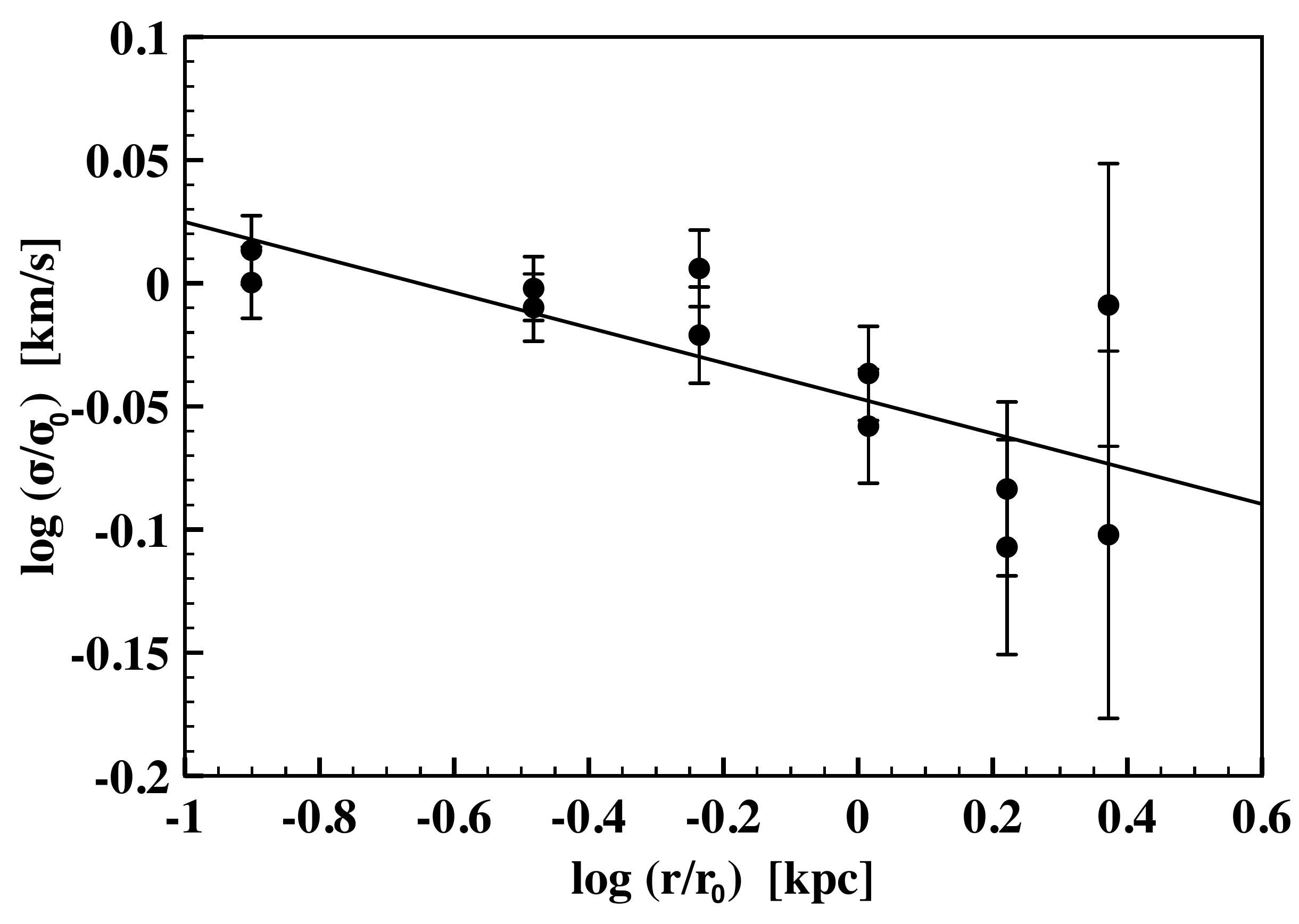}} \\
  \caption{[a] Radial profiles of velocity (V), [b] velocity dispersion ($\sigma$, and [c] power law fit). The grey circles in the velocity dispersion profile of Abell 611 indicate the measurements from \citet{Newman2013a} (see Appendix \ref{kinematics}).}
\label{fig:kin3}
\end{figure*}

\begin{figure*}
\captionsetup[subfigure]{labelformat=empty}
    \subfloat{\includegraphics[scale=0.25]{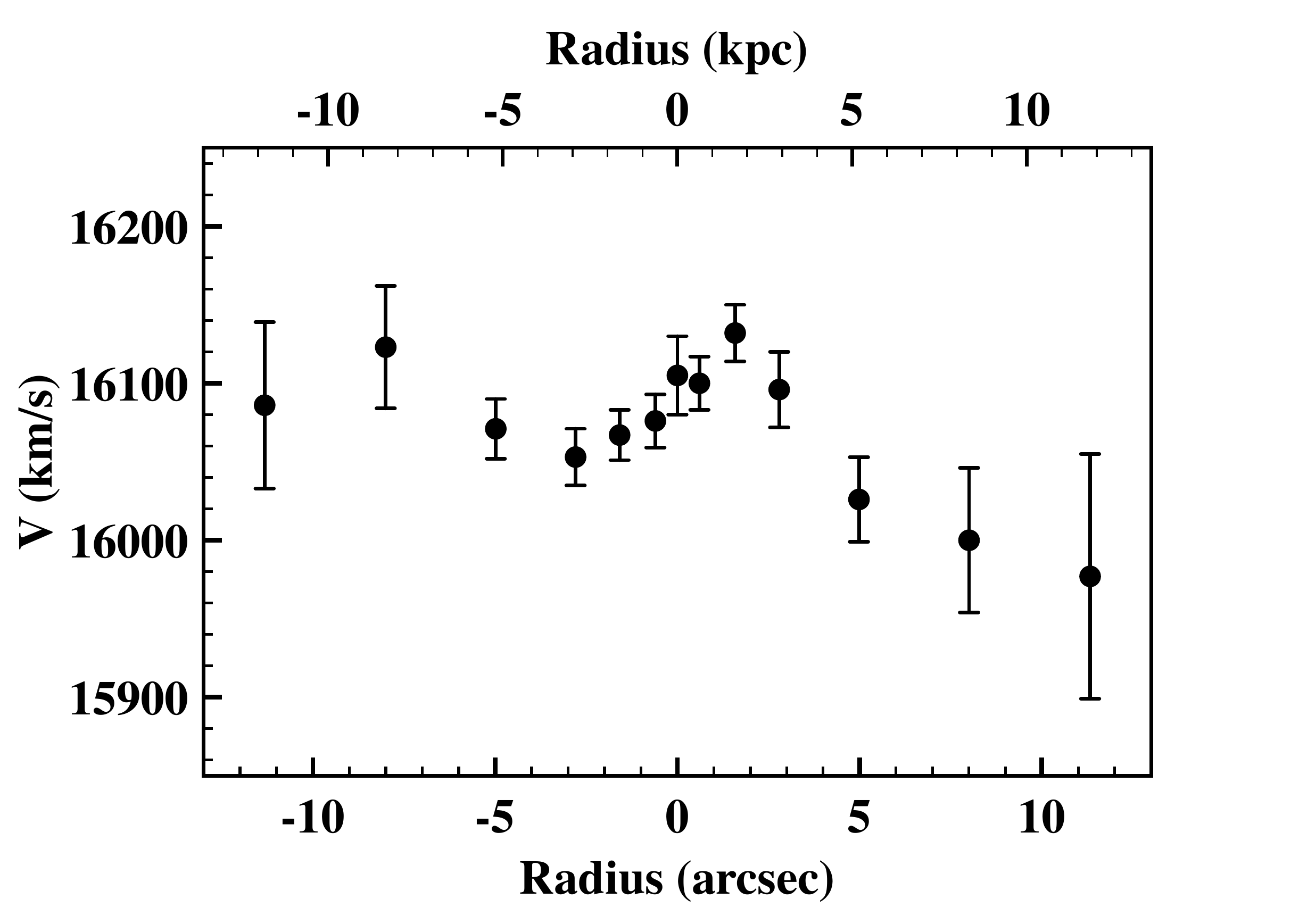}}
         \subfloat[Abell 780]{\includegraphics[scale=0.25]{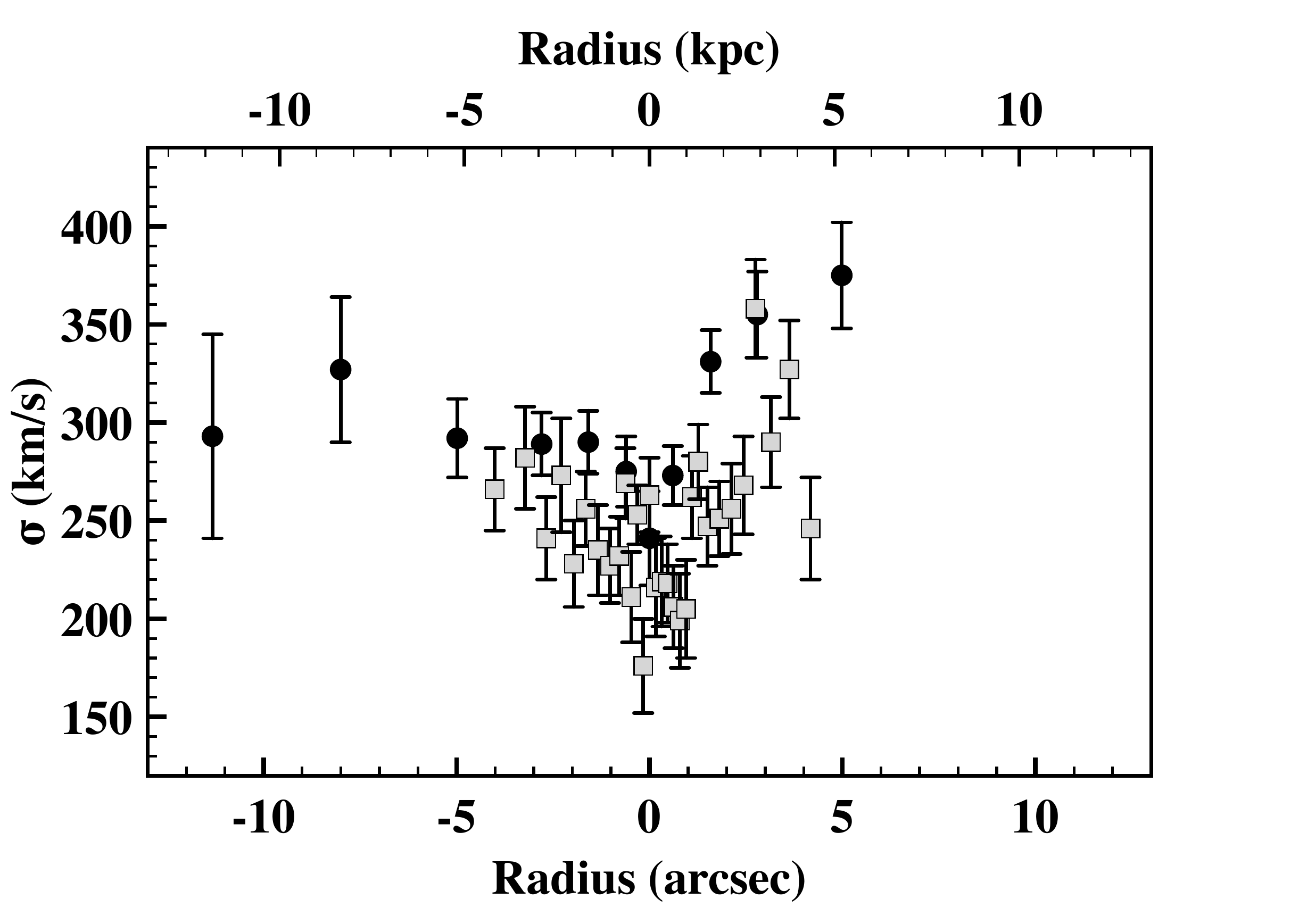}}
   \subfloat{\includegraphics[scale=0.25]{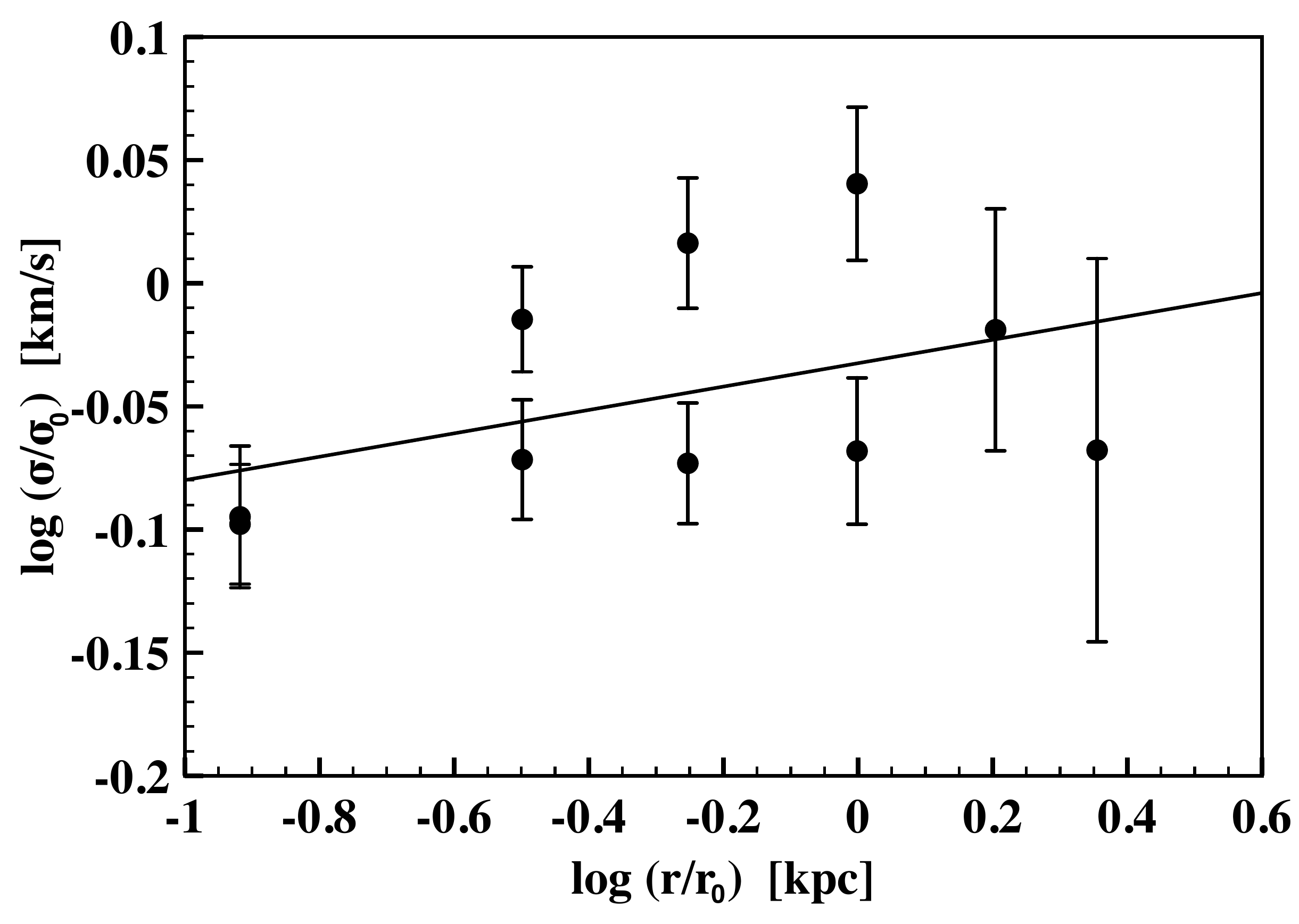}} \\
    \subfloat{\includegraphics[scale=0.25]{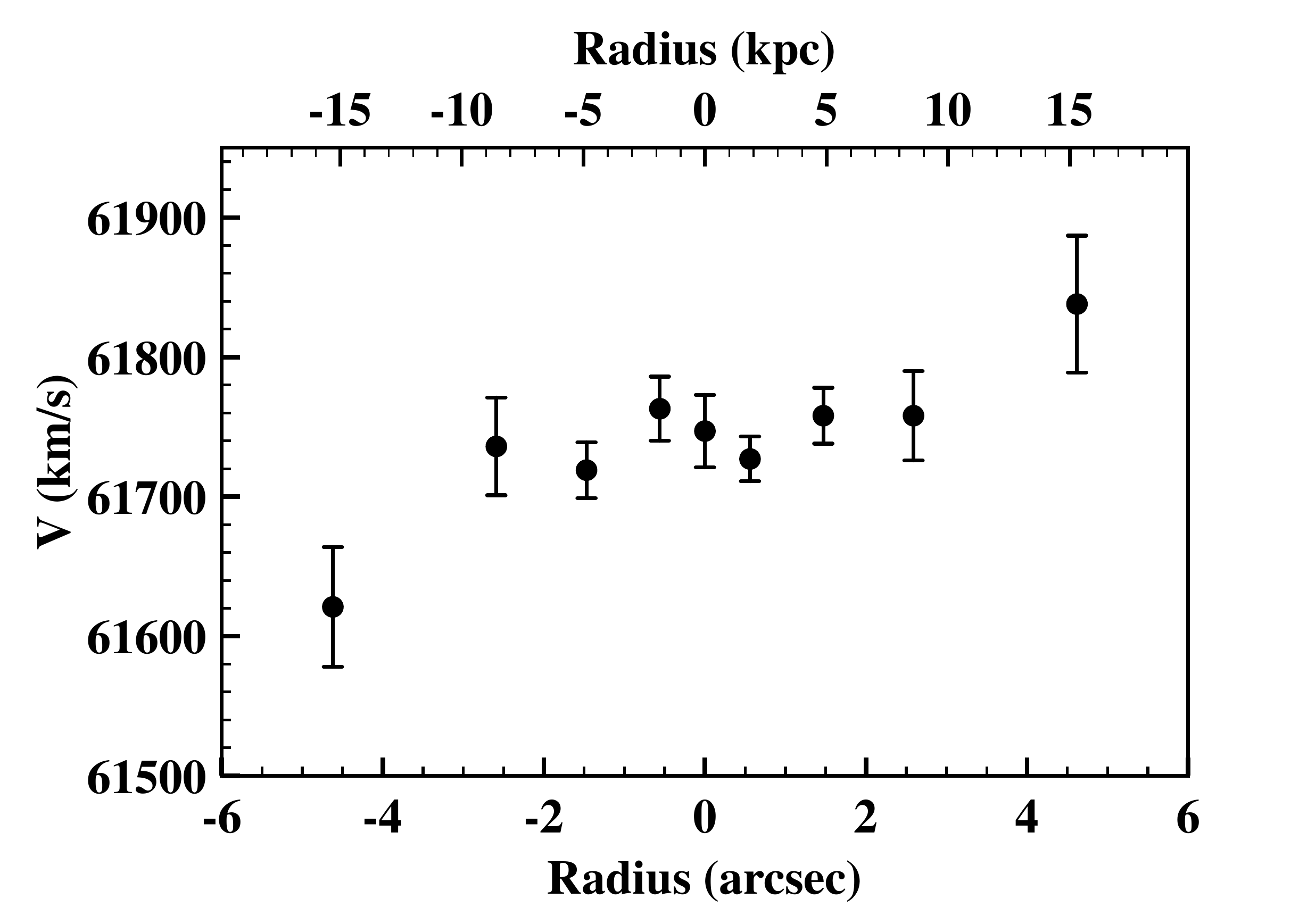}}
         \subfloat[Abell 963]{\includegraphics[scale=0.25]{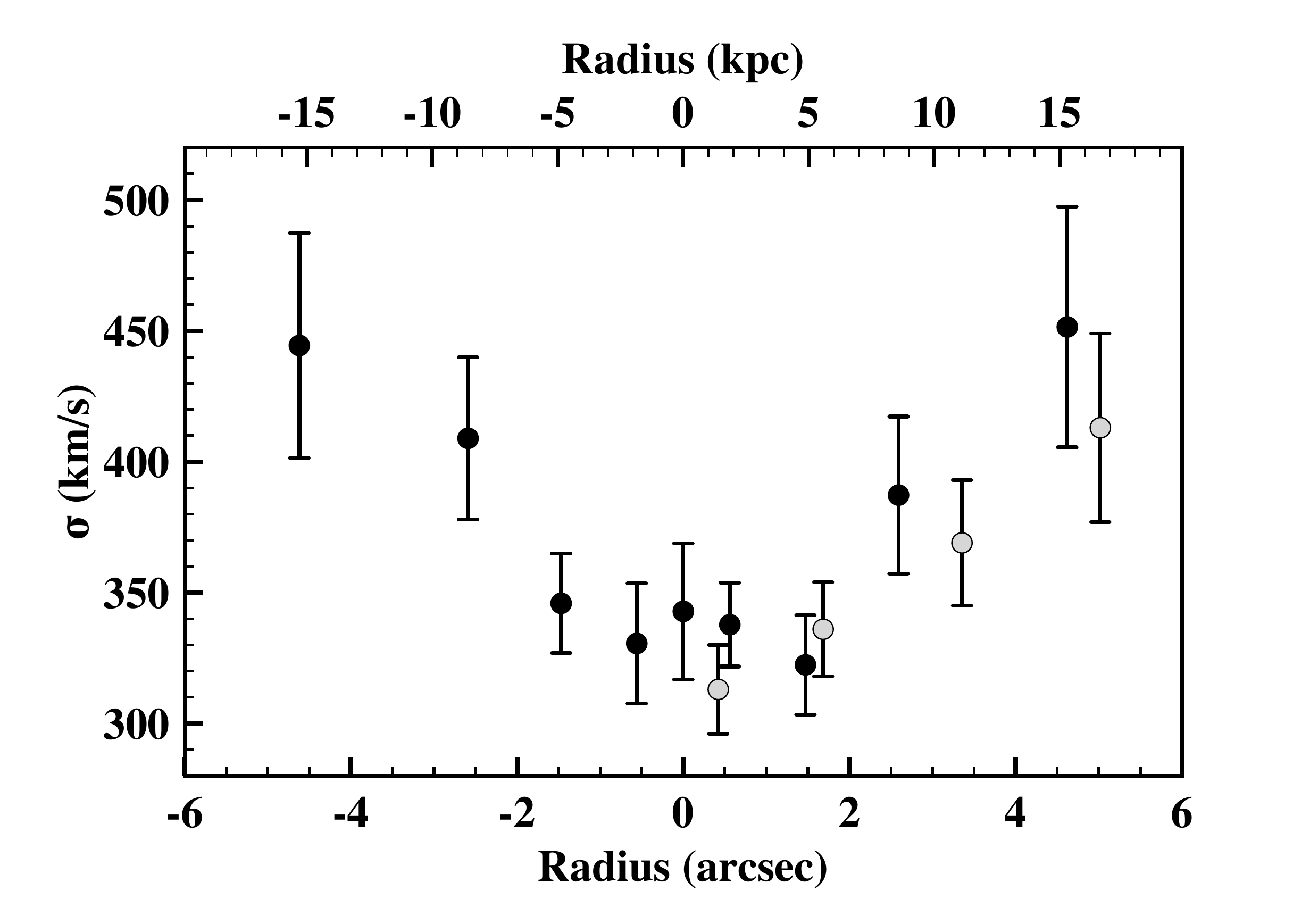}}
   \subfloat{\includegraphics[scale=0.25]{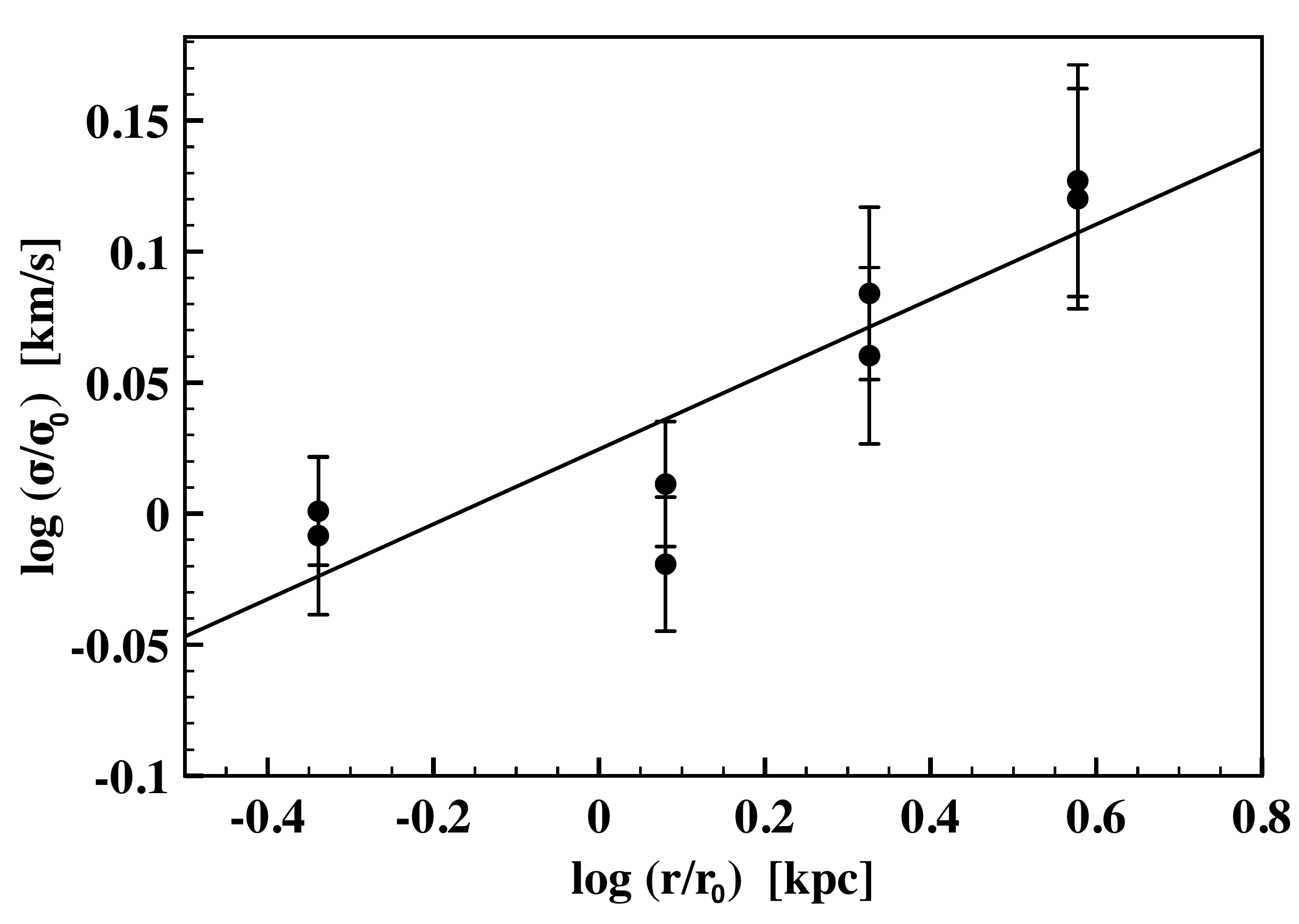}} \\
             \subfloat{\includegraphics[scale=0.25]{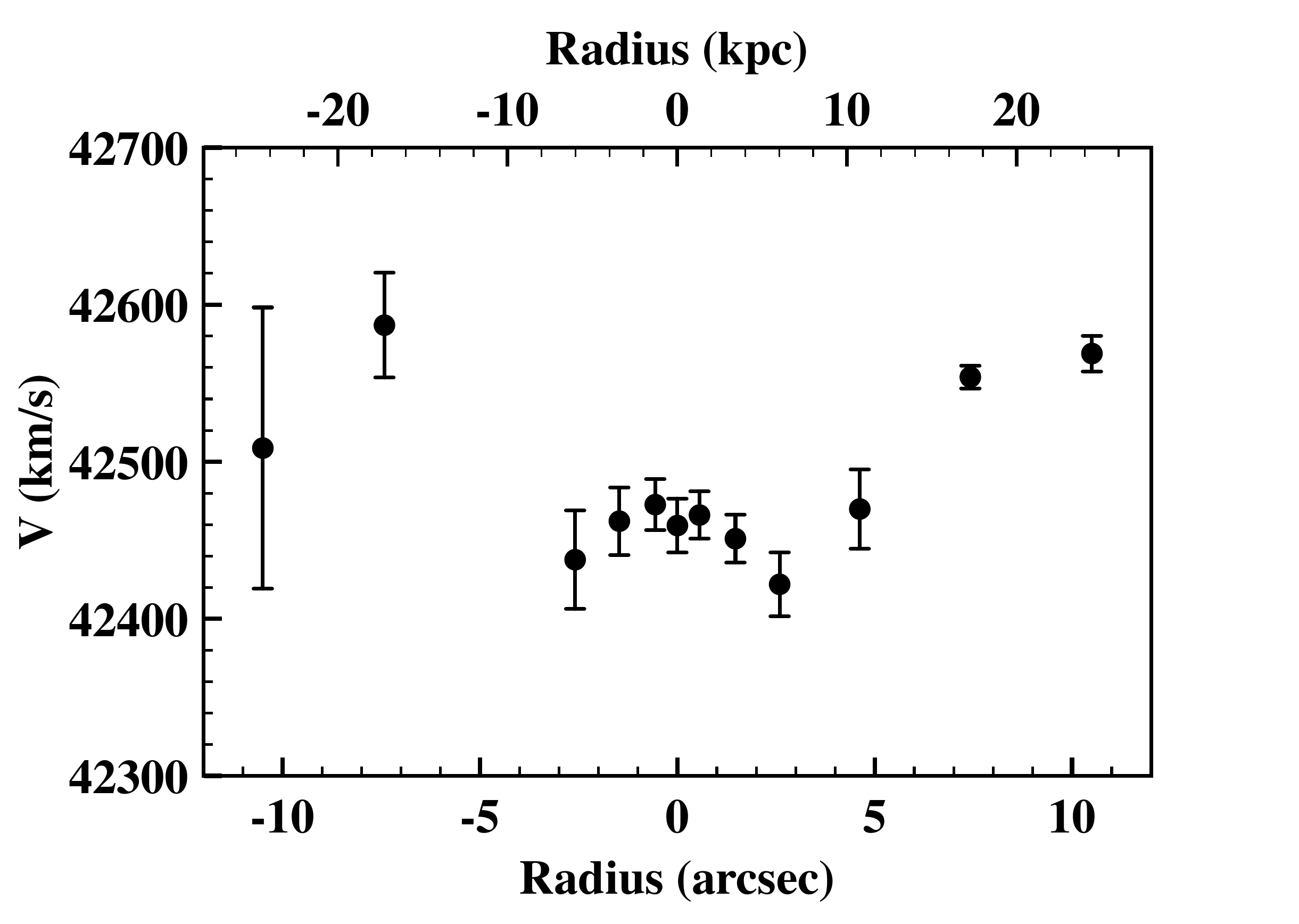}}
         \subfloat[Abell 990]{\includegraphics[scale=0.25]{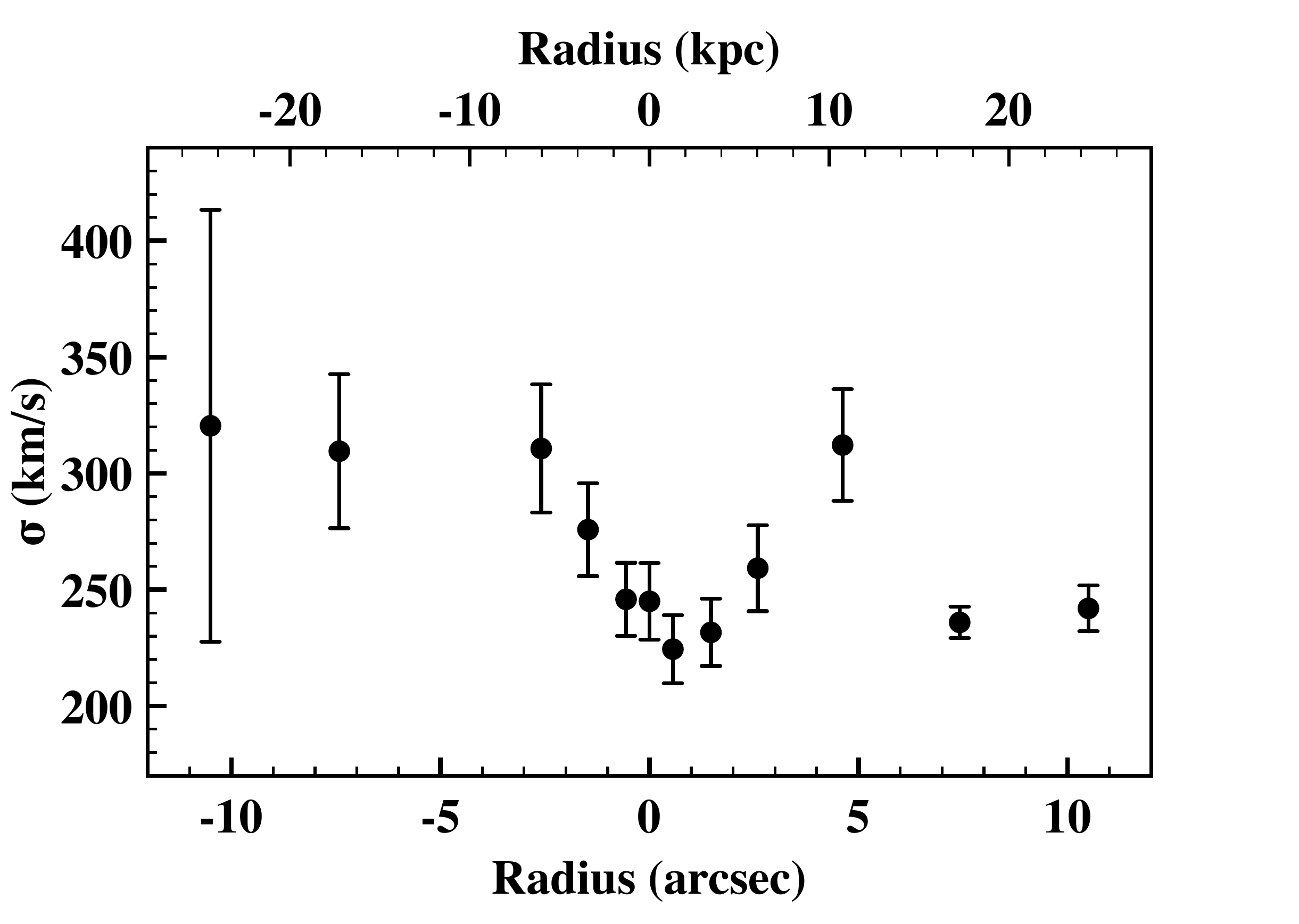}}
   \subfloat{\includegraphics[scale=0.25]{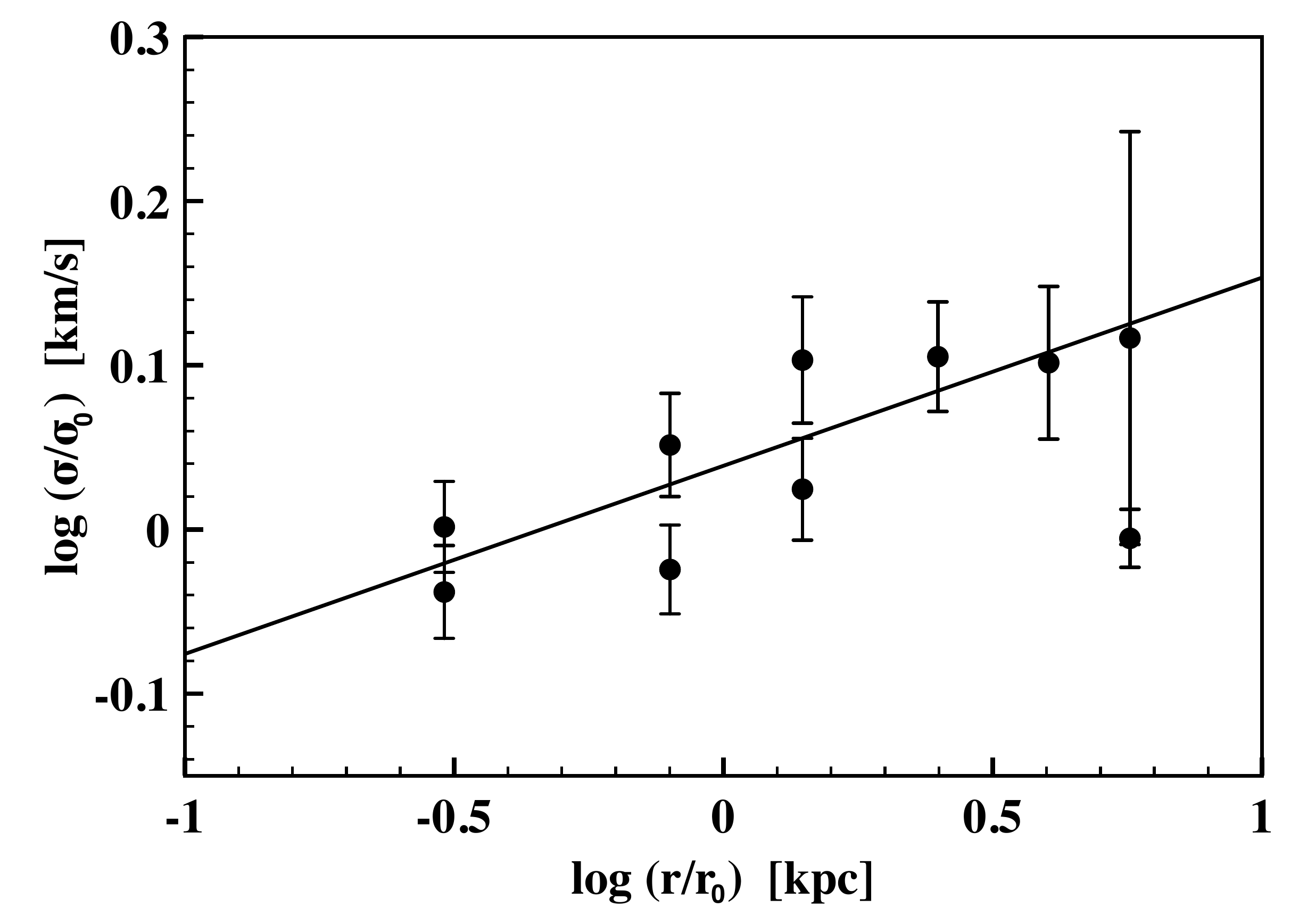}}  \\
    \subfloat{\includegraphics[scale=0.25]{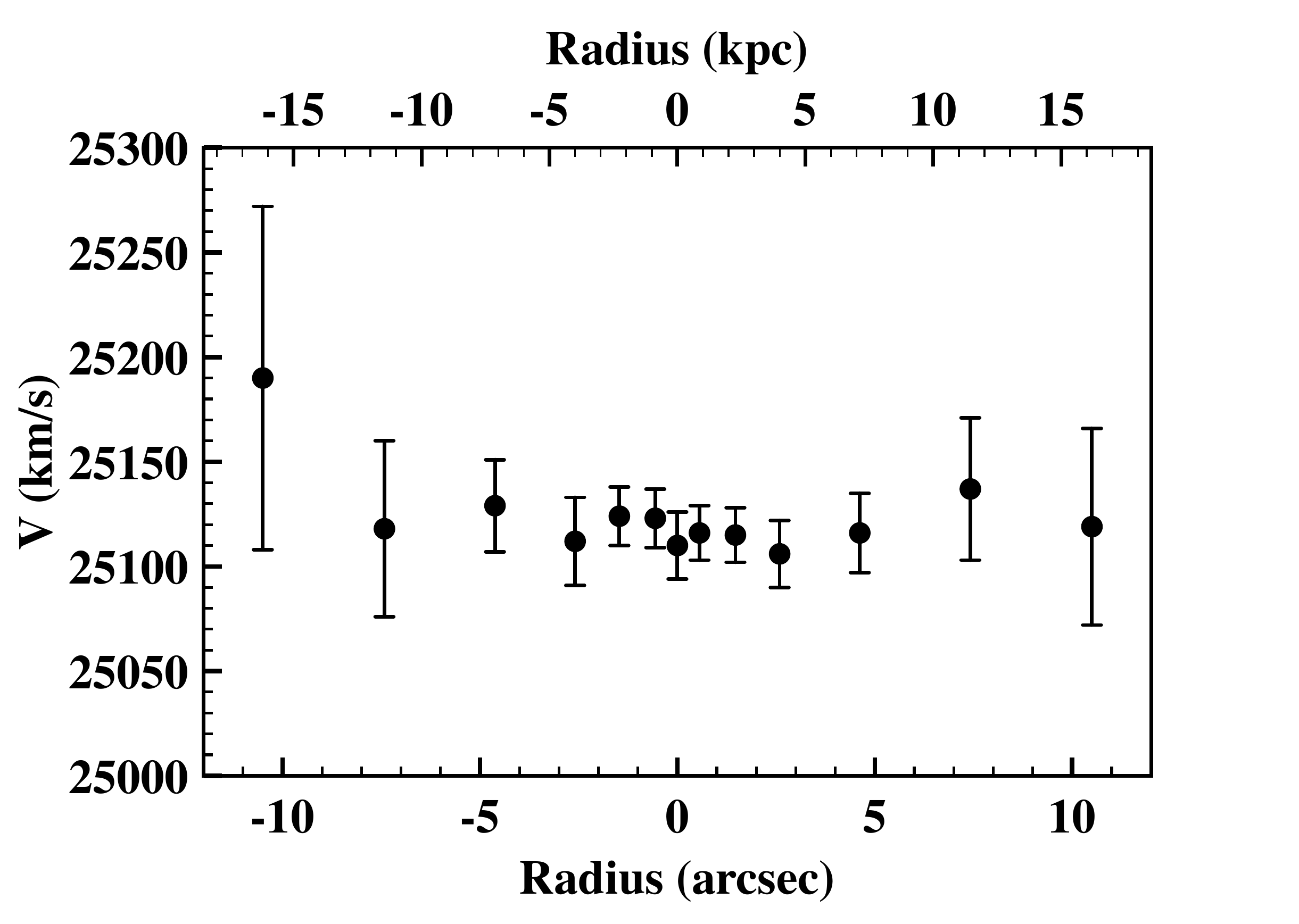}}
         \subfloat[Abell 1650]{\includegraphics[scale=0.25]{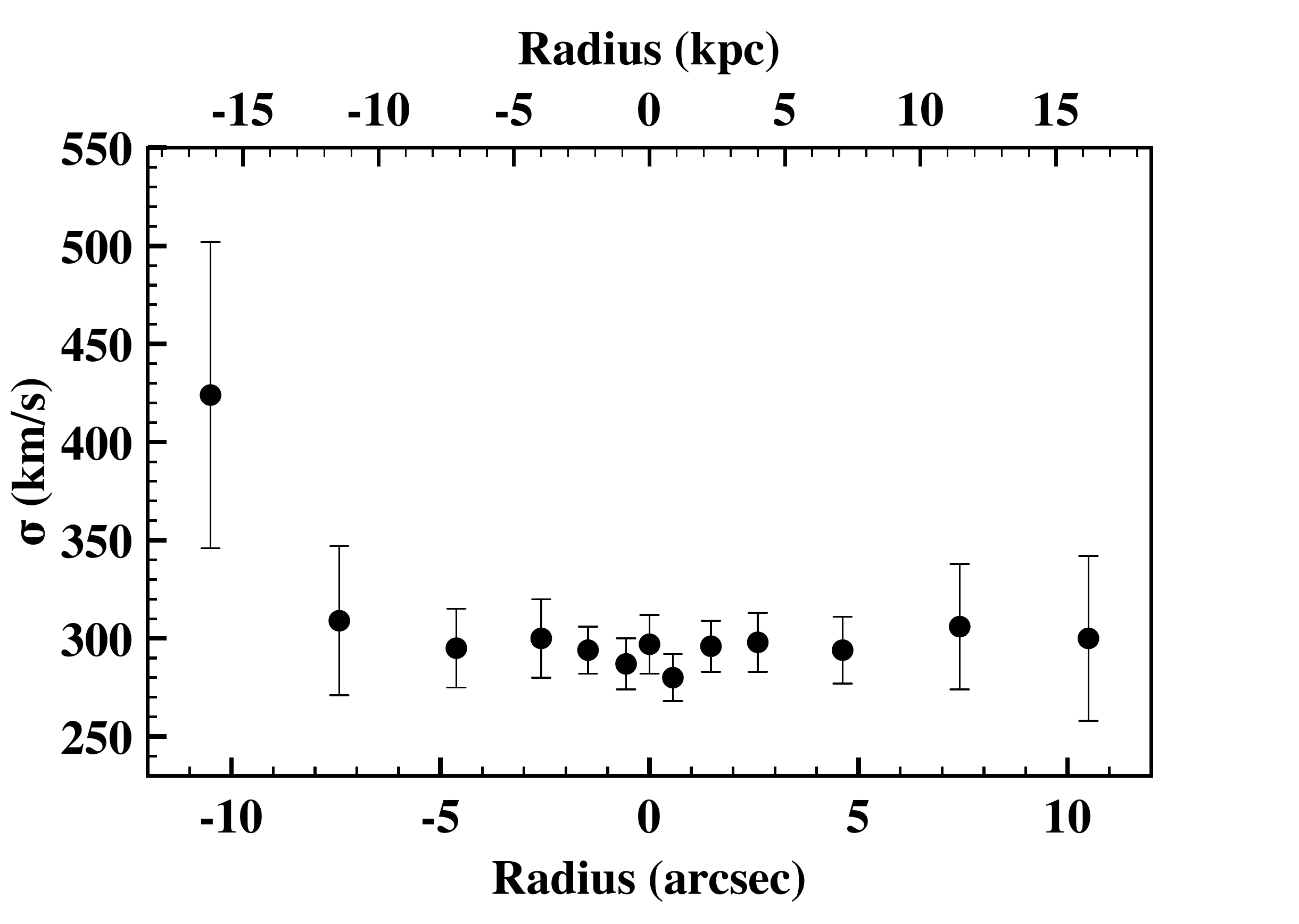}}
   \subfloat{\includegraphics[scale=0.25]{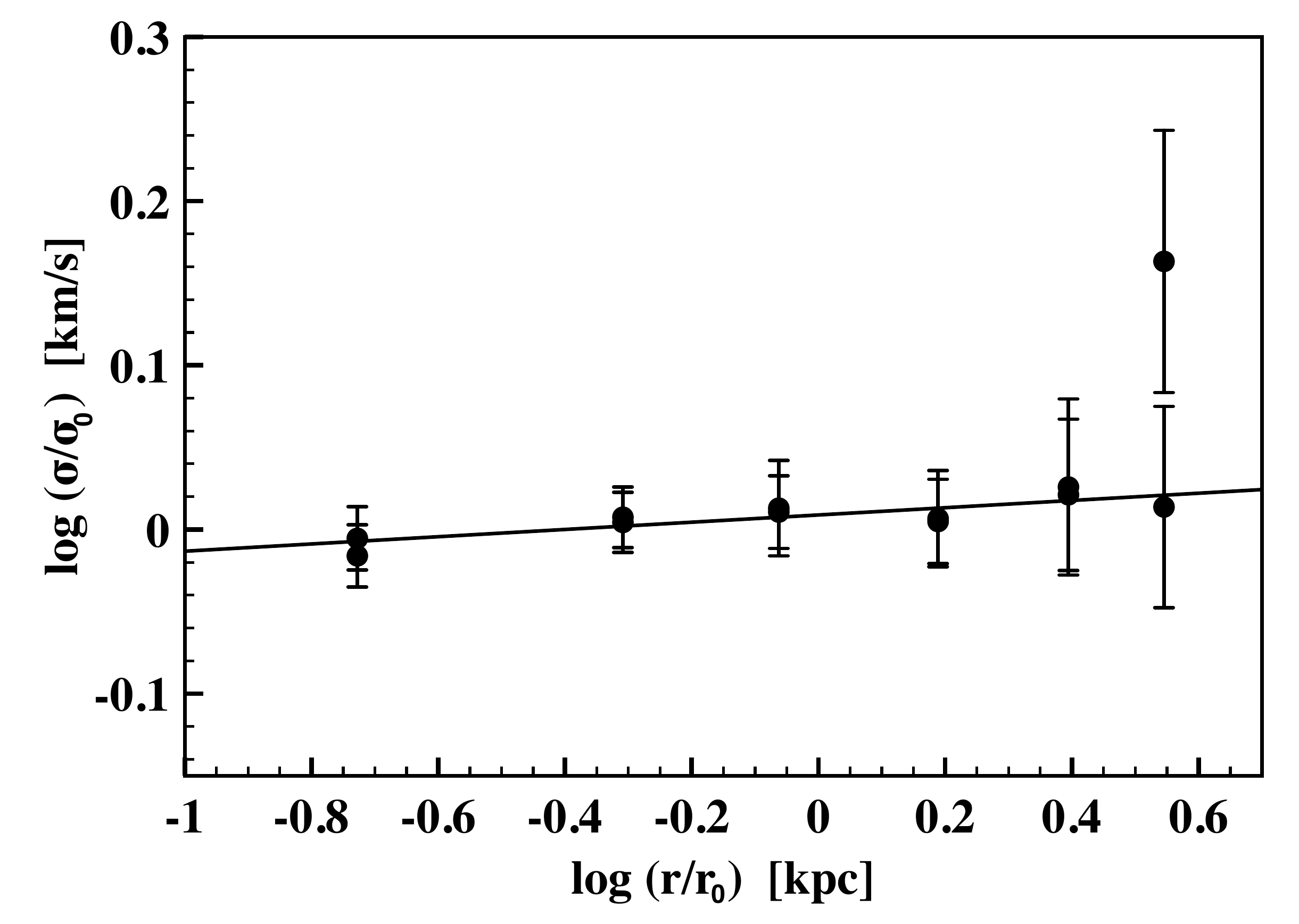}} \\
  \caption{[a] Radial profiles of velocity (V), [b] velocity dispersion ($\sigma$, and [c] power law fit). The grey squares in the velocity dispersion profile of Abell 780 indicate the measurements from \citet{Loubser2008}, and the grey circles in Abell 963 the measurements from \citet{Newman2013a} (see Appendix \ref{kinematics}).}
\label{fig:kin4}
\end{figure*}

\begin{figure*}
\captionsetup[subfigure]{labelformat=empty} 
   \subfloat{\includegraphics[scale=0.25]{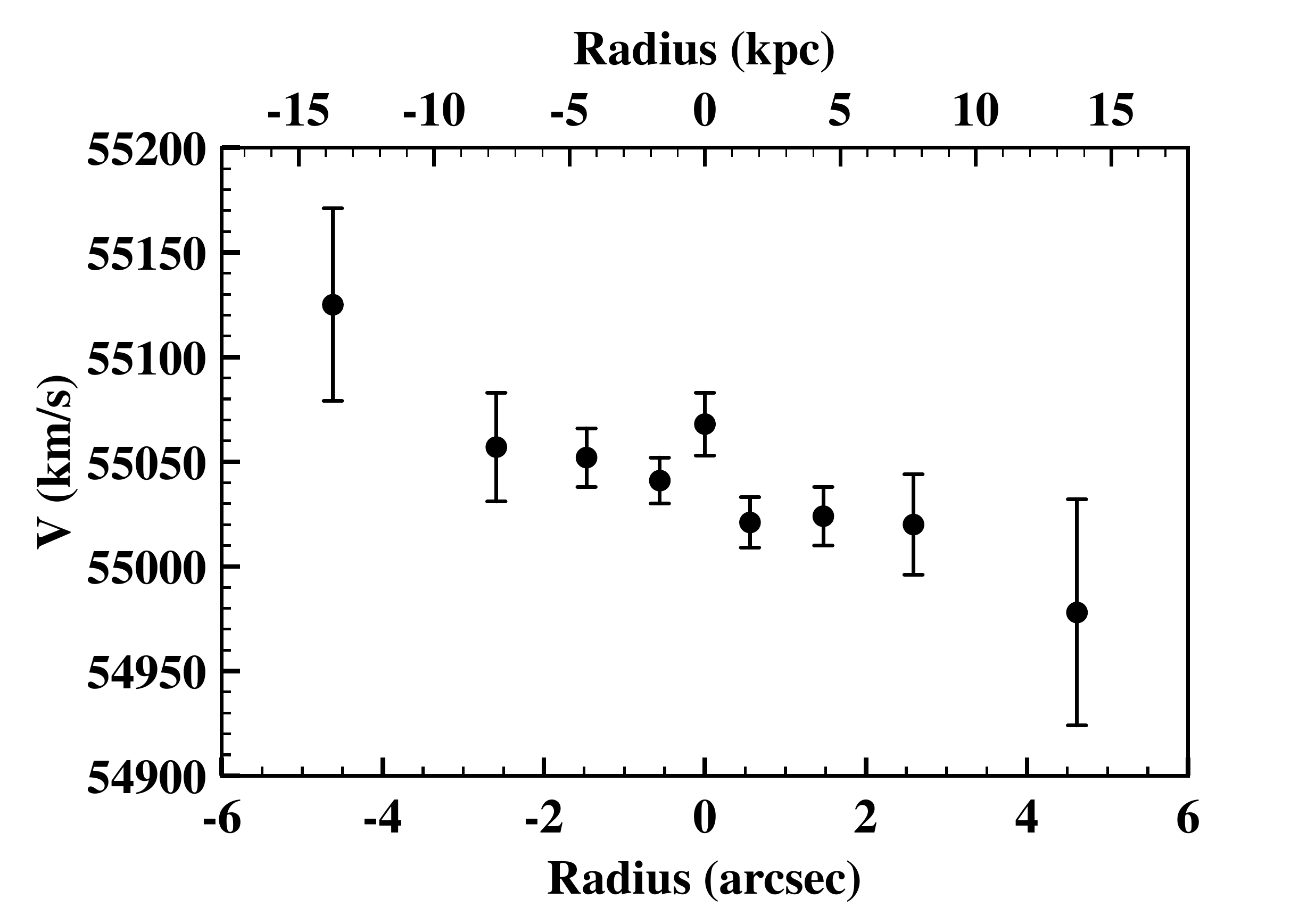}}
         \subfloat[Abell 1689]{\includegraphics[scale=0.25]{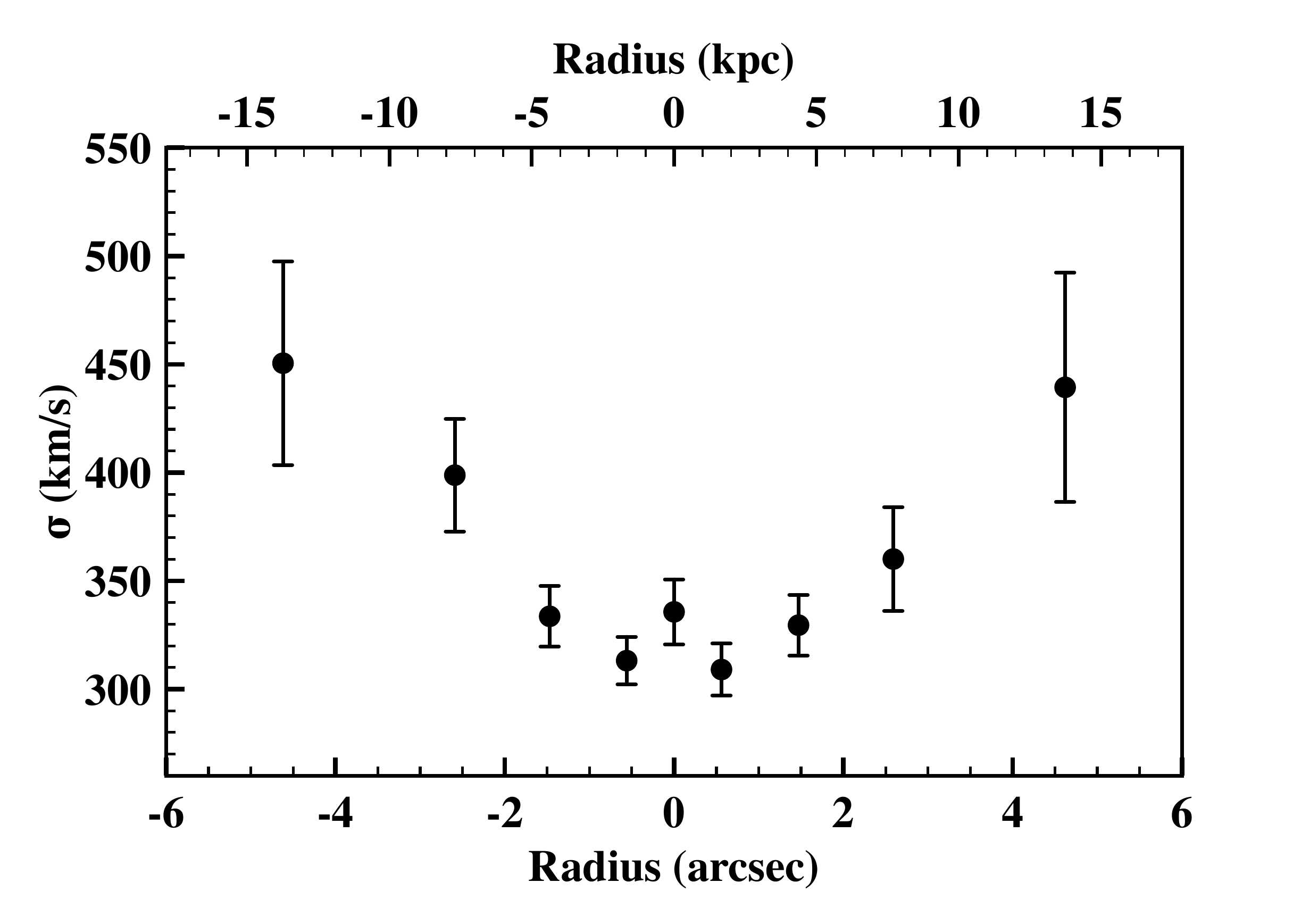}}
   \subfloat{\includegraphics[scale=0.25]{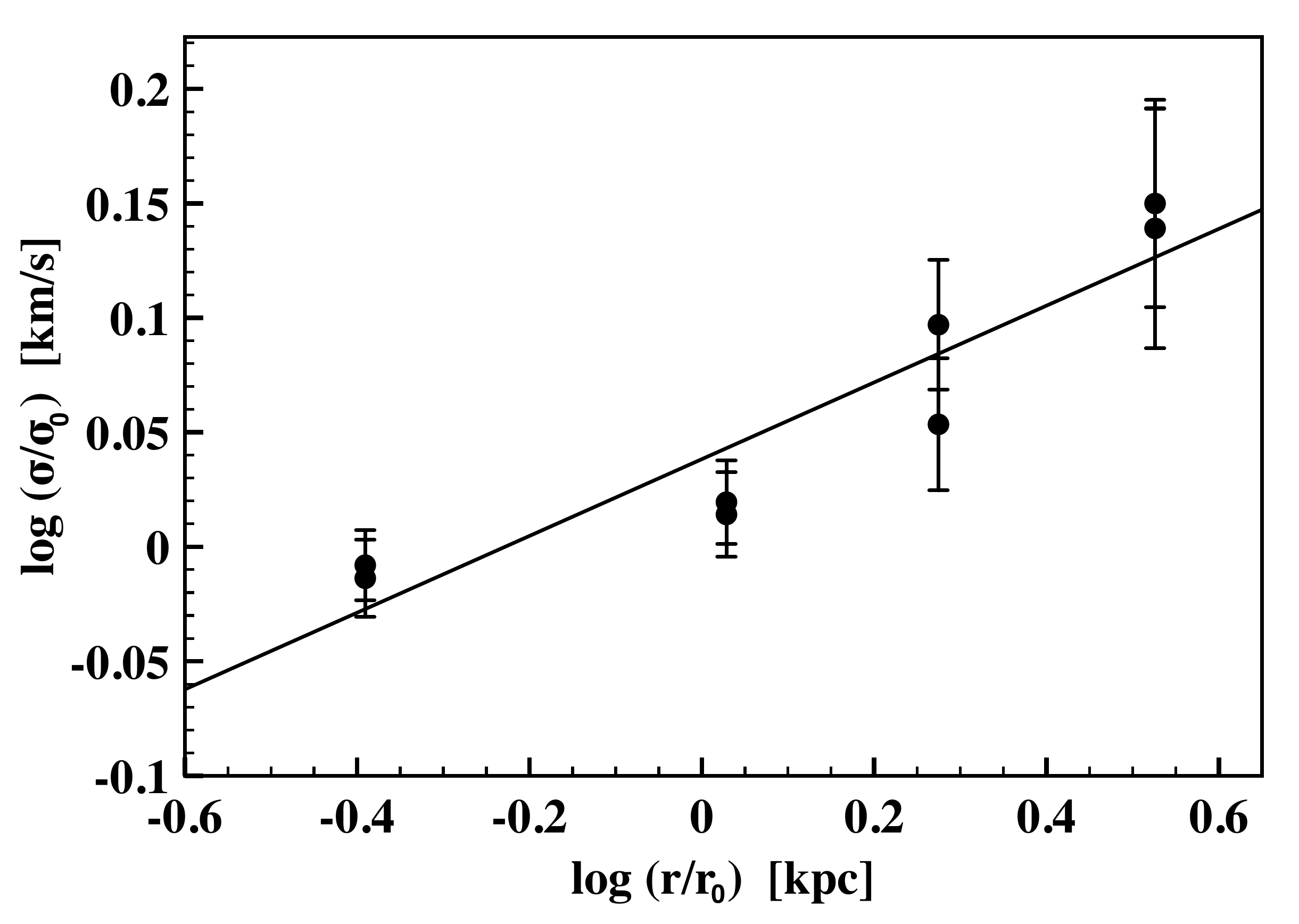}}  \\
     \subfloat{\includegraphics[scale=0.25]{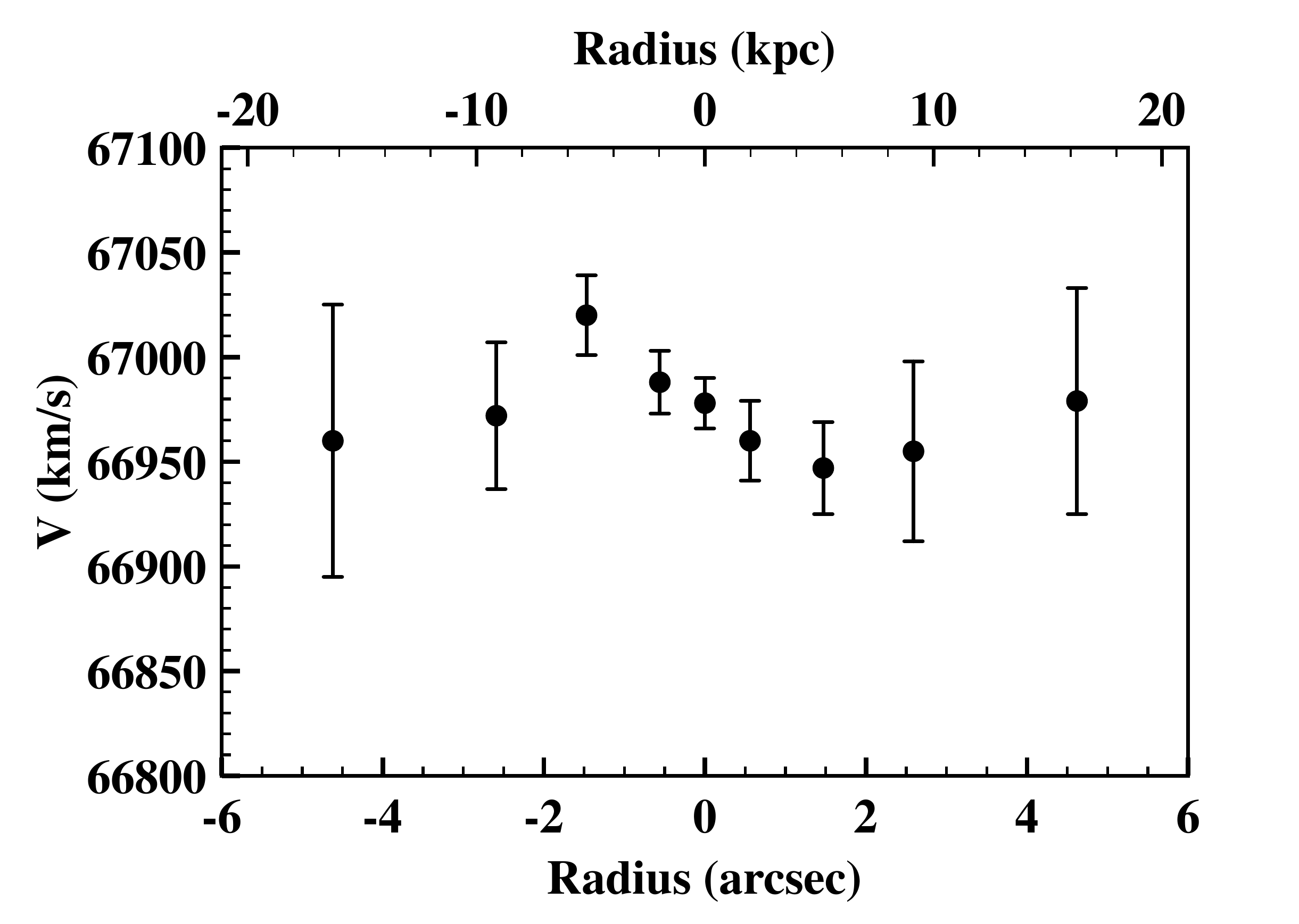}}
         \subfloat[Abell 1763]{\includegraphics[scale=0.25]{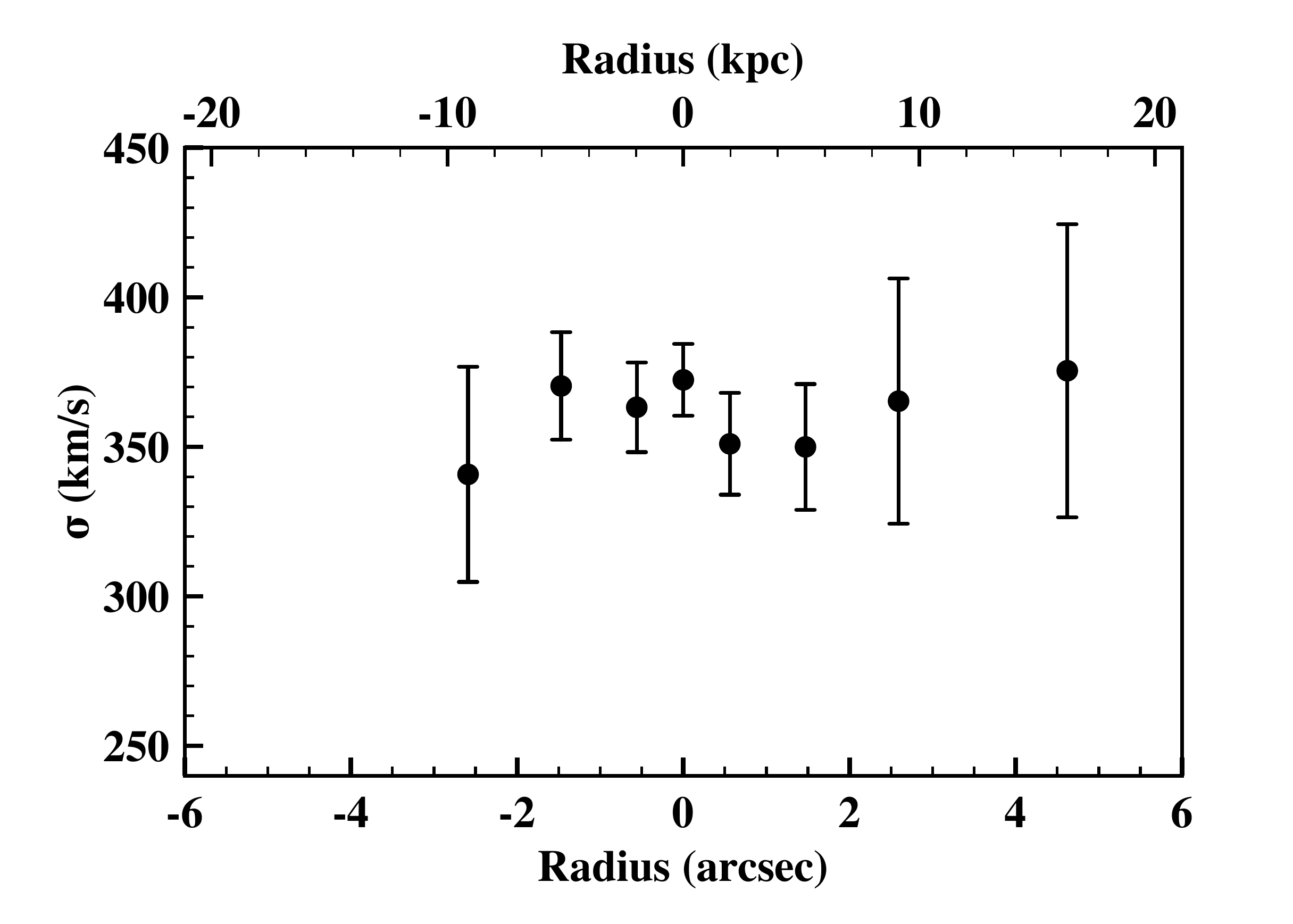}}
   \subfloat{\includegraphics[scale=0.25]{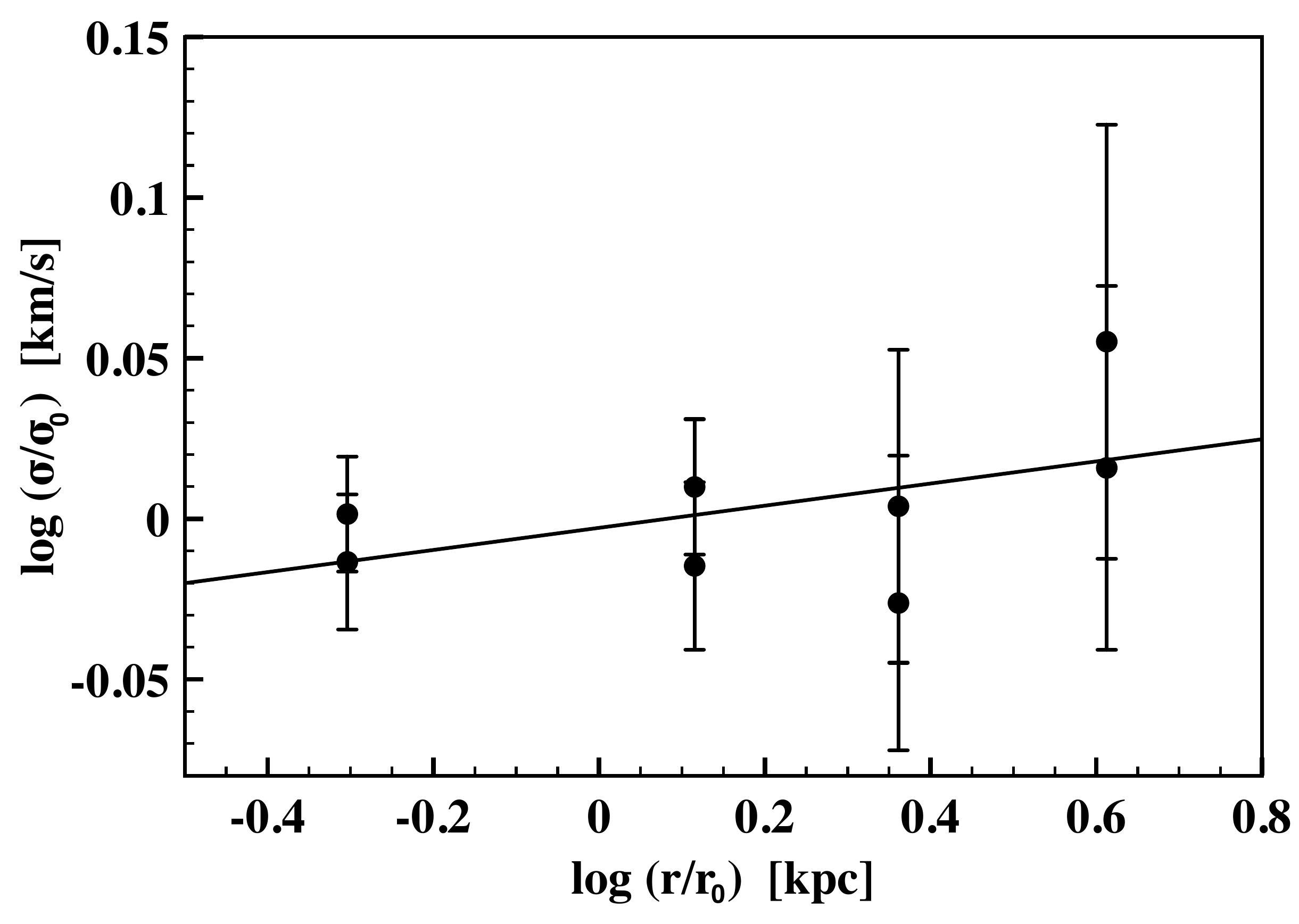}} \\
    \subfloat{\includegraphics[scale=0.25]{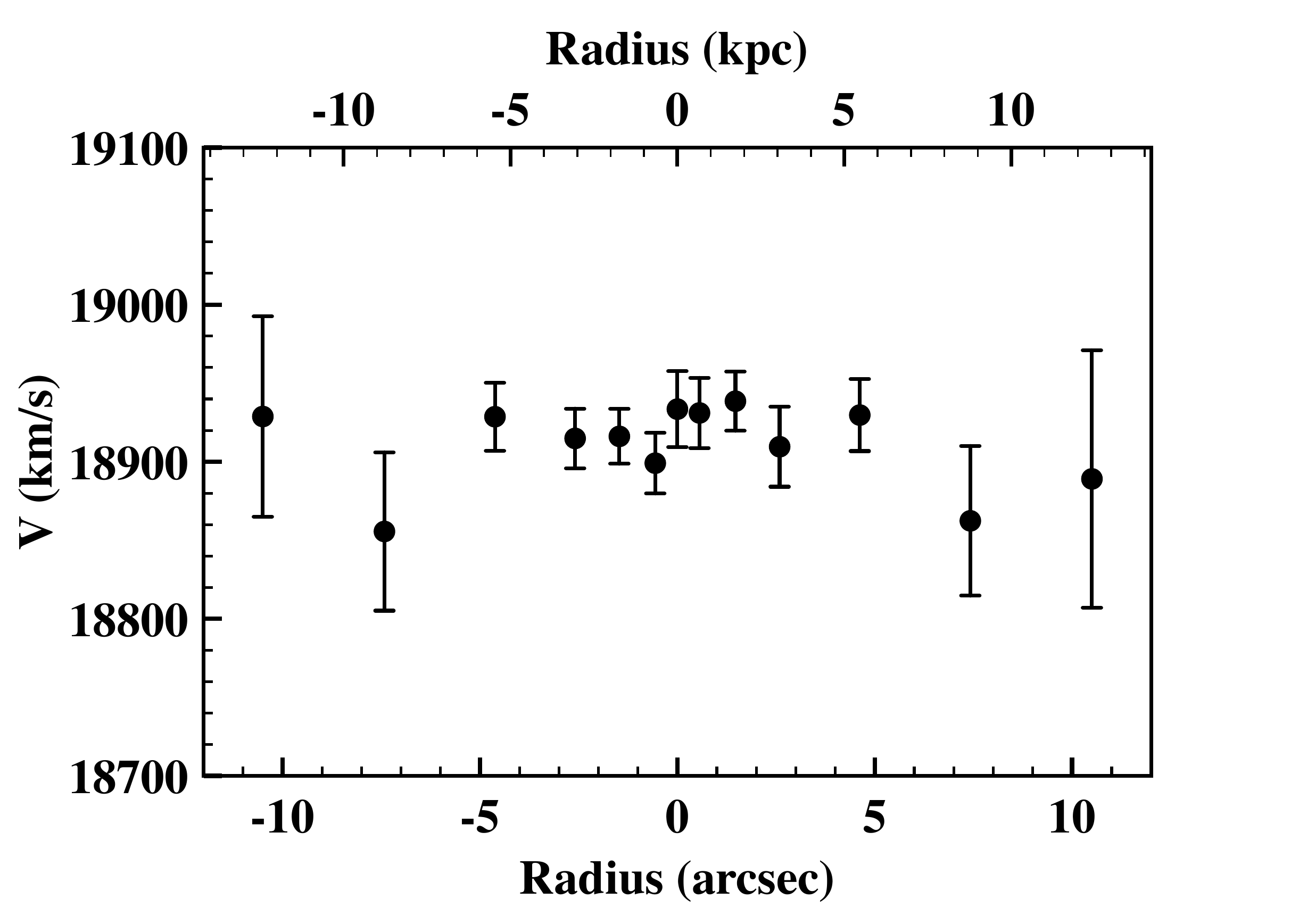}}
         \subfloat[Abell 1795]{\includegraphics[scale=0.25]{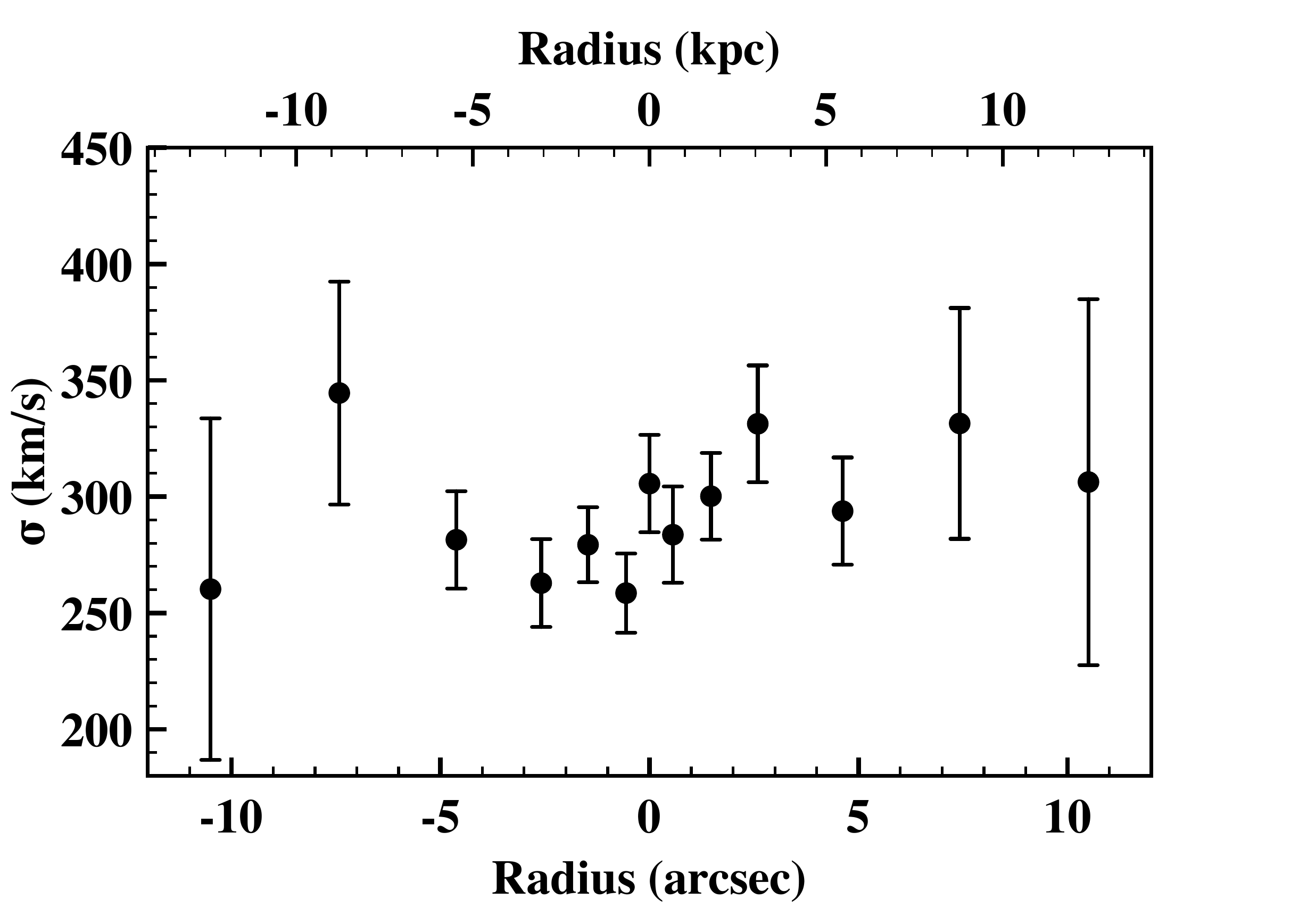}}
   \subfloat{\includegraphics[scale=0.25]{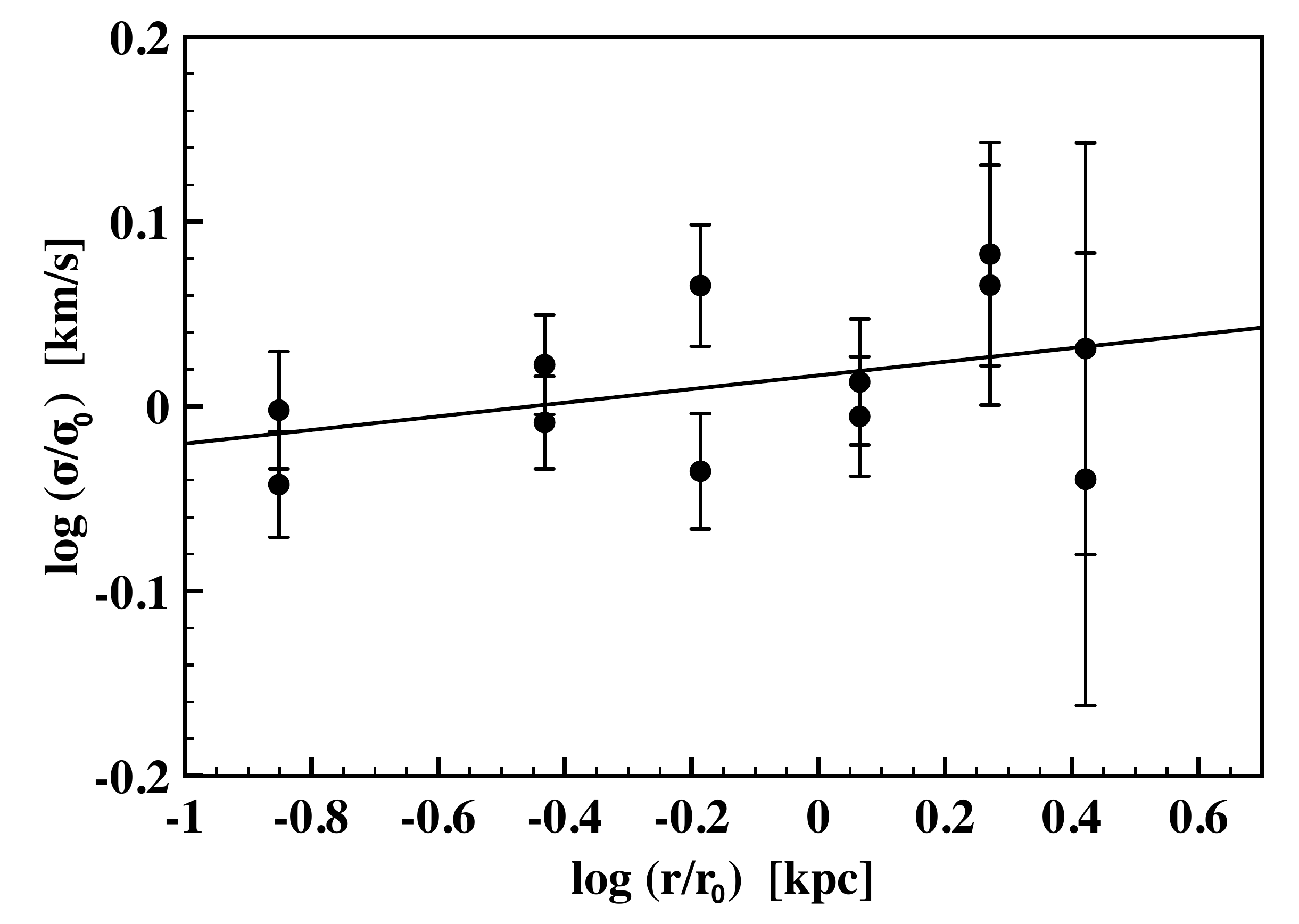}} \\
     \subfloat{\includegraphics[scale=0.25]{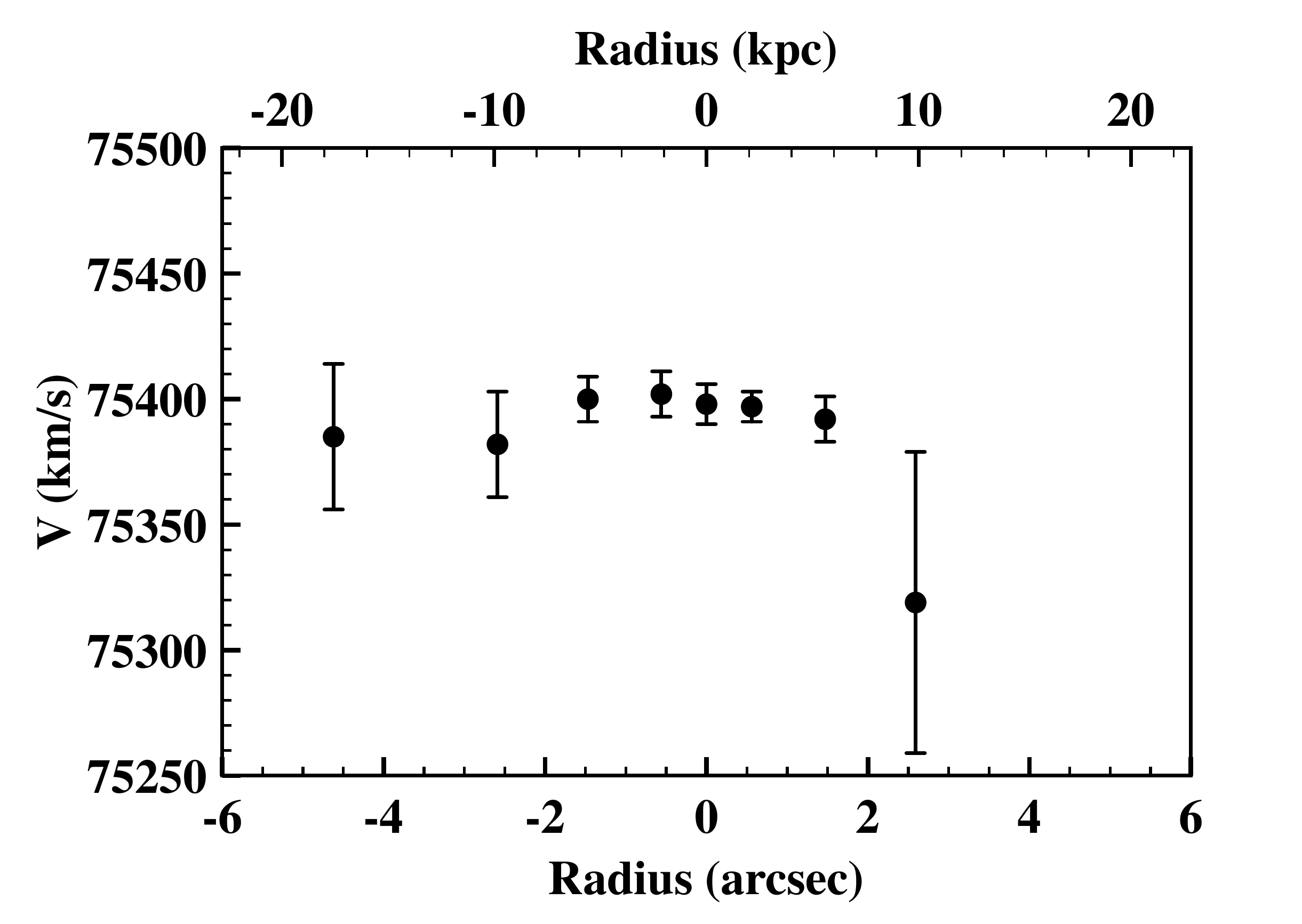}}
         \subfloat[Abell 1835]{\includegraphics[scale=0.25]{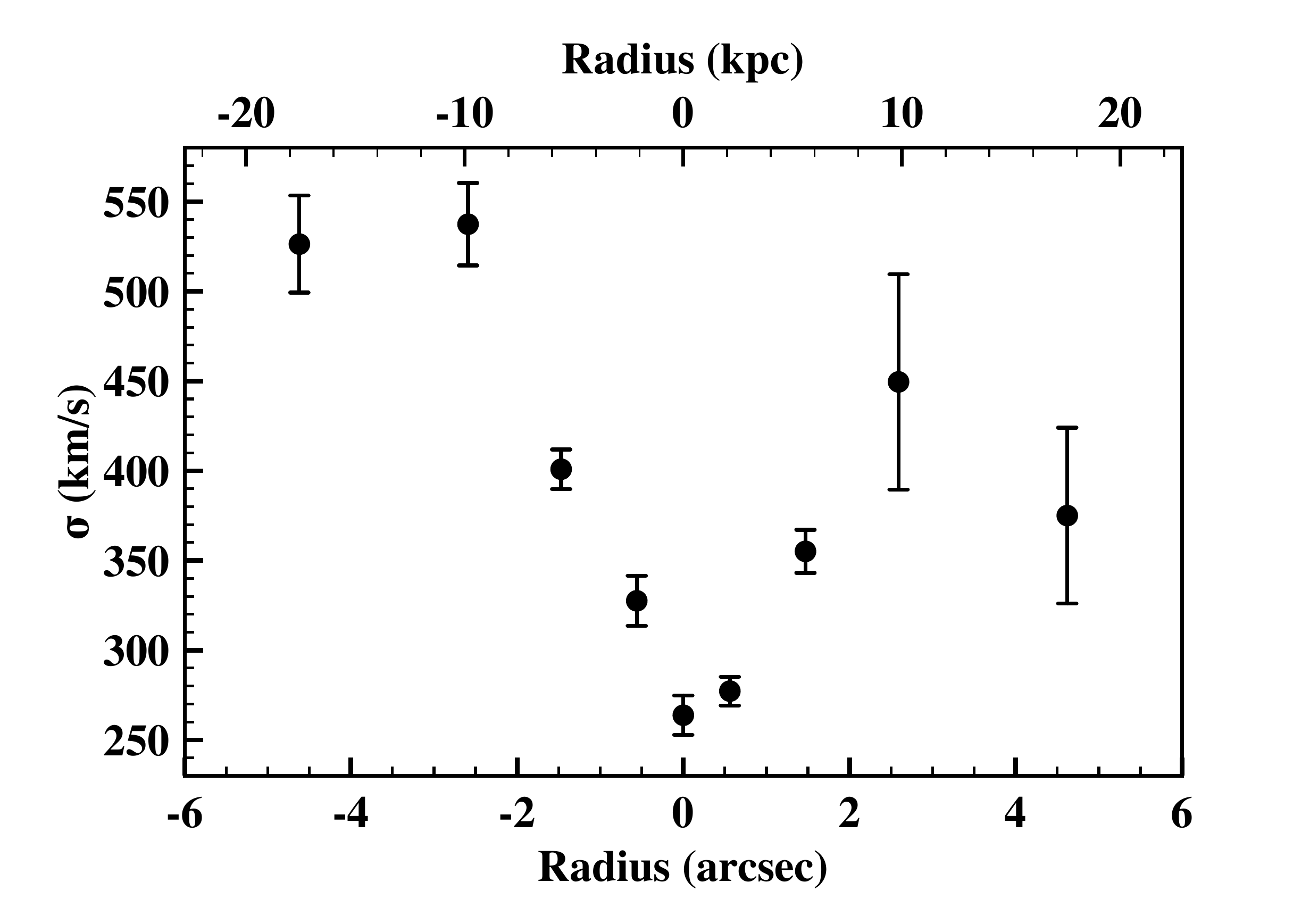}}
   \subfloat{\includegraphics[scale=0.25]{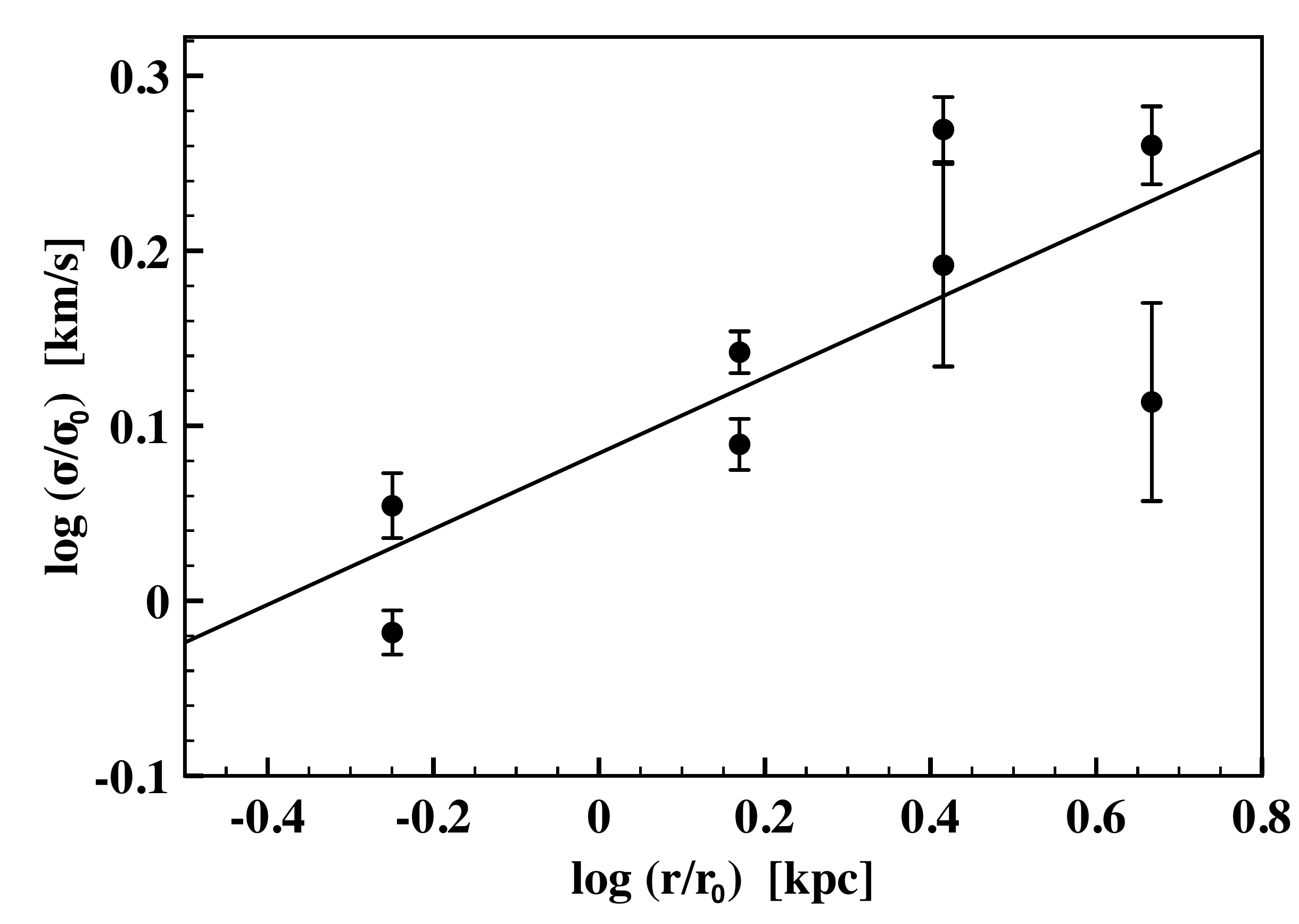}} \\
  \caption{[a] Radial profiles of velocity (V), [b] velocity dispersion ($\sigma$, and [c] power law fit).}
\label{fig:kin5}
\end{figure*}

\begin{figure*}
\captionsetup[subfigure]{labelformat=empty}
      \subfloat{\includegraphics[scale=0.25]{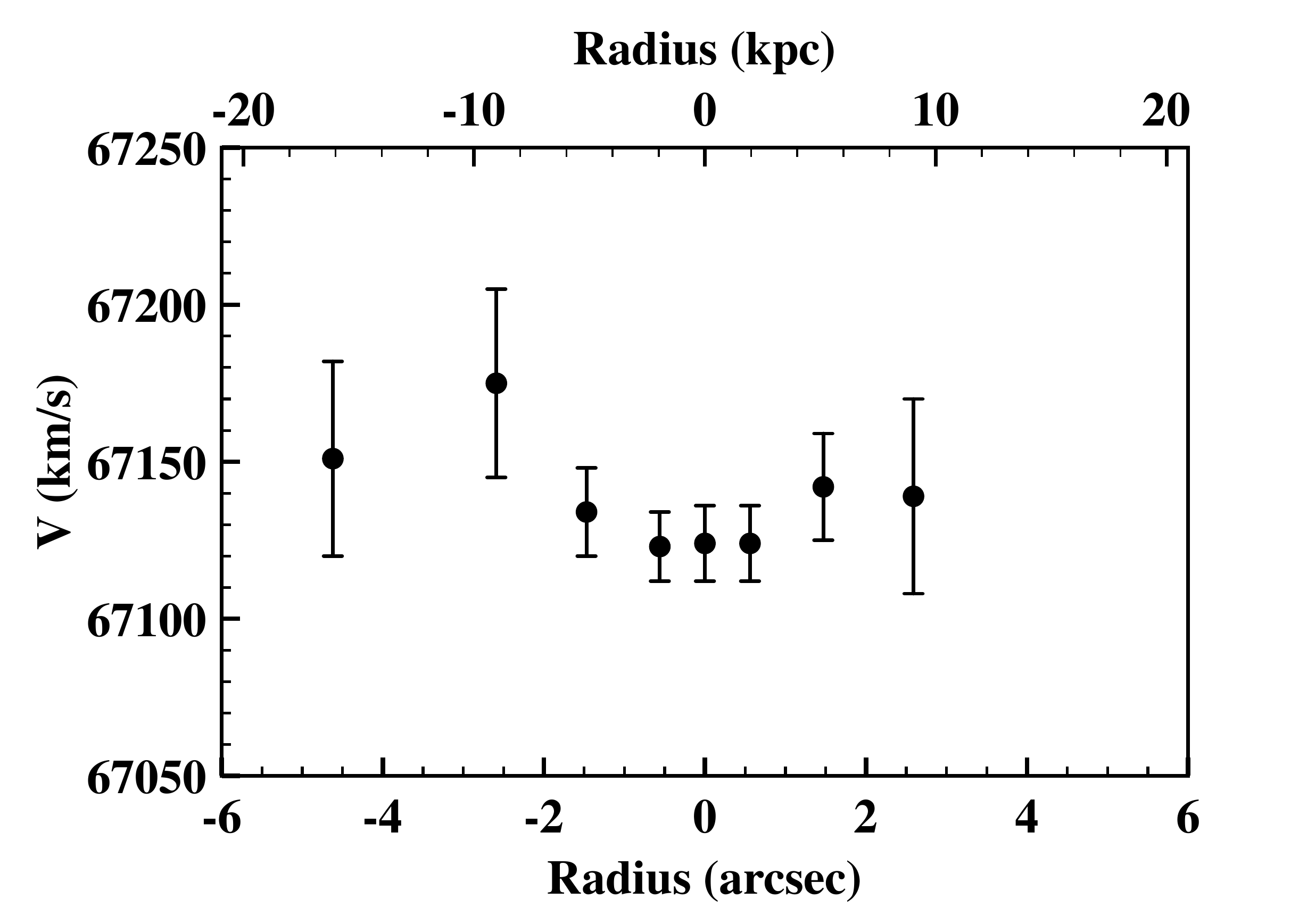}}
         \subfloat[Abell 1942]{\includegraphics[scale=0.25]{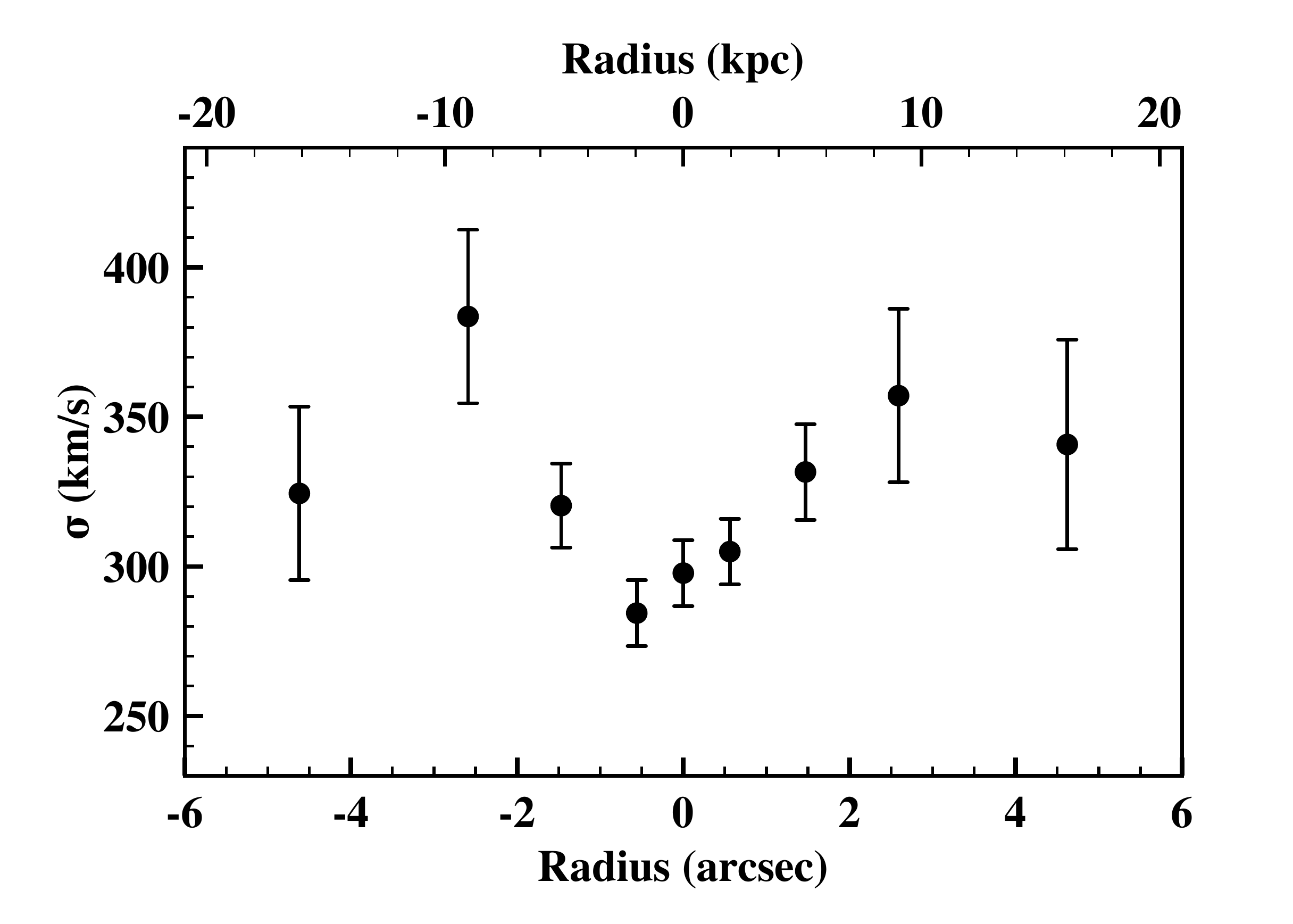}}
   \subfloat{\includegraphics[scale=0.25]{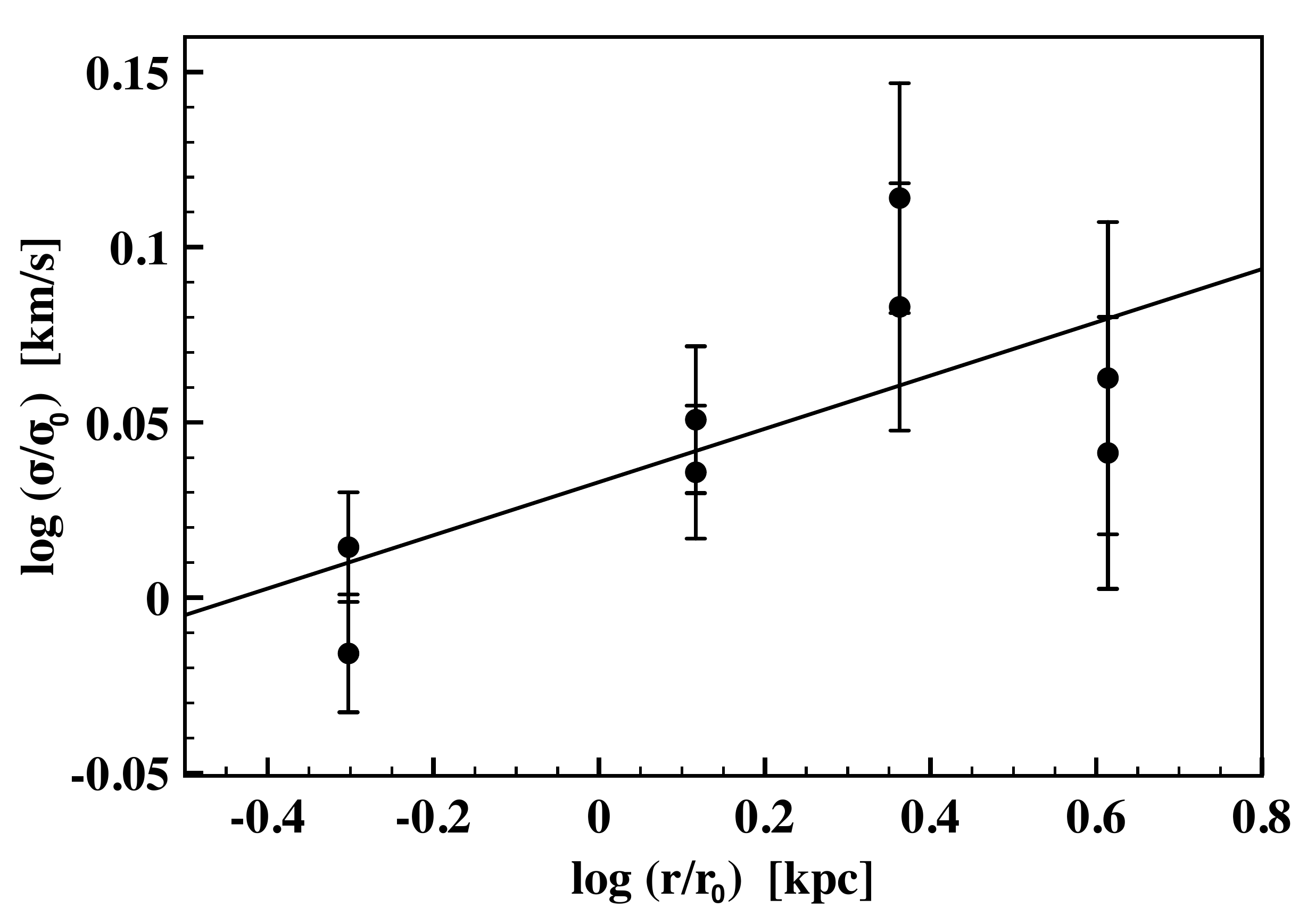}}  \\
    \subfloat{\includegraphics[scale=0.25]{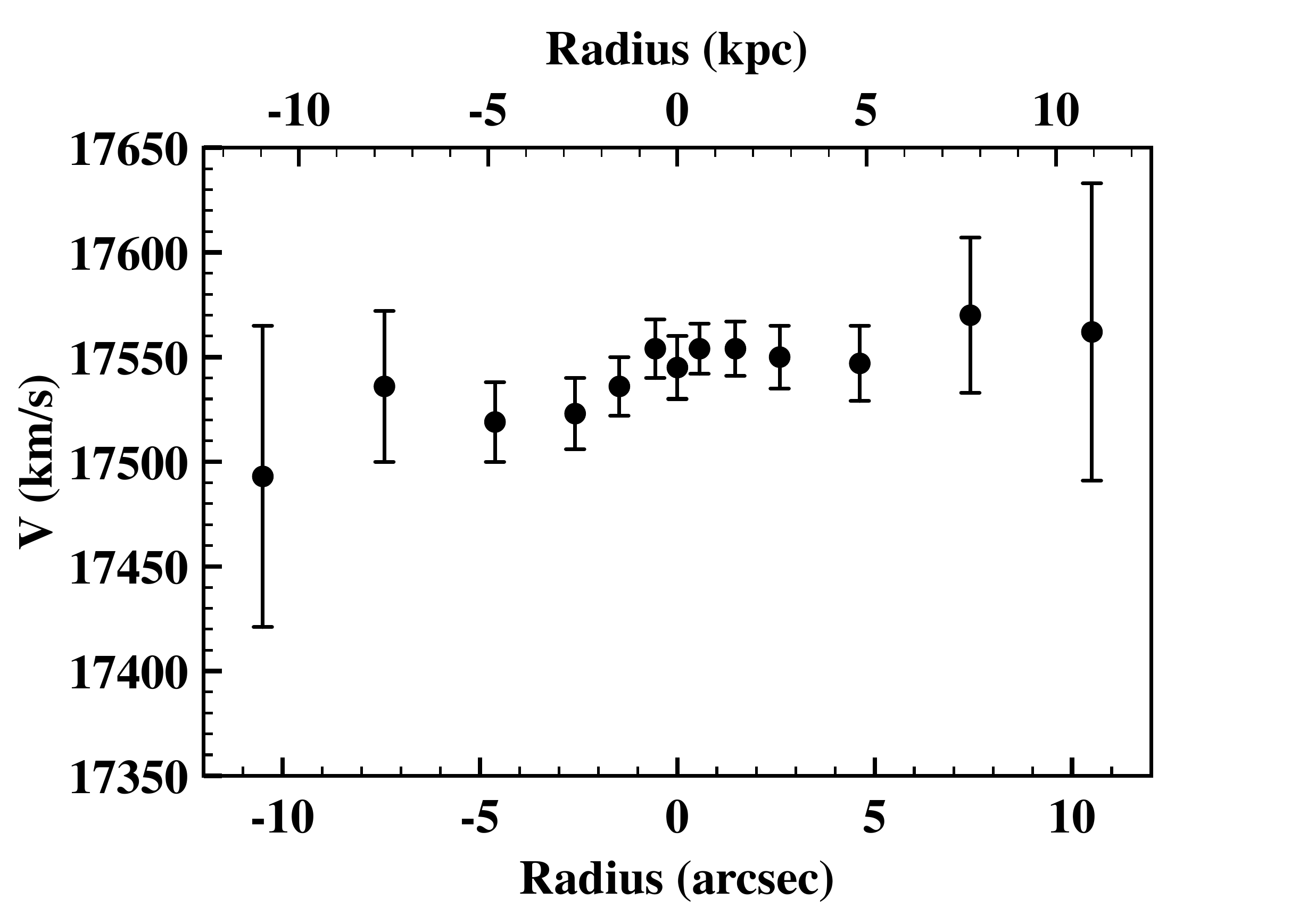}}
         \subfloat[Abell 1991]{\includegraphics[scale=0.25]{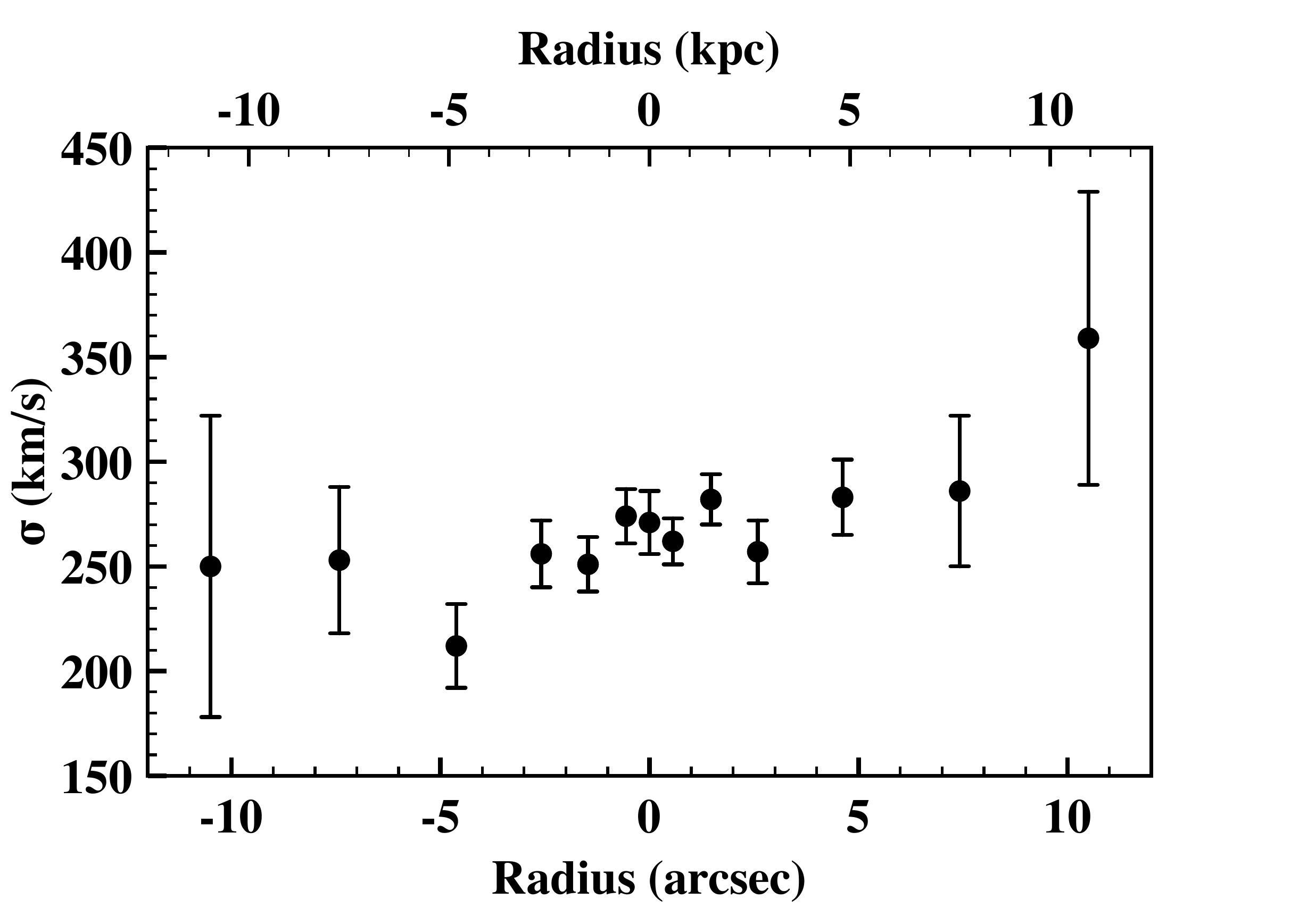}}
   \subfloat{\includegraphics[scale=0.25]{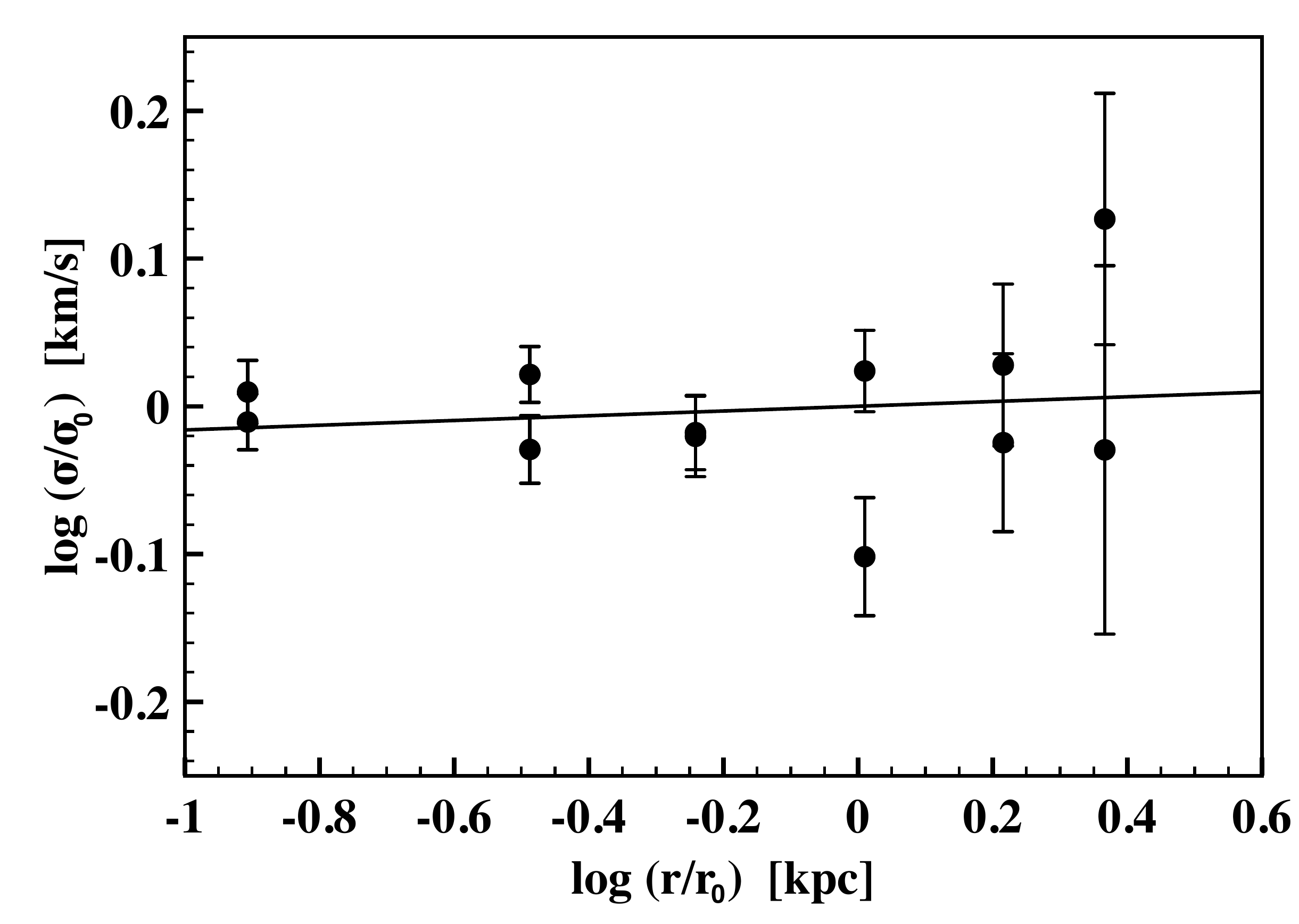}} \\
   \subfloat{\includegraphics[scale=0.25]{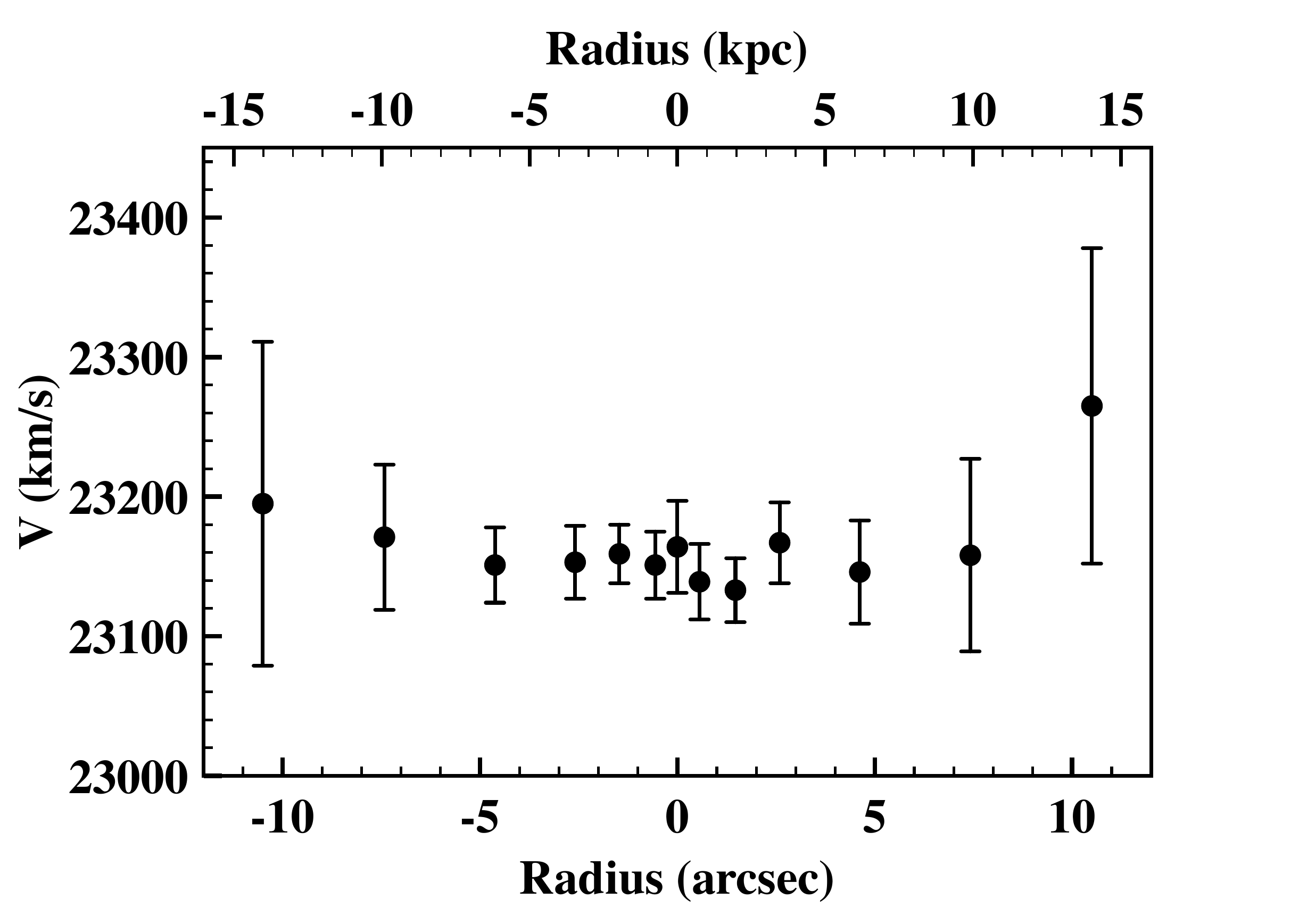}}
         \subfloat[Abell 2029]{\includegraphics[scale=0.25]{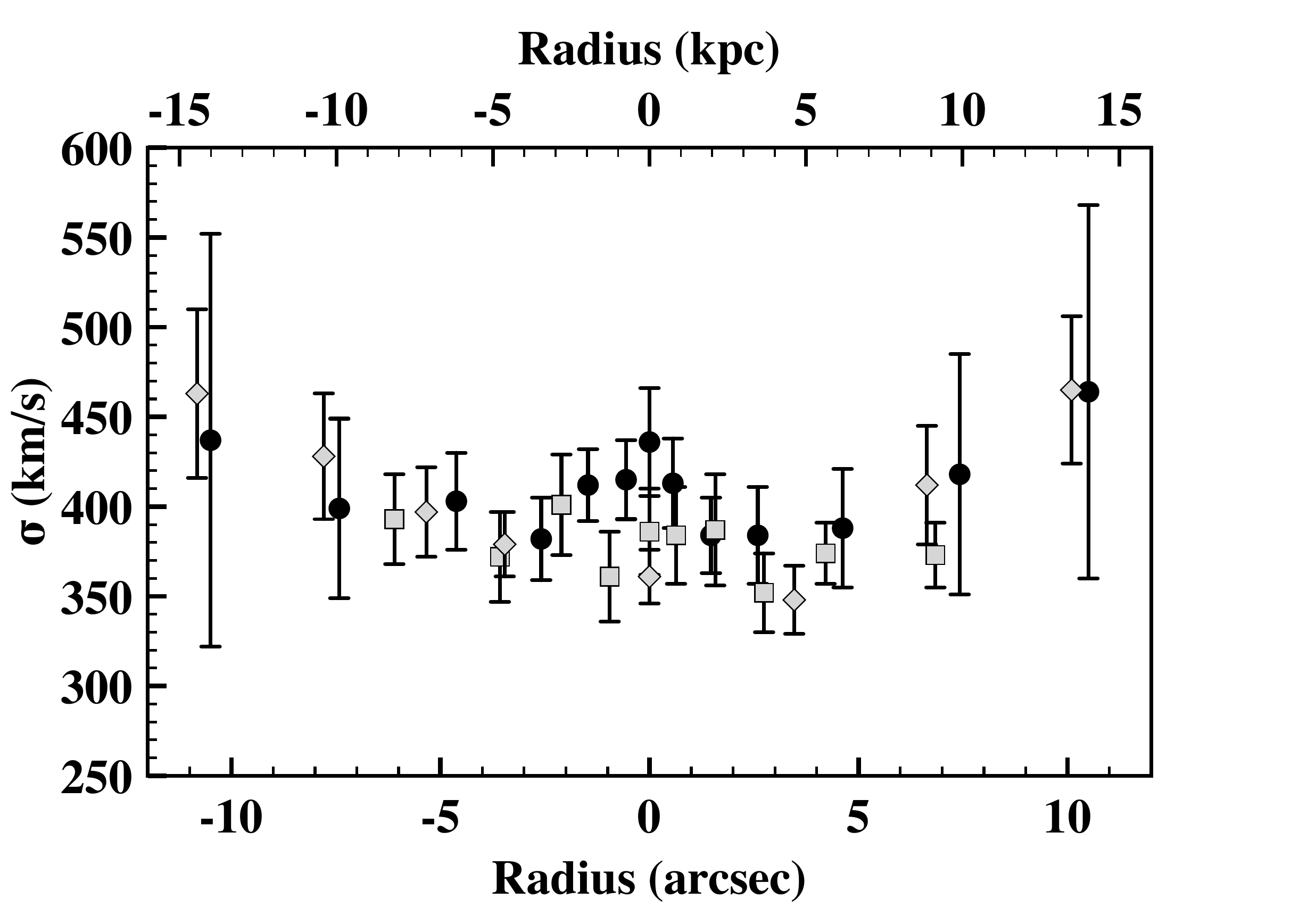}}
   \subfloat{\includegraphics[scale=0.25]{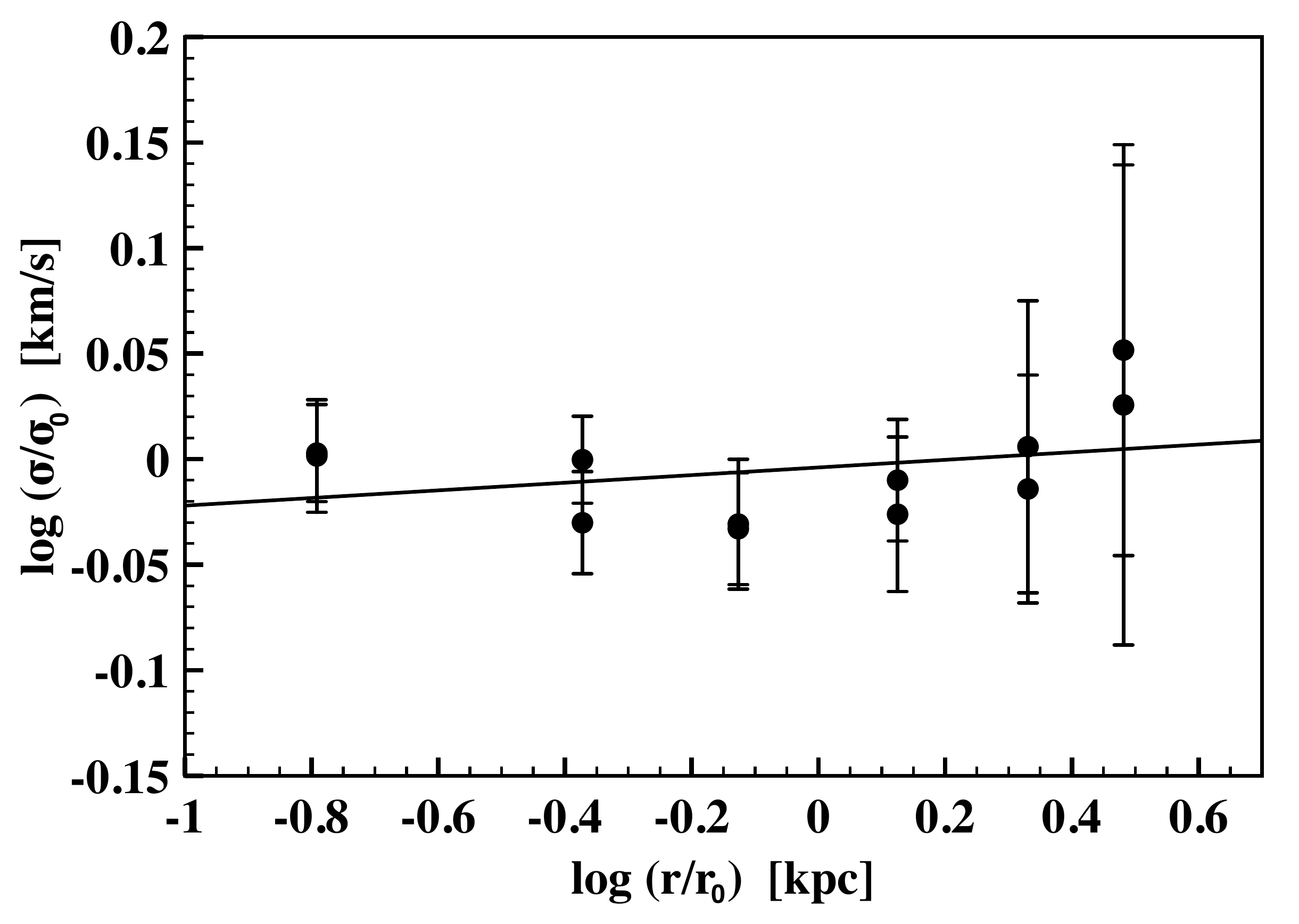}}  \\
     \subfloat{\includegraphics[scale=0.25]{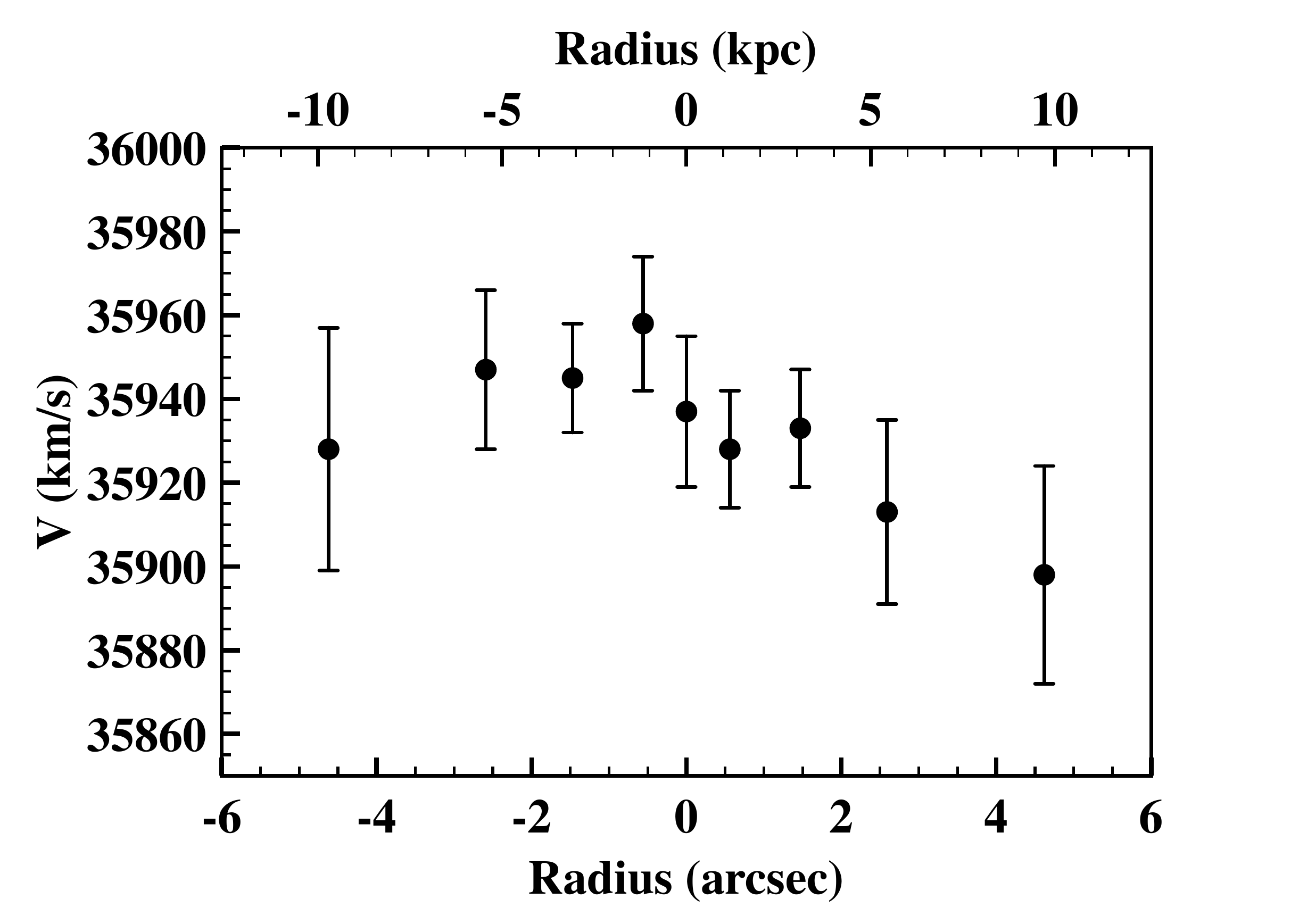}}
         \subfloat[Abell 2050]{\includegraphics[scale=0.25]{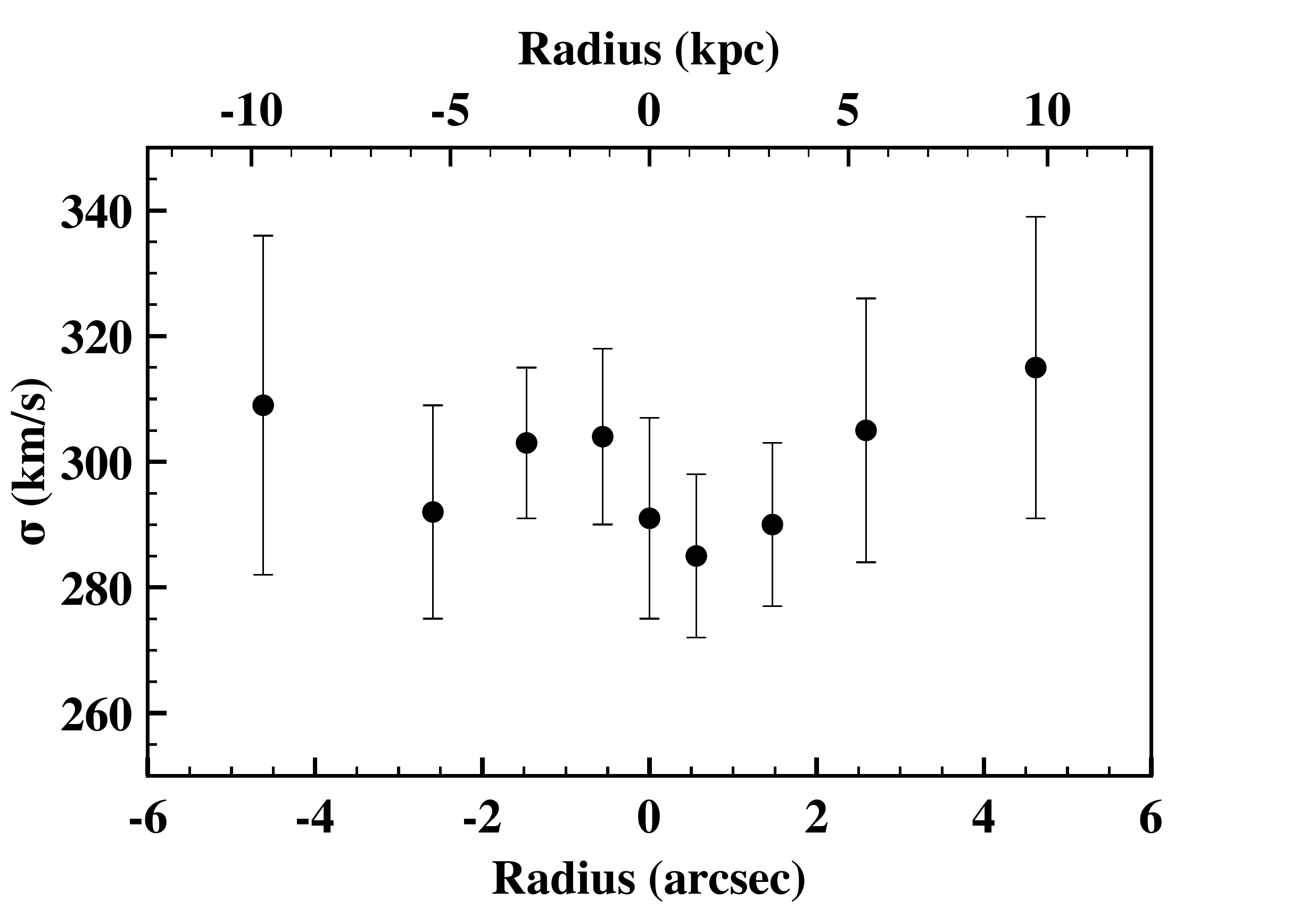}}
   \subfloat{\includegraphics[scale=0.25]{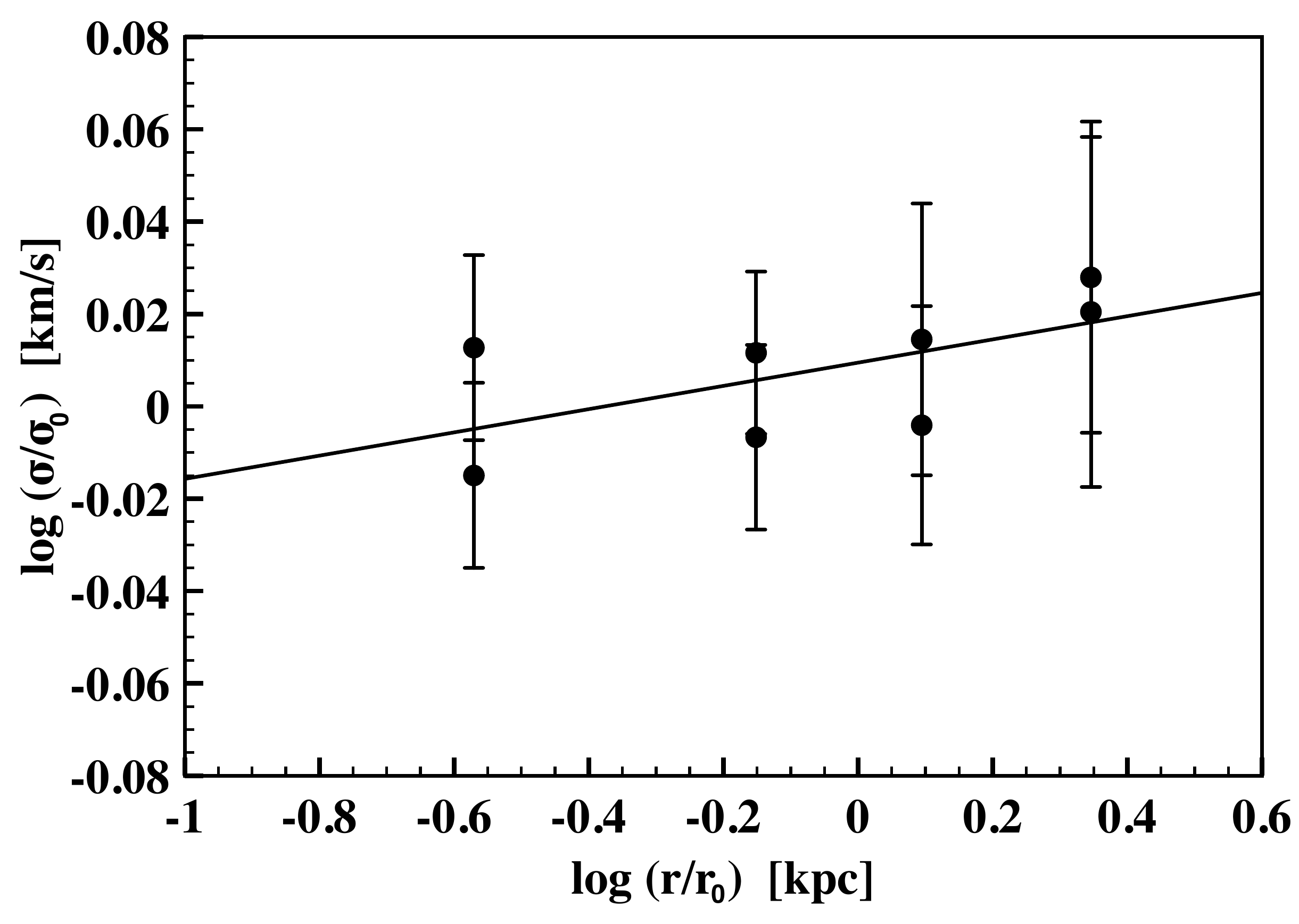}} \\
  \caption{[a] Radial profiles of velocity (V), [b] velocity dispersion ($\sigma$, and [c] power law fit). The grey diamonds and grey squares in the velocity dispersion profile of Abell 2029 indicate the measurements from \citet{Fisher1995} and \citet{Loubser2008}, respectively (see Appendix \ref{kinematics}).}
\label{fig:kin6}
\end{figure*}

\begin{figure*}
\captionsetup[subfigure]{labelformat=empty}
    \subfloat{\includegraphics[scale=0.25]{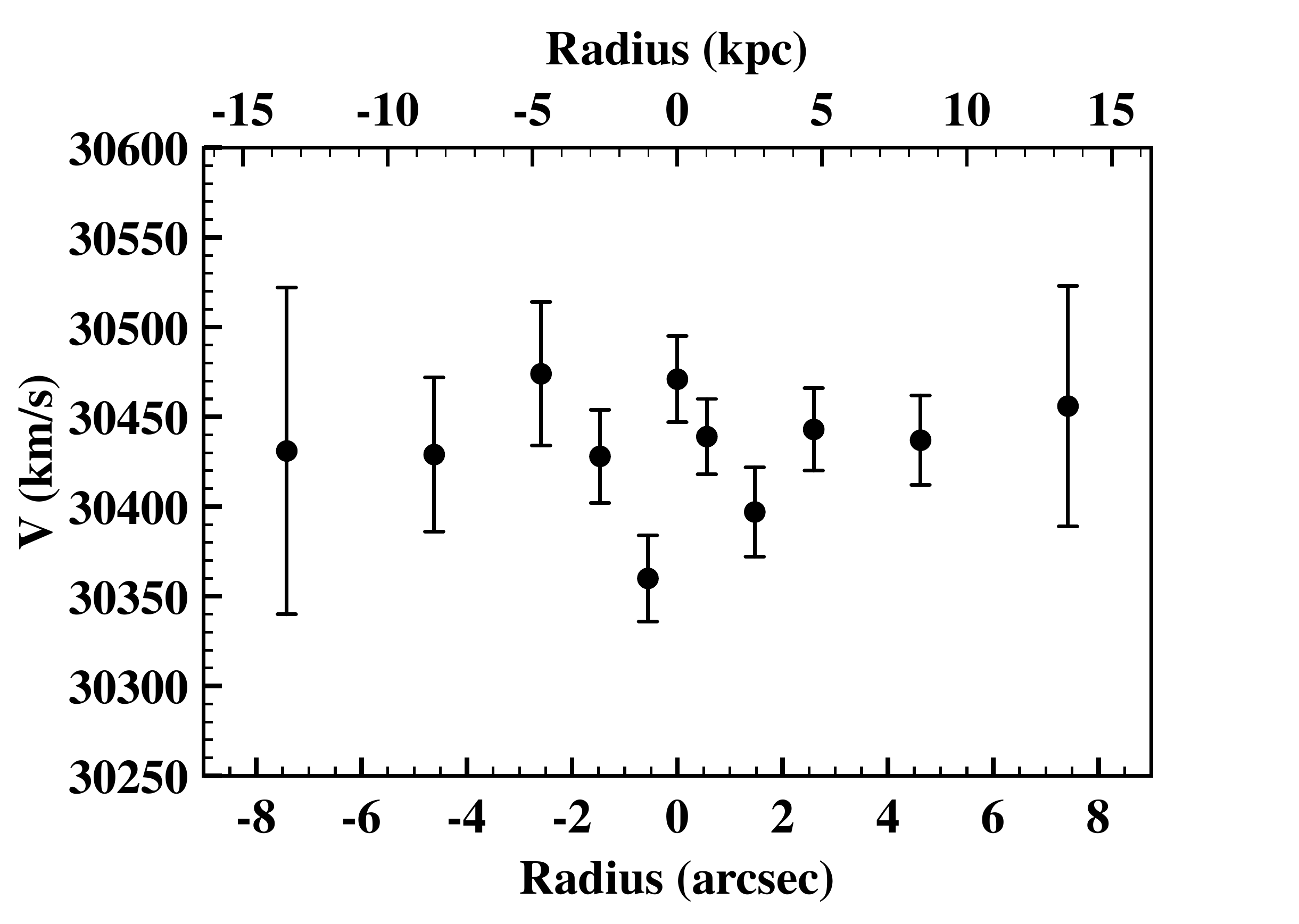}}
         \subfloat[Abell 2055]{\includegraphics[scale=0.25]{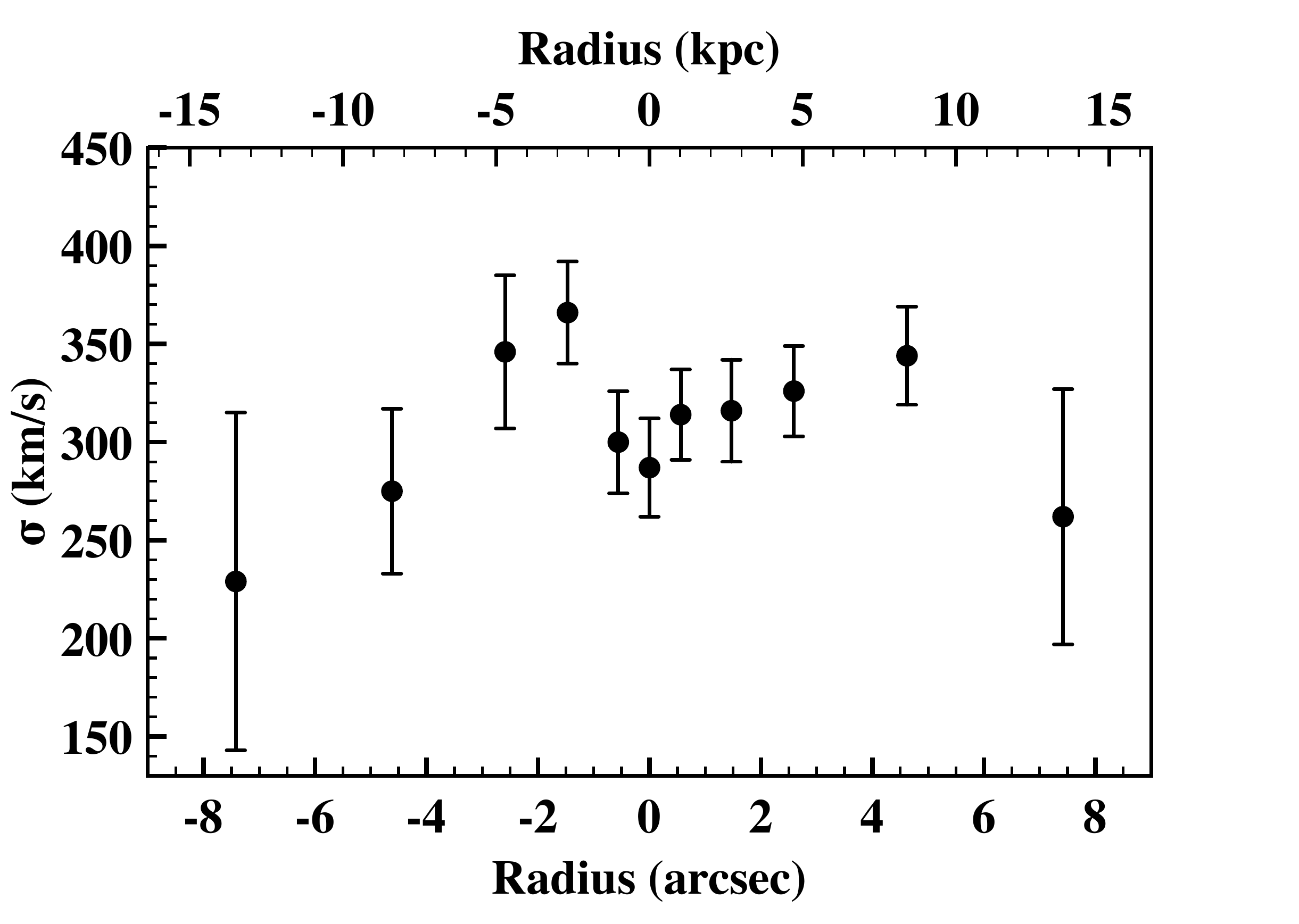}}
   \subfloat{\includegraphics[scale=0.25]{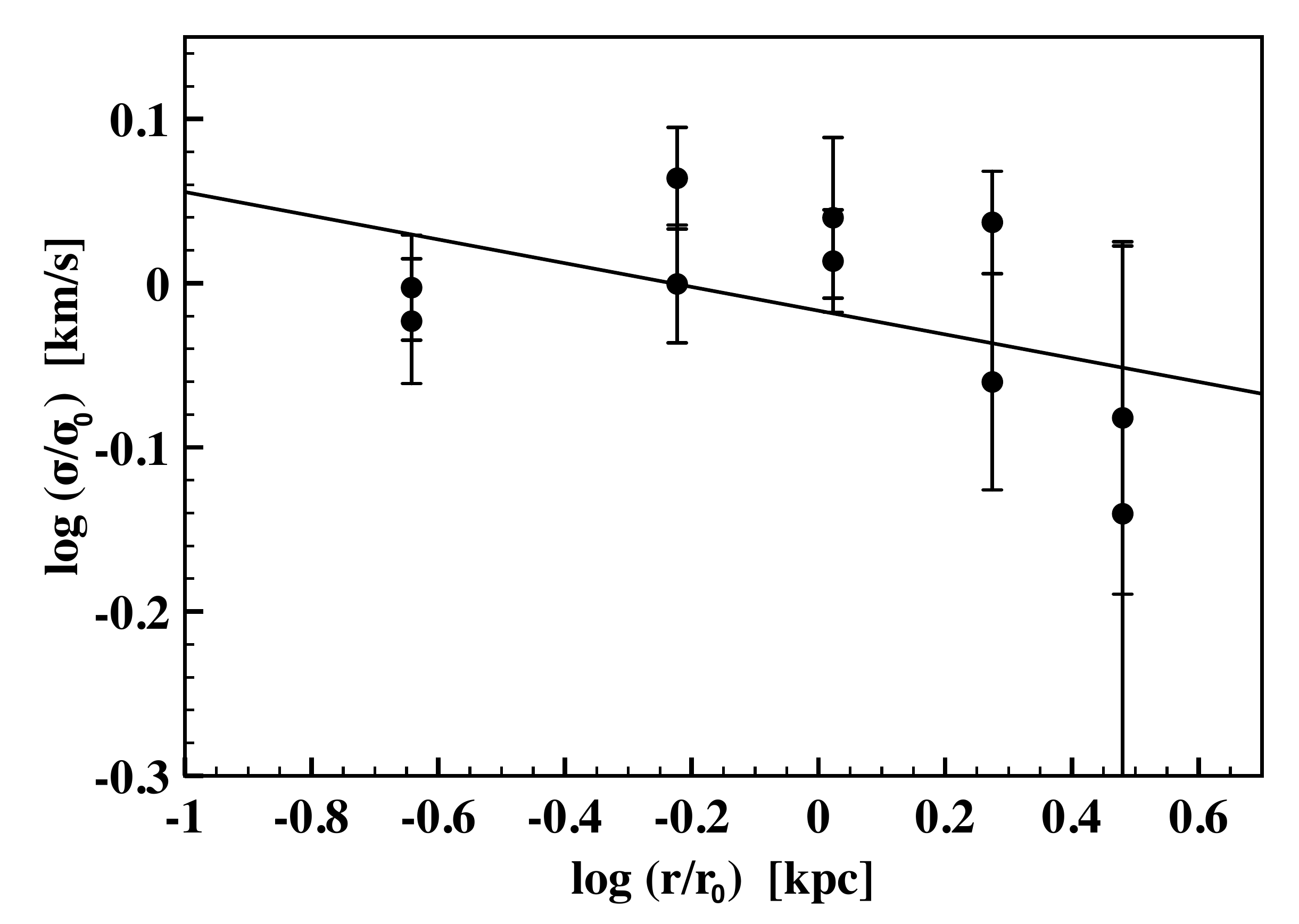}}  \\
   \subfloat{\includegraphics[scale=0.25]{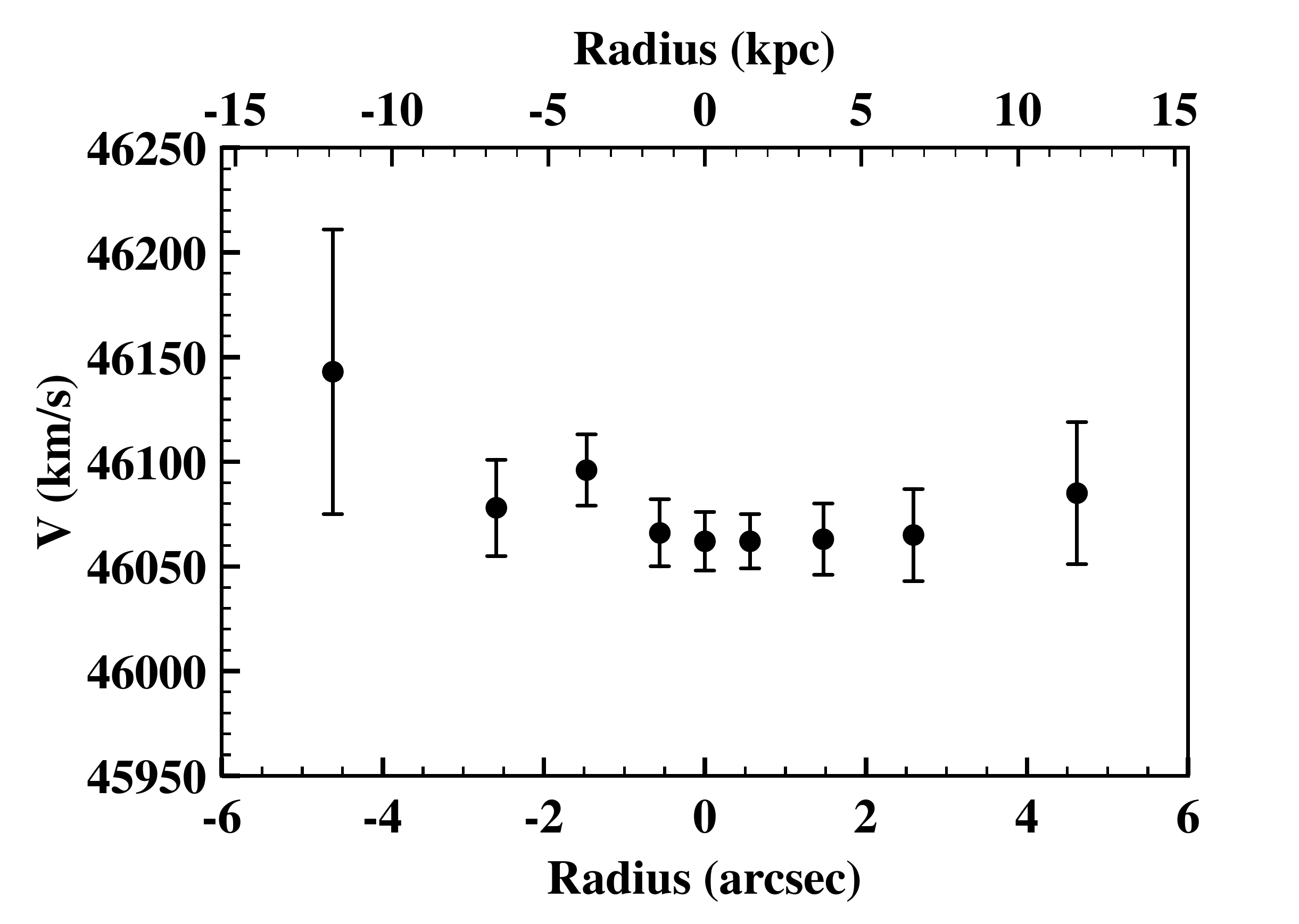}}
         \subfloat[Abell 2104]{\includegraphics[scale=0.25]{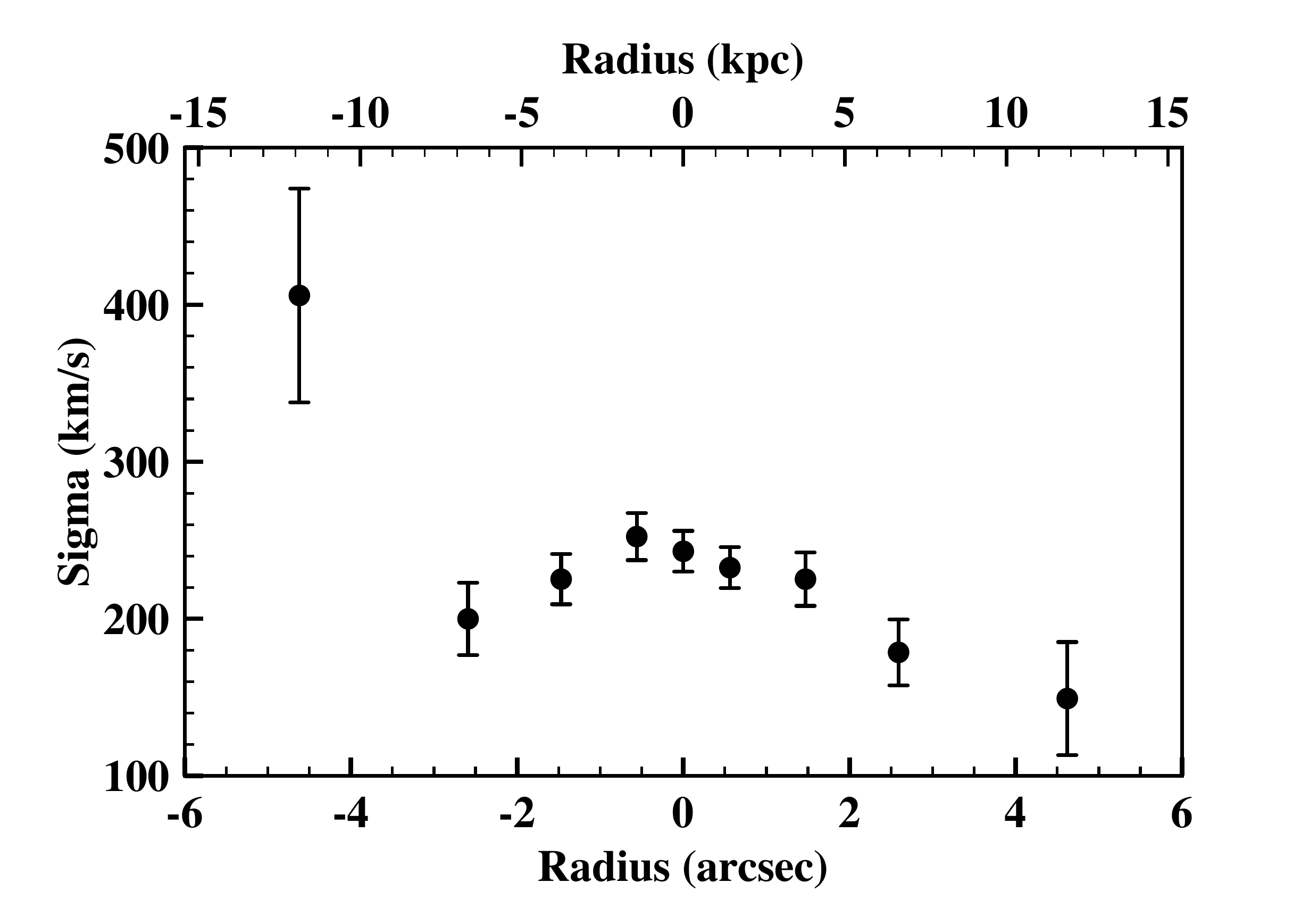}}
   \subfloat{\includegraphics[scale=0.25]{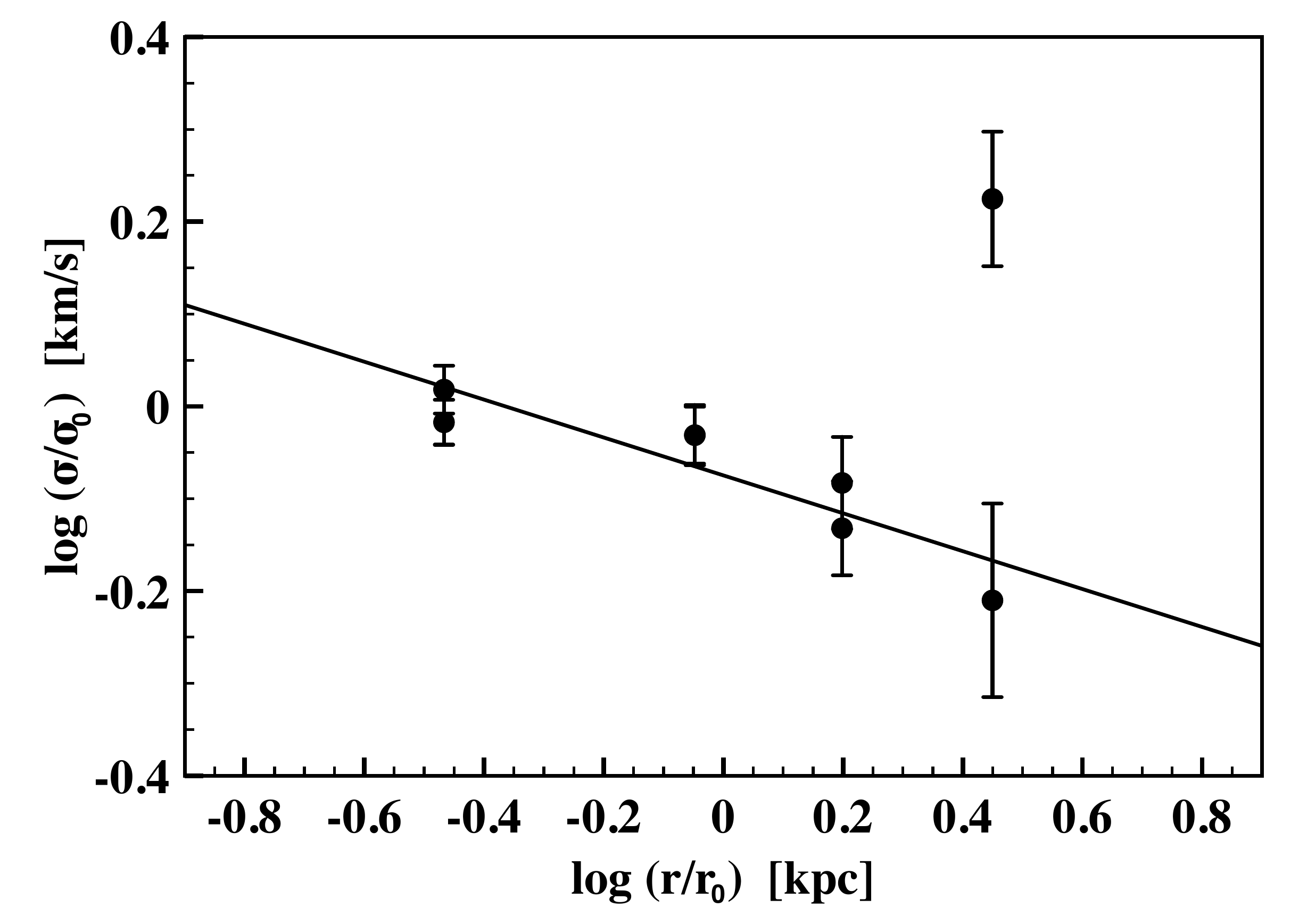}} \\
     \subfloat{\includegraphics[scale=0.25]{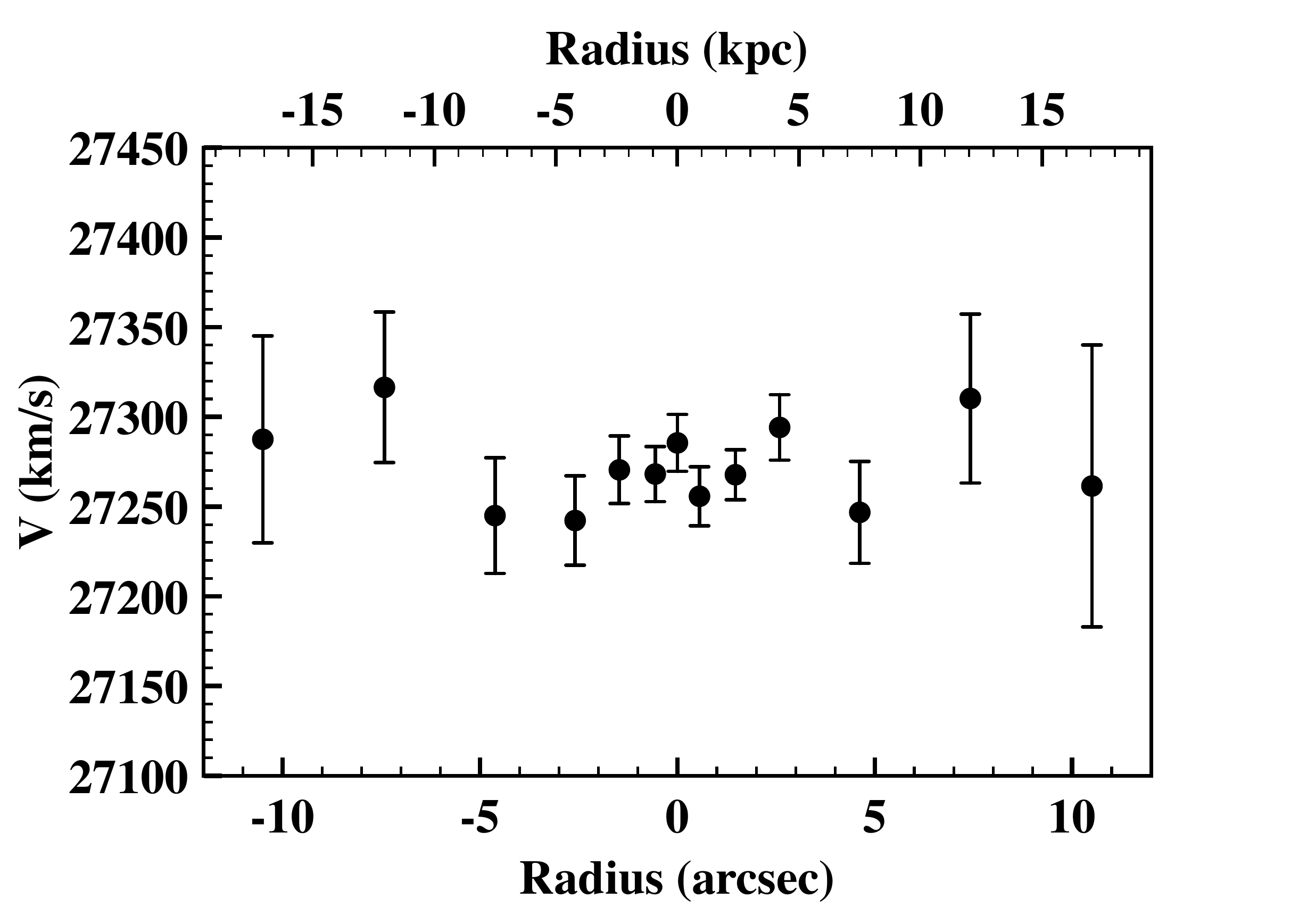}}
         \subfloat[Abell 2142]{\includegraphics[scale=0.25]{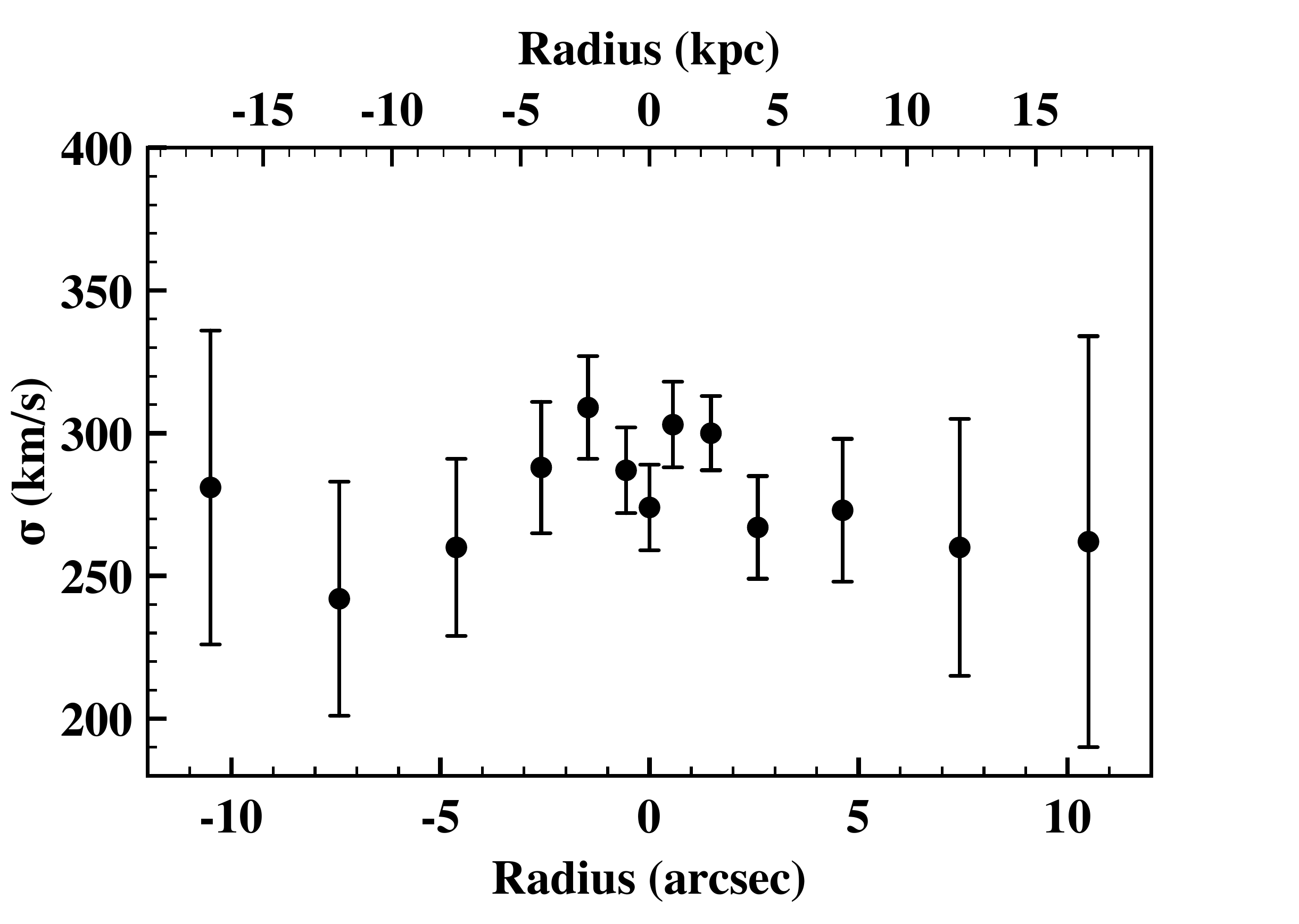}}
   \subfloat{\includegraphics[scale=0.25]{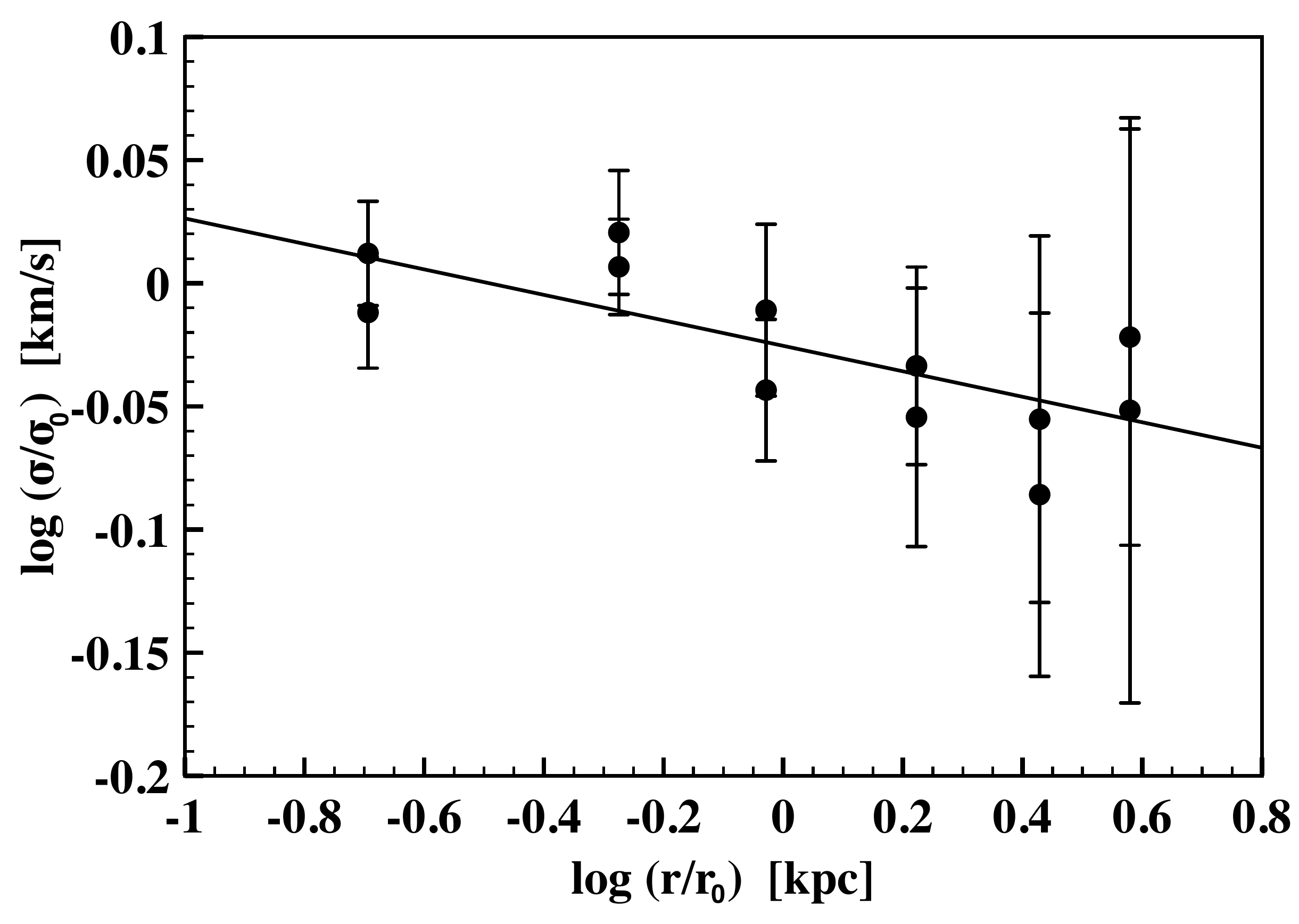}} \\
    \subfloat{\includegraphics[scale=0.25]{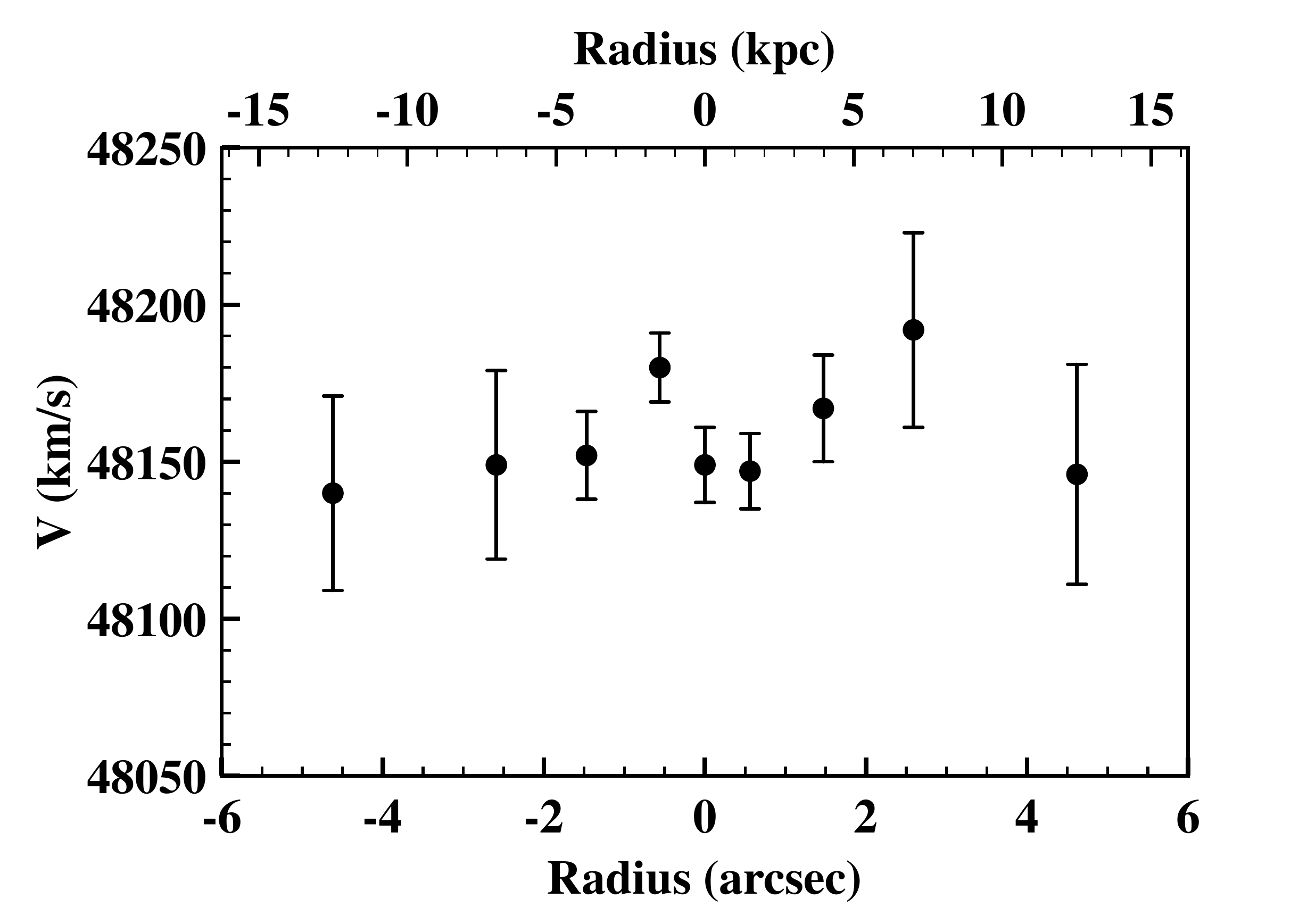}}
         \subfloat[Abell 2259]{\includegraphics[scale=0.25]{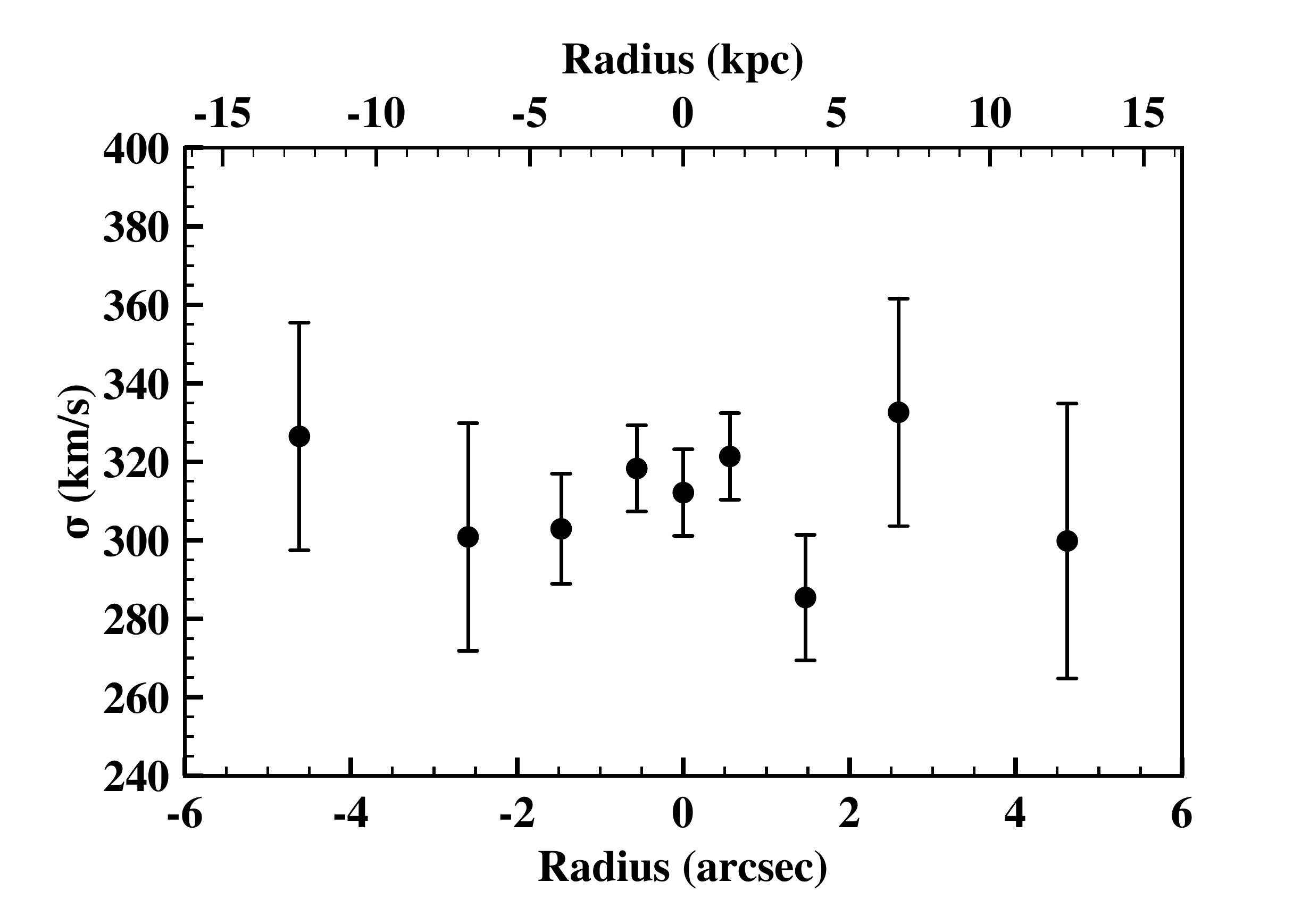}}
   \subfloat{\includegraphics[scale=0.25]{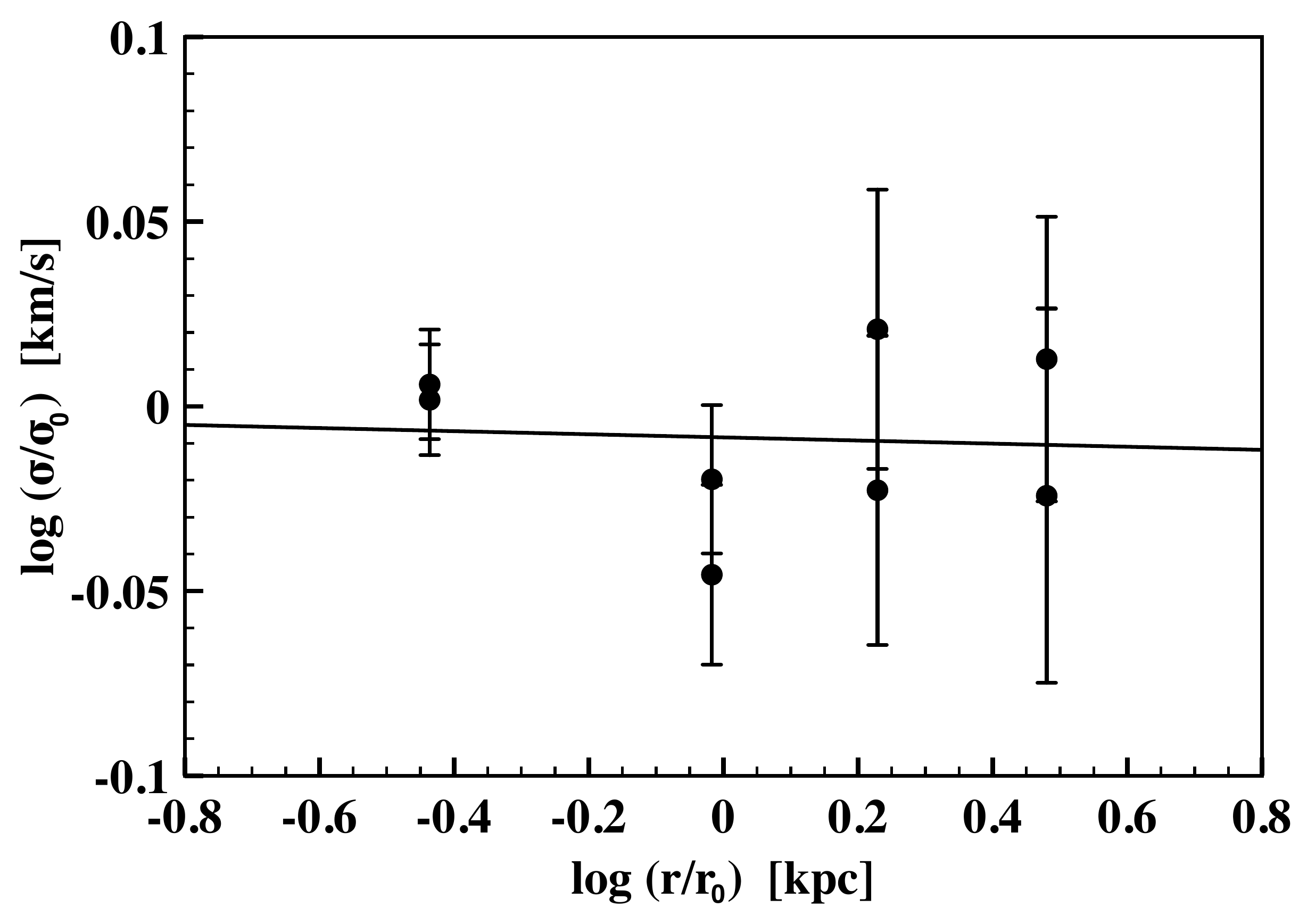}} \\
    \caption{[a] Radial profiles of velocity (V), [b] velocity dispersion ($\sigma$, and [c] power law fit).}
\label{fig:kin8}
\end{figure*}

\begin{figure*}
\captionsetup[subfigure]{labelformat=empty}
 \subfloat{\includegraphics[scale=0.25]{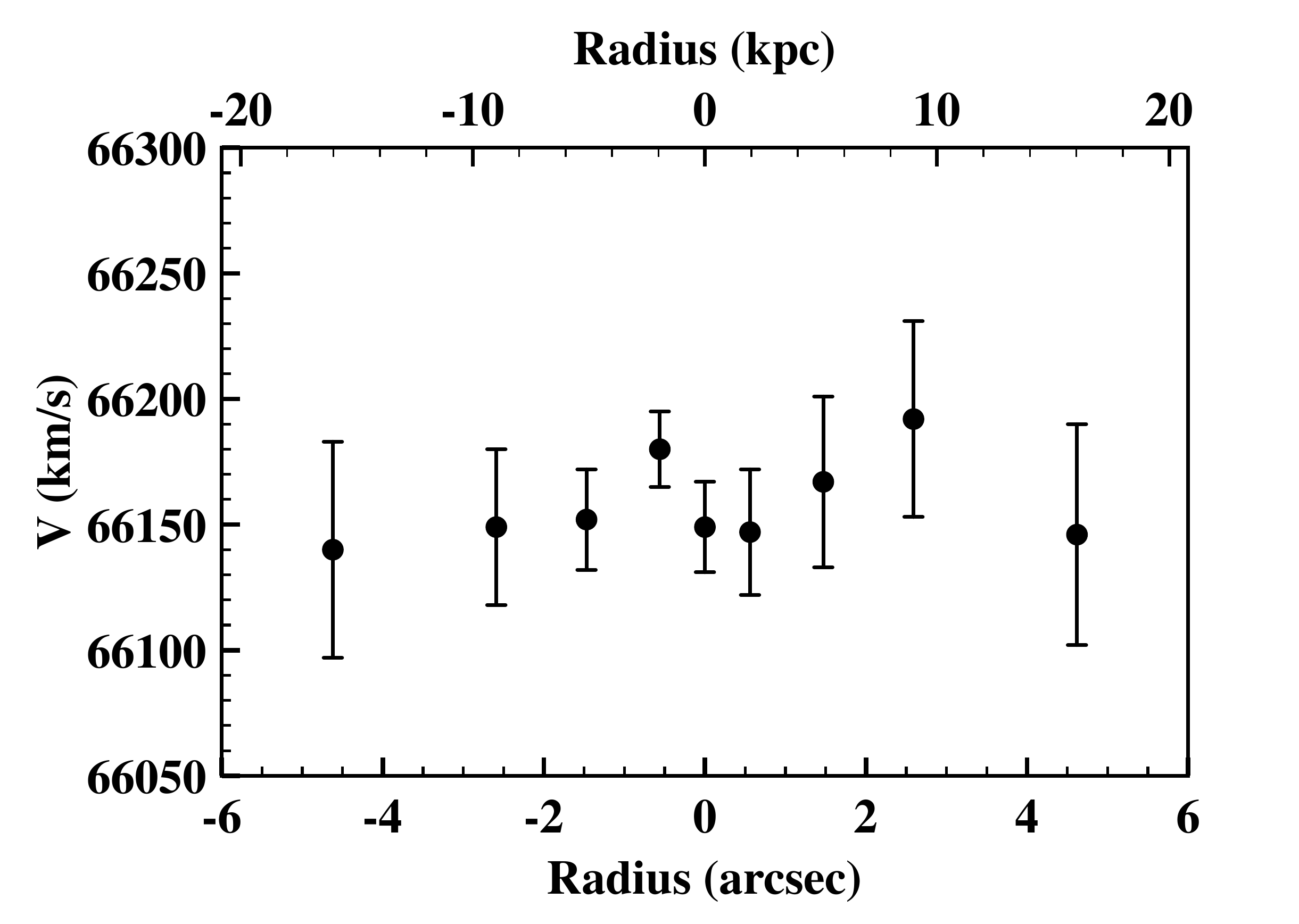}}
         \subfloat[Abell 2261]{\includegraphics[scale=0.25]{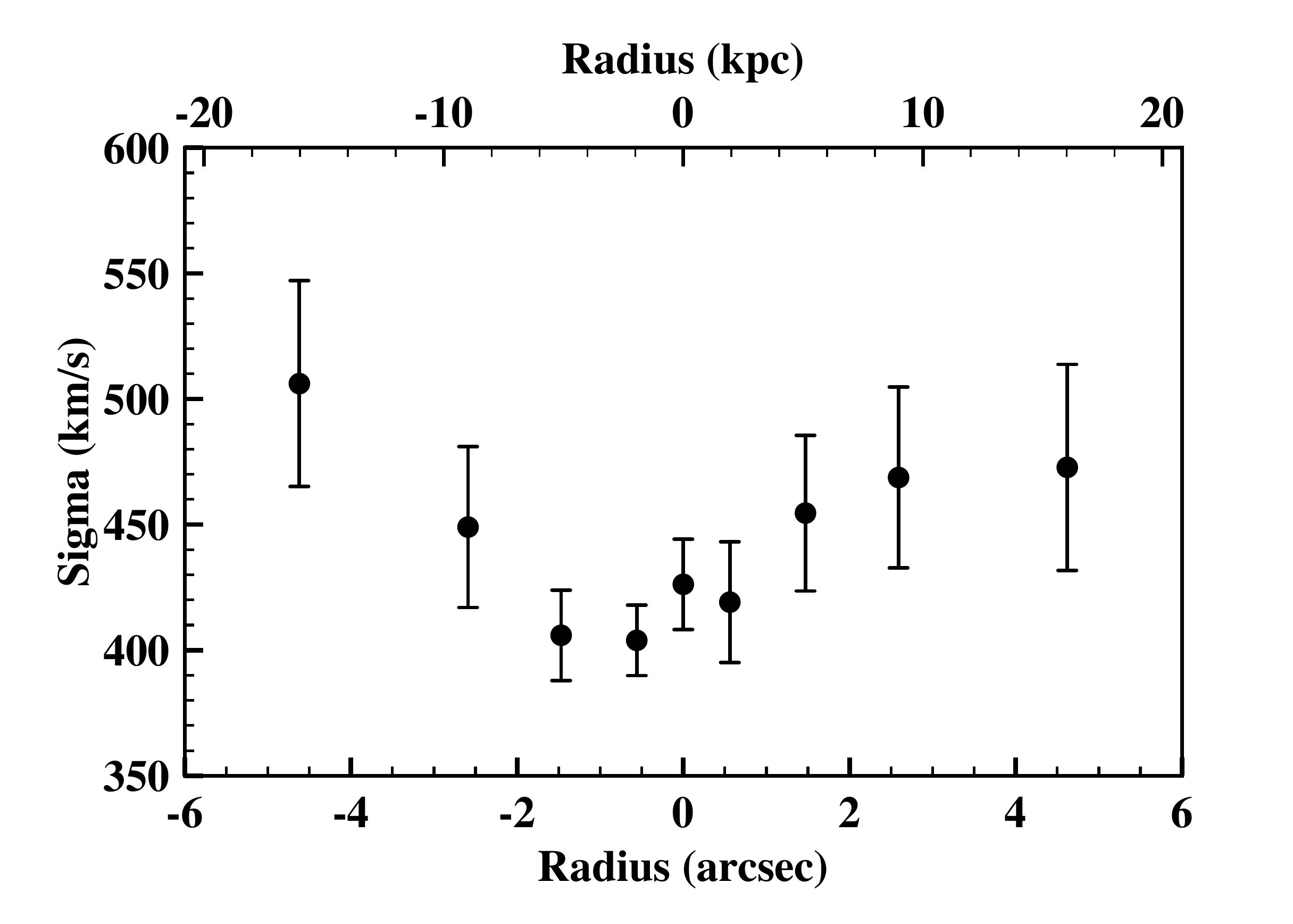}}
   \subfloat{\includegraphics[scale=0.25]{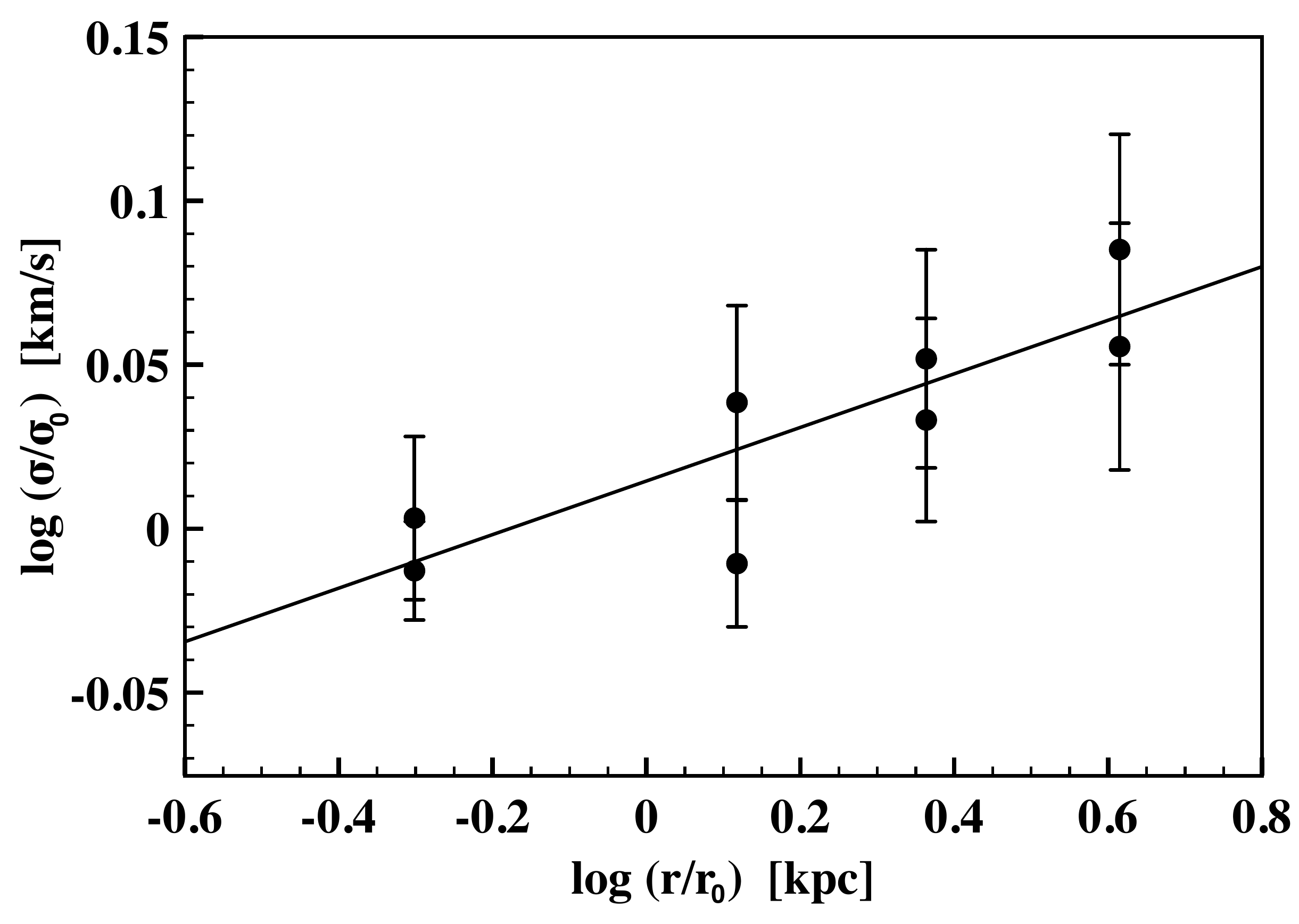}} \\
     \subfloat{\includegraphics[scale=0.25]{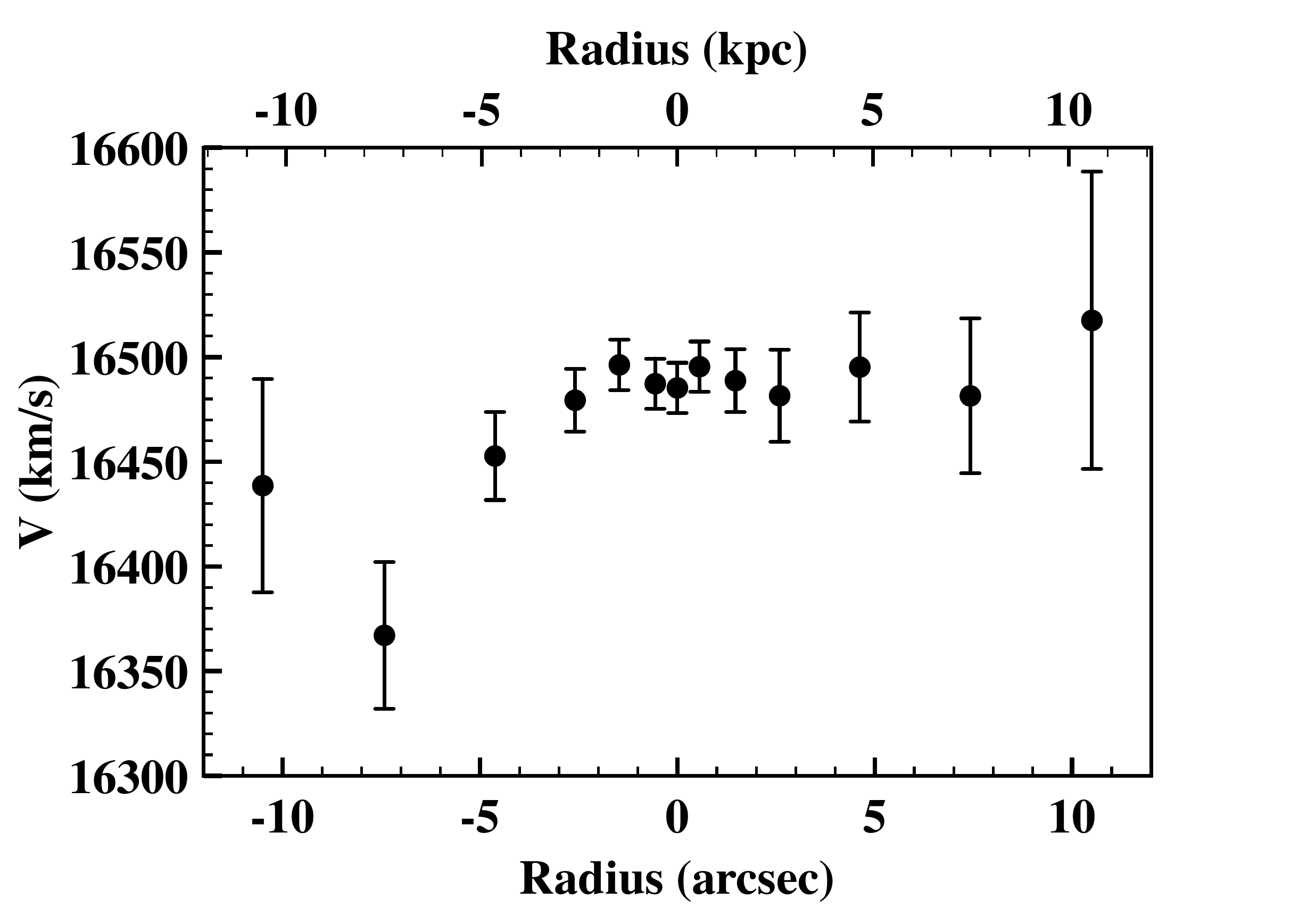}}
         \subfloat[Abell 2319]{\includegraphics[scale=0.25]{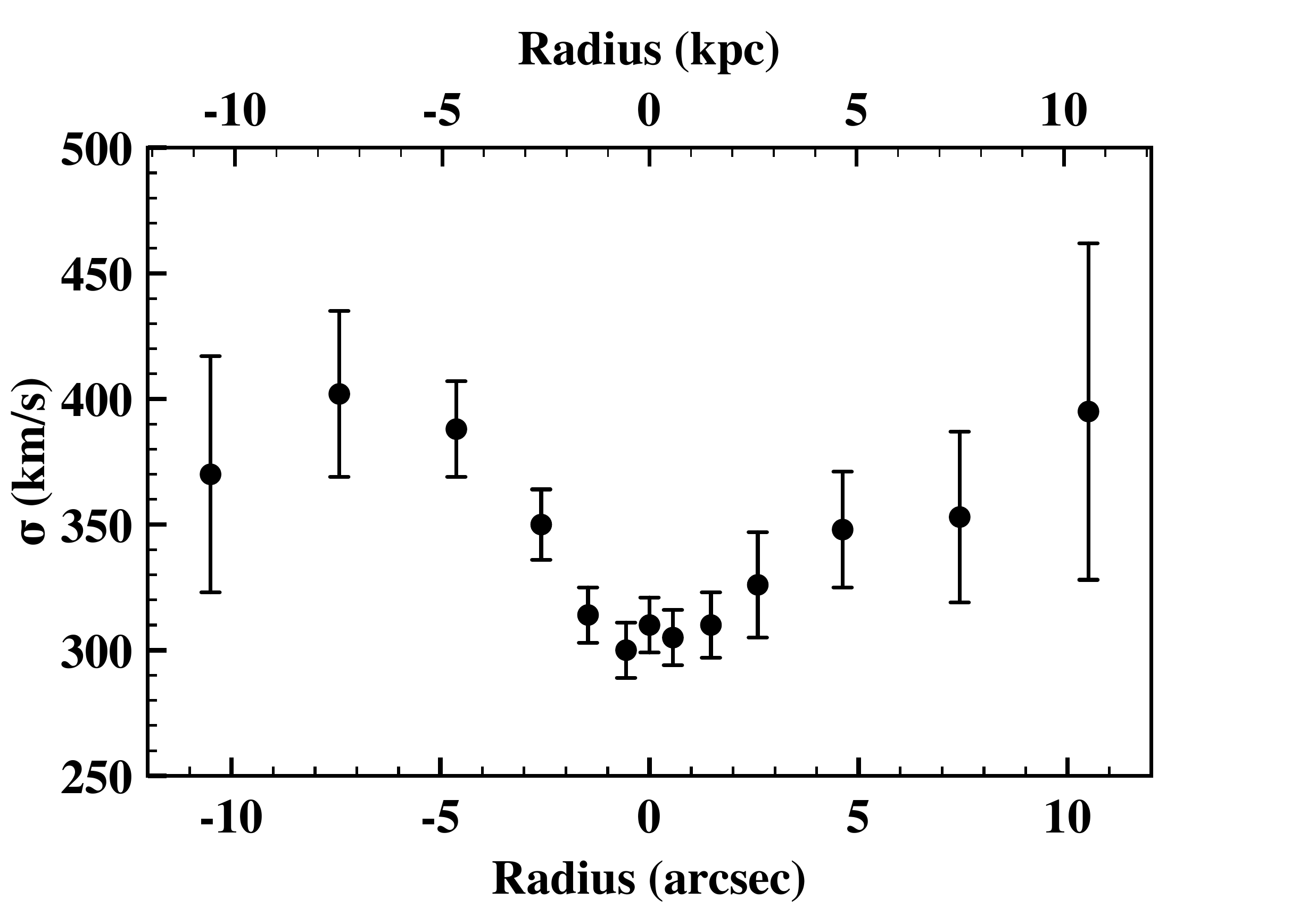}}
   \subfloat{\includegraphics[scale=0.25]{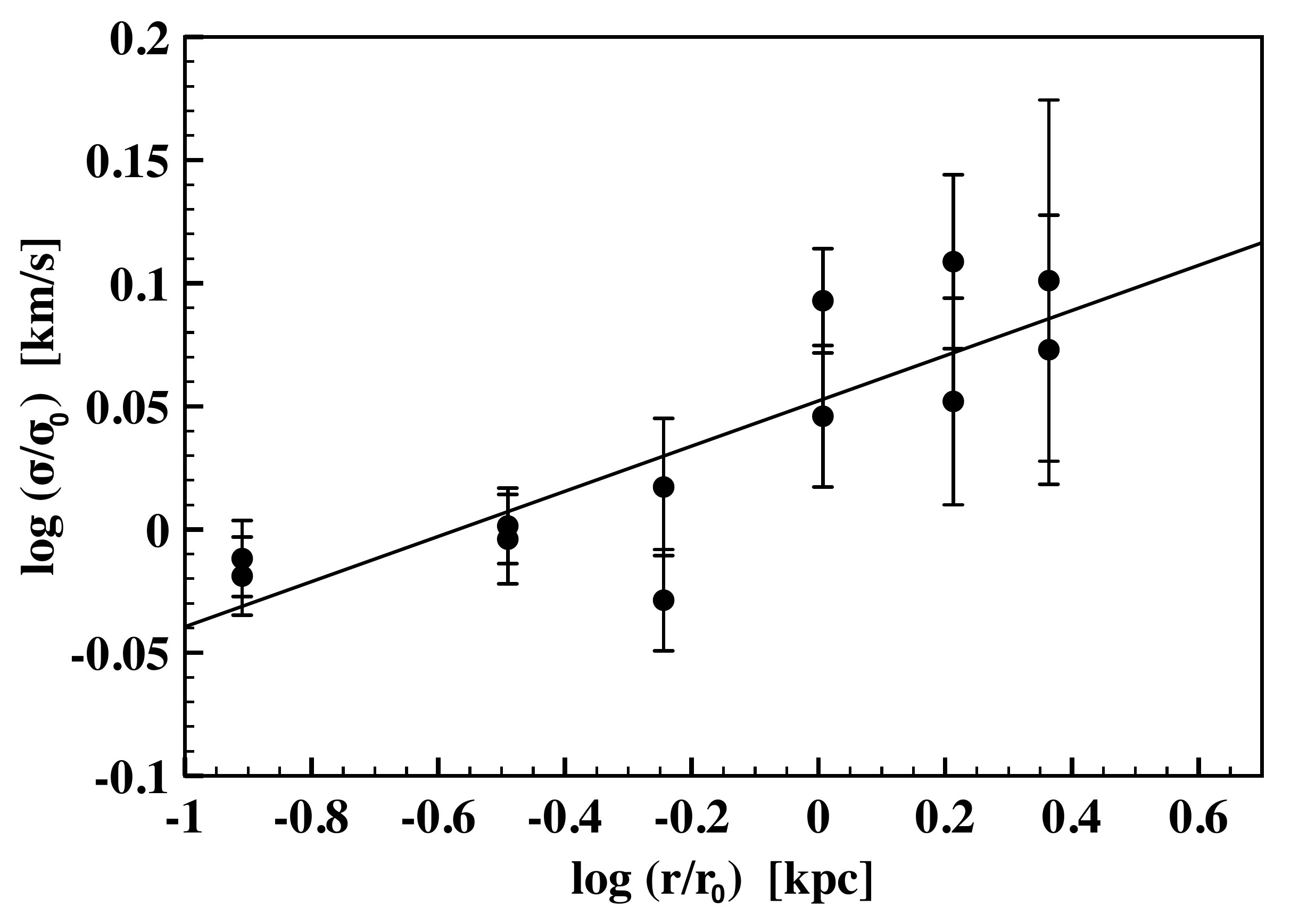}} \\
    \subfloat{\includegraphics[scale=0.25]{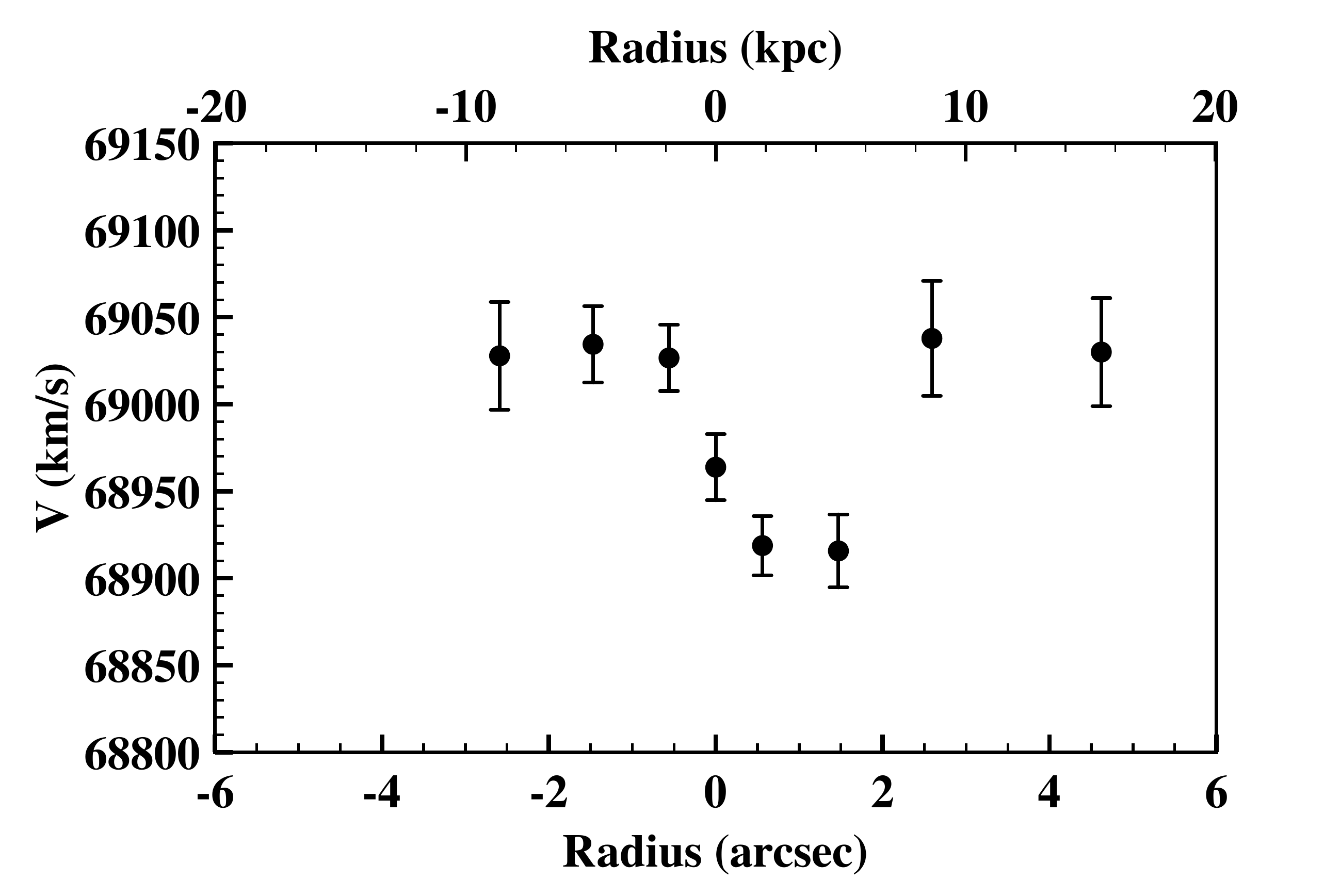}}
         \subfloat[Abell 2390]{\includegraphics[scale=0.25]{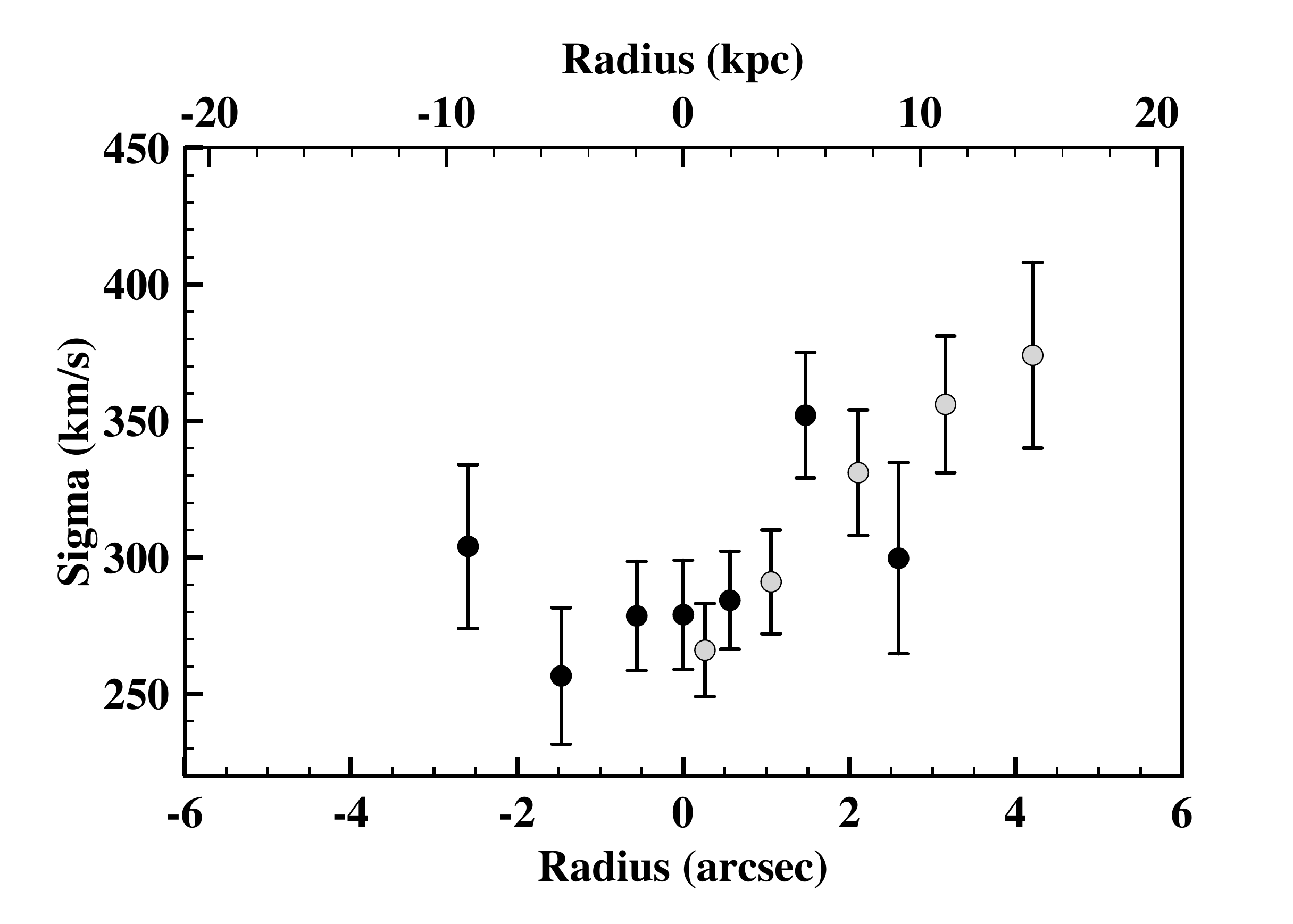}}
   \subfloat{\includegraphics[scale=0.25]{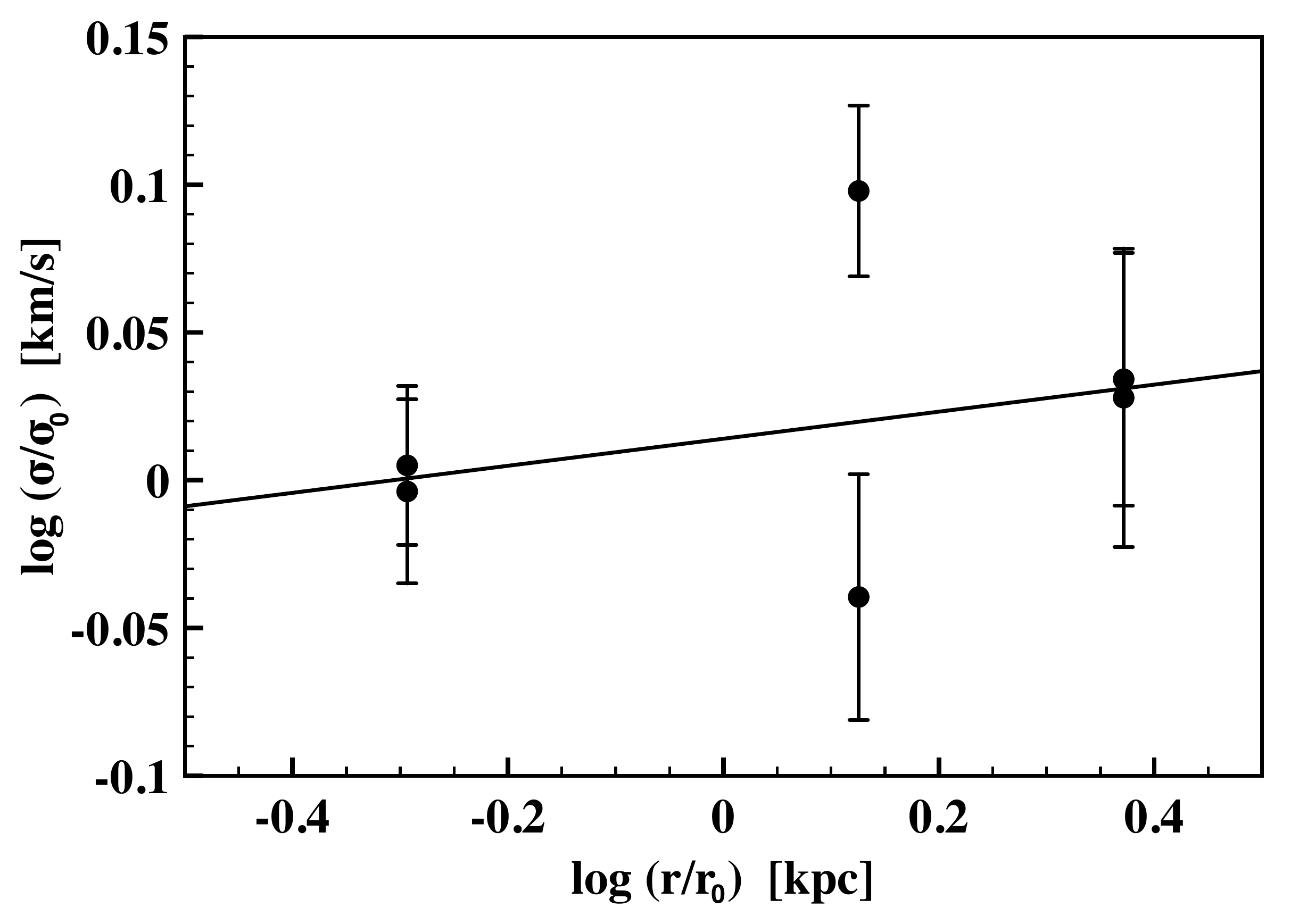}}   \\
   \subfloat{\includegraphics[scale=0.25]{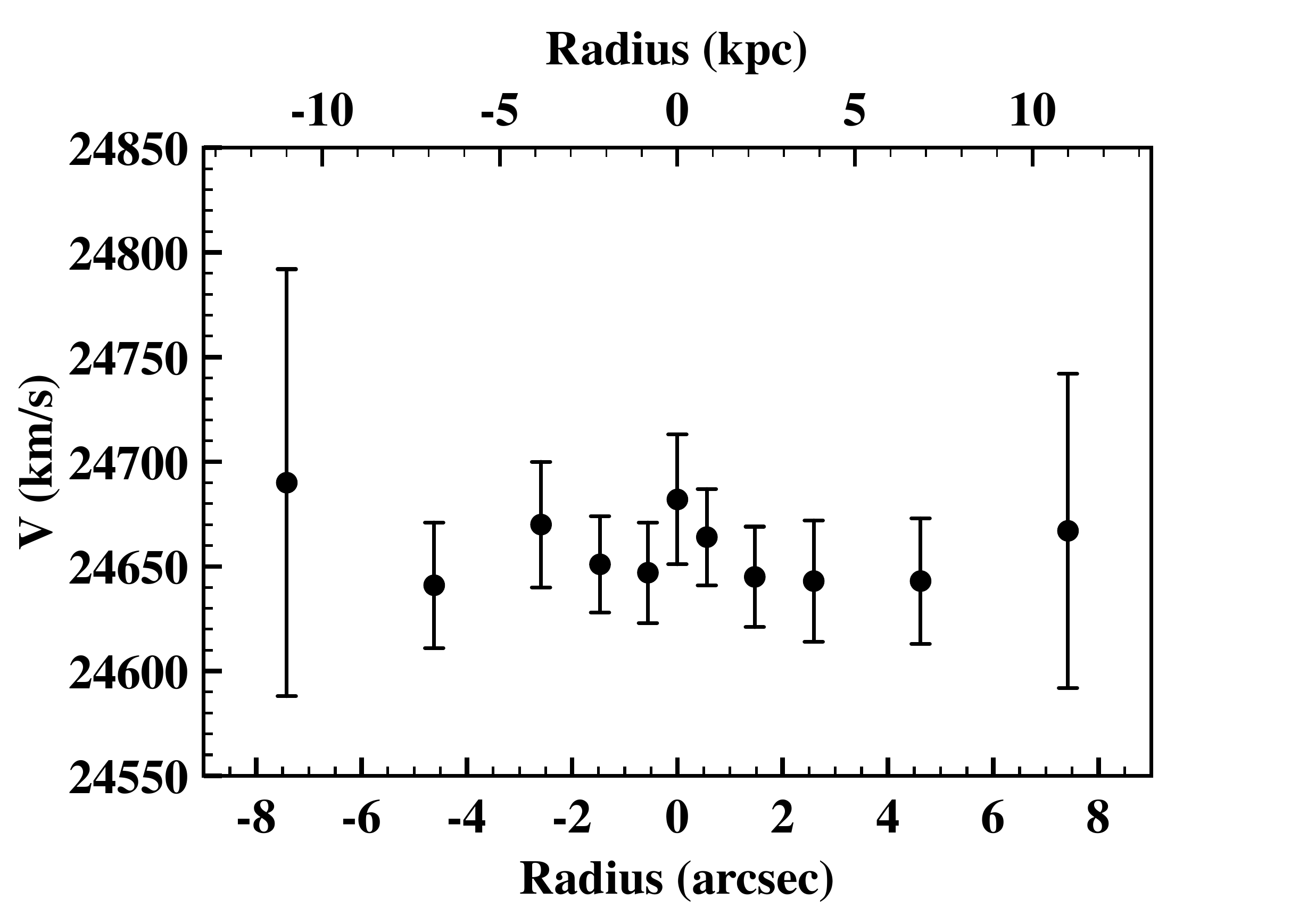}}
         \subfloat[Abell 2420]{\includegraphics[scale=0.25]{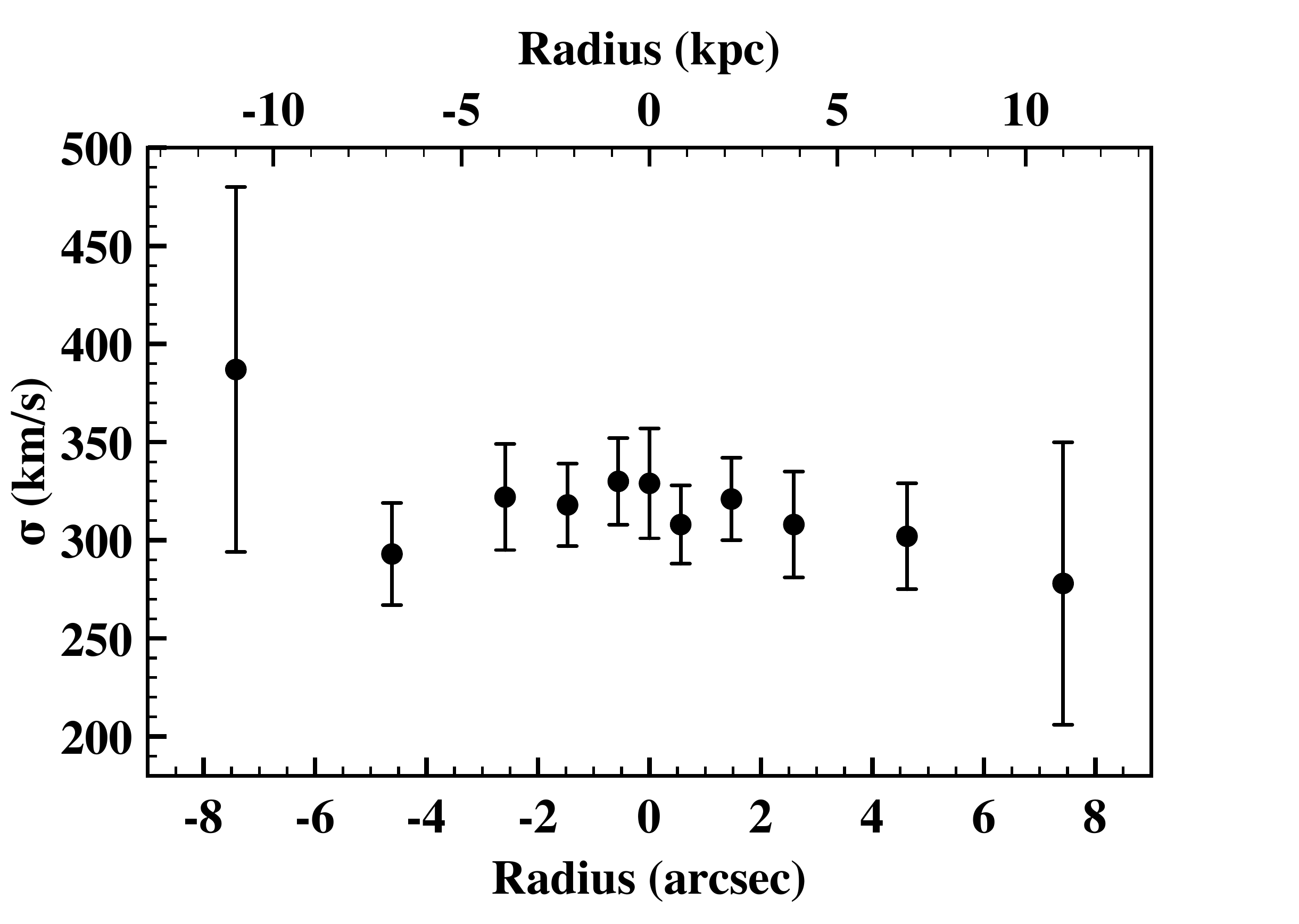}}
   \subfloat{\includegraphics[scale=0.25]{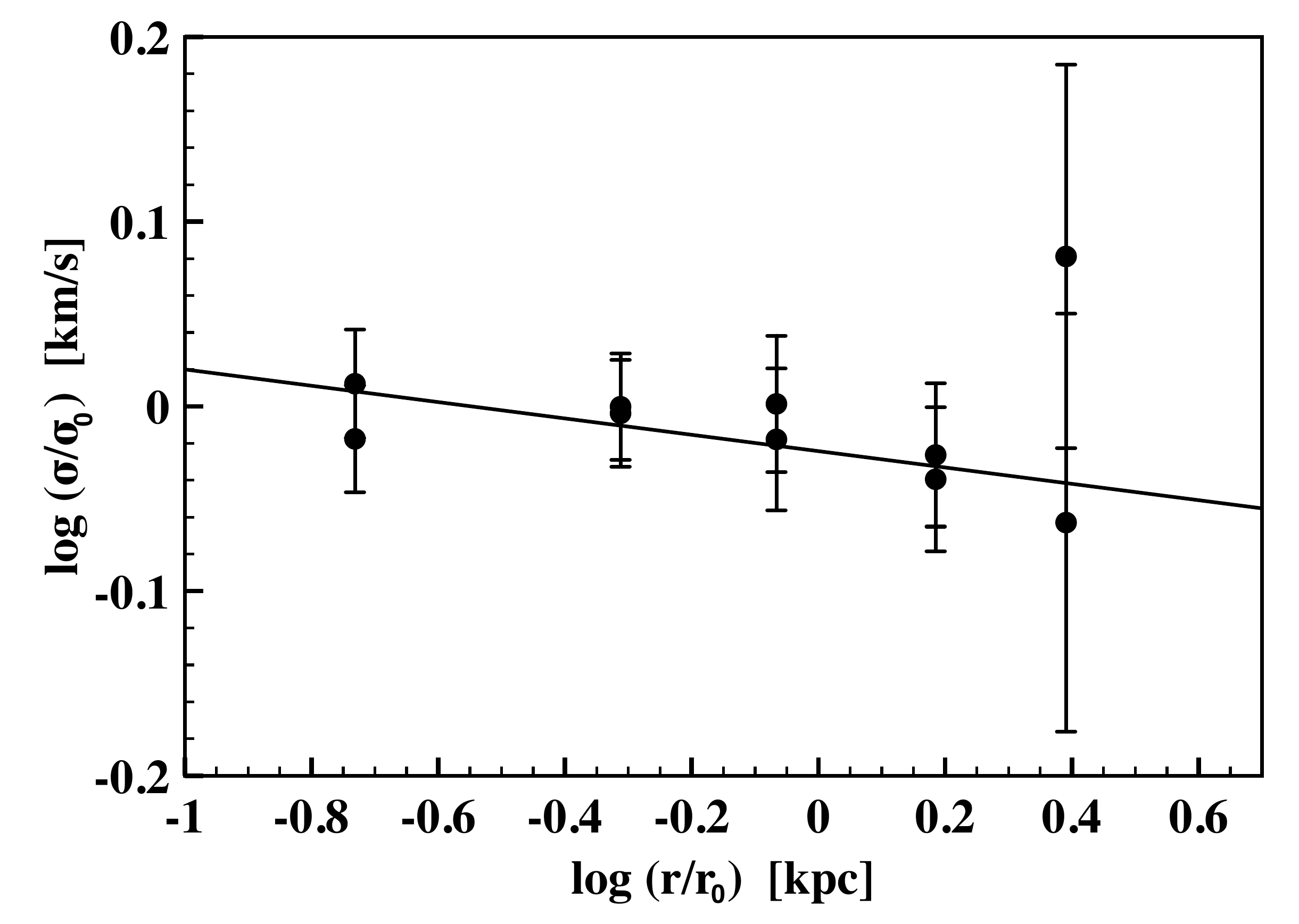}}  \\
  \caption{[a] Radial profiles of velocity (V), [b] velocity dispersion ($\sigma$, and [c] power law fit). The grey circles in the velocity dispersion profile of Abell 2390 indicate the measurements from \citet{Newman2013a} (see Appendix \ref{kinematics}).}
\label{fig:kin9}
\end{figure*}

\begin{figure*}
\captionsetup[subfigure]{labelformat=empty}
     \subfloat{\includegraphics[scale=0.25]{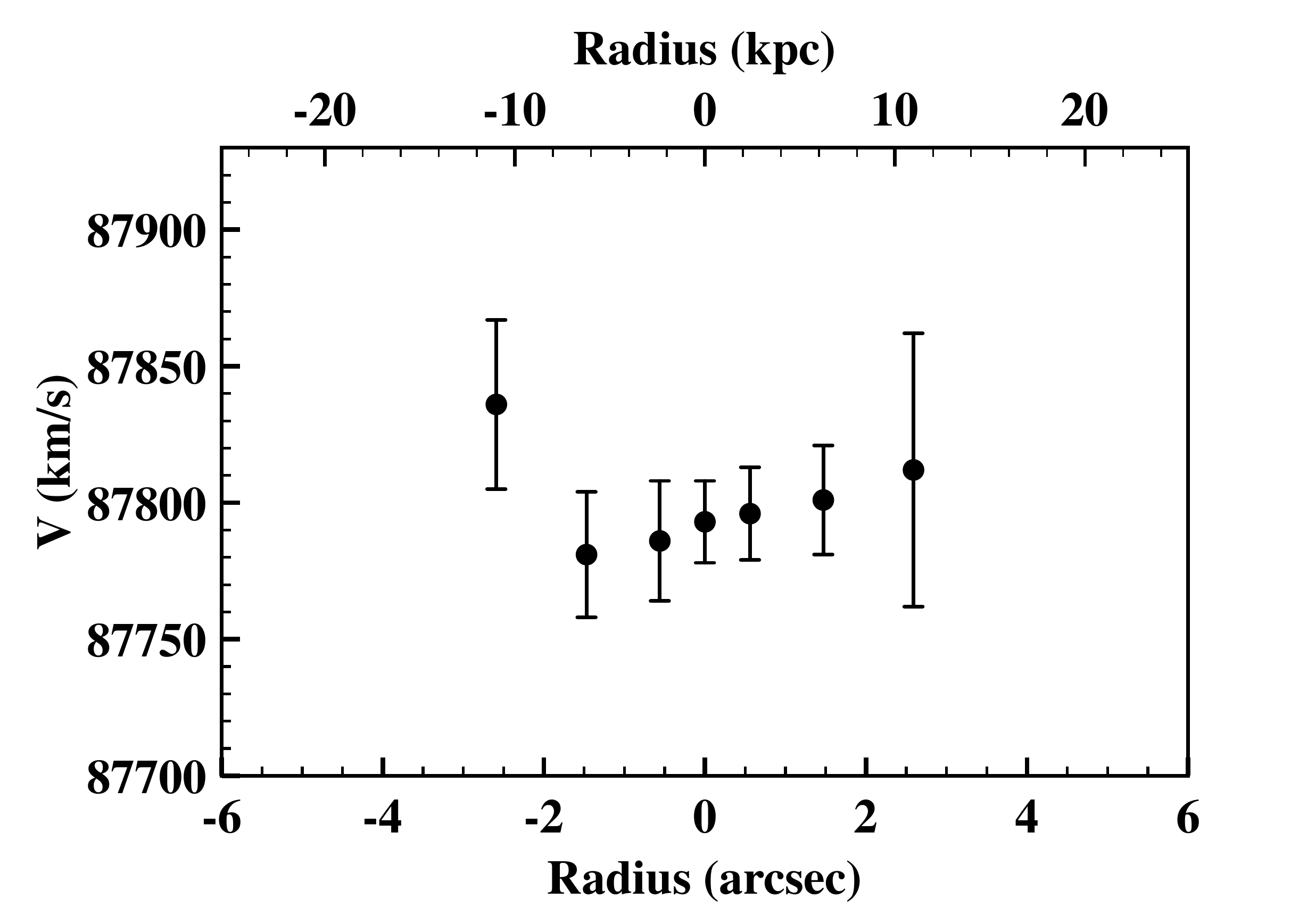}}
         \subfloat[Abell 2537]{\includegraphics[scale=0.25]{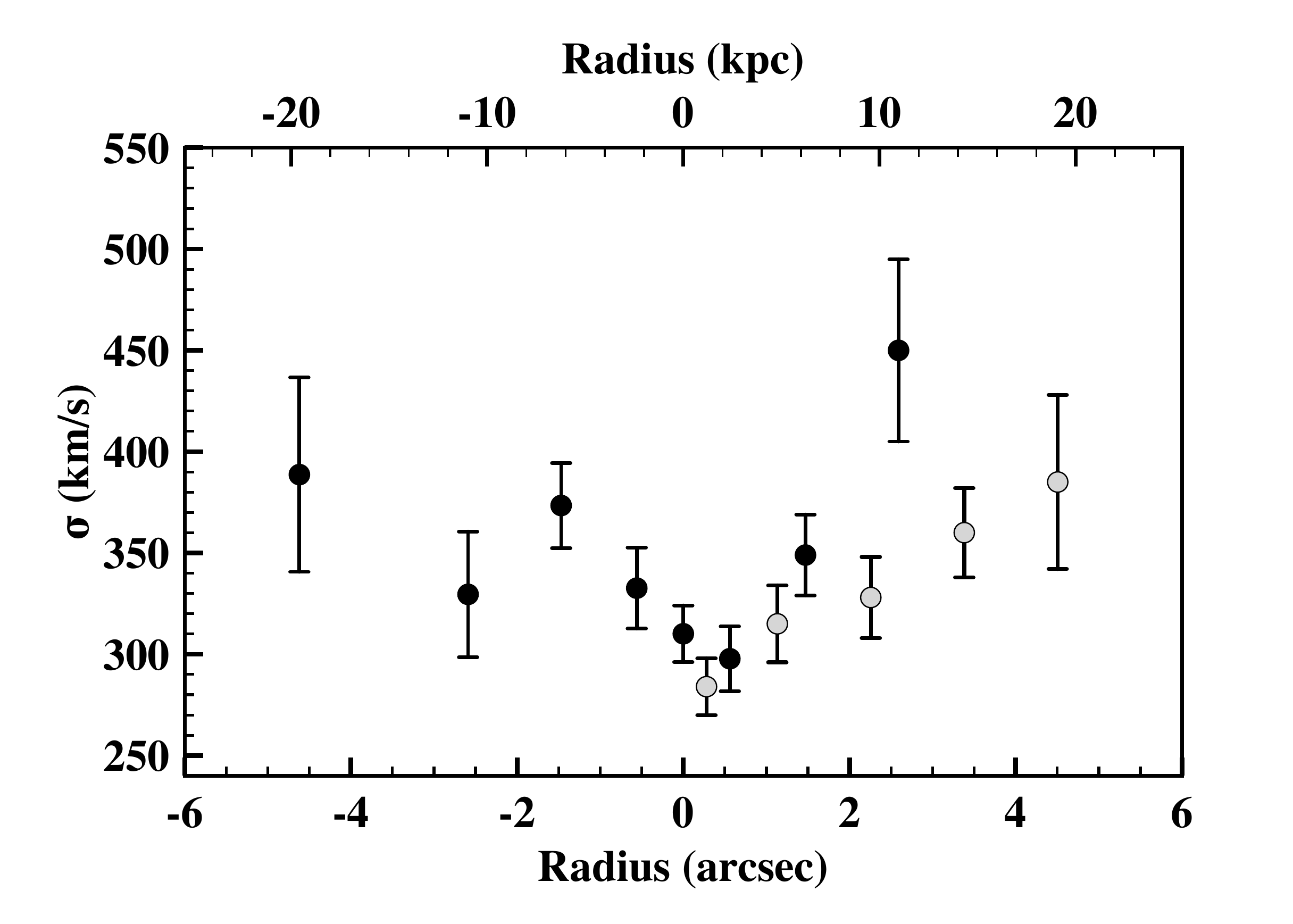}}
   \subfloat{\includegraphics[scale=0.25]{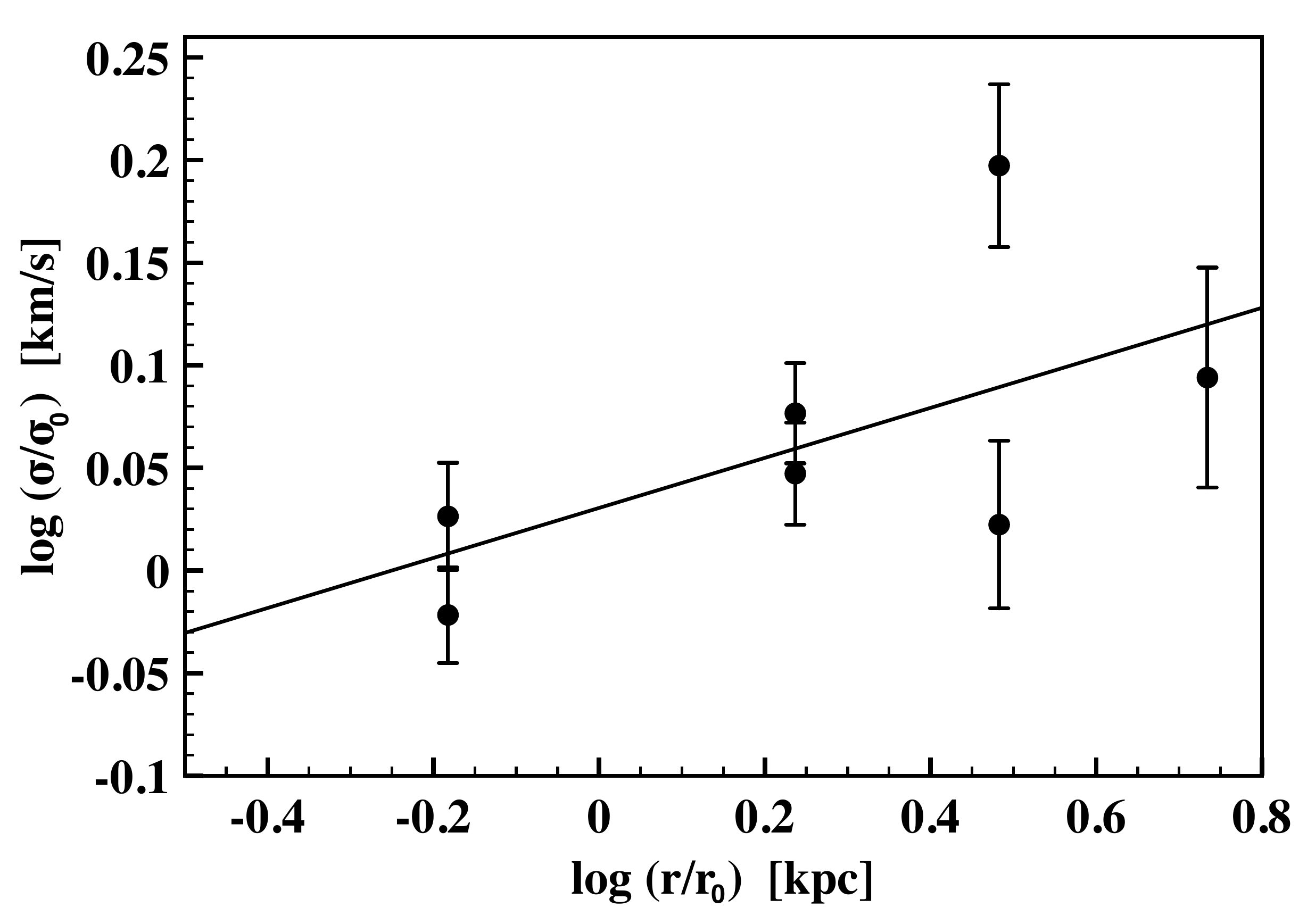}} \\
    \subfloat{\includegraphics[scale=0.25]{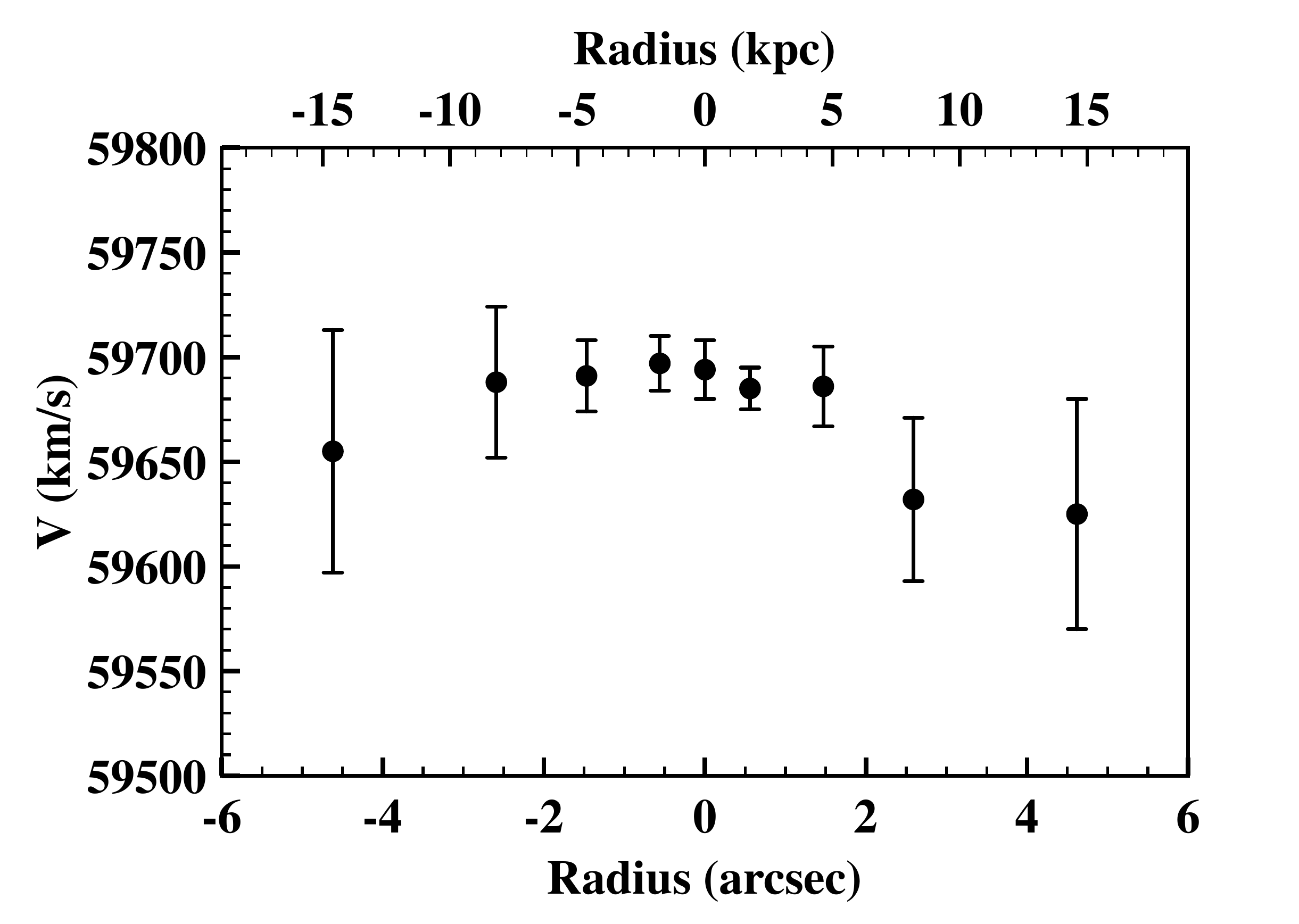}}
         \subfloat[MS 0440+02]{\includegraphics[scale=0.25]{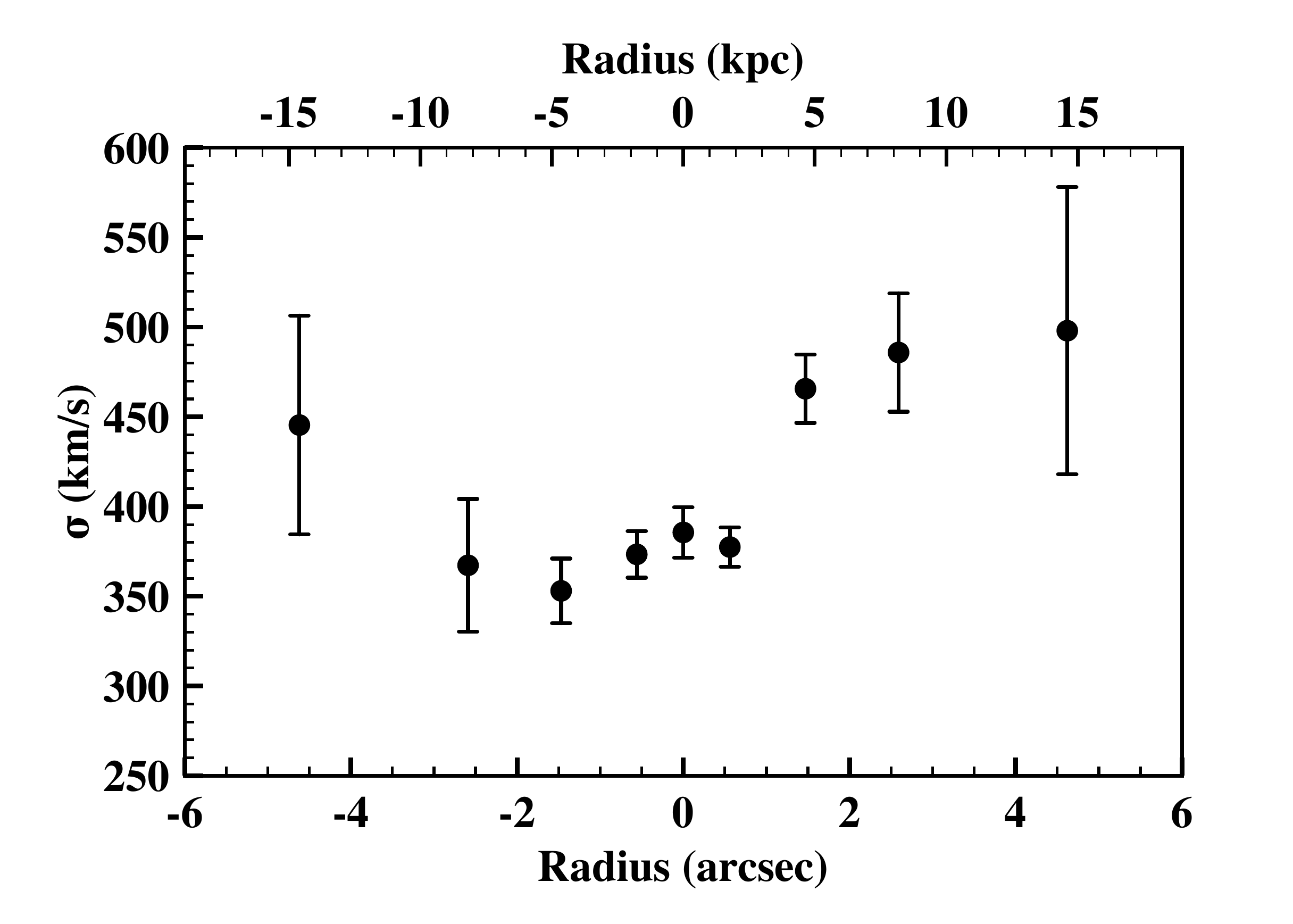}}
   \subfloat{\includegraphics[scale=0.25]{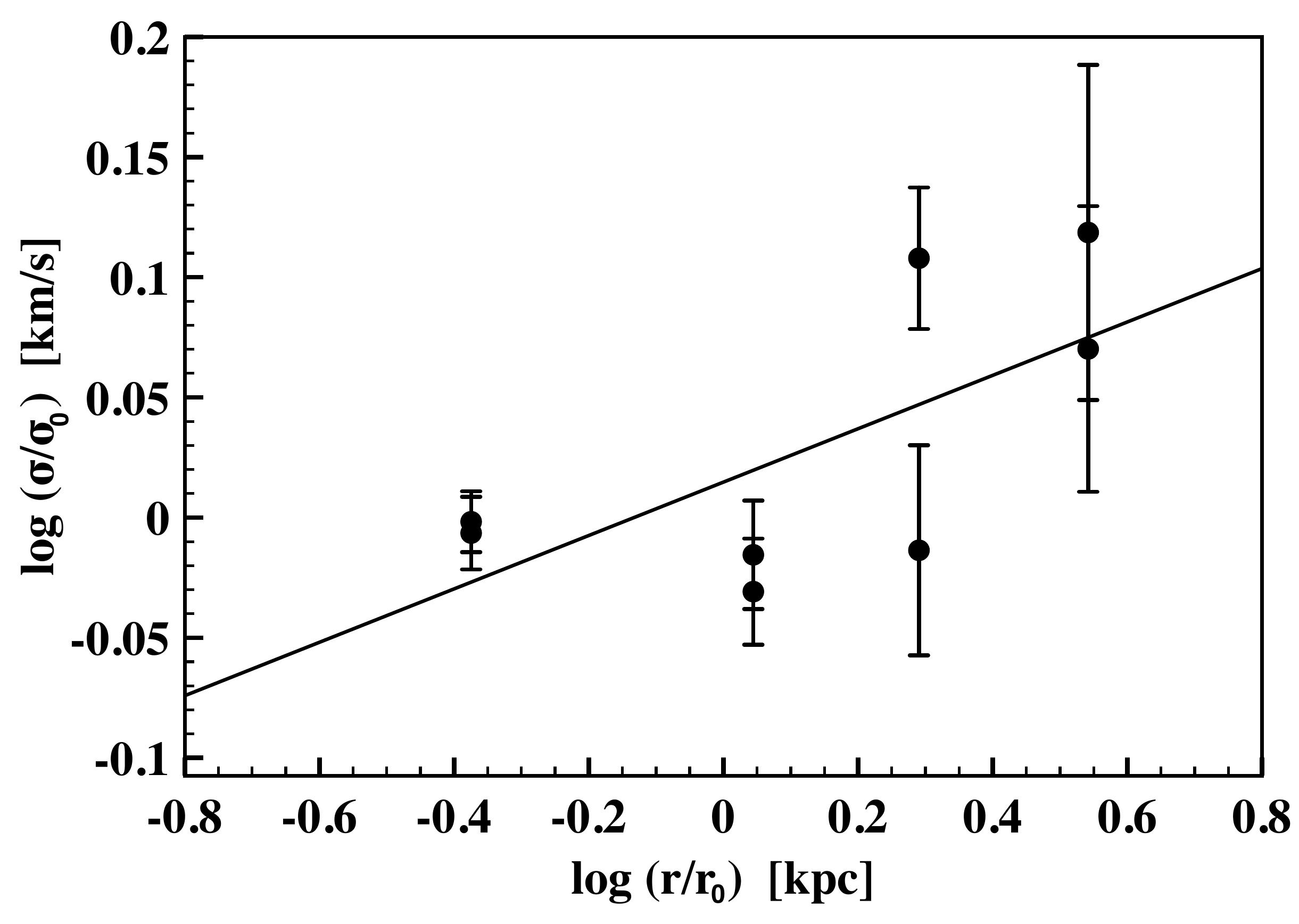}} \\
   \subfloat{\includegraphics[scale=0.25]{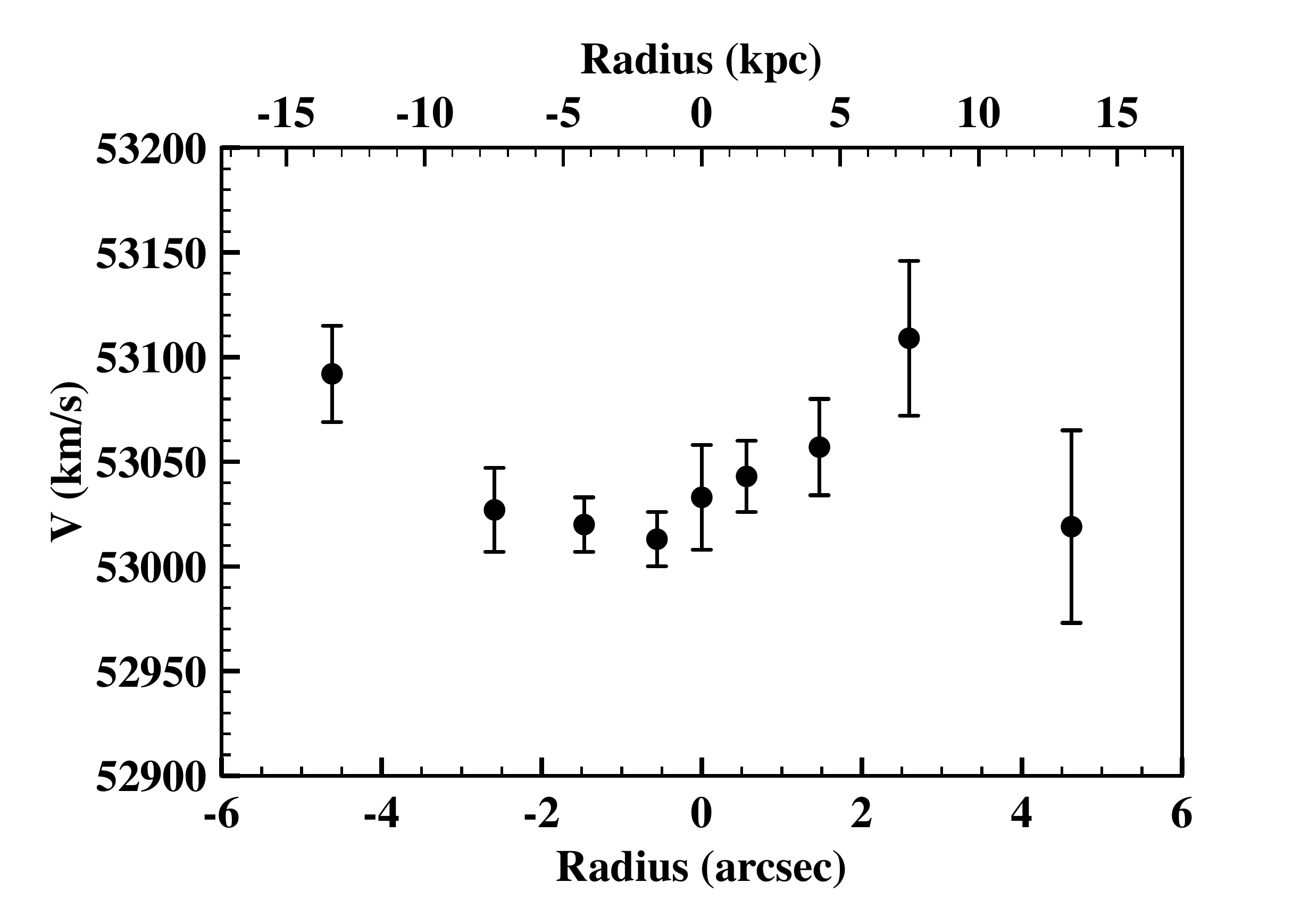}}
         \subfloat[MS 0906+11]{\includegraphics[scale=0.25]{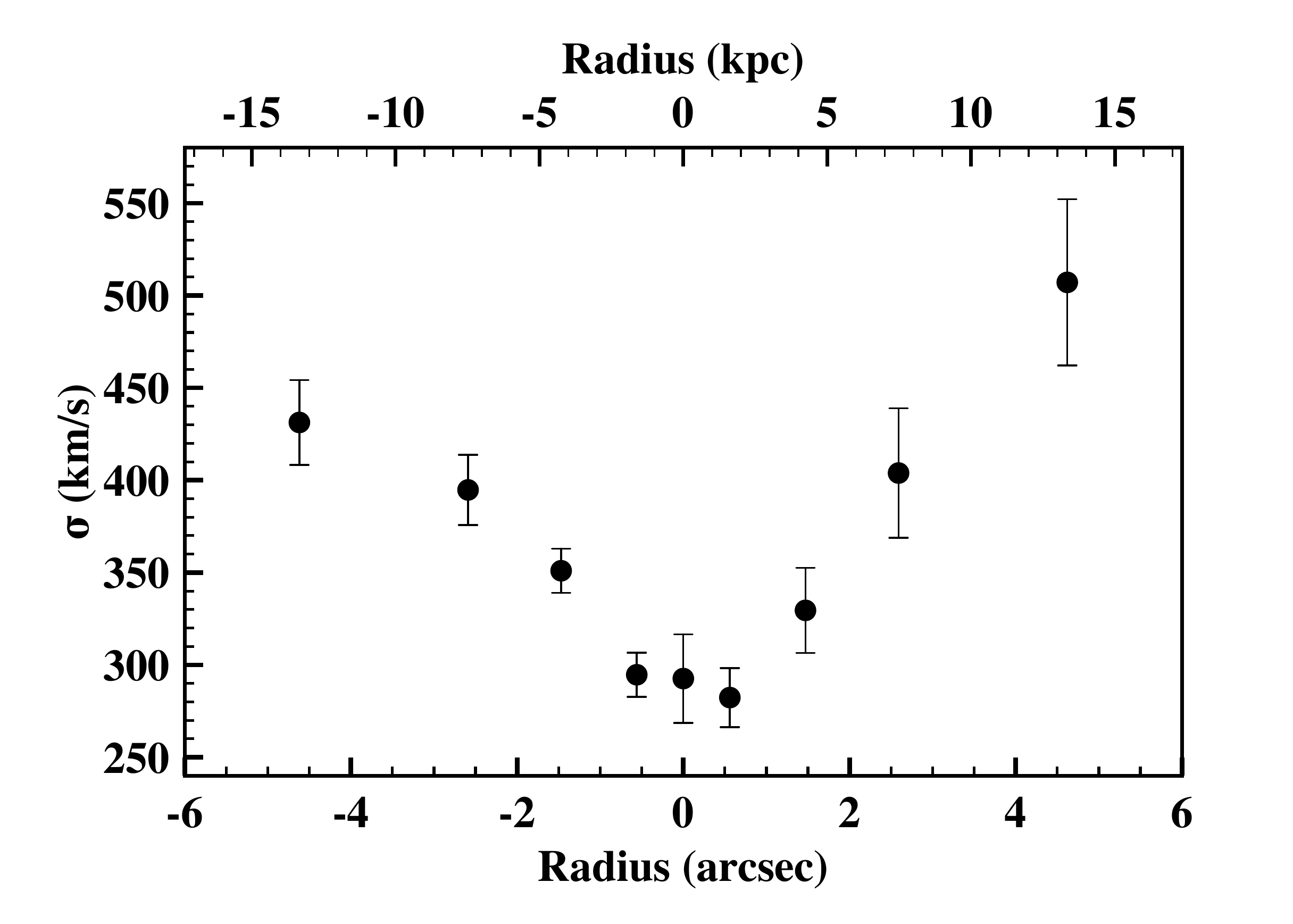}}
   \subfloat{\includegraphics[scale=0.25]{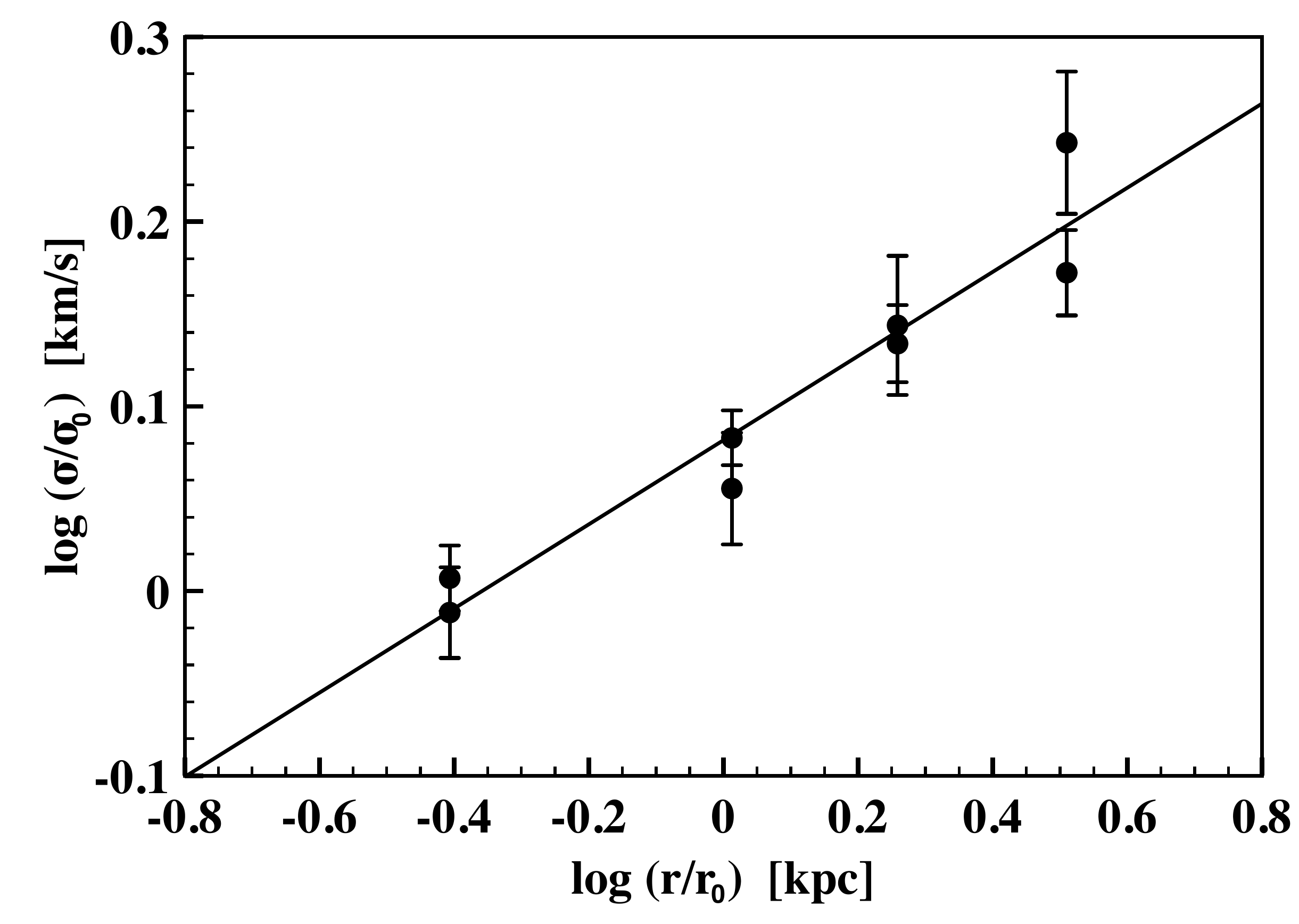}} \\
     \subfloat{\includegraphics[scale=0.25]{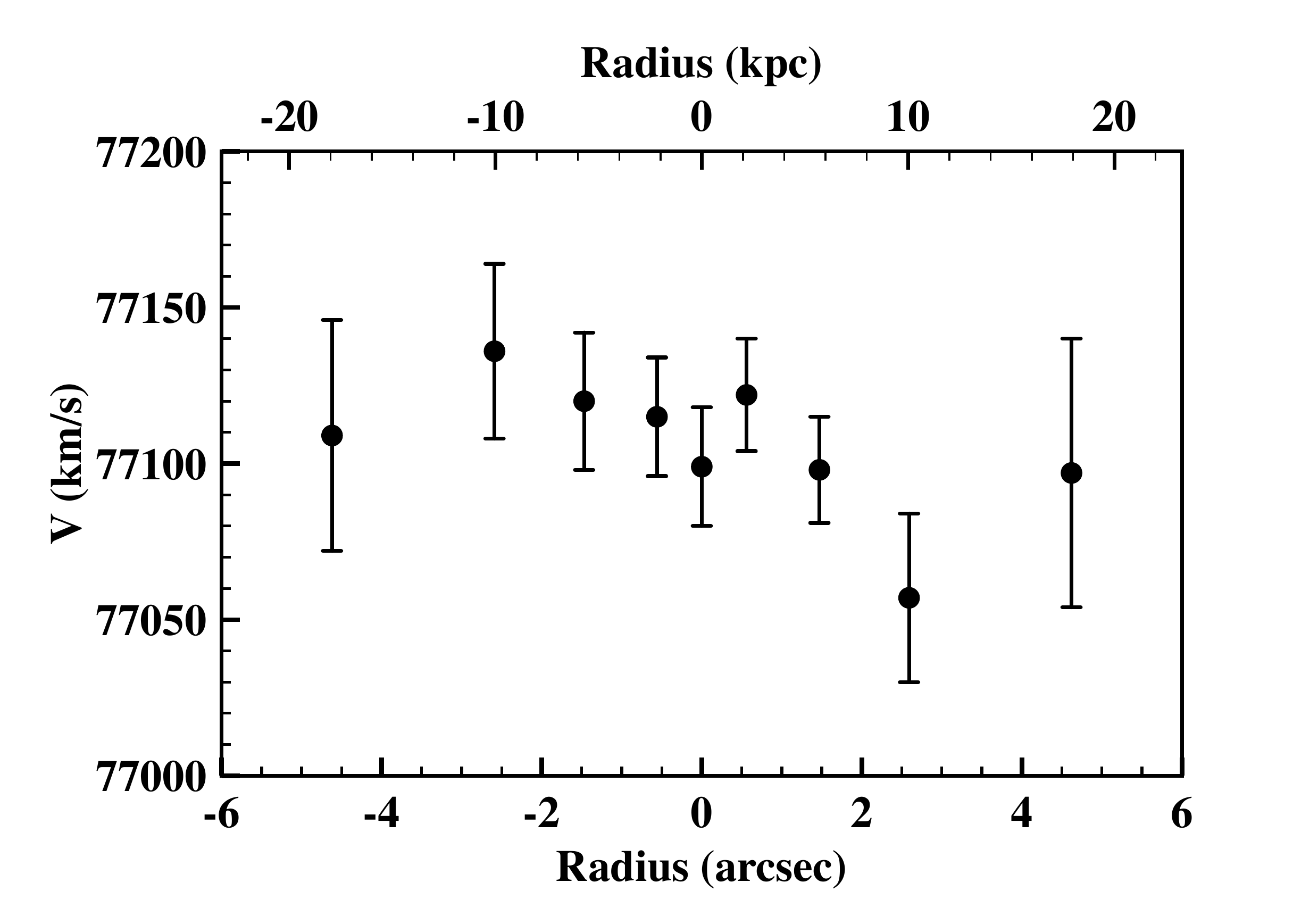}}
         \subfloat[MS 1455+22]{\includegraphics[scale=0.25]{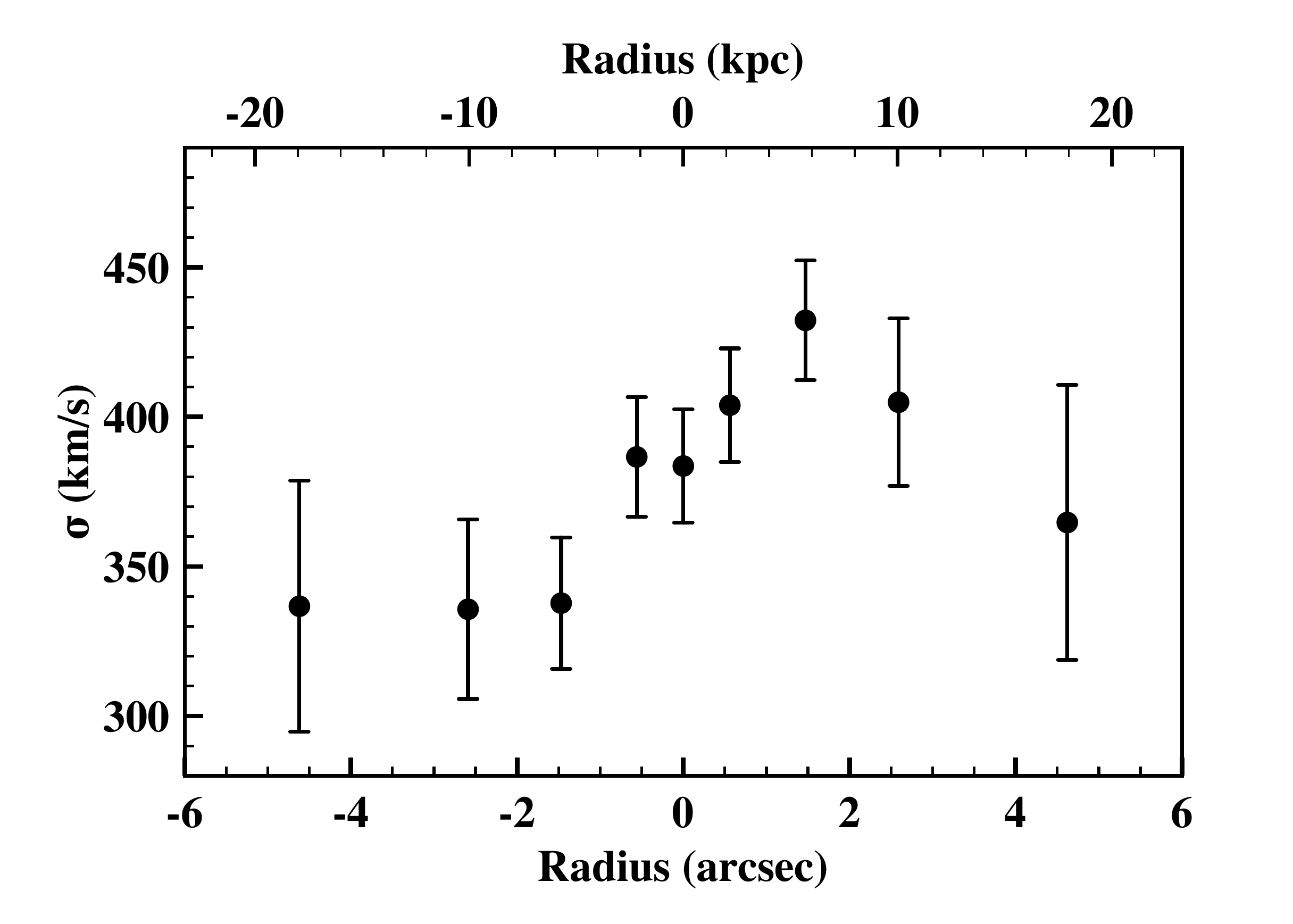}}
   \subfloat{\includegraphics[scale=0.25]{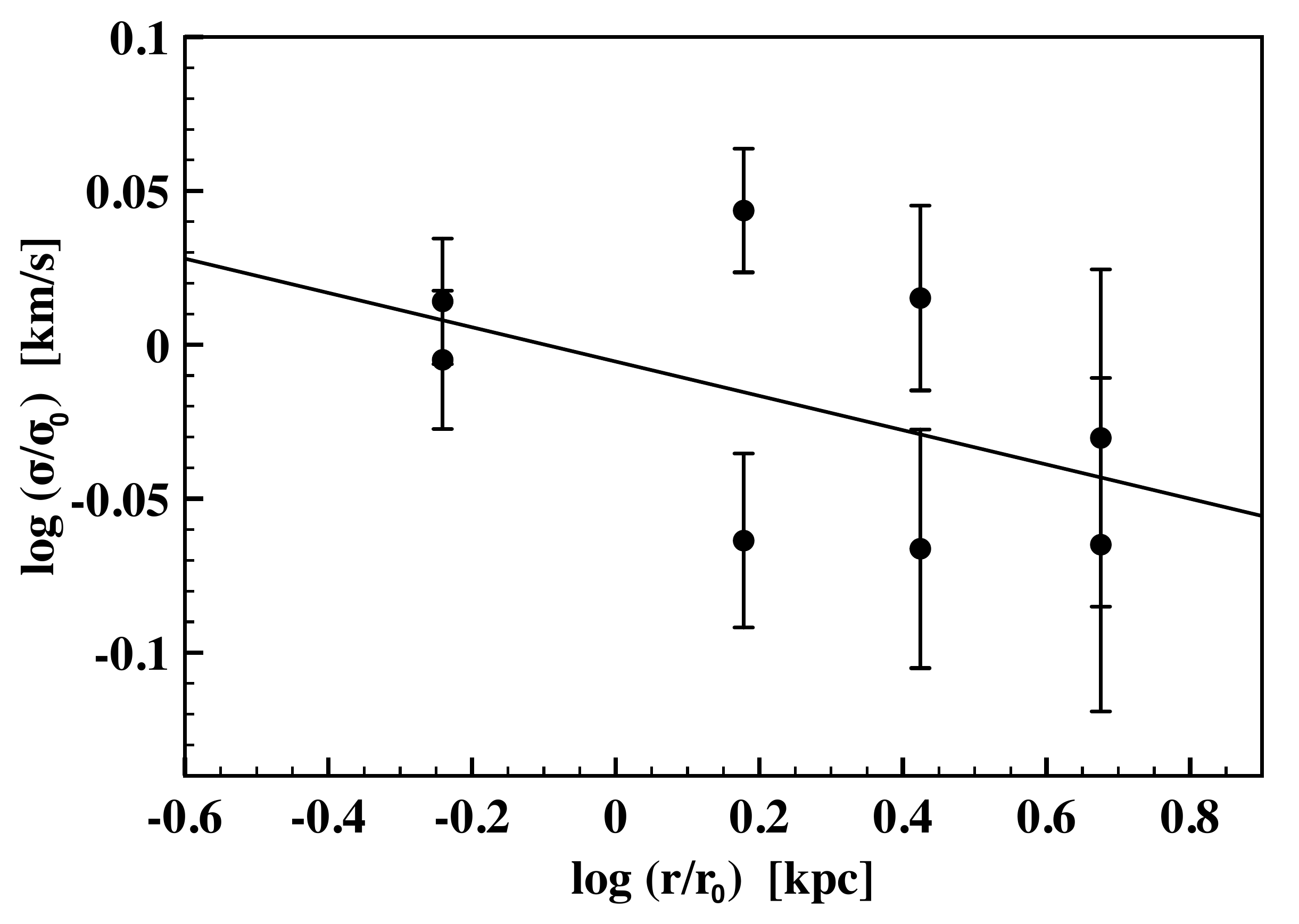}}  \\
  \caption{[a] Radial profiles of velocity (V), [b] velocity dispersion ($\sigma$, and [c] power law fit). The grey circles in the velocity dispersion profile of Abell 2537 indicate the measurements from \citet{Newman2013a} (see Appendix \ref{kinematics}).}
\label{fig:kin11}
\end{figure*}

\begin{figure*}
\centering
   \subfloat[NGC0315]{\includegraphics[scale=0.25]{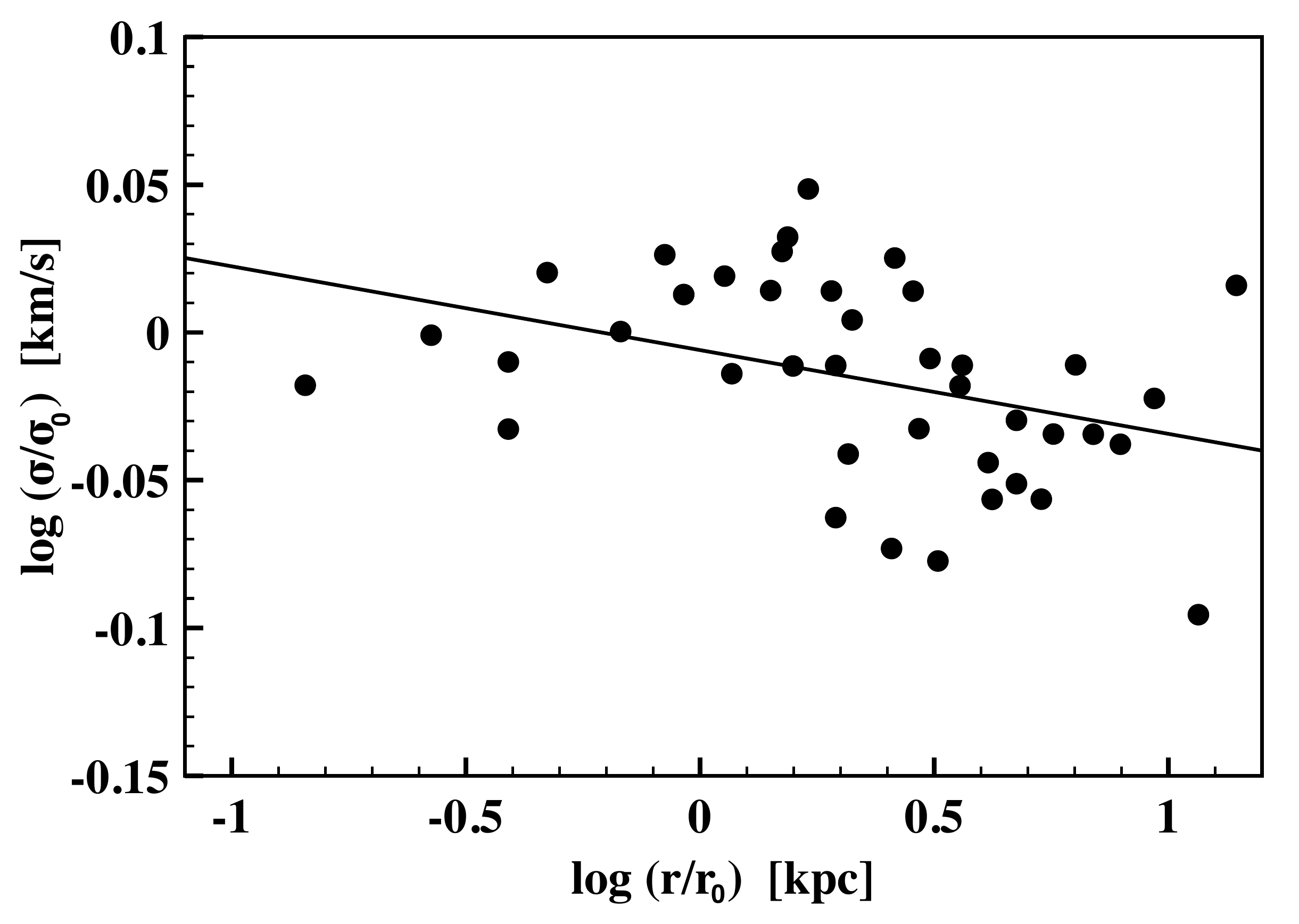}}
   \subfloat[NGC0410]{\includegraphics[scale=0.25]{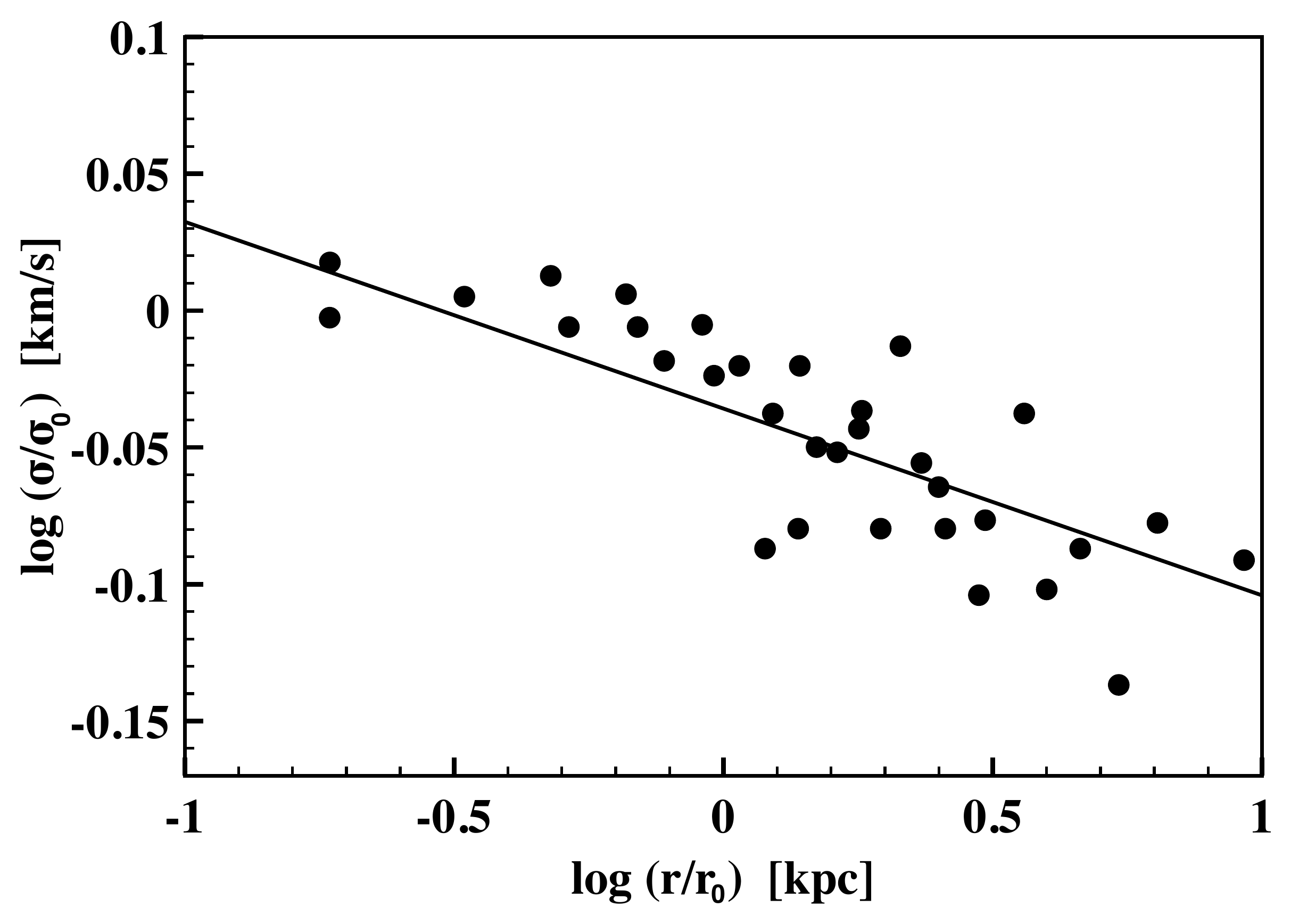}}
         \subfloat[NGC0524]{\includegraphics[scale=0.25]{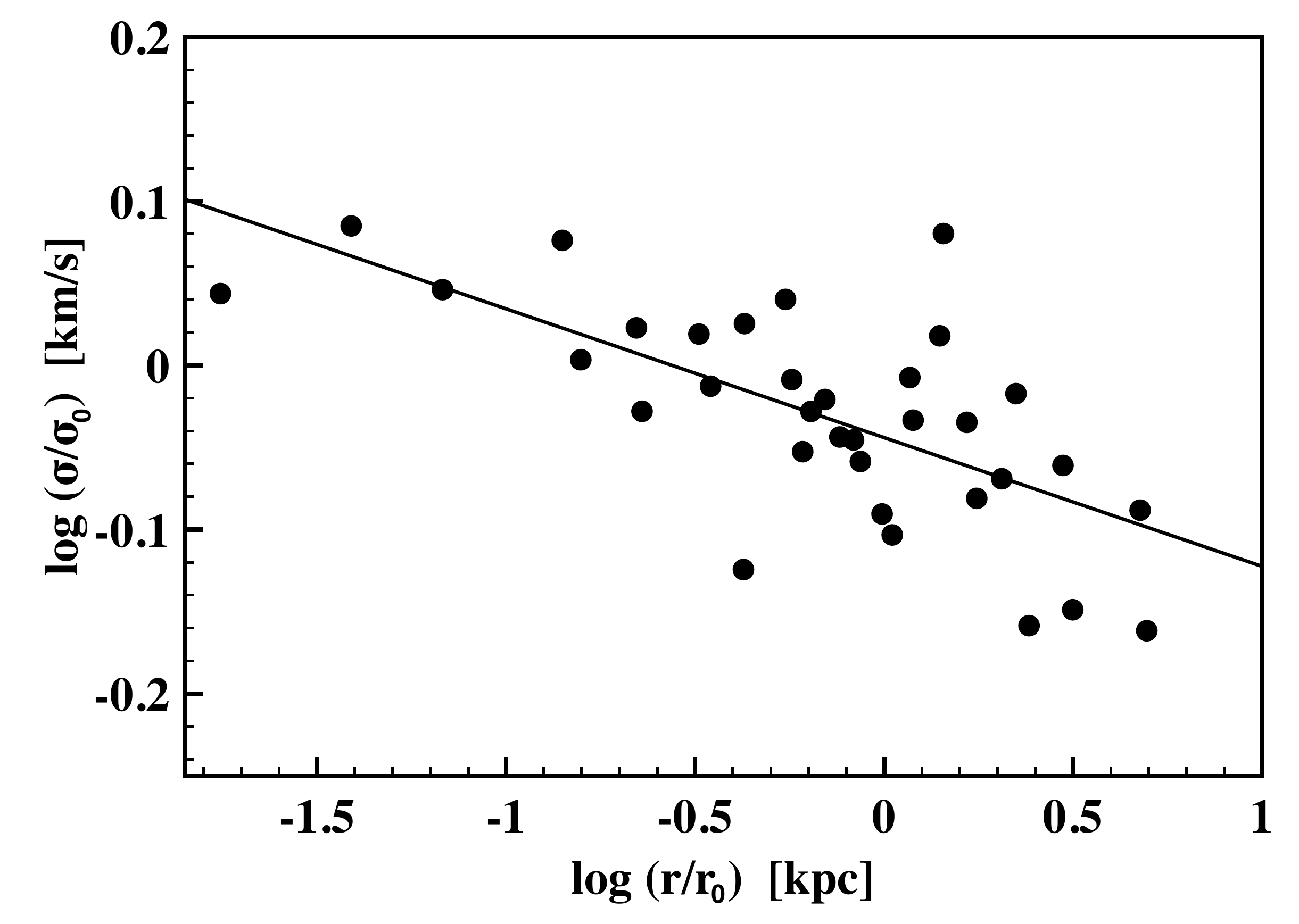}}\\
         \subfloat[NGC0584]{\includegraphics[scale=0.25]{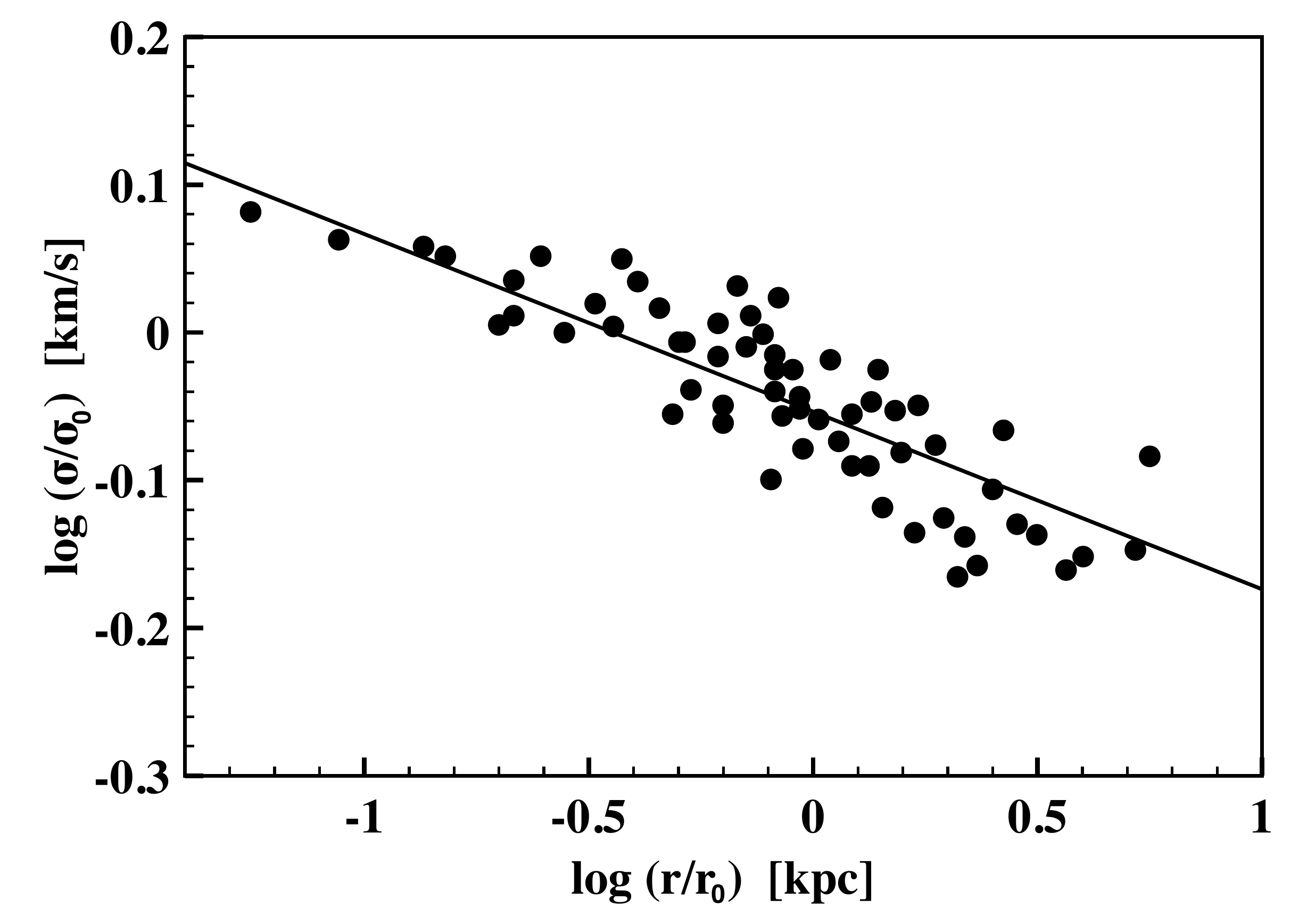}}
         \subfloat[NGC0777]{\includegraphics[scale=0.25]{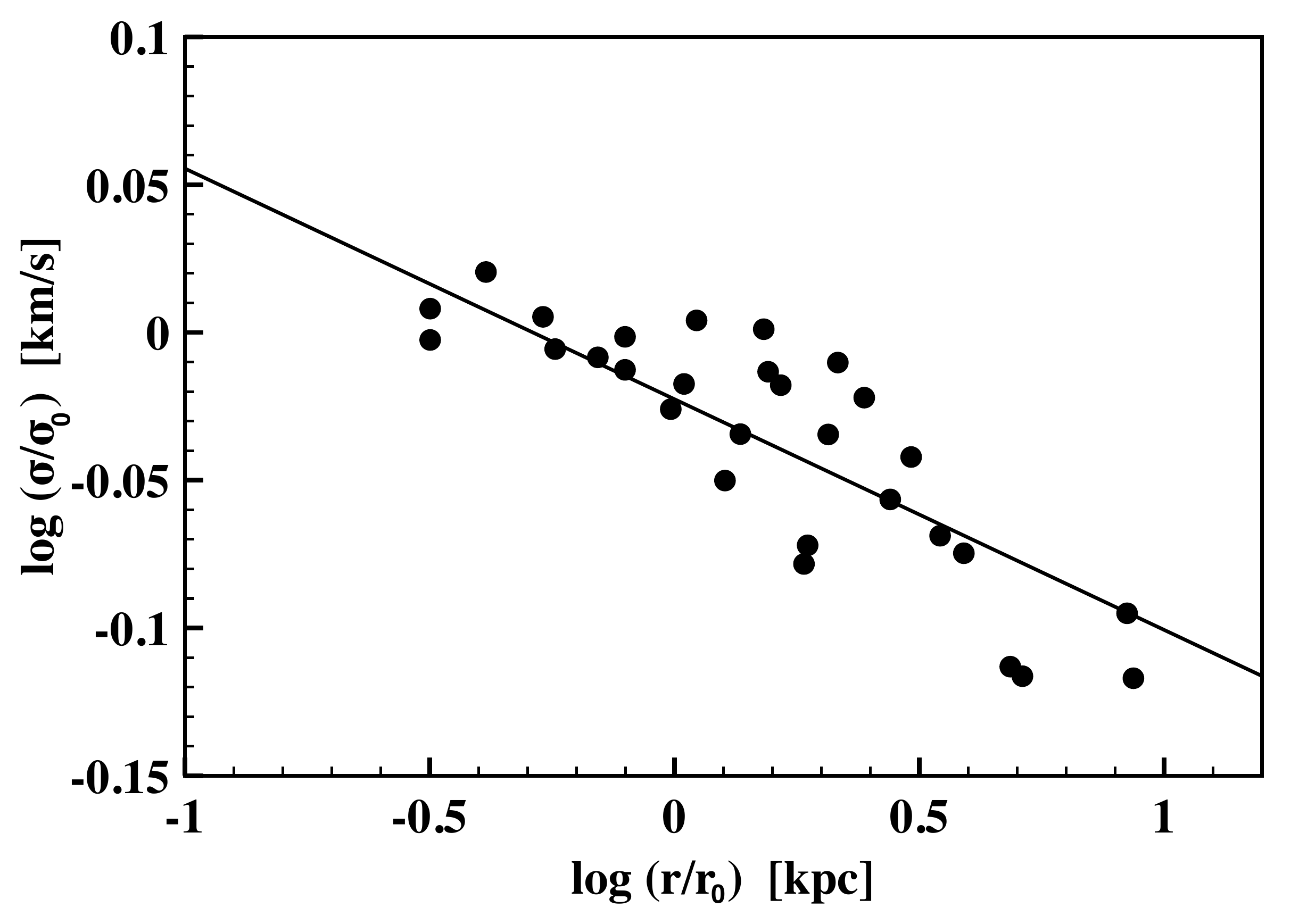}}
      \subfloat[NGC0924]{\includegraphics[scale=0.25]{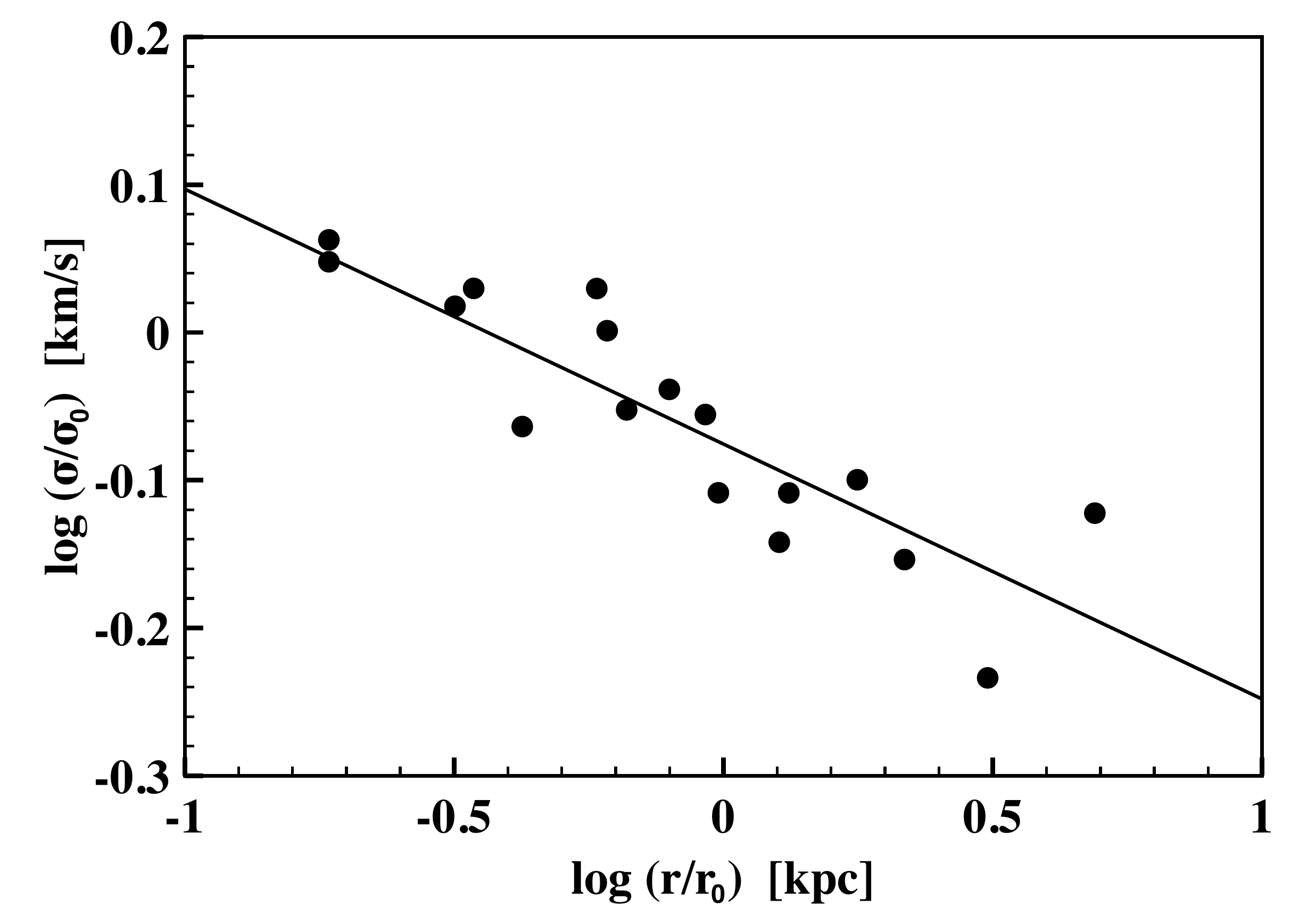}}\\
         \subfloat[NGC1060]{\includegraphics[scale=0.25]{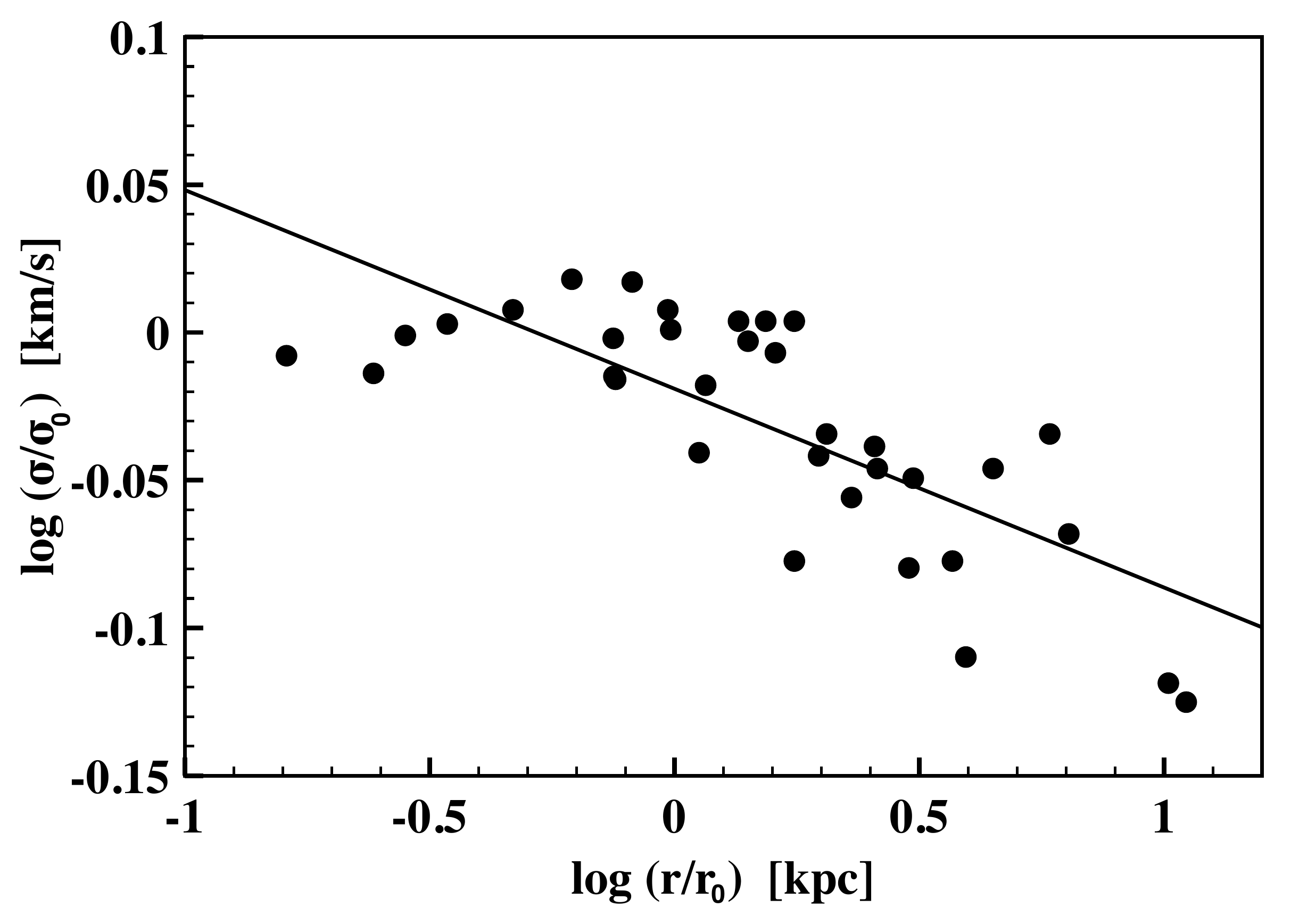}}
         \subfloat[NGC1453]{\includegraphics[scale=0.25]{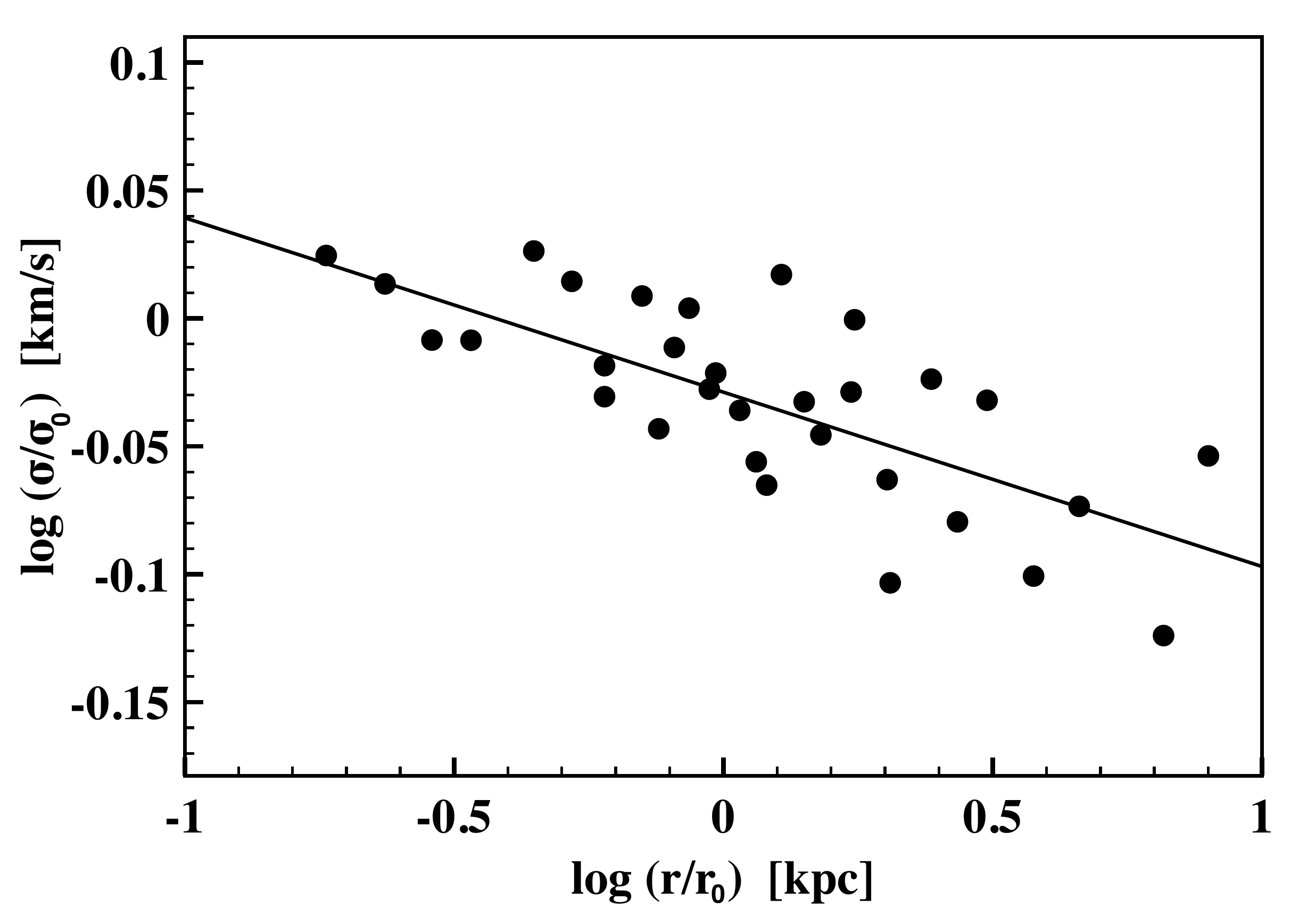}}
         \subfloat[NGC1587]{\includegraphics[scale=0.25]{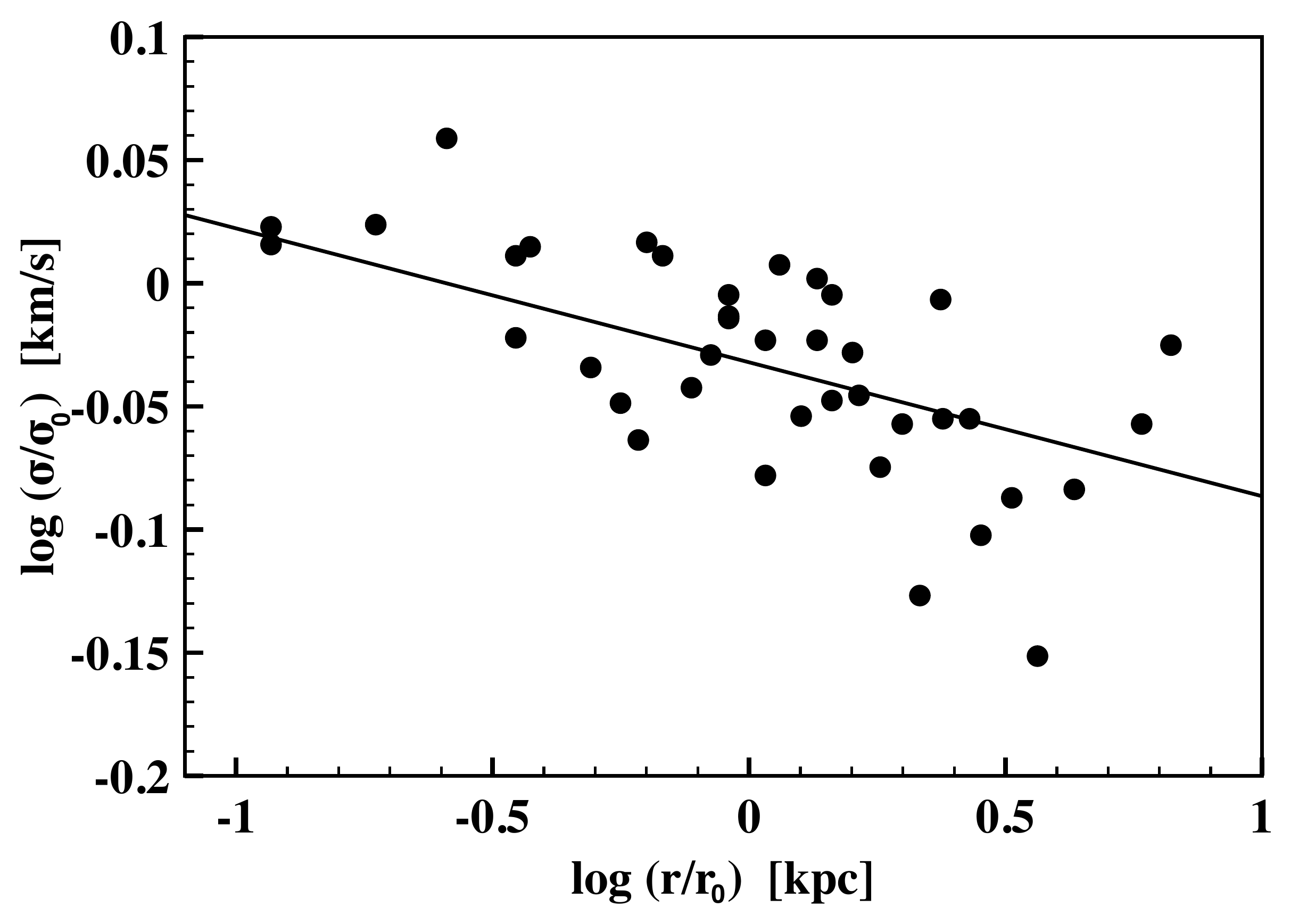}}\\    
          \subfloat[NGC1779]{\includegraphics[scale=0.25]{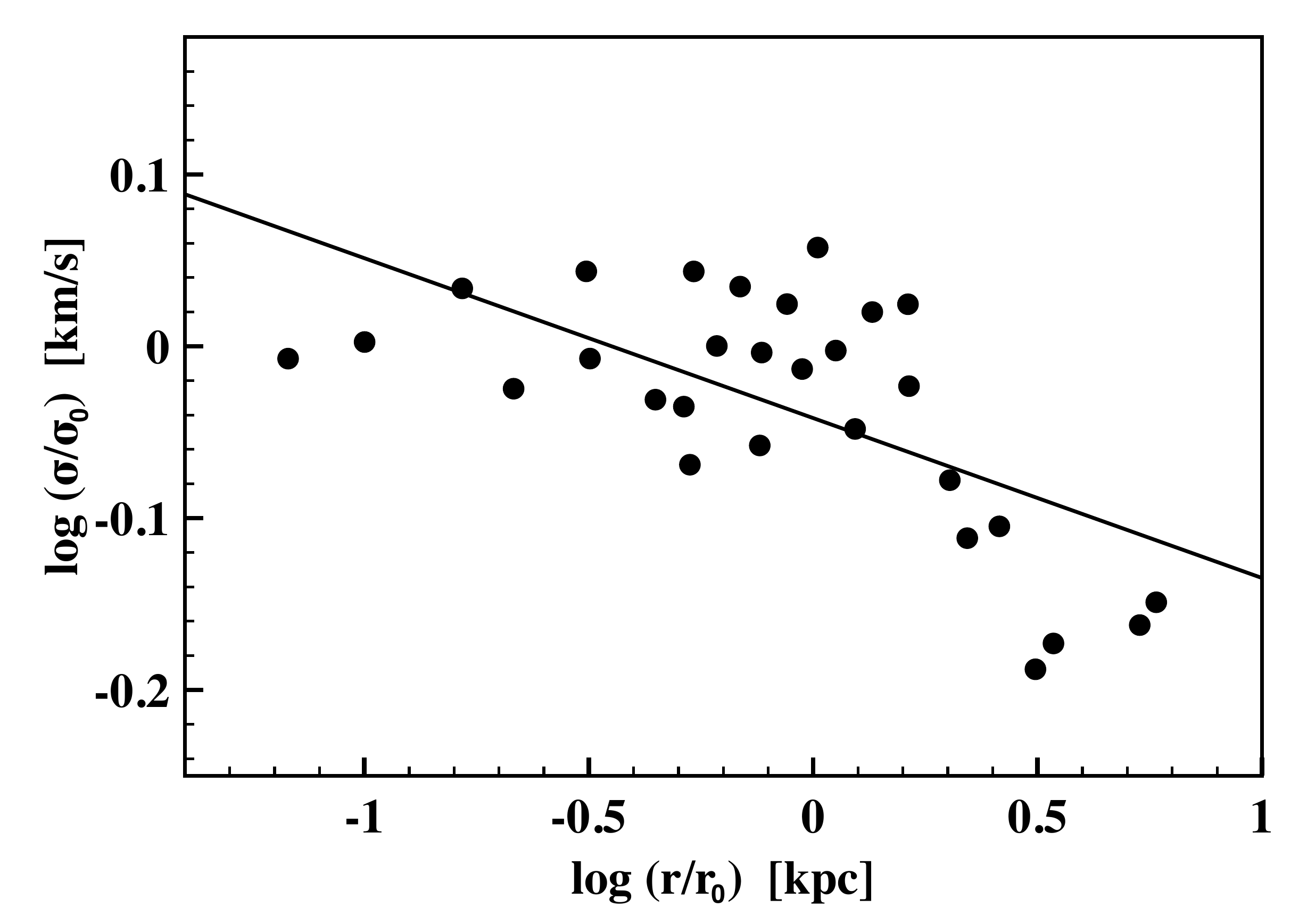}}
         \subfloat[NGC2563]{\includegraphics[scale=0.25]{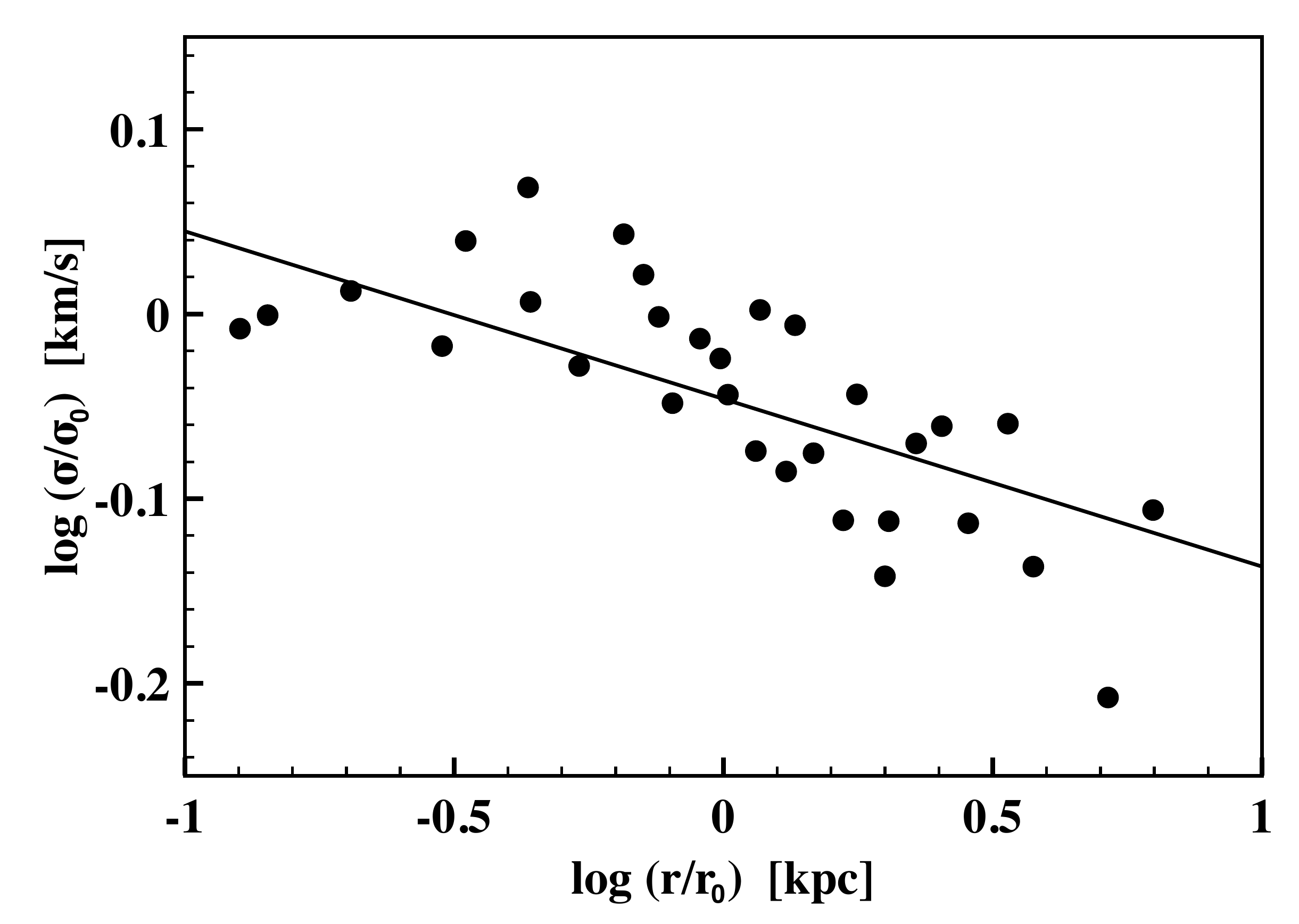}}
         \subfloat[NGC2768]{\includegraphics[scale=0.25]{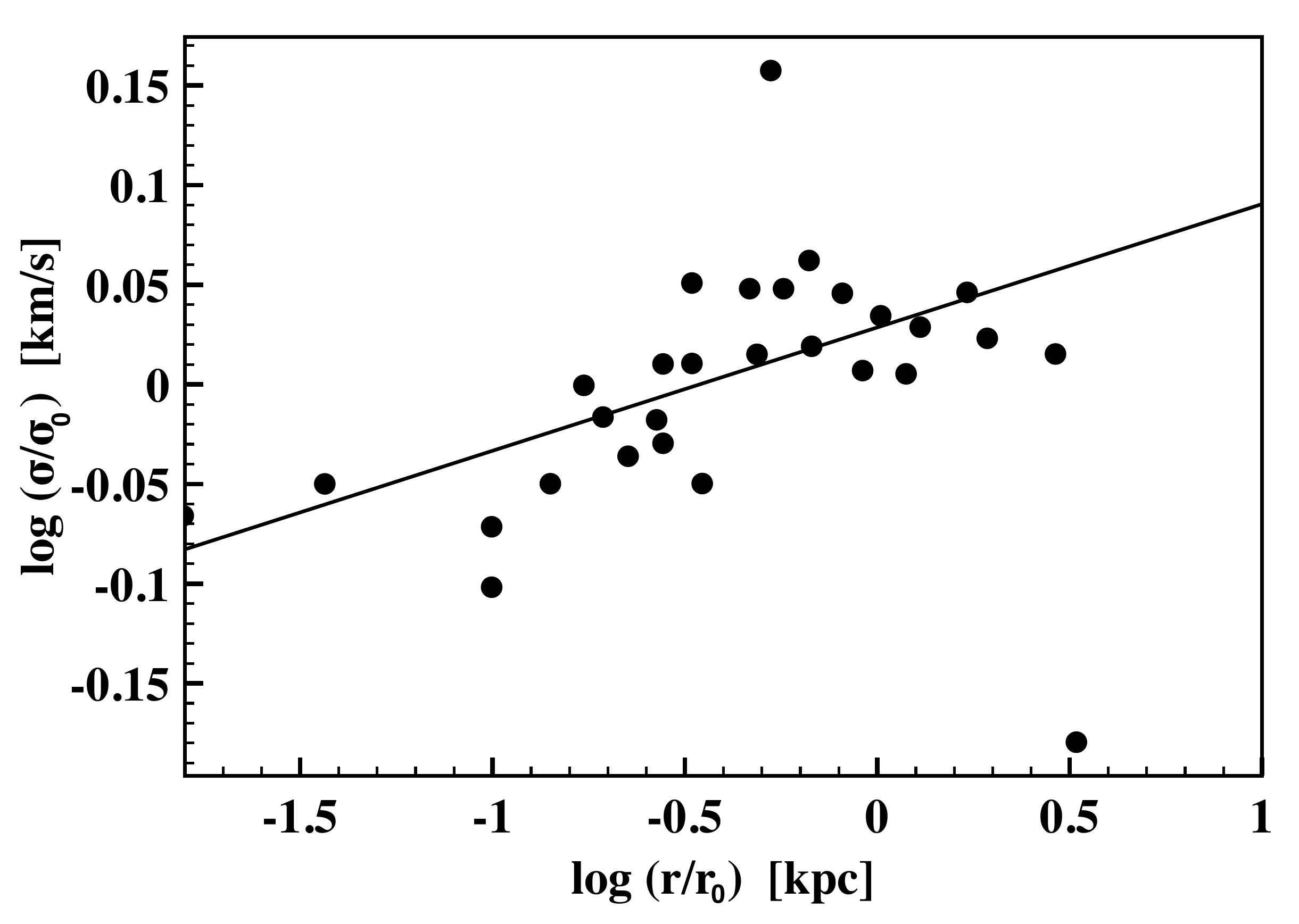}}\\
         \caption{Fits of the velocity dispersion ($\sigma$) profiles of the CLoGS BGGs.}
\label{fig:kin31}
\end{figure*}

\begin{figure*}
\centering
   \subfloat[NGC3613]{\includegraphics[scale=0.25]{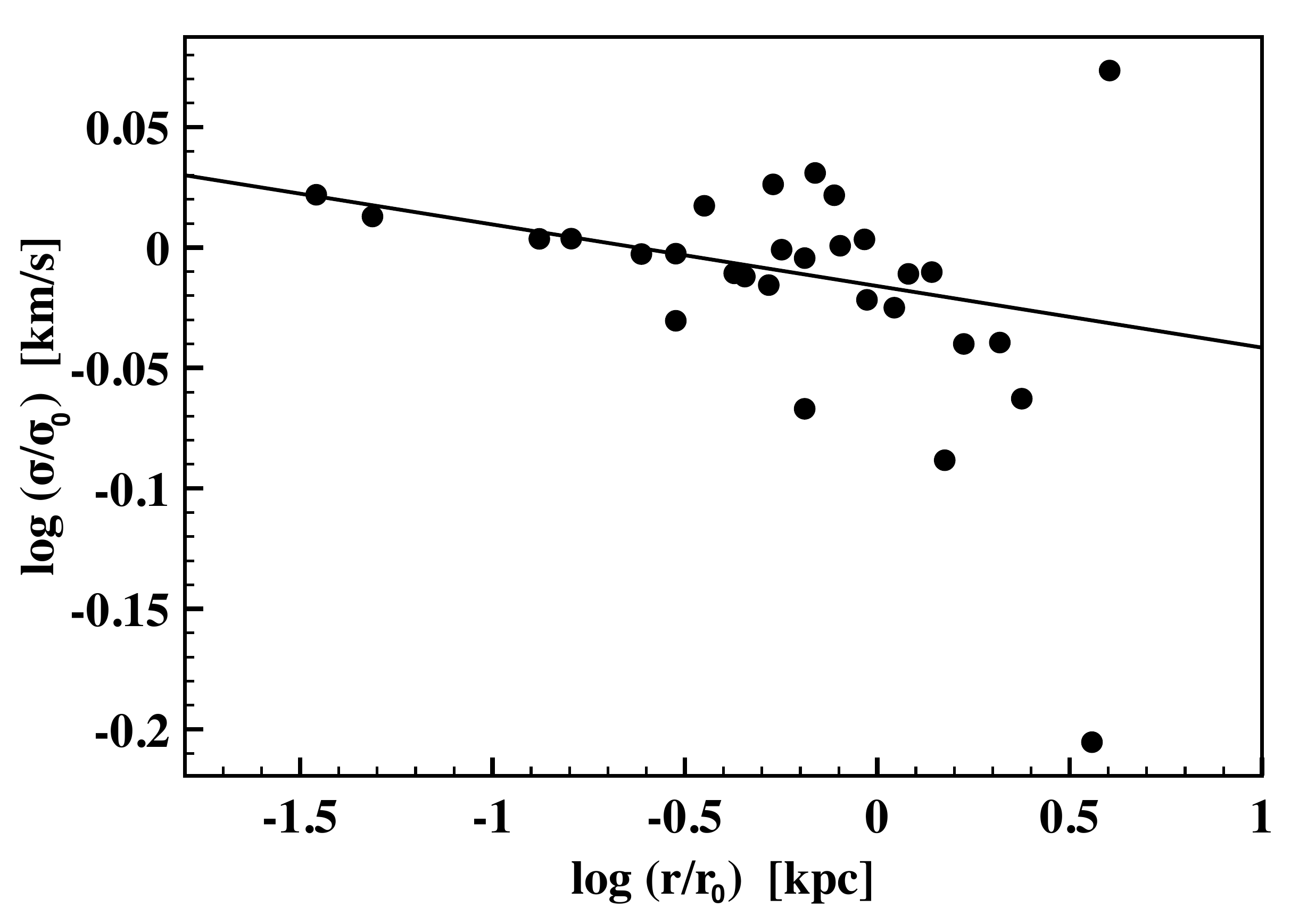}}    
          \subfloat[NGC3665]{\includegraphics[scale=0.25]{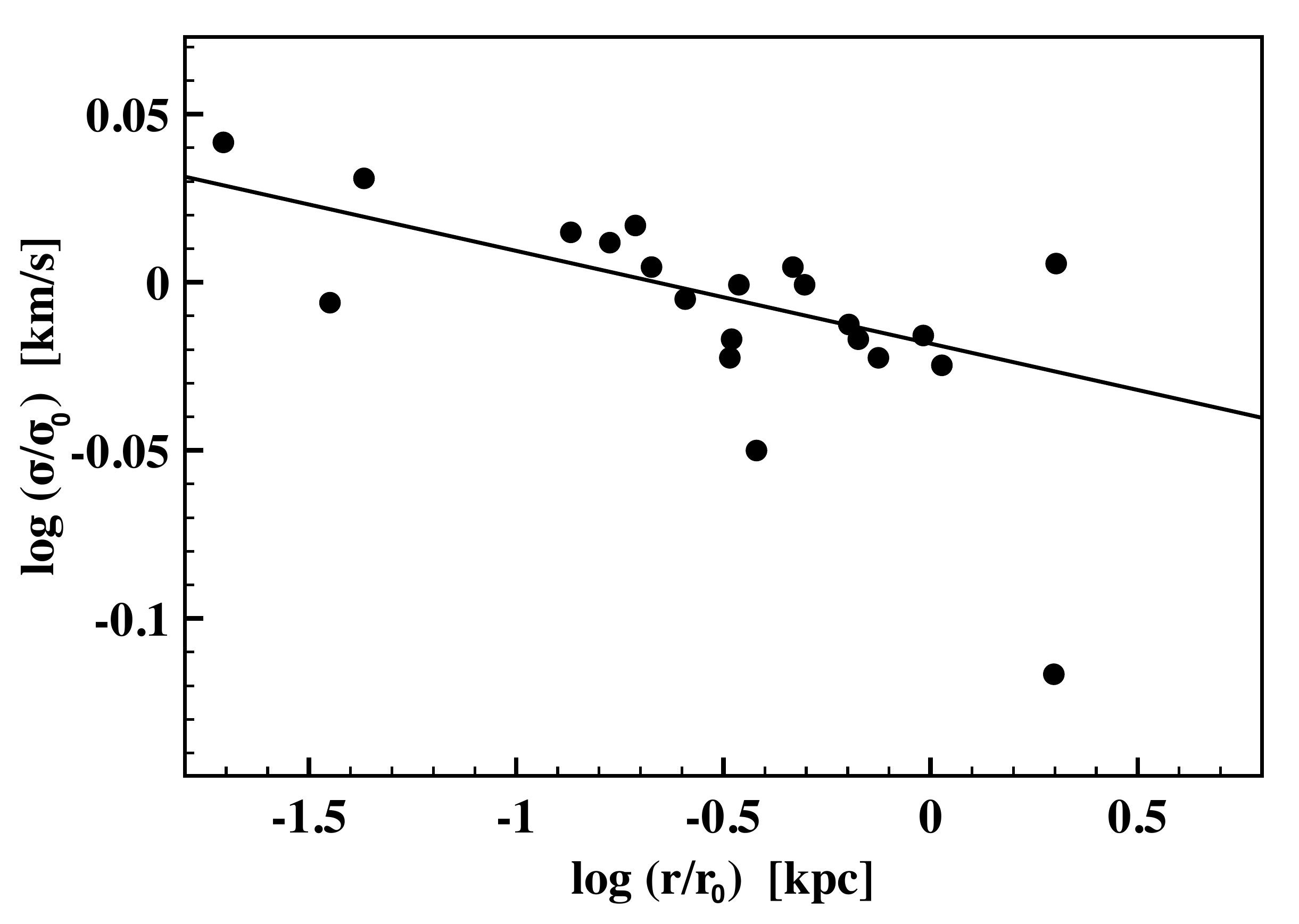}}
         \subfloat[NGC4261]{\includegraphics[scale=0.25]{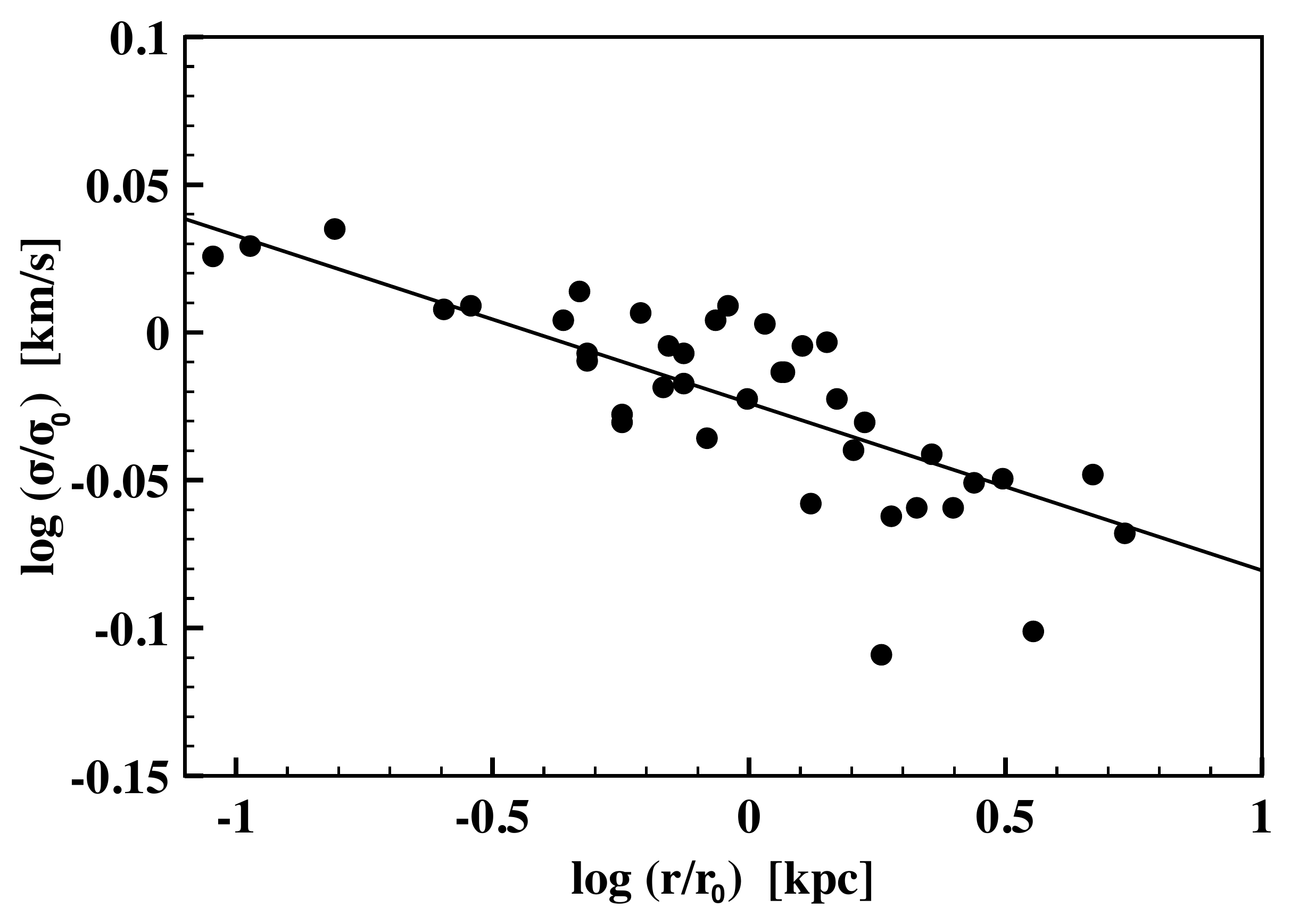}}\\
         \subfloat[NGC5127]{\includegraphics[scale=0.25]{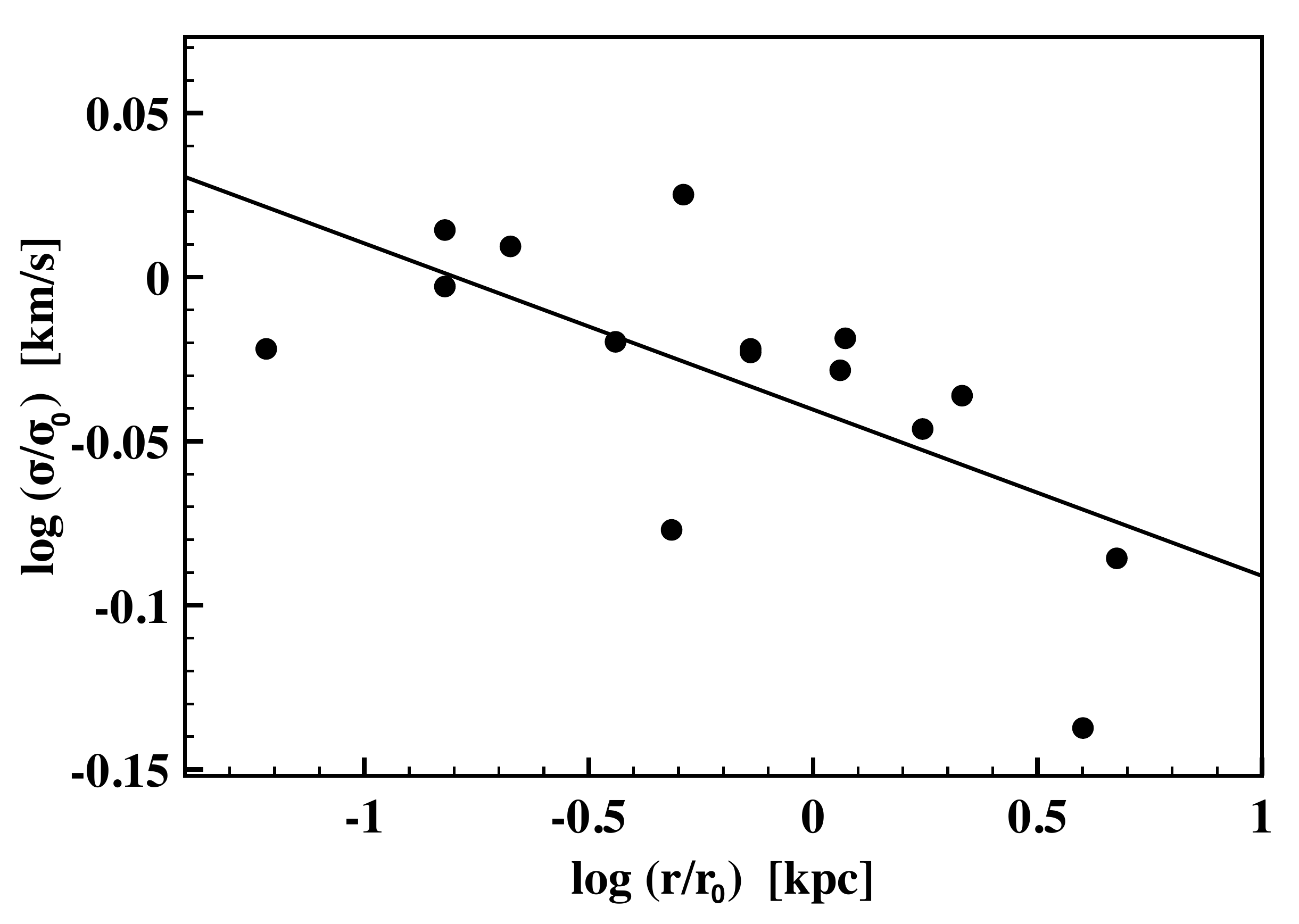}}
         \subfloat[NGC5353]{\includegraphics[scale=0.25]{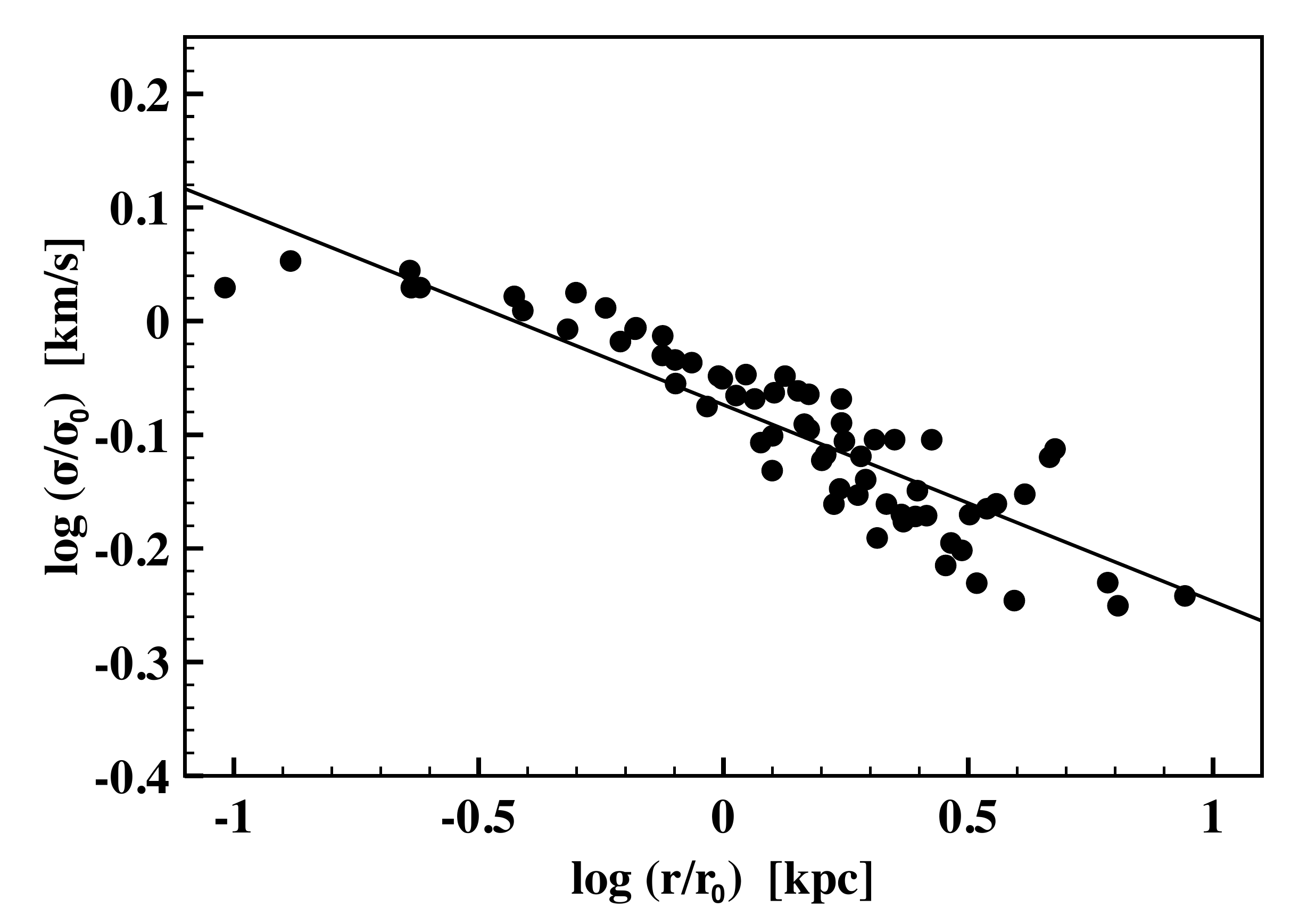}}
         \subfloat[NGC5490]{\includegraphics[scale=0.25]{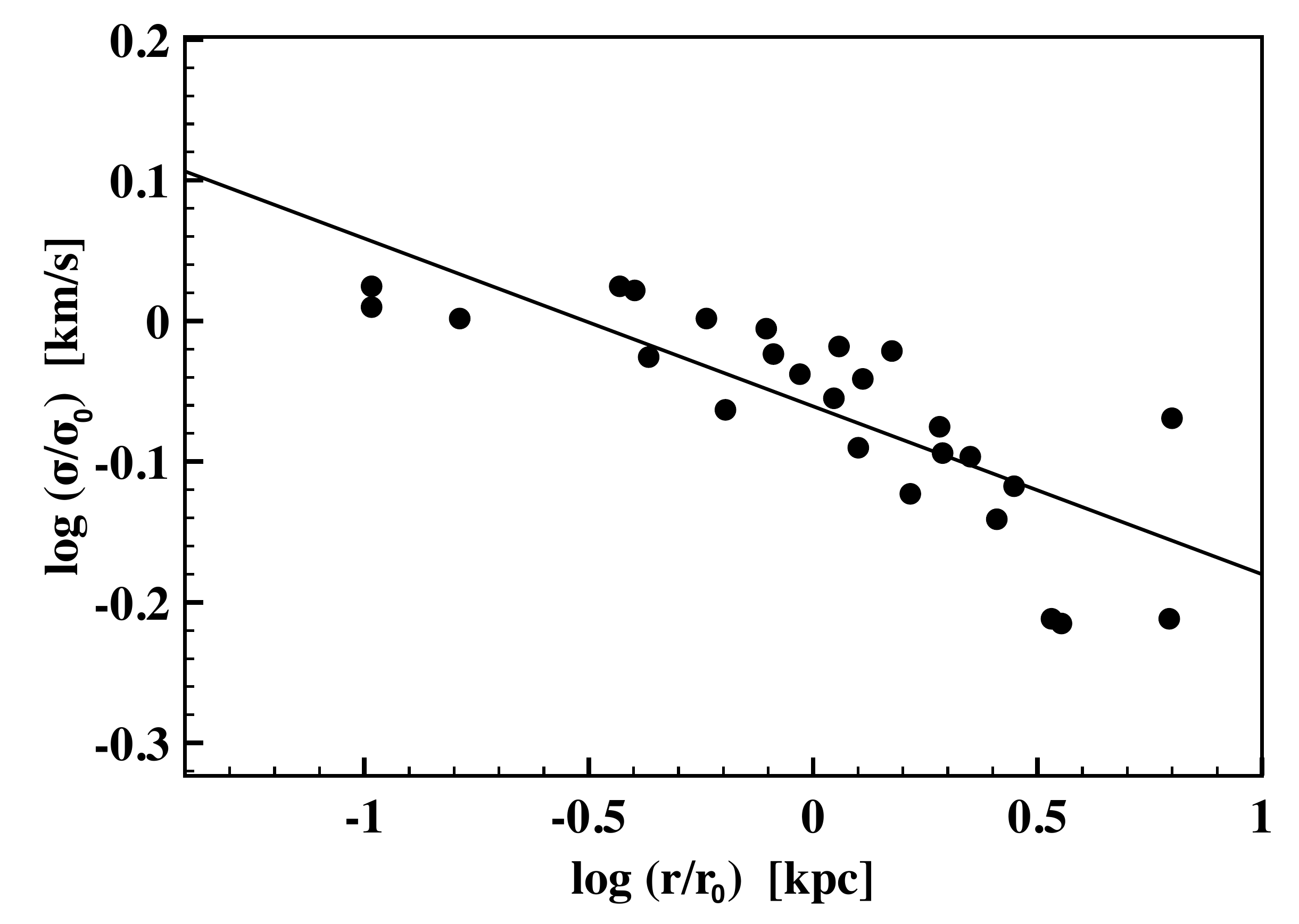}}\\
         \subfloat[NGC5629]{\includegraphics[scale=0.25]{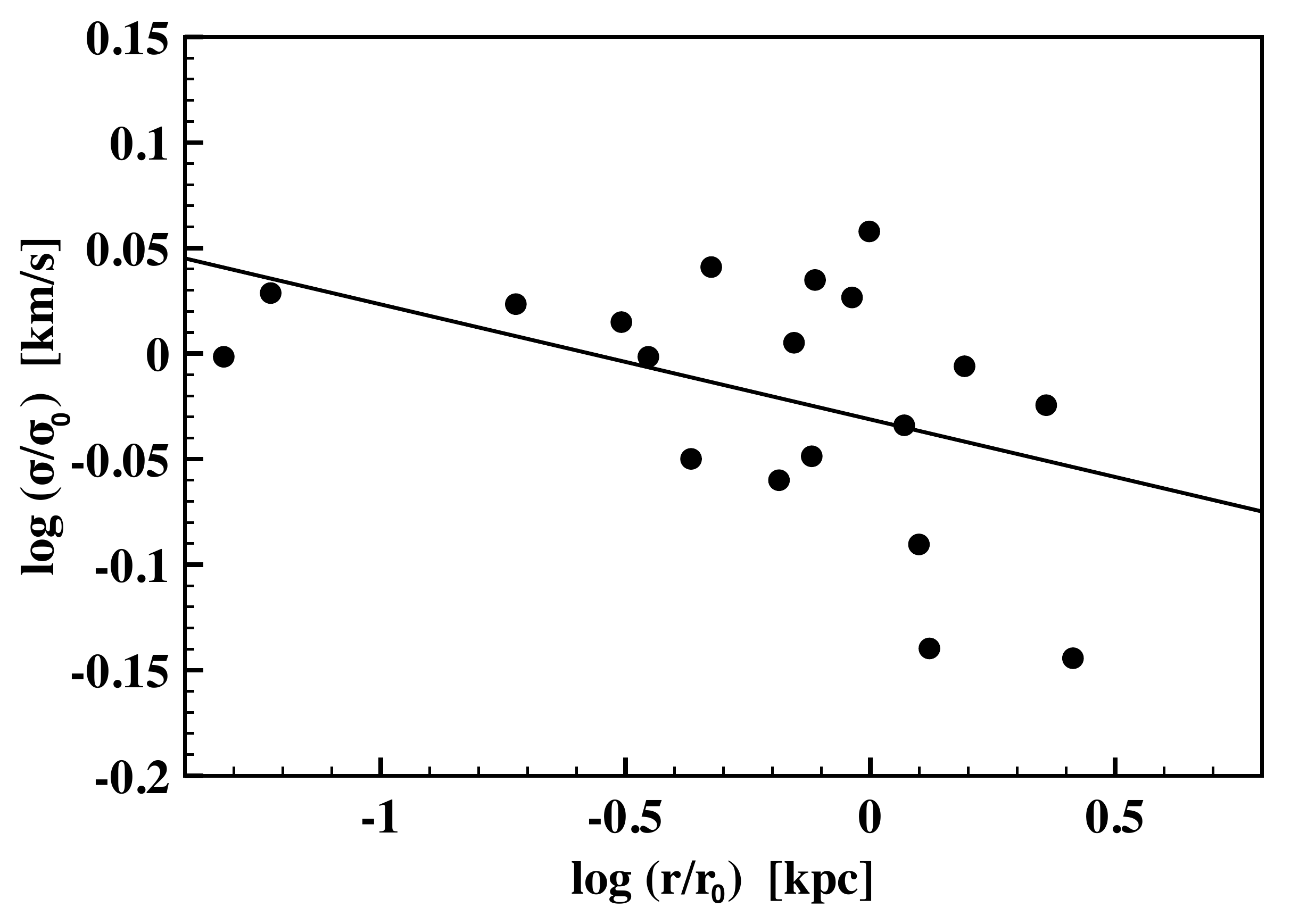}}
         \subfloat[NGC5846]{\includegraphics[scale=0.25]{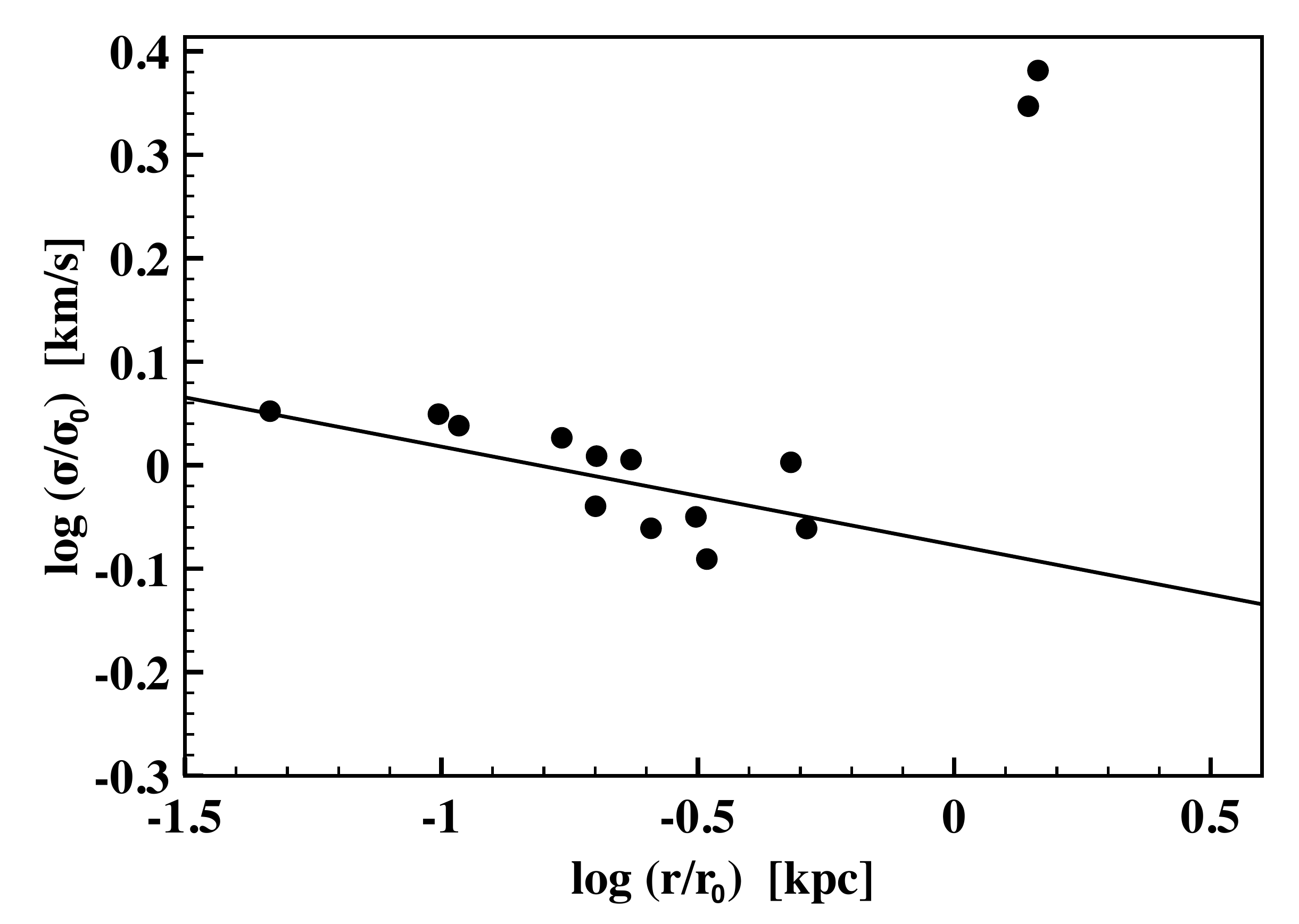}}
         \subfloat[NGC5982]{\includegraphics[scale=0.25]{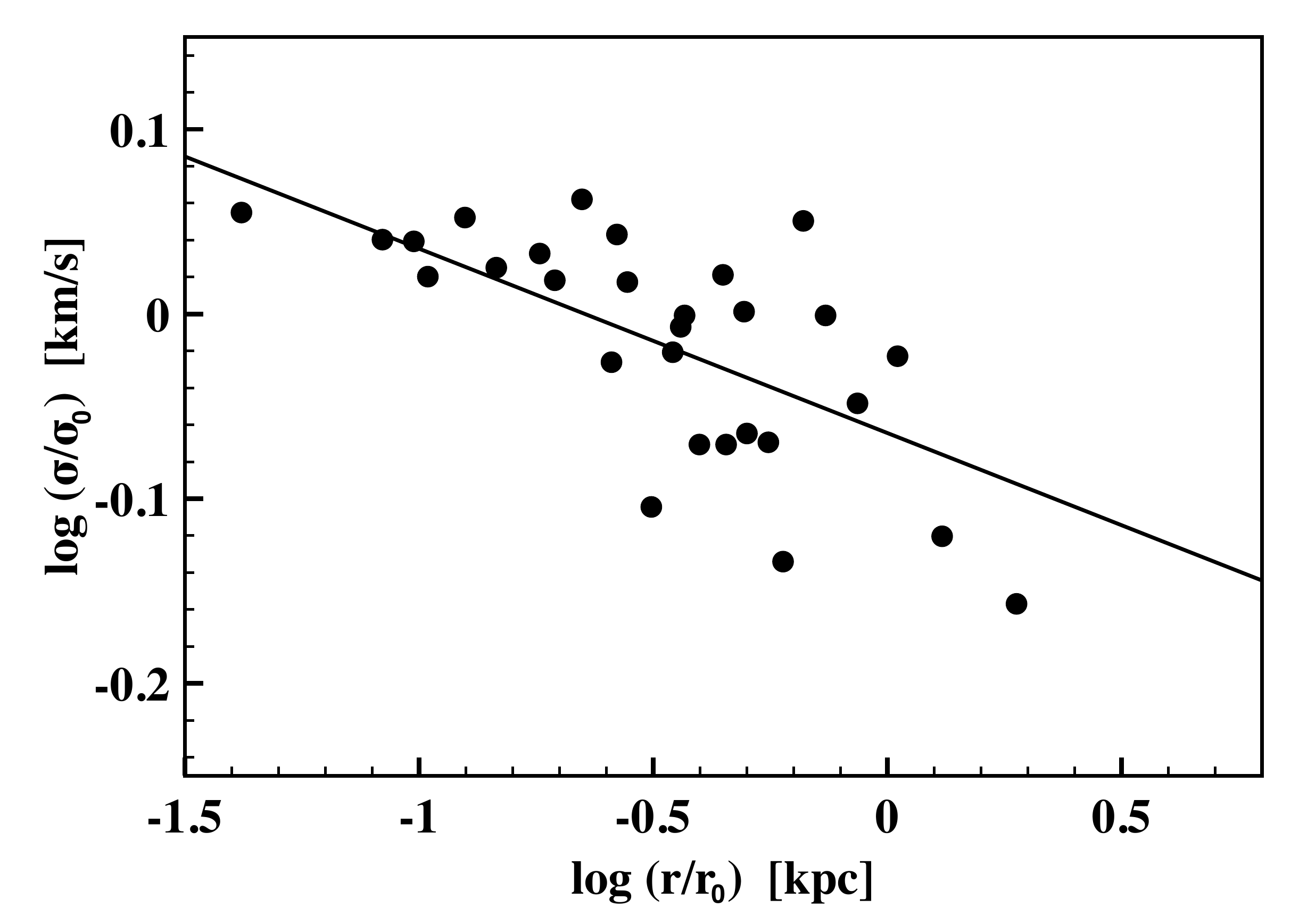}}\\
         \subfloat[NGC6658]{\includegraphics[scale=0.25]{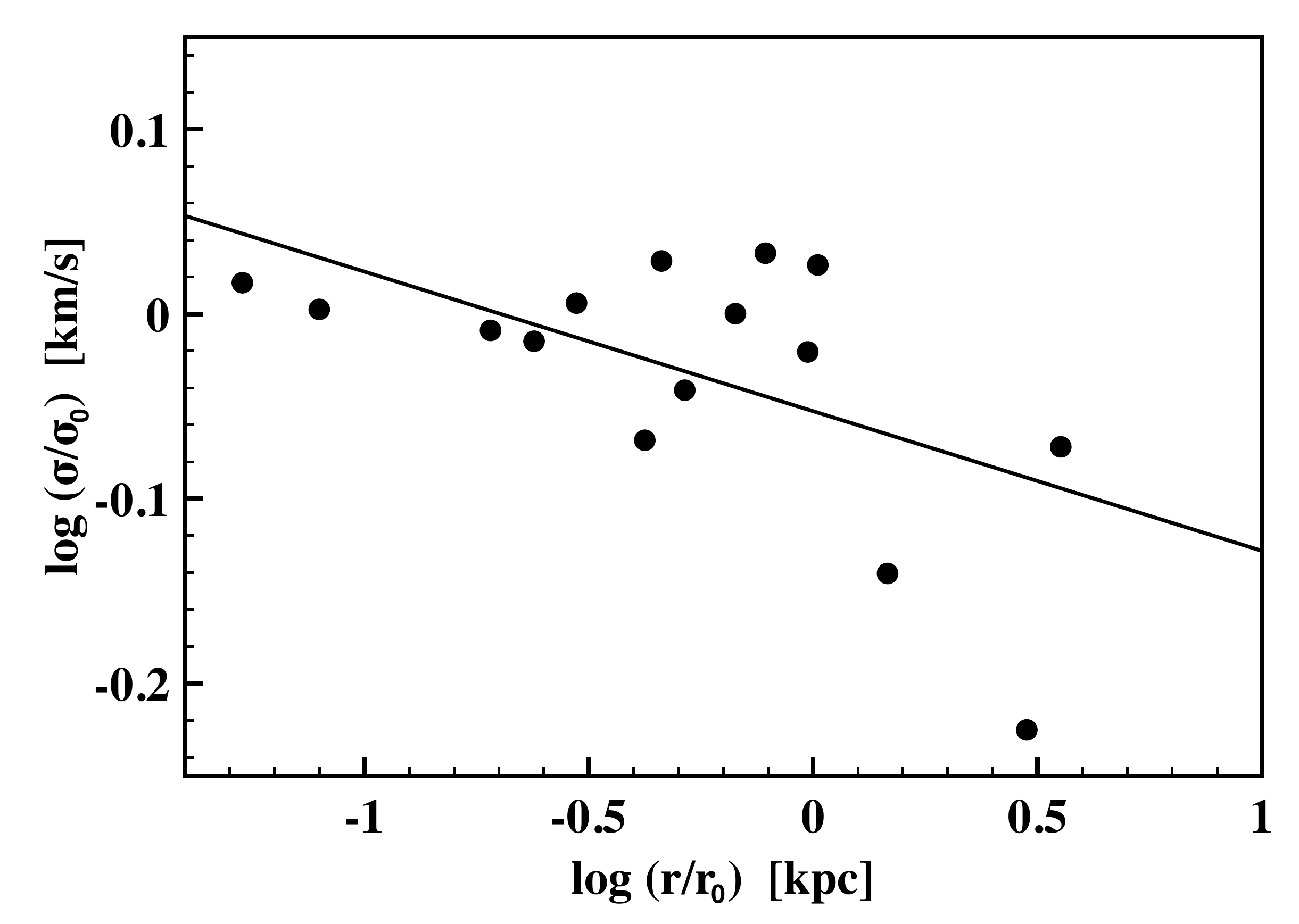}}
          \subfloat[NGC7619]{\includegraphics[scale=0.25]{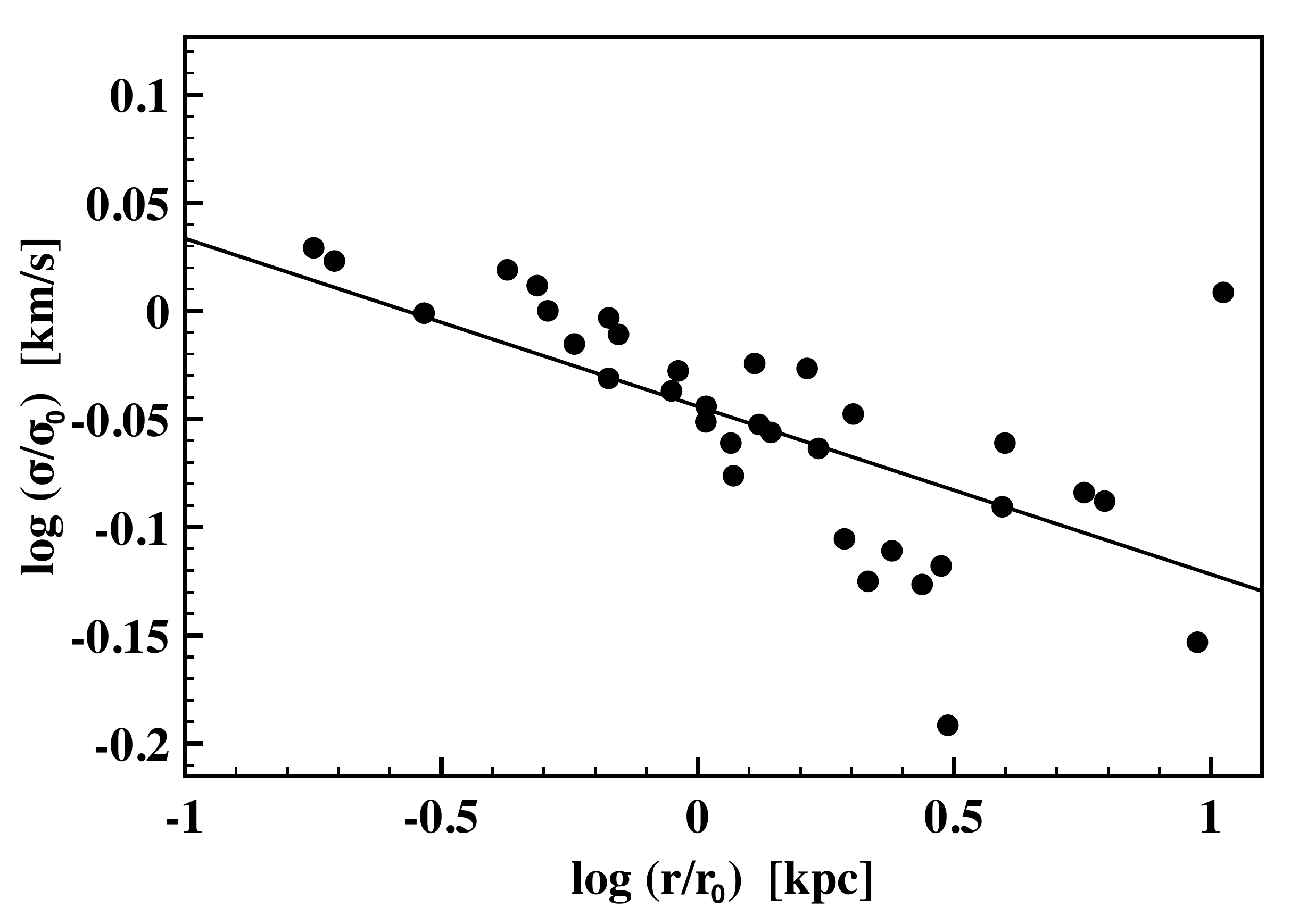}}\\
   \caption{Fits of the velocity dispersion ($\sigma$) profiles of the CLoGS BGGs.}
\label{fig:kin32}
\end{figure*}

The radial profiles of velocity ($V$) and velocity dispersion ($\sigma$, and power law fit) are presented in Figure \ref{fig:kin1} to \ref{fig:kin11}. The power law fits to the velocity dispersion slopes of the CLoGS BGGs are shown in Figure \ref{fig:kin31} and \ref{fig:kin32}.

%%%%%%%%%%%%%%%%%%%%%%%%%%%%%%%%%%%%%%%%%%%%%%%%%%

% Don't change these lines
\bsp	% typesetting comment
\label{lastpage}
\end{document}